\documentclass[oneside]{book}
\usepackage{symbols}
\usepackage{graphics}
\usepackage{feynmf}
\usepackage{epsfig}
\usepackage{latexsym}
\usepackage[figuresright]{rotating}
\begin{document}

\tableofcontents
\listoffigures
\listoftables

\chapter{Introduction}

One of the principal goals of physics is to understand 
the constituents of matter and their interactions
with each other.  Experimental observations provide the basis 
for theoretical models, and these models must in turn be tested by
more experiments.   Theoretical physicists invariably introduce 
certain parameters to simplify these models.  For example, the 
concept of electric charge is introduced to find the relative electric force
on the object.  An object with twice the charge will experience twice the 
force.  Charge is a fundamental quantity and can be used with complete
ignorance of the inner structure of an object. 

We are interested here in another fundamental, yet less-understood, quantity:
polarizability.   The origin of this concept is in the study of atomic spectra.
It was observed that placing atoms in a static electric field  lowers their energy;
the polarizability was introduced to help quantify the amount of this energy shift.
Polarizability is important because it is a basic property of a particle, along with charge and mass, 
and is involved in the description of all electromagnetic interactions.  

Part of the interest in this subject stems from the fact that as an experimental quantity, the 
polarizability is a single number which represents the influence of many effects 
which cannot be directly measured.  However, these effects can be calculated in various
theoretical models and compared with the experimental measurements.  In the quark model,
for instance, a proton is composed of three smaller particles known as quarks.  One of these
is negatively charged and the other two are positive.   
An electric field would exert a force on the positive charges in the opposite direction as the
negative charge, causing the proton to deform slightly.  The amount of this deformation depends 
on, among other things,  how strongly the quarks interact with each other.   The polarizability, on the other hand,
contains information on the deformation, but no knowledge of the quark interactions is needed to 
measure it.  It can therefore be used to independently check the various models which have 
been proposed to describe nuclear matter.

The focus of this paper is on the polarizability of the neutron.  To date, accurate
measurements of the neutron polarizability have not been realized.  This is mainly because
neutrons have no charge, and so the interaction with electromagnetic fields is weak.
The central question to be considered is whether the neutron polarizabililty can be determined in Compton scattering
from a deuteron, which is a particle containing one proton and one neutron.  
This allows the use of a stable target.
Unfortunately, the computation is more difficult for the deuteron because
all of the proton interactions also need to be taken into account.
As discussed in the following section, this is not a new idea and there have been previous calculations.
However, advances on both the theoretical and experimental fronts invite the possibility 
of not only a more accurate calculation, but a more accurate measurement.   
The intent of this work is provide a complete calculation which would
be useful to an experimentalist interested in measuring the neutron polarizability.   

Chapter 2 contains an introduction to polarizability and examines the 
previous theoretical and experimental work which has been done in this area. 
Descriptions of the methods used in the calculation are discussed in Chapter 3.
Checks which were performed to ensure the correctness of the work are
also detailed here.  The actual results of the numerical calculation are
presented in Chapter 4, along with the discussion of the question
of neutron polarizability measurements from deuteron Compton
scattering.  The conclusion can be found in Chapter 5.
Details of calculations omitted in the main text, as
well as  formulas for all of the relevant scattering amplitudes, are in the Appendices.


\chapter{Background and Motivation \label{ch:int} }
\section{Definitions of Polarizability}
Polarizability, like charge, is one of the fundamental properties of a particle.  Roughly speaking,
it is a measure of how easy it is to deform the object in the presence of an electric or magnetic field.
The symbol $\alpha$ is traditionally used to represent the electric polarizability, and $\beta$ is used for the 
magnetic polarizability.  

A more precise definition can be given in terms of induced dipole moments.  The polarizability is the constant of
proportionality between an external field and the average dipole moment that it induces.  We can write
\begin{eqnarray}
\langle \vec{d} \rangle &   = & \alpha \vec{E}, \\
\langle \vec{\mu} \rangle & = & \beta \vec{B},
\end{eqnarray} 
\noindent where $\langle \vec{d} \rangle$ ($\langle \vec{\mu} \rangle$) is the average induced electric
(magnetic) dipole moment.  This induced moment in turn changes the potential energy by an amount
\begin{equation}
\Delta E = -\frac{1}{2} \alpha E^2 - \frac{1}{2} \beta B^2. \label{eq:c1-1}
\end{equation}
\noindent This energy shift is known as the quadratic Stark effect.

We can also formulate a matrix definition of polarizability.  The Hamiltonian for a dipole interacting
with an external field is given by
\begin{equation}
H_{\mathrm{int}} = -d_z E,
\end{equation}
\noindent where we have chosen $\vec{E}$ to be in the $z$-direction for simplicity.  The energy shift can be 
calculated using perturbation theory.  The first-order term is zero, since the dipole operator changes parity.
Restricting ourselves to nucleons, the second order term is
\begin{equation}
\Delta E = \sum_{N'} \frac{ \mx{N}{-d_zE}{N'} \mx{N'}{-d_zE}{N} }{E_N - E_{N'}} ,
\end{equation}
\noindent where $N$ is a ground-state nucleon and $N'$ is some excited state.  An uppercase $N$ 
is used throughout the text to denote a generic nucleon, and $n$ and $p$ 
represent the proton and neutron specifically. Comparing this expression 
with equation~(\ref{eq:c1-1}), we see that the polarizability  can be written as
\begin{equation}
\alpha = 2 \sum_{N'} \frac{ \left| \mx{N}{d_z}{N'} \right|^2 }{E_{N'}-E_N}. \label{eq:c1-2}
\end{equation}
\noindent Similarly,
\begin{equation}
\beta = 2 \sum_{N'} \frac{ \left| \mx{N}{\mu_z}{N'} \right|^2 }{E_{N'}-E_N}.
\end{equation}
\noindent Thus, polarizabilities can be calculated by taking matrix elements of dipole operators between
ground and excited nucleon states.

We can make a rough order-of-magnitude estimate of
$\alpha_N$ by letting $N'$ be an $N^{\ast}$ state.  Then,
\begin{equation} 
\alpha_N \approx \frac {2e^2 |\mx{N}{z}{N^{\ast}}|^2}{ m_{N^{\ast}} - m_N } \approx  20 \times 10^{-4} \mathrm{fm}^3,
\end{equation}
where the square of the matrix element was estimated to be one-third of the mean nucleon radius squared, or 
about $\frac{1}{3} \mathrm{fm}^2$.  This is on the same order as the currently accepted value of
$\alpha_N \approx 12 \times 10^{-4} \mathrm{fm}^3$.   Hereafter the unit of polarizability will be assumed
to be $10^{-4} \mathrm{fm}^3$ .

\begin{figure}
\centering
\parbox{75mm}{\centering\epsfig{file=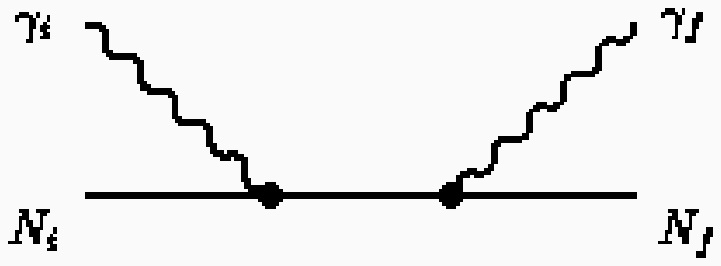} }
\parbox{15mm}{\centering\raisebox{.1in}{(a)} }
\parbox{75mm}{\centering\epsfig{file=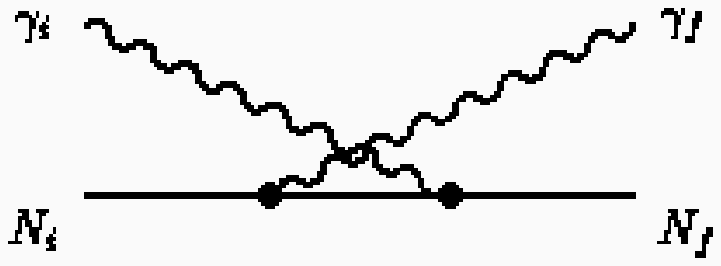} }
\parbox{15mm}{\centering\raisebox{.1in}{(b)} }

\caption[Feynman diagrams for Compton scattering]{Feynman diagrams for Compton scattering, uncrossed (a) and crossed (b) terms. \label{fig:c1-1} }
\end{figure}
 
The above  expressions define the polarizability in terms of either of an energy shift
or an induced dipole moment.  A third definition, which turns out to be the most practical,
involves the scattering amplitude for elastic photon scattering, commonly known as Compton scattering.
The Feynman diagrams for this process are given in Figure~\ref{fig:c1-1}.  The initial  photon has four-momentum
$(\omega_i , \vec{k}_i)$ and polarization vector $\eps$, where $\lambda_i = \pm 1$, while
$(\omega_f , \vec{k}_f)$ and  $\hat{\epsilon}_{\lambda_f}$ describe the final photon.  
The angle between $\kinit$ and $\kfin$ is labeled $\theta$, and the initial and final photon 
energies are related by the kinematic relation
\begin{equation}
\wfin = \frac{ \winit }{ 1 + \frac{\winit}{m_N}(1 - \cos{\theta}) }.
\end{equation}

The scattering amplitude
(to order $\omega^2$) can be written as  \cite{pe81}
\begin{eqnarray}
f(\omega, \winit) & = & -\frac{e^2}{m_N} (\epsdp ) + (\winit + \wfin) g(e,m_N,\kappa_N; \hat{k}_f,\epspr,\hat{k}_i,\eps) + \nonumber\\
& & \ \ \ \ \ \ \ \	\bar{\alpha} \wfin\winit (\epsdp ) + \bar{\beta} \wfin\winit (\hat{k}_f \times \epspr) \cdot
	(\hat{k}_i \times \eps) +  \nonumber\\
& & \ \ \ \ \ \ \ \ \ \ \ \ \ \ \ \ \wfin\winit h(e,m_N,\kappa_N; \hat{k}_f,\epspr,\hat{k}_i,\eps). \label{eq:c1-3}
\end{eqnarray}
\noindent The first term in the amplitude, known as the Thomson amplitude, is the dominant term
at low energies.  It depends on only two parameters: the charge and mass.  Since neither of these quantities assumes any
internal structure for the nucleon, this is consistent with the fact that in scattering at low energies the  
nucleon must behave like a spinless point particle. 

The next order term, represented above by $g$, is a function of not only the charge and mass, but also the
anomalous magnetic   moment $\kappa$.  Therefore, to this order, the nucleon still acts like a point particle,
but one with spin.  Not until the second order does any internal structure become important.  This piece of the
amplitude is divided into two parts.  The terms with the angular dependence $ (\epsdp )$ and
$ (\hat{k}_f \times \epspr) \cdot (\hat{k}_i \times \eps)$ are written separately, and all other combinations
of the four photon unit vectors are lumped together in the function $h$.  We can define the coefficients 
of these two terms to be the generalized polarizabilities $ \bar{\alpha} $ and $  \bar{\beta} $.  This
is the most general definition of polarizability.  By defining 
 $ \bar{\alpha} $ and $  \bar{\beta} $ in this manner, they can be determined experimentally by 
examining the angular dependence of the cross-sections at higher energies.

The Compton scattering definition involves one additional complication: the generalized polarizabilities 
 $ \bar{\alpha} $, $  \bar{\beta} $ are not the same as the static polarizabilities
$\alpha$,$\beta$.  They are related by
\begin{equation}
\bar{\alpha} = \alpha + \Delta\alpha.
\end{equation}
\noindent Since the polarizability is defined as a coefficient in the scattering amplitude, there is no 
way to experimentally distinguish between different terms which have the same energy and angular dependence.
The main contribution to  $ \bar{\alpha} $ comes from the effects of the induced dipole moments, and all other 
contributions are collectively called $\Delta\alpha$.  

 The $\Delta\alpha$ term includes corrections which account for the finite size of the nucleon.  
The first term in the explicit  
expression for $\Delta\alpha$ is \cite{pe63, ma90} :
\begin{equation} 
\Delta\alpha = \frac{e^2r_e^2}{3m} + \frac{e^2(1+\kappa^2)}{4m^3},
\end{equation}
where $r_e$ stands for the charge radius.
$\Delta\alpha$ accounts for about 40\% of $\agen_p$, while it is only about a 5\% effect in the
neutron.  This can be traced to the fact that the neutron has zero charge.  

Corrections also exist for the generalized magnetic polarizability.
The first term in $\Delta\beta$ is \cite{ma90}
\begin{equation}
\Delta\beta = -\frac{e^2 e_e^2}{6m}.
\end{equation}
Additional corrections involve the nucleon form factors $F_1(4m^2), F_2(4m^2)$ and their derivatives \cite{ma90}.

\section{Model Calculations of Polarizability}
While we are mainly interested in determining polarizabilities through experiment, it is useful
to compare these values with those obtained in theoretical calculations.  Many attempts have been 
made to obtain values for $\agen$ and $\bgen$, but all of the results are model-dependent.
We will examine some of these results, including those obtained from dispersion relations 
and chiral perturbation theory.

One of the first to obtain an estimate for the proton polarizabilities was Baldin \cite{ba60}.  His calculation
of 
\begin{equation}
4.0 \le \agen_p \le 15.0
\end{equation}
\noindent is still consistent with recent calculations.  He 
considered only the first excited state ($N' = N + \pi$) in the matrix definition 
of $\alpha_p$ (equation \ref{eq:c1-2}),  
and estimated the matrix elements  using the meson photoproduction data available at the time.

Several useful dispersion relations can be derived for the generalized polarizabilities.  
Gell-Mann and Goldberger \cite{ge54} wrote down the once-subtracted dispersion relation for
the forward scattering amplitude $f(\omega)$:
\begin{equation}
\mathrm{Re} f(\omega) = \mathrm{Re} f(0) + \frac{2\omega^2}{\pi} \int_0^{\infty} 
\frac{ \mathrm{Im} f(\omega') d\omega' }{\omega' (\omega^{\prime 2} - \omega^2)}.
\end{equation}

Looking back at equation (\ref{eq:c1-3}), we see that the actual form of real part of the forward scattering
amplitude is
\begin{equation}
\mathrm{Re} f(\omega) = -\frac{e^2}{m_N} + (\agen + \bgen) \omega^2,
\end{equation}
since the function $g$ is purely imaginary, and contributions from $h$ cancel (this can be seen from 
the explicit expression for $h$ in \cite{pe81}).
After comparing the above equations and then
using the optical theorem,
\begin{equation}
 \mathrm{Im} f(\omega) = \frac{ \omega}{4\pi} \sigma_{\mathrm{tot}}(\omega ),
\end{equation}
where $\sigma_{\mathrm{tot}}(\omega )$ is the total cross-section for photoabsorption, we
can write the dispersion relation \cite{ba60}
\begin{equation}
\agen + \bgen = \int_{m_{ \pi} }^{\infty} \frac{ \sigma_{\mathrm{tot}}(\omega' ) d\omega' }{2\pi^2 \omega^{\prime 2}}.
\end{equation} 

This relationship, the Baldin sum rule,
constrains the possible values of $\agen$ and $\bgen$.  It is model-independent, 
and experimental data alone can be used to evaluate the integral, using some reasonable method
to continue to infinity.  The generally accepted results from the sum rule are \cite{da70,sc80}
\begin{eqnarray}
\agen_p + \bgen_p & = & 14.3 \pm 0.5, \\
\agen_n + \bgen_n & = & 15.8 \pm 0.5. 
\end{eqnarray}
\noindent However, a recent experiment \cite{ba98} predicts the following:
\begin{eqnarray}
\agen_p + \bgen_p & = & 13.69 \pm 0.14, \\
\agen_n + \bgen_n & = & 14.44 \pm 0.69.
\end{eqnarray}
\noindent  The values for the proton are consistent with the older ones, but the new neutron values are 
slightly lower.

A dispersion relation for $\agen - \bgen$ can also be derived \cite{be74, be77},
but it contains a contribution that depends on the amplitudes for
$\gamma\gamma \rightarrow \pi\pi$, which are not well known.  Nevertheless, this
difference has been estimated to be \cite{ho94}
\begin{eqnarray}
\agen_p - \bgen_p & \approx &3.2, \\
\agen_n - \bgen_n & \approx & 3.9. 
\end{eqnarray}
Comparison with the sum rules above yield smaller values for the
individual polarizabilities than the $\agen_p \approx 11$ and $\agen_n \approx 12$ measured 
experimentally.

Various quark models have also been used to determine polarizabilities.  
A simple nonrelativistic model with a harmonic oscillator potential yields
reasonable values for $\bgen$ but predicts $\agen_p > \agen_n$,  so more sophisticated models are needed.
For example, Werner and Weise \cite{we85} calculated $\agen$ and $\bgen$
using a valence quark core surrounded by a pion cloud to be
\begin{eqnarray}
\agen_N & \approx & 7-9, \\
\bgen_N & \approx & 2.
\end{eqnarray}
Again, these values are lower than recent experimental measurements. 

Perhaps the most promising method is using chiral perturbation theory to calculate the polarizabilities.  
Calculations with only the one-loop contribution produce \cite{me92}:  
\begin{equation}
\agen_N = 10 \bgen_N = \frac{f_{\pi N}^2}{4\pi} \frac{5e^2}{24m_N^2m_{\pi} } \approx 13.6.
\end{equation}
\noindent Including the next order as well as the effects of the $\Delta$ resonance gives \cite{me93,me94}
\begin{center}
\begin{tabular}{cc}
$\agen_p = 10.5 \pm 2.0$ & $\agen_n = 13.4 \pm 1.5$ \\
$\bgen_p = 3.5 \pm 3.6 $ & $\bgen_n = 7.8 \pm 3.6$ 
\end{tabular}
\end{center}
If we compare these two calculations with the Baldin sum rule, we find agreement
with the predictions for the proton.  However, the earlier calculation 
gives better agreement with the neutron than the higher-order one, indicating 
a need for more work on this front. 

Both of these methods make predictions about the sources of the various contributions to the polarizability. 
For example, it can be shown that the most important 
contributions to $\agen$ come from polarizing the pion cloud - about 50-70\% \cite{we85}.  Resonances such as the $\Delta$
and the pion polarizability account for the remainder.  The small value of $\bgen \approx 2$ can
be attributed to cancellations between the large paramagnetic contribution from the $\Delta$ resonance ($\beta$) 
and the diamagnetic contribution from the pion cloud ($\Delta\beta$) \cite{mu93}.

\section{Proton Experiments for Determining Polarizabilities}

We now examine some experimental  results.  Several experiments have been performed
over the past few years with the aim of measuring nucleon polarizabilities.  Most of these
have specifically targeted the proton, but some attempts have also been made at measuring 
the neutron polarizabilities.   Since there have been so many more measurements of $\agen_p$ and $\bgen_p$,
we begin with the proton.

The differential cross-section for proton Compton scattering, which is proportional to
the square of the scattering amplitude of equation~(\ref{eq:c1-3}),  can be written to order $\omega^2$ as
\begin{equation}
\frac{d\sigma}{d\Omega}  =  \frac{d\sigma}{d\Omega}_{\mathrm{Born}} 
 - \frac{\alpha}{m_p} \left( \frac{\omega '}{\omega} \right)^2 \omega\omega ' 
\left\{ \frac{1}{2} (\agen_p + \bgen_p ) (1 + \cos\theta)^2 + \frac{1}{2} (\agen_p - \bgen_p ) (1 -\cos\theta)^2 \right\}, \label{eq:c3-2}
\end{equation}
\noindent where $\alpha$ is the fine structure constant, and the Born term, first
written down by Powell \cite{po49}, is
\begin{eqnarray}
\frac{d\sigma}{d\Omega}_{\mathrm{Born}} & = & \frac{1}{2} \left( \frac{\alpha}{m_p} \right)^2
\left( \frac{\wfin}{\winit} \right)^2 \left\{ \frac{\winit}{\wfin} + \frac{\wfin}{\winit} - \sin^2{\theta} \right. + \nonumber\\
& & \kappa_p \frac{2\winit\wfin}{m_p^2} (1-\cos{\theta})^2 + \kappa_p^2 \frac{\winit\wfin}{m_p^2} \left[ 4(1-\cos{\theta}) + \frac{1}{2}
(1-\cos{\theta})^2 \right] + \nonumber\\
& & \left. \kappa_p^3 \frac{\winit\wfin}{m_p^2} \left[ 2(1-\cos{\theta}) + \sin^2{\theta} \right] +
 \kappa_p^4 \frac{\winit\wfin}{2m_p^2} ( 1 + \frac{1}{2}\sin^2{\theta} ) \right\}. \nonumber
\end{eqnarray}
Note that the polarizability terms in equation~(\ref{eq:c3-2}) arise from
interference between the polarizability amplitude of equation~(\ref{eq:c1-3}) and the
Thomson amplitude.

This formula illustrates the sensitivity of the cross-section to $\agen_p$ and 
$\bgen_p$ at different angles.  The cross-section at forward angles is most sensitive to $\agen_p+\bgen_p$,
while at backward angles only $\agen_p-\bgen_p$ can be measured.  At $90^{\circ} $ the $\beta_p$ terms drop out
completely. 

Another point to consider is that there is an optimal range of energies for which
a measurement of the proton polarizability can be done.  The expansion to order $\omega^2$ is 
only valid to about 100 MeV \cite{na94}.  On the other hand, since the polarizability terms are of this order, 
too low an energy means that the cross-section is too insensitive to $\agen_p$ and $\bgen_p$.  
Figure~\ref{fig:c3-1} shows the dependence of the polarizability terms on energy for 
two different angles.  Looking at this graph, it seems that the optimal energy range for an 
experiment would be 70-100 MeV in order to balance sensitivity with expansion validity.  
The cross-section also depends more heavily on the polarizabilities at forward angles;
however, these measurements alone would only test the Baldin sum rule and not yield
individual values for $\agen$ and $\bgen$.

A summary of the data from a few recent proton Compton scattering experiments
is given in Table~\ref{t:c3-1} below.  The Moscow experiment \cite{ba75} is included because it 
was believed to be the most accurate of earlier attempts to measure the polarizabilities,
despite the low energy resolution of the photon detector.  Their value for $\agen_p$
is consistent with theoretical predictions, 
but $\agen_p + \bgen_p$ falls far
below the accepted value from the sum rule, and $\bgen_p$ is negative.   This 
formed part of the motivation for more recent experiments \cite{na94}. The values
listed in the table have actually been generated by a recent re-analysis of the original data \cite{ma95}. 

The Illinois group \cite{fe91} performed one of these latest experiments.  Measurements were 
taken at both a forward and a backward angle, and the energies used fall 
well within the range of validity of the low-energy expansion.  However, 
this was done at the expense at sensitivity to the polarizabilities, which 
explains the high uncertainties.  An improved 1995 experiment by the same group \cite{ma95}
used higher photon energies, but now perhaps too high to permit use
of the low-energy expansion. 

A 1992 Mainz experiment \cite{ze92} is also included in the table.  Since
data was only taken at the backward angle, only the difference 
$\agen_p - \bgen_p$ could be extracted.  Individual values for $\agen_p$ 
and $\bgen_p$ were obtained from the sum rule.  Also, the energy range used
is possibly beyond the limit of validity of the low-energy expansion. 

More extensive reviews of these and other proton Compton scattering experiments
can be found in the literature \cite{pe81, na94}.  However, the point to be made is that, despite
some shortcomings, measurements for the polarizabilities in
these experiments are in reasonable agreement with each other and 
with theoretical predictions.  This is not the case for the neutron. 
\begin{center}
\begin{figure}
\input{figc3-2}
\input{figc3-1}
\caption{Energy dependence of $\agen$ and $\bgen$ terms \label{fig:c3-1} }
\end{figure}
\end{center}
\begin{table}
\caption{Experimental results for $\agen_p$ and $\bgen_p$ \label{t:c3-1} }
\begin{center}
\begin{tabular}{ cccc }
& \underline{$\omega_i$(MeV)} & \underline{$\theta$(degrees)} & \underline{result} \\
Moscow(1975) & 70-110 & 90,150 & $\agen_p + \bgen_p = 5.8 \pm 3.3$ \\
	& & & $ \agen_p - \bgen_p  = 17.8 \pm 2.0$ \\
Illinois (1991) & 32-72 & 60,135 &  $\agen_p = 10.9 \pm 2.2 \pm 1.3$ \\
& & & $ \bgen_p = 3.3 \mp 2.2 \mp 1.3$\\
Mainz(1992) & 98, 132 & 180 & $ \agen_p - \bgen_p = 7.03^{+2.49+2.14}_{-2.37-2.05} $\\
& & & $\agen_p  = 10.62^{ +1.25+1.07}_{-1.19-1.03}$\\
& & & $\bgen_p = 3.58^{ +1.25+1.07}_{-1.19-1.03}$\\
Illinois/SAL(1995) & 70-148 & 90,135 & $\agen_p = 12.5 \pm 0.6 \pm 0.7 \pm 0.5$  \\
& & & $\bgen_p = 1.7 \mp 0.6 \mp 0.7 \mp 0.5$ \\
\end{tabular}
\end{center}
\end{table}

\section{Neutron Experiments for Determining Polarizabilities}

The neutron polarizabilities have not yet been measured with reasonable accuracy.
This is mainly because a neutron Compton scattering 
experiment cannot be directly performed.  Not only is a neutron target unavailable, but 
the cross-section for this process would be extremely small.  Therefore, other methods of 
measuring $\agen_n$ and $\bgen_n$ have been developed.  The two that have produced
results thus far are Coulomb scattering and deuteron photodisintegration.

Coulomb scattering of a low-energy neutron off a heavy nucleus 
is a reasonable means of measuring the electric polarizability because of the   
large dipole moment induced by the Coulomb field.  The interaction energy is given by
\begin{equation}
V = -\frac{1}{2}\alpha_n \frac{Z^2 e^2 }{r^4},
\end{equation}
\noindent where $Ze$ is the charge of the heavy  nucleus.  Note that 
only $\alpha_n$ (rather than $\agen_n$) is involved here, and  
the $\Delta\alpha$ contribution, although small, must be accounted for.

This energy shift turns out
to have only about a 1\% effect on the cross-section \cite{sh89}.  The low sensitivity to $\alpha_n$,
combined with the uncertainty in calculating other small contributions to the cross-section,
makes an accurate determination of the polarizability in this type of experiment very difficult. 
Two separate 1988 experiments, both using $\left.\right.^{208}$Pb as the target nucleus,  
quote large uncertainties \cite{sh88,ko88}:
\begin{eqnarray}
\alpha_n & = & 12.0 \pm 10.0 \ \ \ \ \ \ \ \ \mathrm{          (Vienna/Oak Ridge) }, \\
\alpha_n & = & 8.0 \pm 10.0   \ \ \ \ \ \ \ \ \mathrm{(Munich) }.
\end{eqnarray}

A more recent experiment by Schmiedmayer et al. \cite{sh91}, also using a lead target, produced
the following result:
\begin{equation}
\alpha_n = 12.0 \pm 1.5 \pm 2.0.
\end{equation}
\noindent  This is the value normally quoted for the neutron electric polarizability.  
However, the accuracy of this result has been questioned \cite{en97, ni92}.  It has been claimed that the uncertainties
were underestimated,  and that the best estimate for $\alpha_n$ is really $\sim 7-19$ \cite{en97}.  

The other method used to measure $\alpha_n$ is quasi-free Compton scattering
by the neutron bound in the deuteron.   It has been calculated that the cross-section for
deuteron photodisintegration is sensitive to the neutron polarizability when most of the 
photon's momentum is transferred to the neutron, and the proton remains in the target \cite{lv}.
Such an experiment was carried out by Rose et al. \cite{ro90} in 1990, and yielded the measurement 
\begin{equation}
\alpha_n = 11.7^{+4.3}_{-11.7}.
\end{equation} 
\noindent Again, the uncertainties are large.  However, there have been recent 
arguments that this method should be revisited as it is capable of producing 
more accurate results \cite{wi98}.

It is clear that better experimental measurements of the 
neutron polarizability are needed in order to keep up with recent theoretical developments.  
This is why we have chosen to investigate deuteron Compton scattering as a means 
of obtaining additional information about $\agen_n$.  This idea seems to have been
originally suggested by Baldin \cite{ba60}, who calculated the cross-section in the impulse 
approximation.   A more extensive calculation was undertaken by Weyrauch \cite{we88, we90},
but certain deficiencies in his method to be discussed later, as well 
as a lack of emphasis on the neutron polarizability, suggest that this question
can be reexamined.   

A deuteron Compton scattering experiment was performed in 1994 \cite{lu94}, using 49 and 69 MeV photons, 
and experiments are in progress at Saskatoon \cite{ho98} and Lund \cite{lund98}.
This has spurred some recent activity in this area.  Articles by Arenhovel \cite{ar95}
and L'vov and Levchuk \cite{lv95, lv98} quote new theoretical results for Compton scattering.
A calculation
using effective field theories has also been published \cite{ch98}.
Their results will be compared with the 
present calculation in Chapter 5.


\chapter{Deuteron Compton Scatttering Calculations \label{ch:calc} }
\section{Feynman Diagrams}

\begin{figure}
\centering
\parbox{73mm}{ 
\centering
\epsfig{file=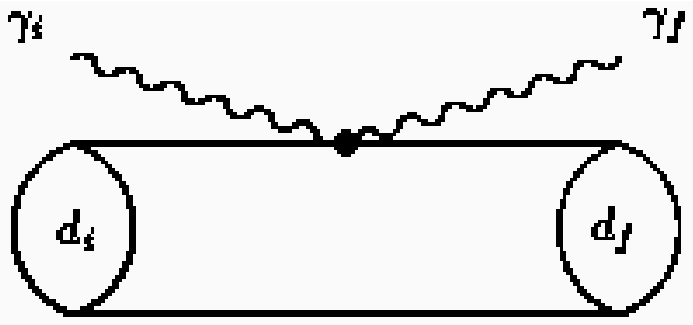}
}
\parbox{35mm}{(a)}
\parbox{73mm}{
\centering
\epsfig{file=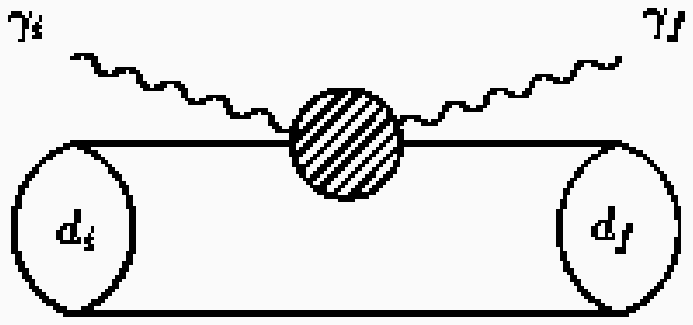}
}
\parbox{35mm}{(b)}
\parbox{73mm}{
\centering
\epsfig{file=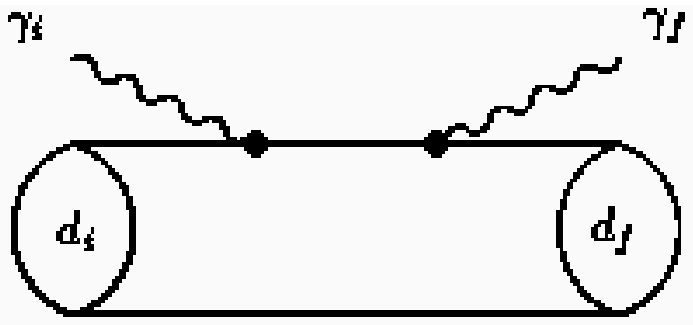}
}
\parbox{35mm}{(c)}
\parbox{73mm}{
\centering
\epsfig{file=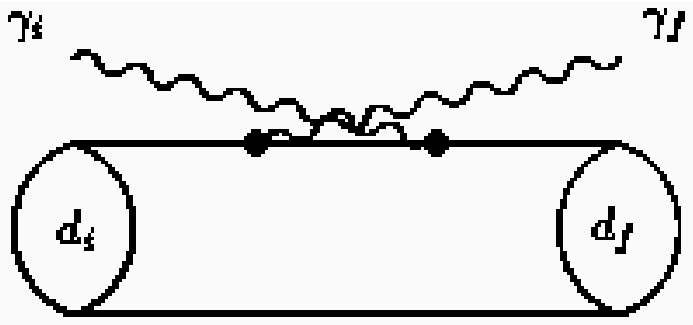}
}
\parbox{35mm}{(d)}
\caption[Feynman diagrams contributing to deuteron Compton scattering]{Feynman diagrams  contributing to deuteron Compton scattering. \label{fig:d1-1} }
\end{figure}

The Feynman diagrams for several of the processes contributing to deuteron 
Compton scattering are shown in  Figure~\ref{fig:d1-1}.  The ovals at either end
represent the deuteron wavefunctions, while the solid lines are the  
individual nucleons with which the photons (wavy lines) interact.  Not all possible combinations
are shown; interactions can occur on either nucleon with either an incoming or outgoing photon.

 Figure~\ref{fig:d1-1}(a) is known as the seagull diagram.  This is the interaction that
arises from the term in the Hamiltonian that is proportional to $A^2$.  It has no overall 
energy dependence, and is an important term at all energies.  All other one-body interactions,
which are at least of order  $\omega^1$, are represented by  Figure~\ref{fig:d1-1}(b).
These include the terms which depend on the polarizabilities $\alpha$ and $\beta$, as well 
as some relativistic corrections.

Dispersive diagrams without meson exchange are depicted in  Figure~\ref{fig:d1-1}(c)-(d).
At lowest order, each vertex can be either an electric or magnetic dipole (E1 or M1) interaction,
which originate in the $\vec{p}\cdot\vec{A}$ and $\vec{\mu}\cdot\vec{B}$ terms
in the Hamiltonian, respectively.  These two diagrams 
account for the bulk of the work in the deuteron Compton scattering calculation.

Three of the meson-exchange effects which are included in this calculation
are shown in  Figure~\ref{fig:d1-1}(e)-(g). They are either second- or third-order
terms, and are rather small at the energies considered here.  The term of Figure~\ref{fig:d1-1}(f)
is called the ``vertex correction''.

\clearpage

\begin{figure}
\centering
\parbox{73mm}{
\centering
\epsfig{file=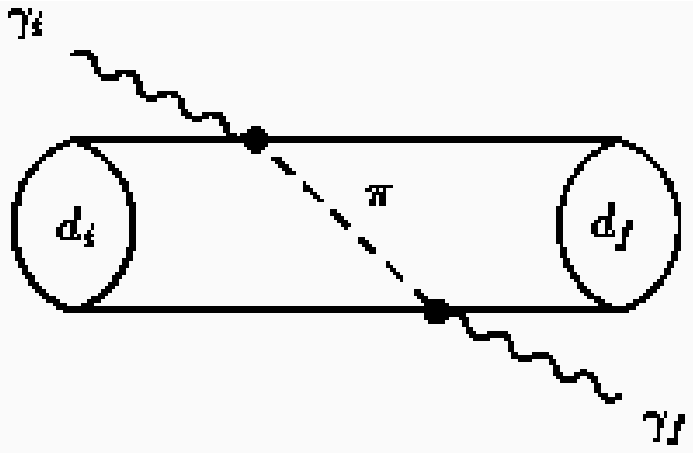}
}
\parbox{35mm}{(e)}
\parbox{73mm}{
\centering
\epsfig{file=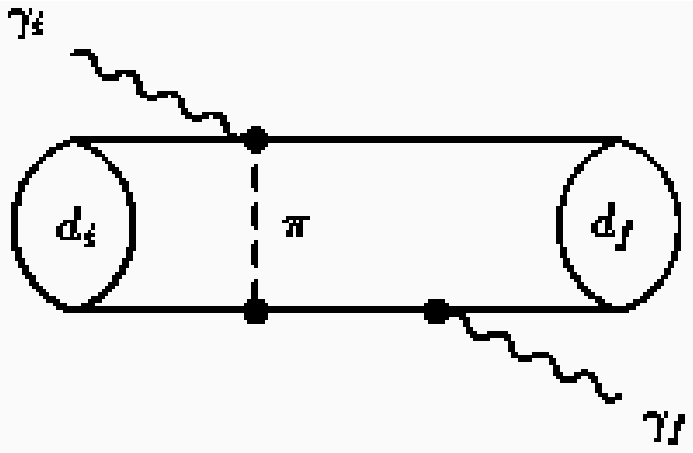}
}
\parbox{35mm}{(f)}

\parbox{73mm}{
\centering
\epsfig{file=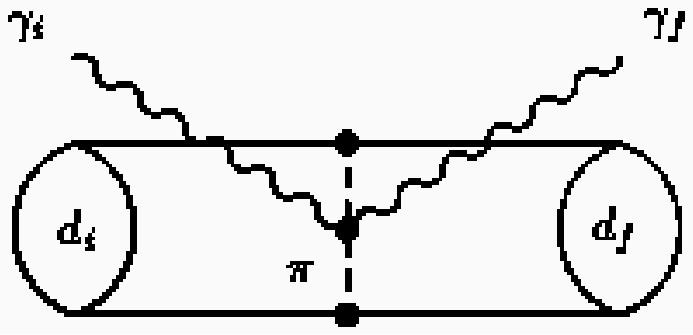}
} 
\parbox{35mm}{(g)}
\end{figure}


\section{Overview of Calculation}

We will begin with an overview of the methods used in the calculation.  
The differential cross-section will be calculated non-relativistically
using Fermi's golden rule, which can be written schematically as
\begin{equation}
d\sigma_{fi} = \frac{2\pi}{\hbar} \delta(\Delta E) \times\frac{1}{\mathrm{initial}\ \mathrm{flux}}
\times (\mathrm{number}\ \mathrm{ of}\ \mathrm{ final}\ \mathrm{ states}) 
\times \left| \mathcal{T}_{fi} \right|^2 \label{eq:d2-1}.
\end{equation}
The subscript $fi$ indicates a dependence on the final and initial energies and polarizations.
We use the convention where the initial photon flux is normalized to be $1$ per
arbitrary volume $V$. A detailed derivation of the phase space factors for this calculation
can be found in Appendix D. We have set $c=1$.

The transition matrix $ \mathcal{T}_{fi}$ is defined by
\begin{eqnarray}
\lefteqn{\mathcal{T}_{fi}  =  \mx{d_f, \vec{P}_f, \gamma_f}{H^{\mathrm{int}} }{d_i, 
\vec{P}_i, \gamma_i} + } \nonumber\\
& & \sum_{C, \vec{P}_C} \frac{ \mx{d_f, \vec{P}_f, \gamma_f}{H^{\mathrm{int}}}{C, \vec{P}_C}
\mx{C, \vec{P}_C}{H^{\mathrm{int}}}{d_i, \vec{P}_i, \gamma_i} }{E_{d_i} + 
P_i^2/2m_d + \hbar\winit - E_C - P_C^2/2m_d + i\varepsilon} + \cdots \label{eq:d2-2}
\end{eqnarray}

The notation $d_{i(f)}$ is meant to represent all quantum numbers needed to describe the
initial (final) deuteron state, except for the center-of-mass momentum $\vec{P}$, which is written
separately so that it can be integrated out and absorbed into a momentum-conserving 
delta function.  Similarly, $C$ is the intermediate $np$ (or possibly deuteron) state, and
$\sum_{C, \vec{P}_C}$ denotes all sums and integrals which are needed to describe this state.
It is also possible for the intermediate state to contain photons and/or mesons, and 
these can be inserted into the above expression as needed.  

In order to calculate the transition matrix, we need to specify a Hamiltonian.  The basic 
non-relativistic Hamiltonian, including polarizability and meson-exchange terms, is
\begin{eqnarray} 
\lefteqn{H^{\mathrm{int}}  =  \sum_{j=n,p} \left[ \frac{e_j^2}{2m_j} A^2(\vec{x}_j) -
\frac{e_j}{m_j} \vec{A}(\vec{x}_j)\cdot\vec{p}_j -  \right.
 \frac{e(1+\kappa_j)}{2m_j} \vec{\sigma}_j\cdot\left( \vec{\nabla}_j \times\vec{A}(\vec{x}_j) \right)- }\label{eq:d2-3} \\      
& &  \frac{1}{2} \bar{\alpha}_j \left( \frac{ \partial\vec{A}(\vec{x}_j) }{\partial t } \right)^2 
-  
\frac{1}{2} \bar{\beta}_j  \left( \vec{\nabla}_j \times\vec{A}(\vec{x}_j)\right)^2
+  \frac{if_{\pi}e_{\pi}}{m_{\pi}} \left( \vec{\sigma}_j \cdot
\vec{A}(\vec{x}_j) \right) \left( \iso{\tau}_j \cdot \iso{\phi}(\vec{x}_j) \right) + \nonumber\\
& & \left.\frac{f_{\pi}\hbar}{m_{\pi}} \left( \vec{\sigma}_j \cdot
\vec{\nabla}_j \right) \left( \iso{\tau}_j \cdot \iso{\phi}(\vec{x}_j) \right) \right]
+ \frac{1}{\hbar^2}\int d^3xe^2 A^2(\vec{x}) \left[ \phi_+(\vec{x})
\phi_-(\vec{x}) + \phi_-(\vec{x})\phi_+(\vec{x}) \right],  \nonumber
\end{eqnarray} 
where
\begin{equation}
 \vec{A}(\vec{x_j}) = \frac{1}{\sqrt{V} } \sum_{\vec{k},\lambda = \pm 1 } \sqrt{\frac{2\pi\hbar }{\omega } } [ a_k
{\hat{\epsilon}}_{\lambda} e^{i\vec{k}
\cdot{\vec{x}}_j } + a_k^{\dagger} {\hat{\epsilon}}_{\lambda}^{\ast} e^{-i\vec{k}\cdot{\vec{x}}_j } ] .
\end{equation}
The operators  $a_{k,\lambda}^{(\dagger )}$ destroy (create) a photon with momentum $\vec{k}$
and polarization $\lambda $.  The explicit time dependence $e^{i\omega t}$ has been suppressed
since it only gives rise to the energy-conserving delta function in Fermi's golden rule
(equation~\ref{eq:d2-1}). Similarly,
\begin{equation}
\phi_{\pm}(\vec{x}_j) = \sum_q \frac{\hbar}{\sqrt{V}} \sqrt{ \frac{2\pi}{E_{\pi}} }
\left( a_{\mp} e^{i\vec{q}\cdot\vec{x}_j} + a_{\pm}^{\dagger} e^{-i\vec{q}\cdot\vec{x}_j} \right).
\end{equation}
Here, $a_{\pm}$ destroys a $\pi_{\pm}$.  Finally, the tildes in the Hamiltonian
indicate vectors in isospin space.

The first term, as discussed before, is known as the Thomson term, and is the interaction represented
by the seagull diagram of  Figure~\ref{fig:d1-1}(a).  The next two terms are the electric and
magnetic dipole terms.  These two expressions will not be actually
substituted into the transition matrix; instead, the  more general observed vector current formalism, to 
be described later, 
will be used to describe the two-body terms.  However, an explicit decomposition of the
vector current expressions yields these dipole terms.   

The polarizability terms are next, and then the last three terms
are the meson-exchange effects  needed to calculate the diagrams in Figures~\ref{fig:d1-1}(e),(f),
and (g), respectively.  
Relativistic corrections 
to this Hamiltonian will be discussed in section~\ref{sec:rel}.


\section{Seagull Term}

The simplest diagram to evaluate is the seagull diagram, so it will be the first one
we will consider.  The basic procedure is to calculate the transition matrix using the
interaction Hamiltonian
\begin{equation}
 H^{SG} = \frac{e^2}{2 m_p} A^2(\vec{x_p}) ,
\end{equation}
Inserting this into the transition matrix gives
\begin{equation}
 \matrixt{SG} = \frac{e^2}{2m_p } \frac{1}{V} \frac{2\pi\hbar}{\sqrt{\winit\wfin}}
\mx{d_f,\vec{P}_f}{2 \eps e^{i\kinit\cdot{\vec{x}}_p } \cdot \epspr 
e^{-i\kfin\cdot{\vec{x}}_p} }{d_i,\vec{P}_i}.
\end{equation}

The next step is to eliminate the individual nucleon variables in favor
of the deuteron variables $\vec{r}$ and $\vec{R}$, where
\begin{eqnarray}
\vec{r} & \equiv & \vec{x}_p - \vec{x}_n, \\
\vec{R} & \equiv & \frac{1}{2} ( \vec{x}_p + \vec{x}_n ).
\end{eqnarray}

We can now insert a complete set of $\vec{r}'$ and $\vec{R}'$ states into
the matrix element.  Defining  $ \vec{q} \equiv \kfin - \kinit $, the transition
matrix becomes
\begin{eqnarray}
 \matrixt{SG} &=  & \int d^3r' d^3R'\frac{e^2}{m_p}  \frac{1}{V} \frac{2\pi\hbar}{\sqrt{\winit\wfin}}(\epsdp)
\times\nonumber\\
& & \ \ \ \ \ \ \ \mx{d_f, \vec{P}_f}{ e^{-\frac{i}{2}\vec{q}\cdot\vec{r'}} 
e^{-i\vec{q}\cdot\vec{R'} } }{ \vec{r'},\vec{R}'}\mxemp{ \vec{r'},\vec{R'} }{d_i, \vec{P}_i}.
\end{eqnarray} 

The integral over $\vec{R}'$ can be performed by assuming that the center-of-mass
wave\-functions for the deuteron are plane waves.  This yields a momentum-conser\-ving 
delta function.  We remove this delta function for convenience by defining the 
scattering amplitude $\matrixm{}$ by
\begin{equation}
 \matrixt{} = \ \frac{1}{V} \frac{2\pi\hbar}{\sqrt{\winit\wfin}} \label{eq:d3-6}
\frac{ {\delta}({\vec{P}}_f + \kfin - {\vec{P}}_i - \kinit)}{V} \matrixm{}  .
\end{equation}   
\noindent Thus,
\begin{equation}
 \matrixm{SG} = \sum_{ll'} \int  d^3r \frac{e^2}{m_p}  (\epsdp) \label{eq:d3-1}
\mx{l' S_f J_f M_f r_f}{ e^{-\frac{i}{2}\vec{q}\cdot\vec{r}} }{\vec{r}}\mxemp{\vec{r} }{l S_i J_i M_i r_i} .
\end{equation}
Here we have explicitly stated what is meant by the label ``$d_i$'', as the radial and angular
parts have been separated.  Isospin is not included as a quantum number, but is
regulated by the generalized Pauli exclusion principle:
\begin{equation}
(-1)^{L+S+T} = -1.
\end{equation}
Also, the orbital quantum number $l$ is not written with an $i$ or $f$ label because
the S and D states of the deuteron are summed at this stage.  The spin and total angular
momenta can only take on the values $S=J=1$ for the deuteron.

\begin{figure}
\centering
\epsfig{file=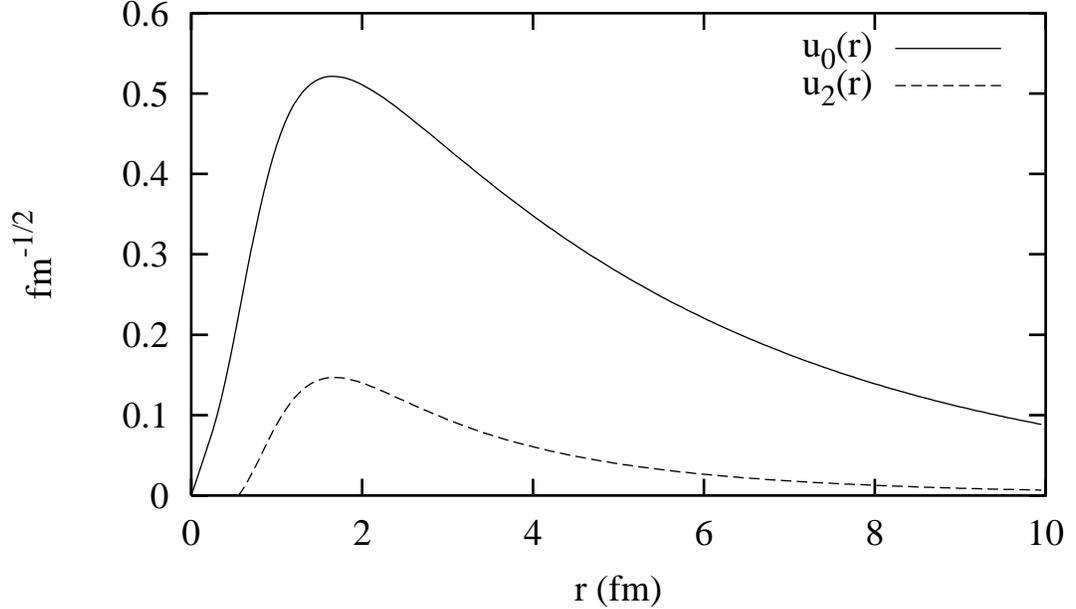}
\caption{Deuteron radial wavefunctions $u_0(r)$ and $u_2(r)$. \label{fig:d3-2} }
\end{figure}

We now turn to a discussion of the deuteron wavefunction.  The form that
will be used here is
\begin{equation}
 \mxemp{\vec{r}}{d_i} =  \frac{u_0(r)}{r}\mxemp{\hat{r}}{011M}  +
\frac{u_2(r)}{r}\mxemp{\hat{r}}{211M} . \label{eq:d3-7}
\end{equation}
 Instead of labeling  
the radial wavefunctions by the conventional $u(r)$ and $w(r)$,
we use the notation $u_0(r)$ and $u_2(r)$ to create an explicit 
index which can be summed over.   We use the  wavefunctions published by Machleidt \cite{ma89}  (the ``B'' wavefunctions,
derived from the Bonn potential).  These are shown in Figure~\ref{fig:d3-2}. 
The angular wavefunctions are of the form $\mxemp{\hat{r}}{LSJM}$;
an algebraic representation will not be needed.
%

The easiest way to evaluate equation~(\ref{eq:d3-1}) is to expand the exponential 
into partial waves using
\begin{equation}
e^{i\vec{k}\cdot\vec{r} } = \sum_{L = 0}^{\infty} 
\sum_{M= -L}^{L} 4\pi i^L Y_{LM}(\hat{r}) Y_{LM}^{\ast}(\hat{k}) j_{\sss L}(kr),
\end{equation}
where $j_{\sss L}(kr)$ is the spherical Bessel function of order $L$.
Inserting this along with the wavefunctions into equation~(\ref{eq:d3-1}) gives
\begin{eqnarray} 
\matrixm{SG} & = &
 \frac{e^2}{m_p}(\epsdp )\sum_{l l' L M}\int dr 4\pi (-i)^L j_L(\frac{qr}{2}) Y_{LM}^{\ast}(\hat{q})
\nonumber\times\\
& & \ \ \ \ \ u_{\sss l}(r) u_{\sss l'}(r) \mxba{ Y_{LM} } \label{eq:d3-3} .
\end{eqnarray}

This radial integral can easily be evaluated numerically; however, the matrix element 
must be simplified further.  We go through this for the seagull term; derivations for
other terms are in the Appendices.  For the seagull term, the operator is a spherical harmonic, but  
other terms in the transition matrix will involve more complicated combinations of spin and angular momentum 
operators.

  The first step is to form tensor products (each with a definite total
 angular momentum, which may be summed over) out of the individual operators.   
In this case, the operator has total angular momentum $L$.  We can then use the Wigner-Eckart theorem
(equation~\ref{eq:formula2}) to create reduced matrix elements which are independent of the azimuthal
quantum number $M$.  The operators in the reduced matrix elements can be uncoupled and evaluated
using standard relations (equations~\ref{eq:formula3}-\ref{eq:formula4}), which can be found in a reference such 
as Rotenberg \cite{ro}.  Further details of this for the seagull term can be found in Appendix A.
The net result is:              
\begin{eqnarray}
\lefteqn{  \mxba{ Y_{LM} }   =  (-1)^{1-M_f+l'} 3 \sqrt{\frac{(2l'+1)(2L+1)(2l+1)}{4\pi } } }
   \nonumber\\
& &\ \ \ \ \ \threej{l}{L}{l'}{0}{0}{0} \threej{1}{L}{1}{-M_f}{M}{M_i} \sixj{1}{l'}{1}{l}{1}{l}
\label{eq:d3-2}.
\end{eqnarray}
Definitions of the $3j$ and $6j$ symbols can also be found in Rotenberg.  One important property to note here is
that for the $3j$ symbol $\threej{l_1}{l_2}{l_3}{m_1}{m_2}{m_3}$, 
$l_1$, $l_2$, and $l_3$ satisfy the triangle inequality, and $m_1+m_2+m_3 = 0.$  This means that,
for the seagull term, $M=M_f-M_i.$  Therefore, the sum over $M$ in equation~(\ref{eq:d3-3}) can be removed and
equation~(\ref{eq:d3-2}) can be inserted to produce the final result:
\begin{eqnarray}
 \matrixm{SG} & = & \frac{\sqrt{12\pi}e^2}{m_p} (\epsdp) (-1)^{1-M_f} \times\label{eq:d3-4}\\
& & \ \ \ \ \ \ \left[ \threej{1}{0}{1}{-M_f}{0}{M_f} Y_{00}^{\ast}(\hat{q}) \right.  (I_0^{00}+I_0^{22}) {\delta}_{M_f, M_i} +
\nonumber \\
& & \ \ \ \ \ \ \ \ \left. \threej{1}{2}{1}{-M_f}{M_f-M_i}{M_i} Y_{2, M_f-M_i}^{\ast}(\hat{q}) (I_2^{02} +
I_2^{20} - \frac{I_2^{22}}{\sqrt{2}}) \right], \nonumber
\end{eqnarray}
where the notation
\begin{equation}
 I_L^{l'l} \equiv \int_0^{\infty} dr u_{\sss l'}(r) j_{\sss L}(\frac{qr}{2}) u_{\sss l}(r) 
\end{equation}
has been used.

The largest term in equation~(\ref{eq:d3-4}) is the one for which $l=l'=L=0$. This is because
the amplitude of the radial wavefunction $u_0(r)$ is several times larger than the amplitude for $u_2(r)$,
as can be seen in Figure~\ref{fig:d3-2}.   Using $\threej{1}{0}{1}{-M_f}{0}{M_f} = (-1)^{1+M_f}/\sqrt{3}$,
we can write the dominant part of the seagull term as
\begin{equation}
 \matrixm{SG}(\mathrm{dominant}) = \frac{e^2}{m_p} (\epsdp) I_0^{00} \delta_{M_f,M_i}.
\end{equation}
\noindent This is the Thomson amplitude for proton Compton scattering reduced by the factor $I_0^{00}$.

The terms that depend on the nucleon polarizabilities are very similar to the
seagull term.  Their transition matrices are calculated in detail in Appendices B and C.

To find the differential cross-section, we insert the scattering amplitude into equation~(\ref{eq:phase4}):
\begin{eqnarray}
\left(\frac{d\sigma}{d\Omega}\right)_{fi} & = &
\left( \frac{\wfin}{\winit} \right)^2 \frac{E_f}{m_d} \mid\matrixm{} \mid^2 , 
\end{eqnarray}
where $E_f =\winit + m_d - \wfin $, and $m_d$ is the deuteron mass.  The spin-averaged cross-section is computed by summing 
over the final and averaging over the initial polarizations.


\section{Dispersive Terms}

Most of the remaining terms to be calculated are known as the dispersive terms, which are
second-order in the interaction Hamiltonian. These terms, like the seagull term,
are of order $\alpha$, the fine structure constant. This
is the small parameter in which we perform our perturbative expansion. 

The transition matrix
has the form
\begin{eqnarray}
\matrixt{} & = & \sum_{C, \vec{P}_C} \left\{ \frac{ \mx{d_f, \vec{P}_f ,{\gamma}_f}{
H^{\mathrm{int} }}{C, \vec{P}_C}
\mx{C, \vec{P}_C}{H^{\mathrm{int}}}{d_i, \vec{P}_i, {\gamma}_i} }{\hbar{\omega}_i+E_{d_i}-E_C-
P_C^2/2m_d +i\varepsilon}
+ \right.  \label{eq:d4-1}\\
& & \left. \frac{ \mx{d_f, \vec{P}_f, {\gamma}_f}{H^{\mathrm{int}}}{C, \vec{P}_C, {\gamma}_i, {\gamma}_f }
\mx{C, \vec{P}_C, {\gamma}_i, {\gamma}_f }{H^{\mathrm{int}}}{d_i, \vec{P}_i, {\gamma}_i} }
{-\hbar{\omega}_f+E_{d_i}-E_C-P_C^2/2m_d+i\varepsilon} \right\} \nonumber .
\end{eqnarray}
 
Instead of inserting the Hamiltonian directly into equation~(\ref{eq:d4-1}), we
make the substitution $H^{\mathrm{int}} = -\int \vec{J}(\vec{\dum})\cdot \vec{A}(\vec{\dum}) d^3\dum$.
To avoid confusion, $\vec{\dum}$ is used as a dummy variable, while $\vec{x}$ is a nucleon variable
and $\vec{r}$ is a deuteron variable.

In addition, $\vec{A}$ is expanded into multipoles using 
\begin{eqnarray}
\lefteqn{ \epsilon_{\lambda} e^{i\vec{k}\cdot\vec{\dum} } =
 \sum_{L = 1}^{\infty} \sum_{M=-L}^{L}\wignerd i^L\sqrt{\frac{2\pi(2L+1)}{L(L+1)}}\times} \label{eq:d4-2}\\
& &  \left\{ -\frac{i}{\omega}\vec{\nabla_{\dum}} \left( 1+\dum\frac{d}{d\dum} \right) j_{\sss L}(\omega\dum) 
Y_{LM}(\hat{\dum}) -
i\omega \vec{\dum} j_{\sss L}(\omega\dum) Y_{LM}(\hat{\dum}) -
\lambda \vec{L}  Y_{LM}(\hat{\dum})  j_{\sss L}(\omega\dum) \right\} \nonumber.
\end{eqnarray}

$\wignerd$ is the Wigner-d function (we use the convention of \cite{rose57}) , which is the overlap of the state $\ket{L\lambda}$, rotated by
the Euler angles $(0,-\vartheta,-\varphi)$, with the state $\ket{LM}$.  
This expression is derived in Appendix E.

To simplify the notation, we will define a functions $\Phi_i$ and $\Phi_f$ by
\begin{eqnarray}
\frac{1}{\sqrt{V}} \sqrt{\frac{2\pi\hbar}{\winit} } \eps
e^{i\kinit\cdot\vec{\dum}} & = & 
\vec{\nabla}\Phi_i(\vec{\dum}) + \cdots, \label{eq:d4-80} \\
\frac{1}{\sqrt{V}} \sqrt{\frac{2\pi\hbar}{\wfin}} \epspr
e^{-i\kfin\cdot\vec{\dum}} 
& = & \vec{\nabla}\Phi_f(\vec{\dum}) + \cdots. \label{eq:d4-81}
\end{eqnarray}
The gradient terms above correspond to the first term in the brackets of equation~(\ref{eq:d4-2}).
In the low-energy limit, $\Phi_i(\vec{\dum}) \rightarrow \vec{\dum}\cdot\eps$, so we
expect $e\Phi$ to be responsible for the electric dipole interaction.

Since $\vec{A}$ contains both a $\Phi$ and a non-$\Phi$ term, each $H^{\mathrm{int}}$ is the sum
of two terms; in addition, there are four different possibilities for each
term of equation~(\ref{eq:d4-1}).  The largest term, which includes an $E1$ interaction at
both $N\gamma$ vertices, is the one for which both matrix elements contain
$\Phi$.  It is calculated in detail in Appendix F, and certain aspects of it will also
be discussed in this section.  All terms in which exactly one of the matrix elements contain $\Phi$ are calculated
and tabulated in Appendix G.  These include all magnetic and spin-induced interactions occurring
at exactly one vertex.
Most of the terms that do not contain $\Phi$ are very small; only the results for the largest of these are listed  
at the end of  Appendix G.  

As stated above, the largest second-order terms are the ones in which 
$-\int \vec{J}(\vec{\dum})\cdot \vec{\nabla}\Phi(\vec{\dum}) d^3\dum$
is substituted for all four $H^{\mathrm{int}}$'s in equation~(\ref{eq:d4-1}).
We want to avoid writing down an explicit expression for $\vec{J}$,  
since this can only be done approximately.   
Instead, we integrate by parts and use current conservation
\begin{eqnarray}
\vec{\nabla}\cdot\vec{J}(\vec{\dum}) & = & -\frac{i}{\hbar}\left[ H,\rho(\vec{\dum}) \right], \\
\rho(\vec{\dum}) & = & \sum_j e_j \delta(\vec{\dum} - {\vec{x}}_j).
\end{eqnarray} 
The integral over $\dum$ can then be performed inside the matrix element 
to give
\begin{eqnarray}
H^{\mathrm{int}} & = & -\int \vec{J}(\vec{\dum})\cdot \vec{\nabla}\Phi(\vec{\dum}) d^3\dum \nonumber\\
& = & -i [H, (e/\hbar) \Phi(\vec{x}_p) ],
\end{eqnarray}
where $H$ is now the full Hamiltonian.  If we examine only the piece of this commutator containing the 
proton kinetic energy, we see that
\begin{equation}
H^{\mathrm{int},p} =
 -i \left[\frac{p_p^2}{2m_p},  (e/\hbar) \Phi(\vec{x}_p) \right] =
-\frac{e}{m_p} \vec{p}_p \cdot \vec{A}(\vec{x}_p),
\end{equation}
which follows from the canonical commutator $[\vec{p}, F(\vec{x})] = \frac{\hbar}{i} \vec{\nabla}  F(\vec{x})$
and the definition of $\Phi$.  This term is identical to one of
the terms of the Hamiltonian in equation~(\ref{eq:d2-3}).  Therefore, the $\Phi$ commutators contain 
the operators responsible for electric dipole (E1) transitions.

We switch to the center-of-mass variables $\vec{p} \equiv (\vec{p}_p - \vec{p}_n)/2$ and
$\vec{P} \equiv \vec{p}_p +\vec{p}_n$ when we evaluate the commutator with
the full Hamiltonian.  Defining the ``internal'' Hamiltonian $H^{np} \equiv \frac{p^2}{2m_p} + V,$ 
we get for the $\Phi_i$ term 
\begin{equation}
H^{\mathrm{int}} 
 =  -\frac{e}{m_p}\vec{P}\cdot\eps e^{i\vec{k}_i\cdot\vec{r}/2} e^{i \vec{k}_i\cdot\vec{R}}  -i 
 \left[ H^{np} , (e/\hbar)\Phi_i(\vec{r}/2) \right]  
e^{i \vec{k}_i\cdot\vec{R}}.\label{eq:d4-3}
\end{equation}

The term with the $\vec{P}$ operator will be called the ``recoil correction''; its matrix 
elements are calculated in Appendix H.  It is a small effect at the energies
of interest (exactly how small will be discussed in section 3.3). 
The effects of all other recoil operators which occur in higher-order terms will be assumed to be negligible.  

The substitution 
$\vec{x}_p = \vec{r}/2 + \vec{R}$ must  generate an exponential of the form $e^{i\vec{P}\cdot\vec{R}}$,
which will combine will other similar exponentials to create the momentum-conserving delta function.
This is the delta function which is removed in the matrix element $\matrixm{}$ that was defined
in equation~(\ref{eq:d3-6}).  The net result of ignoring the recoil operators, therefore,  is to replace
$\vec{x}_p$ by $\vec{r}/2$ and switch from $\matrixt{}$ to $\matrixm{}$.

The interesting physics is contained in the second term of equation~(\ref{eq:d4-3}).
Plugging this into equation~(\ref{eq:d4-1}), we get
\begin{eqnarray}
\matrixm{a, \mathrm{uncr}}  & = &  -\sum_{C} \left\{\frac{ \mx{d_f}{[H^{np},
\hat{\Phi}_f] }{C}
\mx{C}{[H^{np}, \hat{\Phi}_i] }{d_i} }{\hbar\winit+E_{d_i}-E_C-
P_C^2/2m_d+i\varepsilon} \right.\nonumber - \\
& &  \left.\frac{ \mx{d_f}{[H^{np}, \hat{\Phi}_f] }{C}
\mx{C}{[H^{np}, \hat{\Phi}_i] }{d_i} }{\hbar\winit+E_{d_i}-E_C- P_C^2/2m_d+i\varepsilon} \right\}.
\end{eqnarray}
The dimensionless operators  $\hat{\Phi}_i \equiv (e/\hbar)\Phi_i(\vec{r}/2)$
and $\hat{\Phi}_f \equiv (e/\hbar)\Phi_f(\vec{r}/2)$  have been introduced, and the
switch from $\matrixt{}$ from $\matrixm{}$ has already been made. 
We have also labeled this term with the superscript ``$a$''.

Since $\ket{d_{i,f}}$ and $\ket{C}$ are eigenstates of $H^{np}$,
the commutators can be expanded and some cancellations with the
denominator can be made.  This generates four terms (after adding the crossed term of equation~\ref{eq:d4-1}:
\begin{equation}
\matrixm{a}  =  \matrixm{a1} + \matrixm{a2} + \matrixm{a3} + \matrixm{a4},
\end{equation}
\noindent where
\begin{eqnarray}
\matrixm{a1} & = & \left[ \frac{(\hbar\winit)^2}{2m_d} - \hbar\winit \right]^2 \sum_C
\frac{ \mx{d_f}{\hat{\Phi}_f}{C}
\mx{C}{ \hat{\Phi}_i}{d_i} }{\hbar\winit- \frac{(\hbar\winit)^2}{2m_d}+E_{d_i}-E_C+i\varepsilon}
, \label{eq:d4-4}  \\
\matrixm{a2} & = & \left[ \hbar\wfin + \frac{(\hbar\wfin)^2}{2m_d} \right]^2  \sum_C \label{eq:d4-5}
\frac{ \mx{d_f}{\hat{\Phi}_i}{C} \mx{C}{ \hat{\Phi}_f}{d_i}}{-\hbar\wfin- \frac{(\hbar\wfin)^2}{2m_d}+
E_{d_i}-E_C+i\varepsilon},  \\
\matrixm{a3} & = & \left[\hbar\wfin + \frac{(\hbar\wfin)^2}{2m_d} - 
\hbar\winit +  \frac{(\hbar\winit)^2}{2m_d} \right]
\mx{d_f}{\hat{\Phi}_f\hat{\Phi}_i}{d_i} \label{eq:d4-6},  \\
\matrixm{a4} & = & \frac{1}{2} \mx{d_f}{\left[ \ [H^{np},\hat{\Phi}_i],  \hat{\Phi}_f \right] +
\left[\ [ H^{np},\hat{\Phi}_f  ] ,  \hat{\Phi}_i \right] }{d_i} \label{eq:d4-7}.
\end{eqnarray}

The first two terms correspond to the Feynman diagram Figure~\ref{fig:d1-1}(b) and (c), and includes the case where both vertices
are E1 interactions.  The third term is a small correction, calculated in Appendix F.  
The final term is responsible for compliance to the low-energy theorems  which result from
demanding gauge invariance.  This is discussed at length in Section~\ref{sec:let}.

The calculation of these terms, while more complex than the one-body terms, is straightforward except
for the additional complications of the intermediate states and energy denominators.  
We use Green's functions to handle this problem.  Defining $E_0 \equiv \hbar\winit- 
\frac{(\hbar\winit)^2}{2m_d}+E_{d_i}$, the $\matrixm{a1}$ term can be rewritten as 
(neglecting overall constants)
\begin{equation}
\matrixm{a1} = \sum_C  \mx{d_f}{\hat{\Phi}_f \frac{1}{E_0-H^{np}_{L_C}+i\varepsilon} }{C}
\mx{C}{ \hat{\Phi}_i}{d_i},
\end{equation} 
since we know that $E_C$ is an eigenvalue of the $H^{np}_{L_C}$ operator.  
The dependence of the Hamiltonian on the orbital angular momentum is now explicitly
shown in order to remove the temptation to collapse the complete set of $C$ states.
However, if we split these states into angular and radial parts, the radial part
can be removed, leaving a sum over the angular quantum numbers $(L_C S_C J_C M_C)$, which
we denote collectively by $\hat{C}$.  Therefore, we split all parts of the transition
matrix into radial and angular parts.   
The initial and final deuteron states can be separated 
according to the form of the wavefunction, equation~(\ref{eq:d3-7}).  
Looking at the algebraic representation of the $\Phi$ operator in the appendix, equations~(\ref{eq:mainterm10})
and (\ref{eq:mainterm11}),
we see that it also contains distinct angular and radial pieces, which we call 
$O$ and $J$, respectively (suppressing the sums and related indices here for simplicity).
Lastly, we insert two complete set of radial
states to be able to write a meaningful expression:
\begin{eqnarray}
\matrixm{a1} & = & \sum_{\hat{C}} \sum_{l=0,2} \sum_{l'=0,2} \mx{l'11M_f}{O_f}{\hat{C}} \mx{\hat{C}}{O_i}{l11M_i} \times\nonumber\\
& & \int_0^{\infty} \int_0^{\infty}  r r' 
u_{l'}(r') J_f(r') \green J_i(r) u_l(r) dr dr'. \label{eq:d4-8}
\end{eqnarray} 
The Green's function is defined by
\begin{equation}
\green \equiv \mx{r'}{\frac{1}{E_0-H^{np}_{\hat{C}}+i\varepsilon} }{r}.
\end{equation}
By applying the operator $E_0-H^{np}_{\hat{C}}$ to both sides of this equation, the Green's
function can also be seen to satisfy the following differential equation:
\begin{equation}
\left[ E_0 + \frac{\hbar^2}{m_p}\frac{1}{ r} \frac{d^2}{dr^2} r - \frac{\hbar^2L_C(L_C+1)}{m_p r^2} -
V_{\hat{C}}(r) \right] \green =
\frac{\delta(r'-r)}{r^2}. 
\end{equation}
For $V_{\hat{C}}(r) $, we use the Reid93 $np$ potential as published in \cite{sto94}.

All of the second-order terms can be reduced to the form of equation~(\ref{eq:d4-8}), so
we need to be able to evaluate this double integral.   Computing this integral directly
would be a formidable task.  Therefore, we evaluate it in two steps.  First,
we compute the function $\chi_f^{l'\hat{C}}(r)$ defined  by
\begin{equation}
\chi_f^{l'\hat{C}}(r) \equiv \int r' dr' u_{l'} (r') J_f(r') \green, \label{eq:d4-9}
\end{equation}
and then evaluate the single integral $I_{fi}^{ll'\hat{C}}$:
\begin{equation}
I_{fi}^{ll'\hat{C}} \equiv \int r dr u_l(r) J_i(r) \chi_f^{l'\hat{C}}(r).
\end{equation}

The more difficult step, of course, is the determination of $\chi_f^{l'\hat{C}}(r).$ If we apply the operator
$E_0 - H^{np}_{\hat{C}}$ to both sides of equation~(\ref{eq:d4-9}), we get
\begin{equation}
\left[ \frac{d^2}{dr^2} + \frac{m_p}{\hbar^2}\left[ E_0 -
V_{\hat{C}}(r) \right]- \frac{L_C(L_C+1)}{r^2}\right] r \chi(r) =  \frac{m_p}{\hbar^2}\frac{u_{l'} (r) J_f(r)}{r} \label{eq:d4-10}.
\end{equation}

This is an ordinary, inhomogeneous, second-order differential equation.  We use the Numerov method \cite{al70}
to obtain solutions for both the homogeneous and inhomogeneous solutions.  However, the correct linear
combination of these solutions still needs to be determined.  The boundary condition is that $\chi(r)$
must be an outgoing spherical wave at large distances.  Assuming that $J(r) \rightarrow 0$
as $r$ goes to infinity,  a condition always satisfied here, the boundary condition can be written as 
\begin{equation}
\lim_{r \rightarrow\infty} \chi(r) = h^{(1)}_{L_C}(Qr),
\end{equation}
where $h^{(1)}_{L_C}(Qr)$ is the spherical Hankel function defined by
\begin{equation}
h^{(1)}_{L_C}(Qr) = j_{L_C}(Qr) + in_{L_C}(Qr),
\end{equation}
and where  
\begin{equation}
Q^2 = \frac{m_p E_0}{\hbar^2}.
\end{equation}
$Q$ can be either real or imaginary.
Therefore, to solve our problem, we must find a constant $\lambda$ such that
\begin{equation}
\lim_{r \rightarrow\infty}  \{ \chi^{\mathrm{inhom}}(r) + \lambda   \chi^{\mathrm{hom}}(r) \}=
 h^{(1)}_{L_C}(Qr).
\end{equation}

Since the asymptotic solution of the differential equation~(\ref{eq:d4-10}) must be a linear combination
of  $h^{(1)}_{L_C}(Qr)$ and $j_{L_C}(Qr)$, we can write the solutions for all $r$ as
\begin{equation}
\chi(r) = C_{L_C}(r) [ j_{L_C}(Qr) + t_{L_C}(r) 
h^{(1)}_{L_C}(Qr) ], \label{eq:d4-11}
\end{equation}
where the functions $C_{L_C}(r)$ and $t_{L_C}(r)$ approach the constants $C_{L_C}$ and
$t_{L_C}$ as $r$ approaches infinity.   The choice of $\lambda$
which satisfies the boundary condition is then
\begin{equation}
\lambda = -\frac{  C_{L_C}^{\mathrm{inhom}} }{ C_{L_C}^{\mathrm{hom}} }.
\end{equation}

Our task is thus reduced to determining these coefficients.  This must be done in the region
where $C_{L_C}(r)$ and $t_{L_C}(r)$ are both constant, so that their derivatives vanish. We numerically evaluate $\chi$ and
its derivative, and then use the logarithmic derivative $D \equiv \frac{d}{dr} \ln \chi(r)$ to 
solve for $t_{L_C}$, since
\begin{equation}
t_{L_C} = \frac{ D  j_{L_C}(Qr) - \frac{d}{dr} j_{L_C}(Qr) }{  D h^{(1)}_{L_C}(Qr) -
\frac{d}{dr} h^{(1)}_{L_C}(Qr) }.
\end{equation}
After calculating $t_{L_C}^{\mathrm{inhom}}$ and $t_{L_C}^{\mathrm{hom}}$, equation~(\ref{eq:d4-11})
can be used to determine $C_{L_C}^{\mathrm{inhom}}$ and $C_{L_C}^{\mathrm{hom}}$.  The value of $\lambda$ 
which satisfies the boundary condition can then be found. 

We have sketched the method used to calculate the integral in equation~(\ref{eq:d4-9}).
In terms of numerical computation, this integral is the most demanding part of the calculation,
so we would like to make a quick check on this algorithm.  Since the Green's function is symmetric in
$r$ and $r'$, the entire integral must be invariant under this interchange.  Therefore, if $l=l'$, making
the switch $J_i \leftrightarrow J_f$ cannot change the integral, even if $J_i \ne J_f$.  However, if it is
true that $J_i \ne J_f$, the intermediate functions $\chi(r)$ created by the differential equation
routine will \emph{not} be the same.  There are several pairs of interference terms which have $J_i \leftrightarrow J_f$
and $J_i \ne J_f$, so that comparing the values of the integrals for these terms would be a good check
of the algorithm described above.  This check has been performed on several pairs of terms, and the integrals
are found to agree to 6 significant figures.


\section{Relativistic Corrections \label{sec:rel} }

We now discuss relativistic corrections to our non-relativistic formulation.
The major correction here is the spin-orbit effect. Corrections
arising from boosting the final deuteron wavefunction into  a moving frame
are assumed to be small and therefore are neglected.

The origin of the relativistic spin-orbit effect can be understood by looking at 
the magnetic term in the Hamiltonian of equation~(\ref{eq:d2-3}):
\begin{equation}
H^M = \sum_{j=n,p} \frac{e(1+\kappa_j)}{2m_j} \vec{\sigma}_j\cdot\vec{B}(\vec{x}_j). 
\end{equation}
The relativistic correction to the magnetic field for an object moving with velocity $\vec{v}$ is
\begin{equation}
\vec{B} \rightarrow \vec{B} - \vec{v}\times\vec{E},
\end{equation}
This produces a relativistic term in the Hamiltonian:
\begin{equation}
H^{RC} = -\sum_{j=n,p} \frac{e(1+2\kappa_j)}{4m_j} \vec{\sigma}_j\cdot\left(\vec{v}_j\times\vec{E}_j\right).
\end{equation}
The effects of Thomas precession have also been included.  Since $\vec{v} \approx \vec{p}/m$ and  
$\vec{E} = \vec{\nabla}V(r) \sim \vec{r}$ for a central potential, the dot product above goes as 
$\vec{\sigma}\cdot\vec{L}$, where $\vec{L}$ is the orbital angular momentum. 

If we make the gauge invariant substitution
\begin{equation}
\vec{p} \rightarrow \vec{p} - e\vec{A},
\end{equation}
and use
\begin{equation}
\vec{E} = -\frac{d\vec{A}}{dt} ,
\end{equation}
$H^{RC}$ becomes
\begin{equation}
H^{RC} =\sum_{j=n,p} \frac{e(1+2\kappa_j)}{4m_j^2} e_j \vec{\sigma}_j\cdot \left[ \vec{A}(\vec{x}_j) \times
\frac{d}{dt} \vec{A}(\vec{x}_j)  \right].
\end{equation}

The leading relativistic correction to the scattering amplitude can easily be calculated from this 
Hamiltonian since it is a contact term, like the seagull term.  Only the proton gives a correction
because of the factor $e_j$.  The result is 
\begin{equation}
\matrixm{RC} = \bra{d_f} \frac{-ie^2}{4m_p^2} \vec{\sigma}_p\cdot(\epspr\times\eps)
 (\hbar\wfin + \hbar\winit ) (2\kappa_p+1) e^{-i\vec{q}\cdot\vec{r}/2} \ket{d_i}. \label{eq:d5-1}
\end{equation}

Our calculation, however, does not use the Hamiltonian of equation~(\ref{eq:d2-3}) ;
instead, the substitution $H = -\int \vec{J}(\vec{\dum})\cdot\vec{A}(\vec{\dum}) d^3\dum$ was made and charge density
and current operators were defined.  To be consistent with this, we make a correction
to the charge density:
\begin{equation}
\rho = \sum_j e_j \delta(\vec{\dum} - \vec{x}_j) + \rho^{R},
\end{equation}
where
\begin{equation}
\rho^{R} = e\hbar \left[ \frac{2\kappa_p + 1}{4m_p^2} \vec{\nabla}_{\dum} \delta(\vec{\dum} - \vec{r}/2) \cdot
(\vec{\sigma}_p \times \vec{p}) - \frac{\kappa_n}{m_p^2} \vec{\nabla}_{\dum} \delta(\vec{\dum}+ \vec{r}/2) \cdot
(\vec{\sigma}_n \times \vec{p}) \right].
\end{equation}
This is derived from the non-relativistic reduction of the Dirac operators in the relativistic formulation. \cite{b:ar91}
The corrections to the scattering amplitude from this new operator are calculated in 
Appendix I.   The leading-order term is identical to equation~(\ref{eq:d5-1});
higher-order terms, including one proportional to $\kappa_n$, are  neglected.


\section{Checking the Calculation \label{sec:let} }

There are checks which can be performed to help ensure that certain
parts of the calculation are correct.  One of these involves the use of
the optical theorem to calculate the total cross-section for deuteron 
photodisintegration, the process in which a photon and a deuteron interact
to form an unbound $np$ state.    These cross-sections have been more extensively studied, both
theoretically and experimentally, so that the comparison of the value of the photodisintegration
cross-section extracted from this larger calculation with the known values would be a good check.

The differential cross-section for deuteron Compton scattering is given by (see equation \ref{eq:phase4}):
\begin{eqnarray}
\left(\frac{d\sigma}{d\Omega}\right)_{fi} & = &\left( \frac{\wfin}{\winit} \right)^2 
\left[ 1 - \frac{\winit - \wfin}{m_d} \right] \mid\matrixm{} \mid^2 .
\end{eqnarray} 
Combining the definitions of the
transition matrix $\matrixt{}$ (equation~\ref{eq:d2-1}), the relationship between $\matrixt{}$
and $\matrixm{}$ (equation \ref{eq:d3-6}), and the mathematical identity
\begin{equation}
\frac{1}{x + i\varepsilon} = \frac{1}{x} + i \pi \delta(x),
\end{equation}
we get a relation for the imaginary part of the forward scattering amplitude:
\begin{equation}
\imag \matrixm{}(\theta = 0) =  \sum_C \frac{V\omega}{2\hbar} \delta(\Delta E) \mid 
\mx{C}{H^{\mathrm{int} }}{d} \mid^2, \label{eq:d6-1}
\end{equation}
where $\omega$ is now the initial or final photon energy. 
To find the total photodisintegration cross-section $\sigma_{\mathrm{tot}}$,
we can use Fermi's golden rule.  The result is
\begin{equation}
\sigma_{\mathrm{tot}}  =  \frac{2\pi}{\hbar} \frac{1}{1/V}  \sum_C \delta(\Delta E) \mid 
\mx{C}{H^{\mathrm{int}} }{d} \mid^2 , \label{eq:d6-2}
\end{equation}
where $C$ (as above) is an unbound $np$ state.  This can be combined with equation~(\ref{eq:d6-1}) to yield
\begin{equation}
\sigma_{\mathrm{tot}}  =  \frac{4\pi}{\omega} \imag \matrixm{}(\theta = 0).
\end{equation}
This is the optical theorem.  Since all of the $\matrixm{}$s have already been calculated,
determining the photodisintegration cross-sections is basically a matter of summing the imaginary parts
of the forward scattering amplitudes of all relevant terms, and over all possible polarizations (of which 
there are only six, since the initial and final states are identical here).  

The only diagram that needs to be considered here is the uncrossed second-order term of Figure~\ref{fig:d1-1}.  
However, all combinations of interactions at the two vertices must be included.  In a 1964 paper, 
Partovi \cite{pa64} calculated the contributions from each of these interactions at 3 different energies.  
The dominant term is the one in which
there is an E1 interaction at each vertex.  He called this cross-section ``approximation A'', and added the 
other terms one by one in successive approximations, until every term was included (``approximation I'').
These results are reproduced in Table~\ref{t:d6-1}, along with the corresponding
cross-sections extracted from this calculation.  There is reasonable agreement at all energies;
any differences can be attributed to improvements in the wavefunctions and potentials 
(the short-range parts, in particular) since the time that Partovi's work was published. 

The meanings of the approximations in Table~\ref{t:d6-1}:

Approximation A: E1 only.  

Approximation B: A + singlet M1.

Approximation C: B + E2.

Approximation D: C + triplet M1.

Approximation E: D + spin induced triplet M1.

Approximation F: E + spin induced triplet M2.

Approximation G: F + spin induced triplet E1.

Approximation H: G + retardation corrections to E1.

Approximation I: H + all other terms.

\begin{sidewaystable}
\caption[Comparison of photodisintegration cross-sections at 20, 80, and 140 MeV calculated in several approximations]{ \label{t:d6-1} 
Comparison of photodisintegration cross-sections calculated
in several approximations.  $\sigma_P$ is the cross-section published by Partovi \cite{pa64}, while $\sigma_K$
is from this calculation.}

\begin{tabular}{ccccccccc}
	\multicolumn{1}{c}{Approximation} &
	\multicolumn{2}{c}{20 MeV} &
	\multicolumn{1}{c}{ } &
	\multicolumn{2}{c}{80 MeV} &
        \multicolumn{1}{c}{ } &
	\multicolumn{2}{c}{140 MeV} \\ 

&  $\sigma_{P}$  ($\mu b/sr$) & $\sigma_{K}$ ($\mu b/sr$) & 
&  $\sigma_{P}$  ($\mu b/sr$) & $\sigma_{K}$ ($\mu b/sr$) &
&  $\sigma_{P}$  ($\mu b/sr$) & $\sigma_{K}$ ($\mu b/sr$) \\ 
A & 579.1 & 583.3 & & 77.15 & 80.54 & & 34.04 & 34.56 \\
B & 589.2 & 593.2 & & 83.50 & 85.83 & & 39.77 & 38.59 \\ 
C & 591.5 & 595.0 & & 84.55 & 86.86 & & 40.38 & 38.99 \\  
D & 591.7 & 595.1 & & 84.85 & 87.06 & & 40.67 & 39.16 \\ 
E & 591.6 & 594.7 & & 84.73 & 86.61 & & 40.55 & 38.70 \\ 
F & 592.1 & 595.3 & & 85.43 & 87.68 & & 41.31 & 40.26 \\ 
G & 591.9 & 594.6 & & 90.45 & 90.22 & & 47.07 & 43.33 \\ 
H & 588.2 & 591.2 & & 87.52 & 86.36 & & 44.64 & 39.44 \\ 
I & 588.2 & 591.2 & & 87.40 & 86.40 & & 44.53 & 39.52 \\

\end{tabular}

\end{sidewaystable}

\begin{figure}
\centering
\epsfig{file=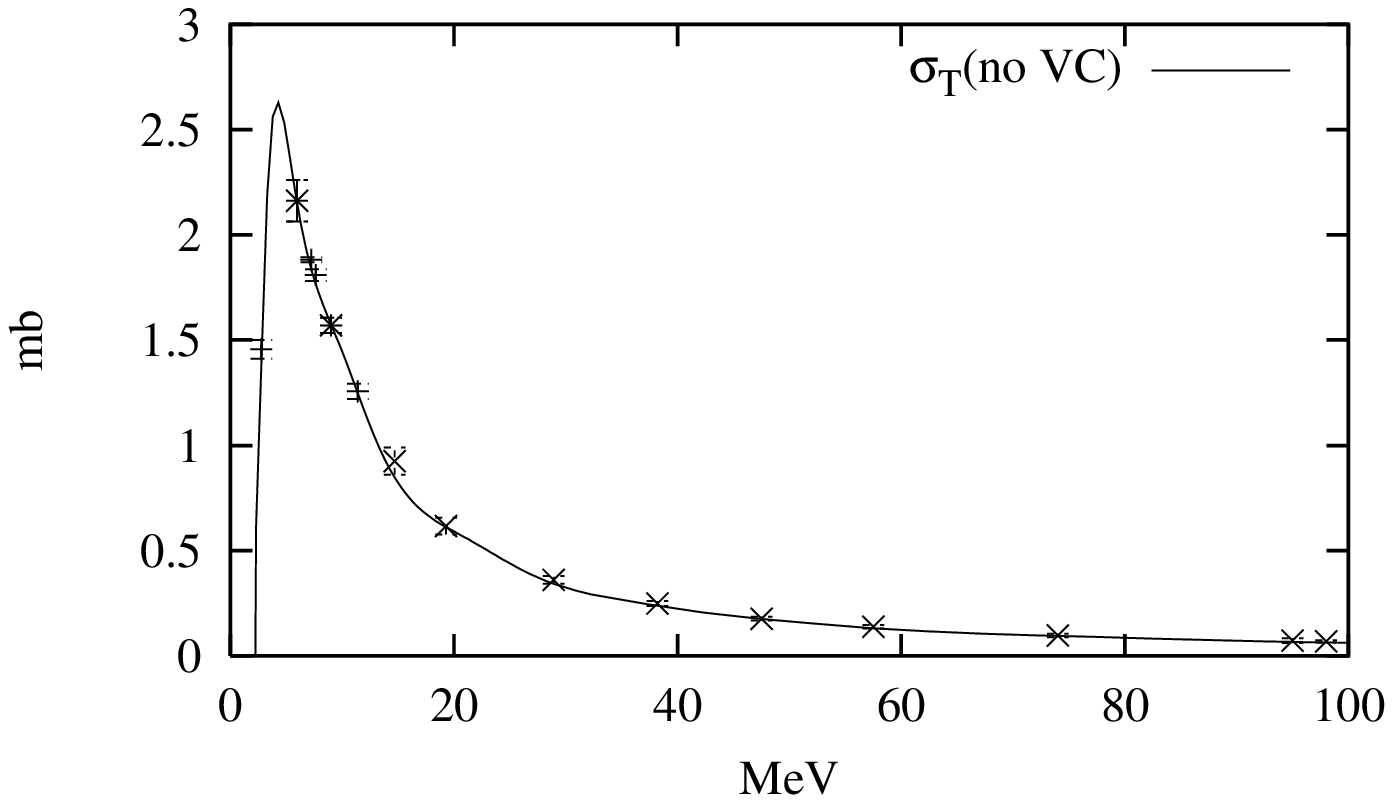}
\epsfig{file=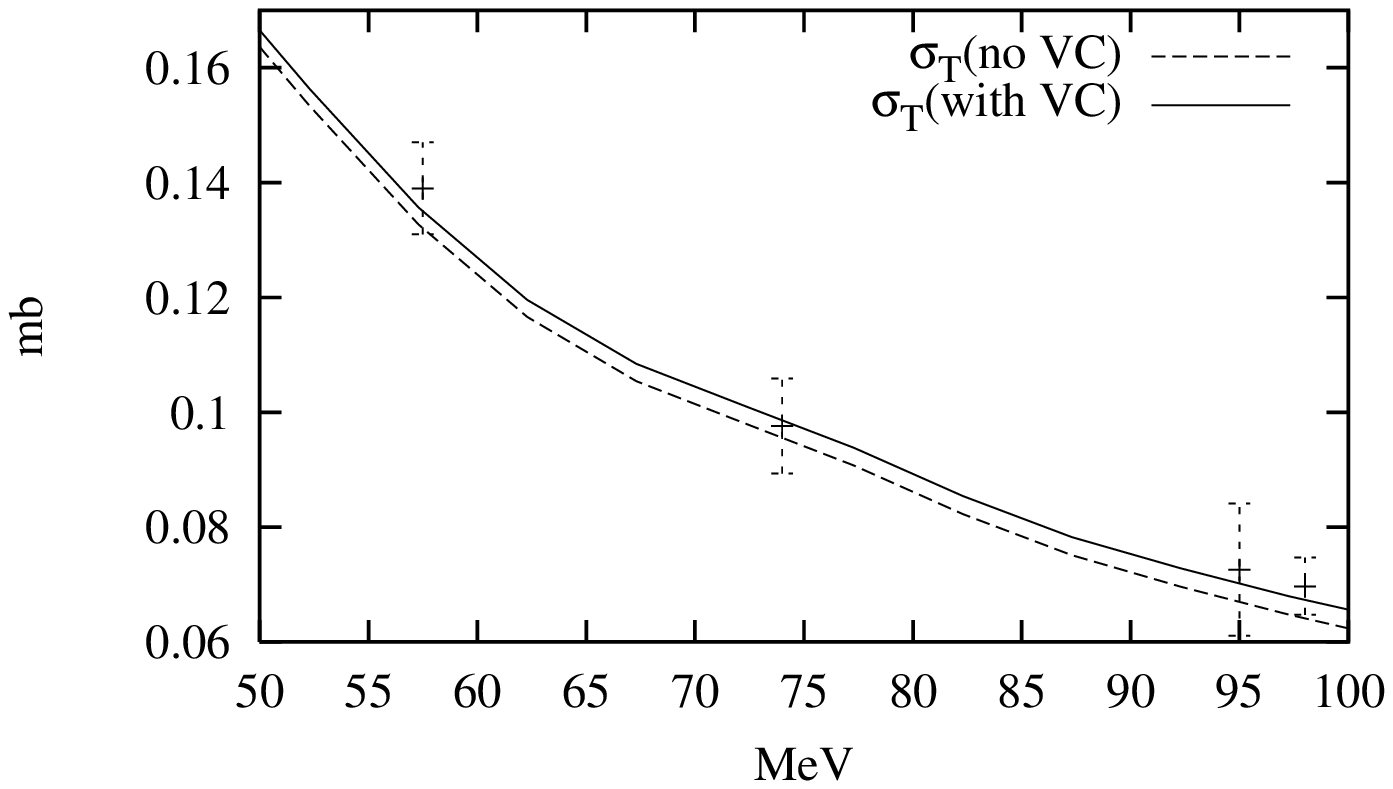}
\caption[Total photodisintegration cross-sections vs. energy compared with experimental data]{
\label{f:d6-5} Total photodisintegration cross-sections $\sigma_T$ vs. energy compared with experimental data.
The range 50-100 MeV is magnified and a vertex correction (VC) is added. The experimental data come from \protect\cite{sa89} 
and \protect\cite{bi85}.}
\end{figure}

We also compare these photodisintegration cross-sections with experimental data up to 100 MeV.
This is shown in Figure~\ref{f:d6-5}.  A vertex correction (see figure~\ref{fig:d1-1}f) that
contributes to this process has also been included in a second graph of the range 50--100 MeV.
There is good agreement with the data.  Our photodisintegration cross-section also reporoduce the
correct threshold behavior as described by Brown \cite{br}.

Another important check is ensuring  that the calculation is gauge invariant.  In
electrodynamics, gauge invariance arises because Maxwell's equations remain
unchanged under the transformation
\begin{equation}
\vec{A} \rightarrow \vec{A} + \vec{\nabla}\Lambda,
\end{equation}
where $\Lambda$ can be any scalar function.  Since the relativistic two-photon scattering amplitudes 
must have the form $\epsilon_i^{\mu}T_{\mu\nu}\epsilon_f^{\nu}$, it follows that the condition imposed 
by gauge invariance is that 
\begin{equation}
k_i^{\mu}T_{\mu\nu}\epsilon_f^{\nu} = \epsilon_i^{\mu}T_{\mu\nu} k_f^{\nu} = 0.
\end{equation}

An important consequence of this is what is usually referred to as the low-energy theorem.
Friar \cite{fr75} showed that, as the photon energy goes to zero, the full Compton amplitude must go as 
\begin{equation}
\mathcal{M} \rightarrow \frac{e^2}{m_t} (\hat{\epsilon}_f\cdot\hat{\epsilon}_i).
\end{equation}
This was derived assuming only that the calculation be gauge invariant.  It is 
non-trivial because it involves the target(deuteron) mass; at low-energy
the seagull amplitude depends on the \emph{proton} mass :
\begin{equation}
\mathcal{M}_{\mathrm{sg}} \rightarrow \frac{e^2}{m_p} (\hat{\epsilon}_f\cdot\hat{\epsilon}_i).
\end{equation}
Therefore, all other terms, including meson-exchange terms, must exactly cancel half of the seagull
amplitude at threshold.

If we ignore meson-exchange terms, this is not as difficult as it sounds.  The only term which survives in the low-energy
limit, from equation~(\ref{eq:d4-7}), is
\begin{equation}
\mathcal{M}^{a4}  =\frac{1}{2} \mx{d_f}{\left[ [ \frac{p^2}{m_p} ,\hat{\Phi}_i],  \hat{\Phi}_f \right] +
\left[\ [ \frac{p^2}{m_p},\hat{\Phi}_f  ] ,  \hat{\Phi}_i \right] }{d_i}, \label{eq:d6-3}
\end{equation}
where we have only included the kinetic energy part of $H^{np}$; the potential energy term involves
meson-exchange currents and will be discussed shortly.

This can easily be shown to satisfy Friar's low-energy theorem. Since $\vec{A} \rightarrow \hat{\epsilon}$ at threshold (the 
dipole approximation), the definition of $\hat{\Phi}_i$ (equations \ref{eq:d4-80} and \ref{eq:d4-81}) implies that 
\begin{eqnarray}
\hat{\Phi}_i & \rightarrow & \frac{e}{\hbar} \frac{\vec{r}}{2}\cdot\hat{\epsilon}_i, \\
\hat{\Phi}_f & \rightarrow & \frac{e}{\hbar} \frac{\vec{r}}{2}\cdot\hat{\epsilon}_f. 
\end{eqnarray}
Evaluating the commutators of equation~(\ref{eq:d6-3}) then yields
\begin{equation}
\mathcal{M}^{a4} \rightarrow -\frac{e^2}{m_d} (\hat{\epsilon}_f\cdot\hat{\epsilon}_i),
\end{equation}
which cancels half of the seagull term as required.  

\begin{figure}
\centering
\epsfig{file=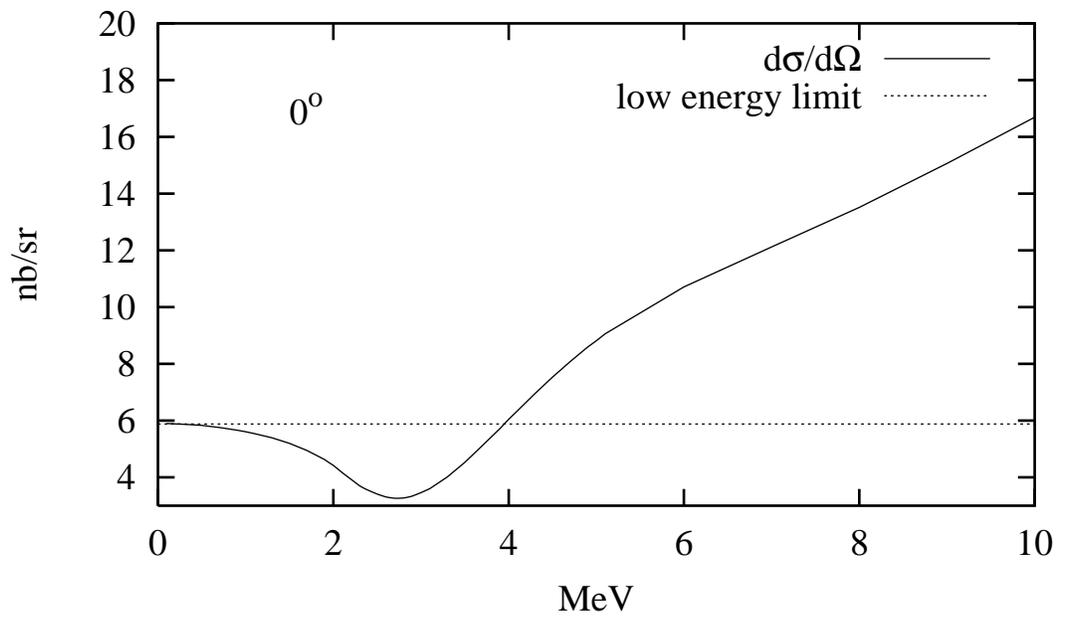}
\caption{ \label{fig:d6-10}  Forward differential cross-section vs. energy  compared with the low-energy limit. }
\end{figure}

Figure~\ref{fig:d6-10} further emphasizes this point by demonstrating that our numerical calculation
reproduces this result.  Since the cross-section goes like the square of the amplitude, we expect
the total cross-section at low energy to be one-fourth of what the cross-section would be if only the
seagull term were included.  The differential cross-section in this figure includes all terms except meson-exchange terms,
and is seen to approach the correct limiting value.  
This plot shows only the forward cross-section, but the low energy theorem is satisfied at all angles. 

\begin{figure}
\centering
\parbox{75mm}{\centering\epsfig{file=pions1} }
\parbox{30mm}{(a)}
\parbox{75mm}{\centering\epsfig{file=pions13} }
\parbox{30mm}{(b)}
\parbox{75mm}{\centering\epsfig{file=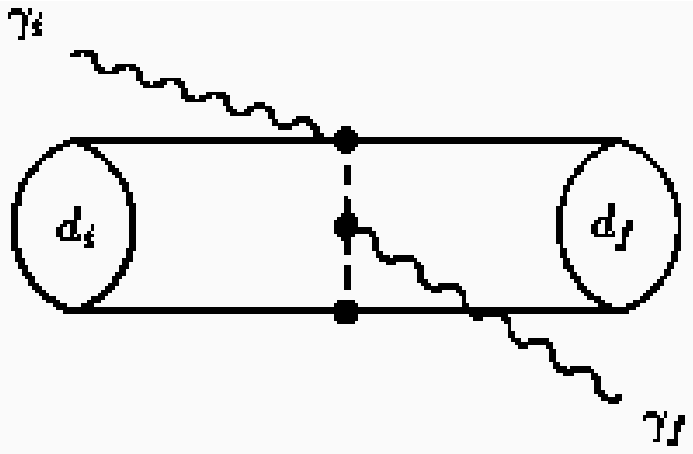} }
\parbox{30mm}{(c)}
\parbox{75mm}{\centering\epsfig{file=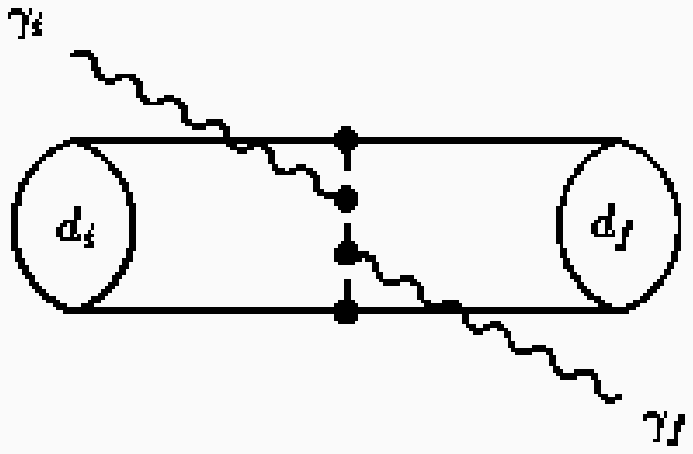} }
\parbox{30mm}{(d)}
\caption{Gauge-invariant set of pion-exchange diagrams \label{fig:d6-2}  }
\end{figure}

We now examine the effects of the meson-exchange currents that were previously neglected.
Since the potential energy portion of the double commutator term (equation~\ref{eq:d6-3})
has not yet been accounted for, we suspect that this must cancel the explicit meson-exchange
terms at low energy.  This is indeed the case.  Figure~\ref{fig:d6-2} shows the four pion-exchange diagrams 
which can contribute, according to Arenhovel \cite{ar80}.  He showed that these terms analytically
cancel the $V^{\mathrm{OPEP}}$ term at threshold. 


\begin{figure}
\centering
\epsfig{file=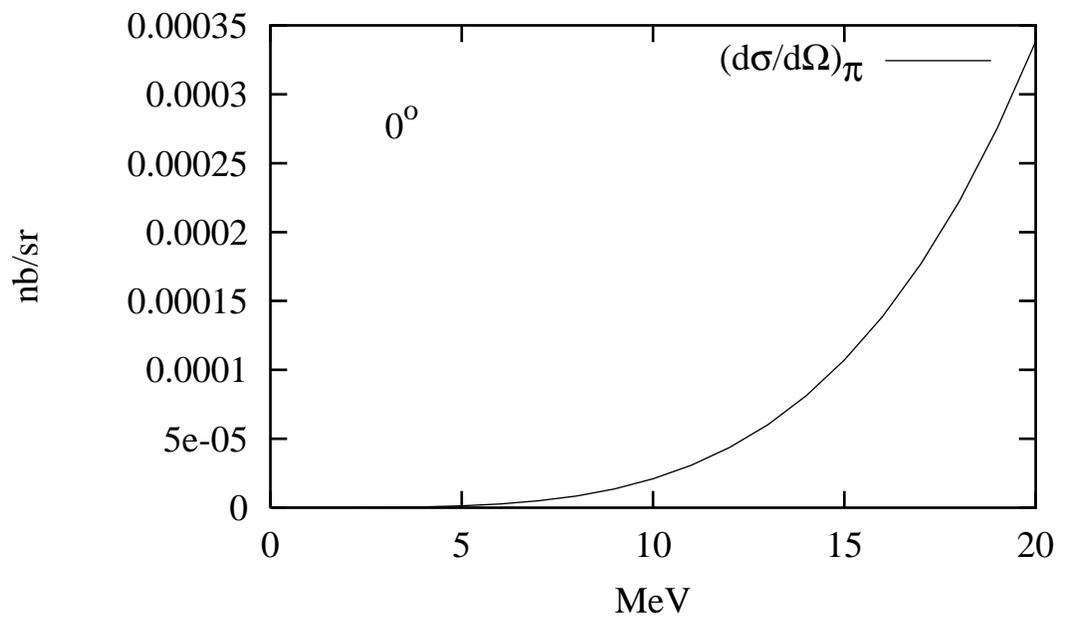}
\caption{ \label{fig:d6-15} Forward differential cross-section including only pion-exchange and $V^{\mathrm{OPEP}}$ contributions vs. energy}
\end{figure}

These four terms and the $V^{\mathrm{OPEP}}$ term in calculated in Appendix I.  
Figure~\ref{fig:d6-15} shown a graph of a differential cross-section at $0^{\circ}$ which includes only these 5 terms.
It is exactly zero at threshold.  This is true at all angles, not just in the forward direction.
The cross-section can be seen to have an $\omega^4$ dependence, which means that the amplitude goes like $\omega^2$. 

This is a very non-trivial check.  Gauge invariance has been satisfied without explicitly demanding
it in the formulation of the problem.   This check, along with the photodisintegration comparisons, inspires
confidence in the correctness of our calculation.    

\chapter{Results and Discussion \label{ch:res} }
\section{Effects of Various Terms in Cross-Section}

We are now ready to calculate the deuteron Compton scattering cross-sections.  We concentrate on three energies: 
49 MeV, 69 MeV (the energies of an Illinois experiment \cite{lu94}) and 95 MeV (the energy of 
a new Saskatoon experiment \cite{ho98}).  

First, we study the contributions to the cross-section from the major interactions.  They will be labeled as
follows:

SG: The Seagull Term (Appendix A).

EM: All interactions arising from the multipole decomposition of the vector potential as described in Appendices F and G, 
except for recoil corrections.  
These include all electric and magnetic dipoles, as well as  the double commutator term needed to satisfy the low-energy theorem.

$\pi$:  All pion interactions pictured in Fig~\ref{fig:d6-2} which are needed to satisfy the low-energy theorem.
These are calculated in Appendix J.

$\alpha + \beta$:  The polarizability terms, calculated in Appendices B and C.  Unless
otherwise specified, the values for the polarizabilities are $\alpha_p = 10.9$, $\alpha_n = 12.0$, $\beta_p = 3.3$, $\beta_n = 2.0$,
as determined in \cite{fe91,sh91}.

\begin{figure}
\centering
\epsfig{file=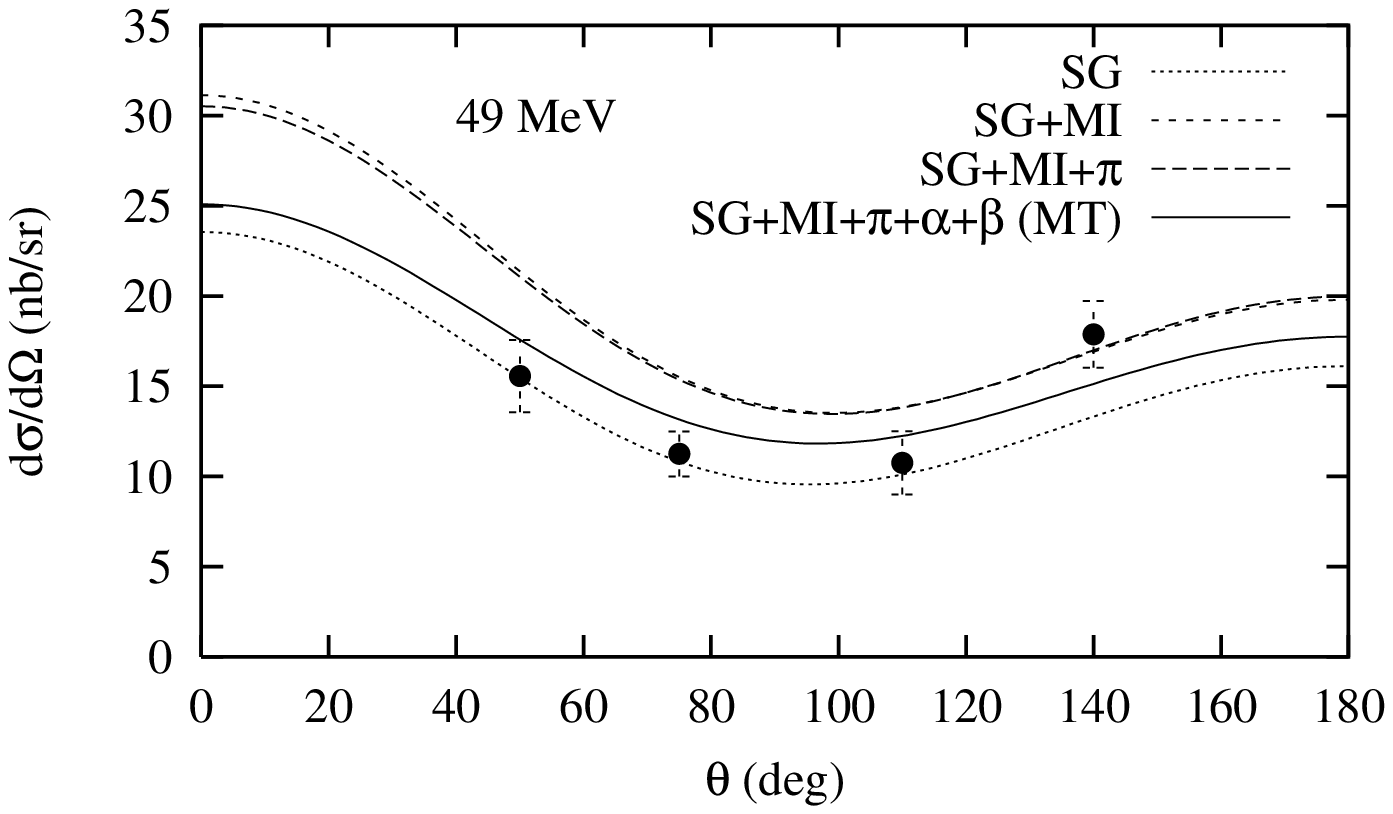}
\epsfig{file=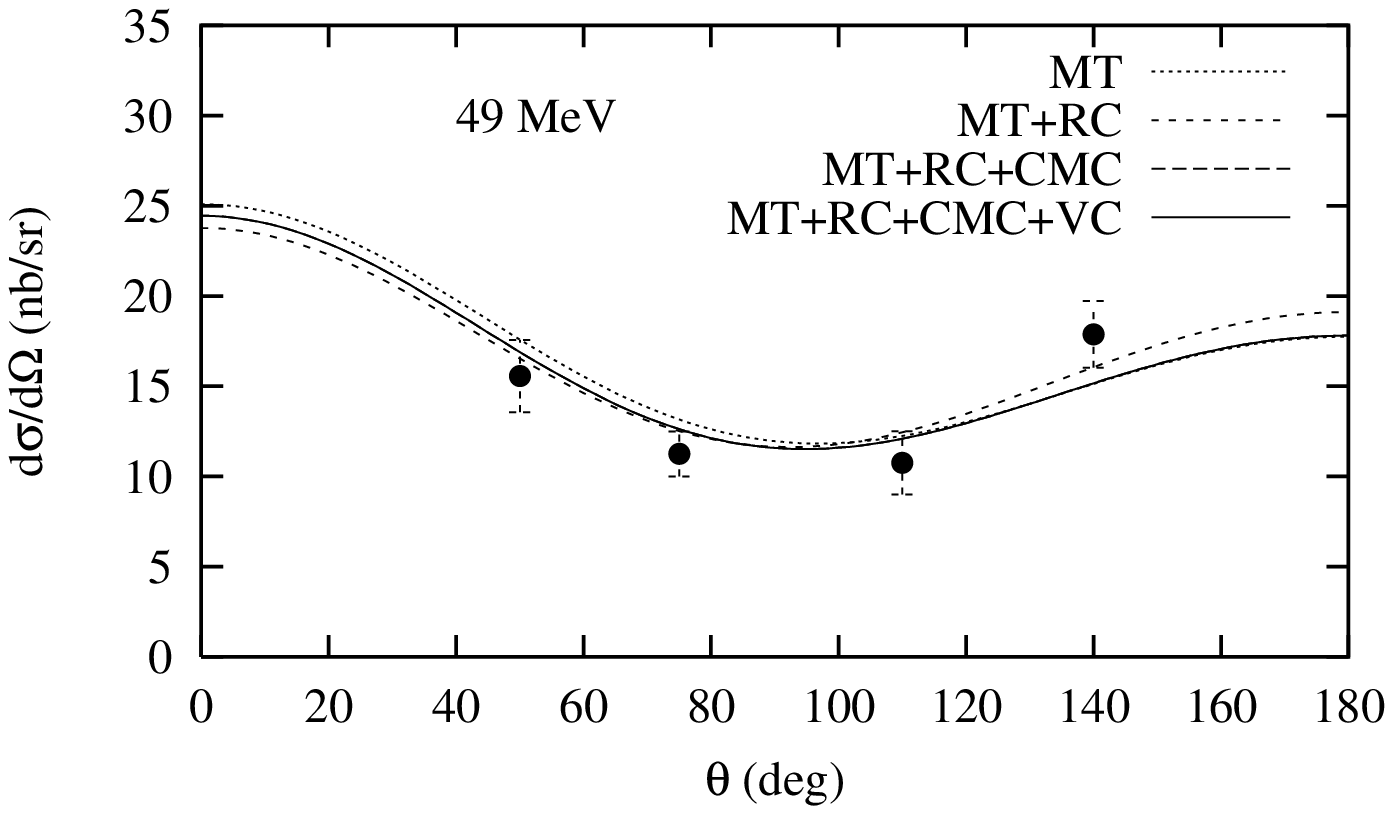}
\caption[Contributions of different terms to Compton cross-section at 3 different energies]{ 
\label{fig:e1-1} Contributions of different terms to Compton cross-section at 3 different energies. The upper graphs include the seagull (SG), 
electromagnetic multipole interactions (MI), pion ($\pi$) , and polarizability terms.  These terms are collectively called the 
``major terms'' (MT).  The second graph in each set includes relativistic (RC), recoil (CMC), and vertex (VC) corrections.}
\end{figure}

These terms are displayed on the top graph in each set of graphs of Figure~\ref{fig:e1-1}.  
The data from \cite{lu94} is displayed on all graphs at 49 and 69 MeV; no data has yet been
published for 95 MeV.

The seagull term alone provides a reasonable description of the data at 49 MeV, and does no worse than all the terms together
at 69 MeV.  At low energy, the seagull term has a $(1 +\cos^2{\theta})$ dependence.  As the energy becomes higher,
this dependence is maintained at the forward angles, while the cross-section gets increasingly smaller at the backward angles.
It is the dominant term at all energies under consideration.

The EM terms raise the seagull cross-section and become increasingly important as the photon energy rises.  
The $\pi$ terms, on the other hand, do not have a very large effect, even at 95 MeV.  These terms exactly cancel at threshold and
continue to nearly cancel at higher energies.  The amplitude of these terms has an $\omega^2$ dependence, as was mentioned in 
section 3.6.
This is borne out in the figure as the contribution at 95 MeV is approximately four times that at 49 MeV.

The polarizability amplitudes also have an $\omega^2$ dependence.  They are nearly as large as the 
EM terms at energies as low as 49 MeV, and  have a major effect on the cross-section at 95 MeV.  This 
large effect is necessary if there is to be any chance of measuring the polarizabilities at these energies.

These terms are collectively called the Major Terms (MT).  They describe the data reasonably well without introducing
the smaller corrections detailed in the second graph of each set of Figure~\ref{fig:e1-1}.   These corrections are:

RC: The Relativistic Correction, as calculated in Appendix I.  This includes only
the first-order correction.

CMC: Recoil Corrections (from the  Center-of-Mass operator), as calculated in Appendix H.  This 
also includes only the largest correction.

VC:  $\gamma N$ Vertex Corrections, such as the one pictured in Figure~\ref{fig:d1-1}(f).  These are calculated at the end of
 Appendix J.

The RC term has an $\omega^2$ dependence. The scattering amplitude itself goes like $\omega$, but does not interfere with
the seagull term because of its spin operator.  It is a small effect at 49 MeV in that in cannot be detected because of the
size of the experimental error bars.  However, at 95 MeV it is important.  This term decreases the forward cross-section and raises it
at backward angles.  Somewhere around $90-110^{\circ}$  it has no effect at all.

The CMC terms are very small at all energies here.  Their presence cannot be detected within the error bars
and do not affect the calculation.
 
The VC terms tend to counteract the RC terms.  Their effect is largest at backward angles.  However, they
do not increase as rapidly in energy and therefore become less important relative to the RC and polarizability terms.
They also cancel at about the same angle as the RC terms.

\begin{figure}
\centering
\epsfig{file=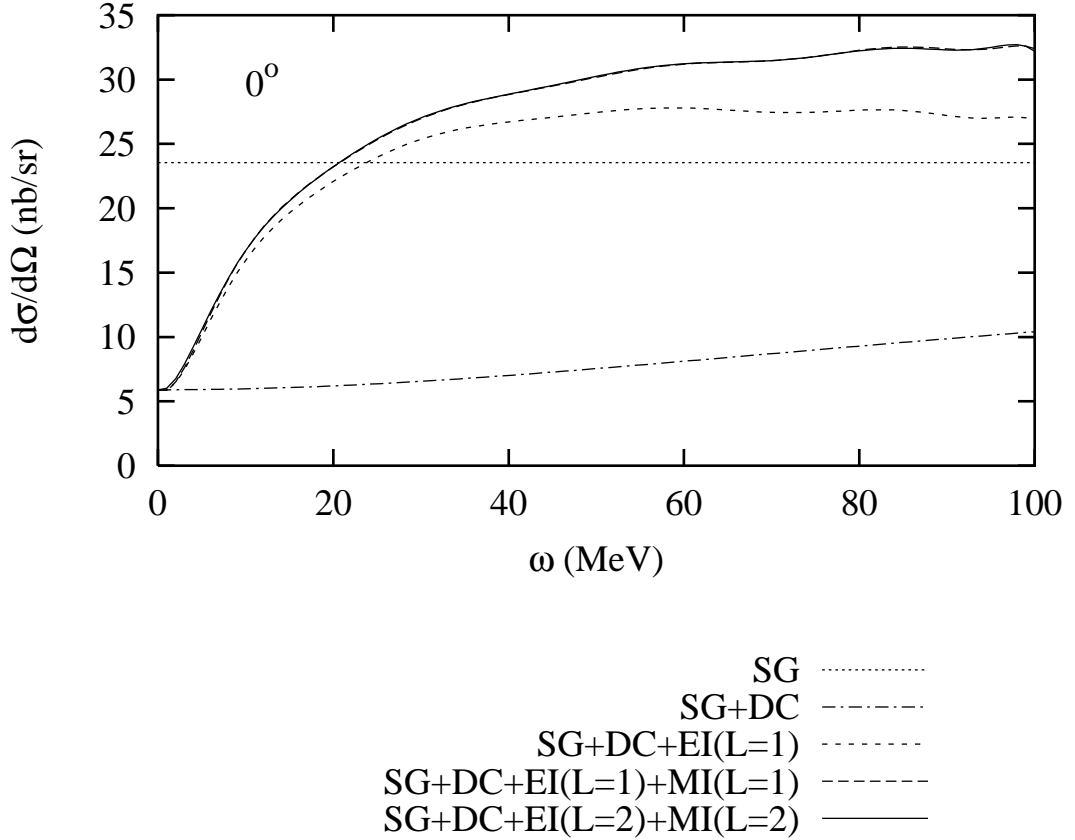}
\caption[Contributions of different multipoles to the electromagnetic terms vs. energy]{ 
\label{fig:e1-2} Contributions of different multipoles to the electromagnetic terms (MI) vs. energy.  Included are the seagull (SG),
double commutator (DC), electric multipole (EI), and magnetic multipole (MI) terms.}
\end{figure}

Next, we examine the EM terms in more detail.  A plot of the cross-section versus energy at $\theta = 0^{\circ}$ is shown
in Figure~\ref{fig:e1-2}.   The following terms are included:

SG: The Seagull Term, as above.  

DC:  The Double Commutator term of equation~(\ref{eq:d4-7}), calculated in Appendix F.

EI:  Electric Interactions arising from the $J^{(p)}$ operator in equation~(\ref{eq:nextterm5}), calculated in Appendix G.
This also includes the second-order terms of equation~(\ref{eq:d4-4}) and (\ref{eq:d4-5}). 
The $L=1$ term is the first term in the multipole expansion.  

MI: Magnetic Interactions arising from the $J^{(\sigma)}$ operator in equation~(\ref{eq:nextterm4}),also calculated in Appendix G.

The seagull term is independent of energy at the forward angle. It depends on $\vec{q}$ and not
$\wfin$ or $\winit$ individually.   The DC term is seen here to give the correct low-energy limit, and
slowly decreases with energy.  This term contains pieces of both magnetic and electric interactions, because
current conservation was used to derive it.  This is the reason we are not using the labels $E1$ and $M1$; these
terms in the usual sense cannot be extracted from the DC term.

The $L=1$ terms for the EI and MI  terms are added next.  These terms are zero at low energy.  The EI term for $L=1$ is important at low energies,
increasing rapidly until above 40 MeV, after which it remains approximately constant.    The MI term with $L=1$ becomes noticeable only around
20 MeV, and increases steadily.  Adding the $L=2$ terms produces no noticeable effect; we conclude that multipoles higher
than $L=1$ are unimportant below 100 MeV.

\clearpage

\begin{figure}
\centering
\epsfig{file=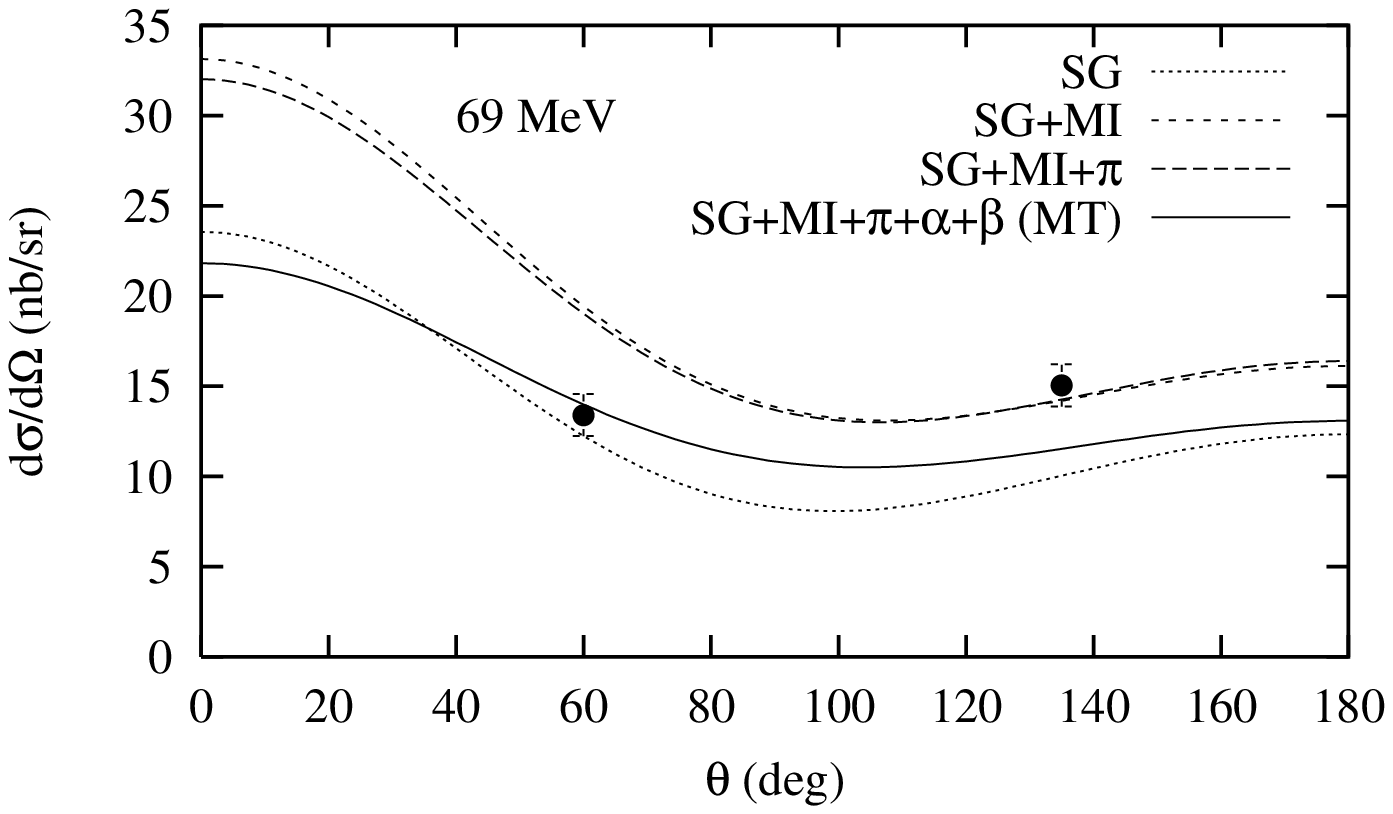}
\epsfig{file=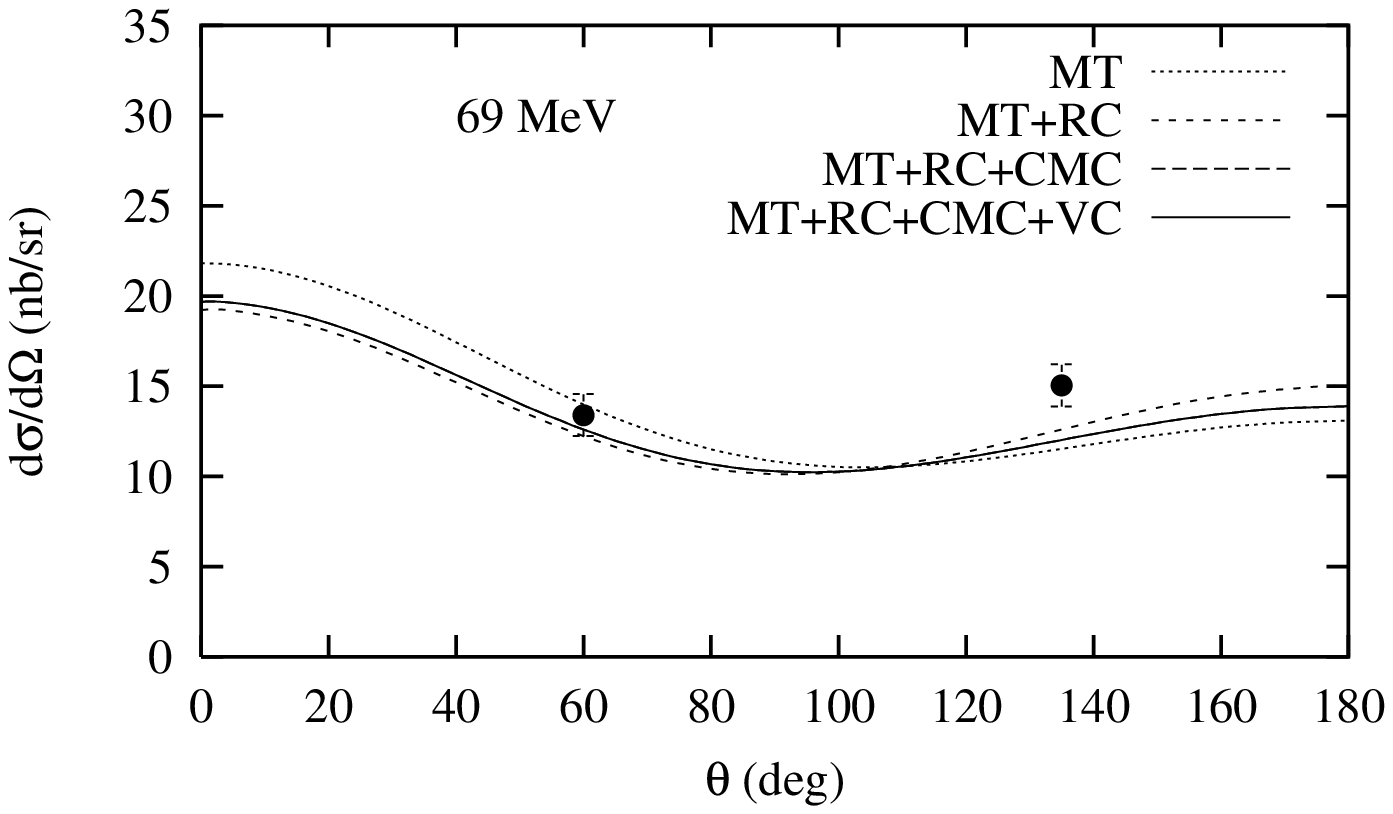}
\end{figure}
\clearpage

\begin{figure}
\centering
\epsfig{file=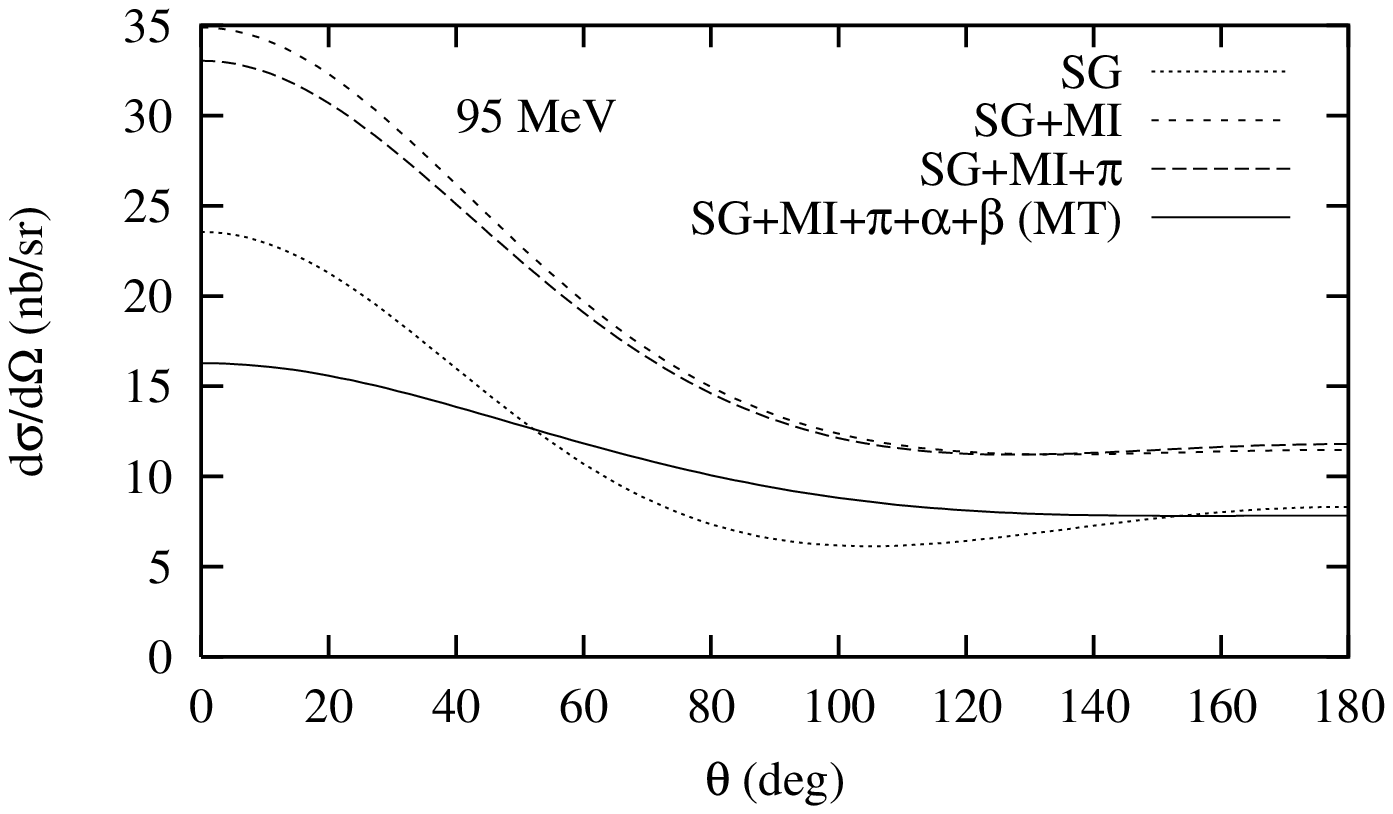}
\epsfig{file=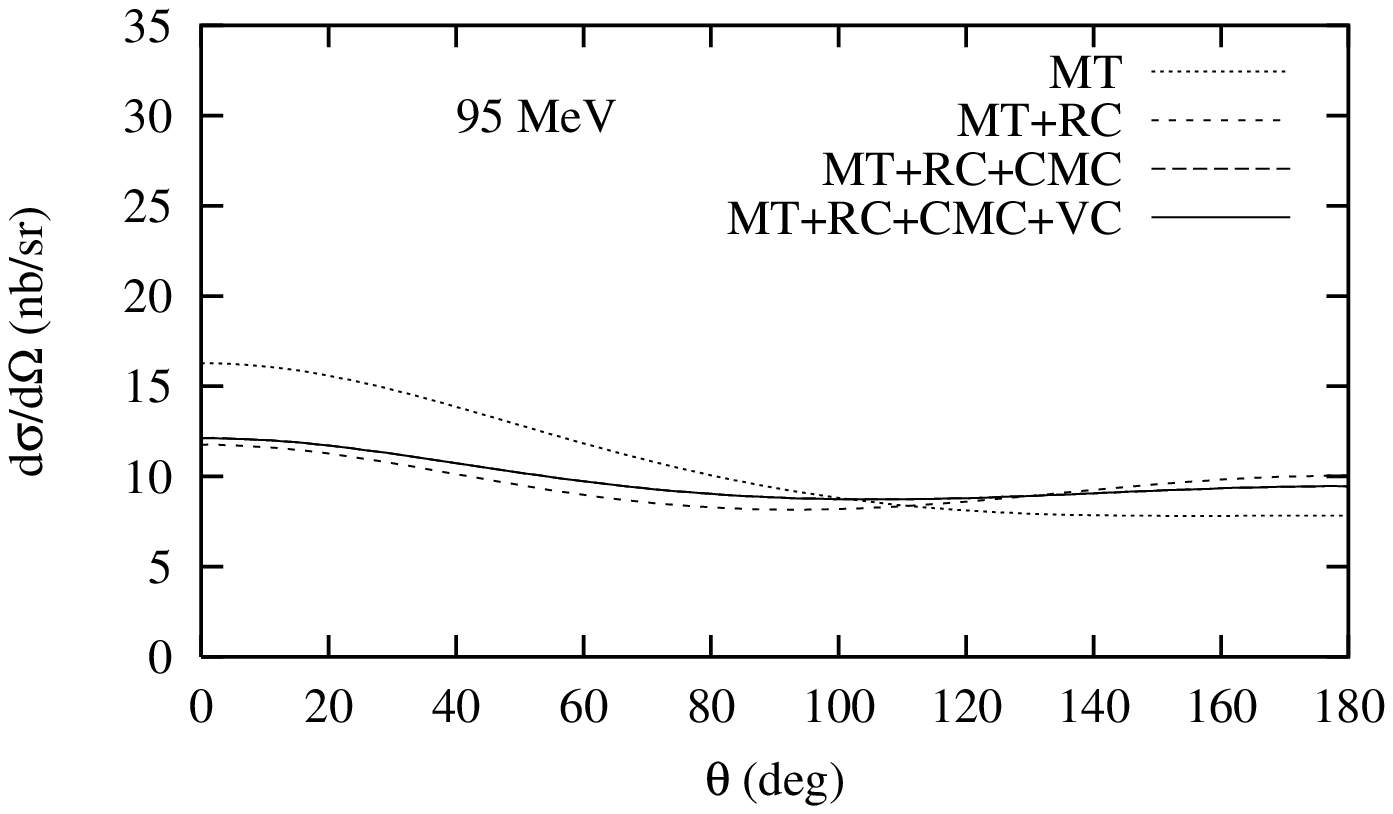}
\end{figure}
\clearpage

\section{Determining Polarizabilities from Deuteron Cross-Sections}
\begin{figure}
\centering
\epsfig{file=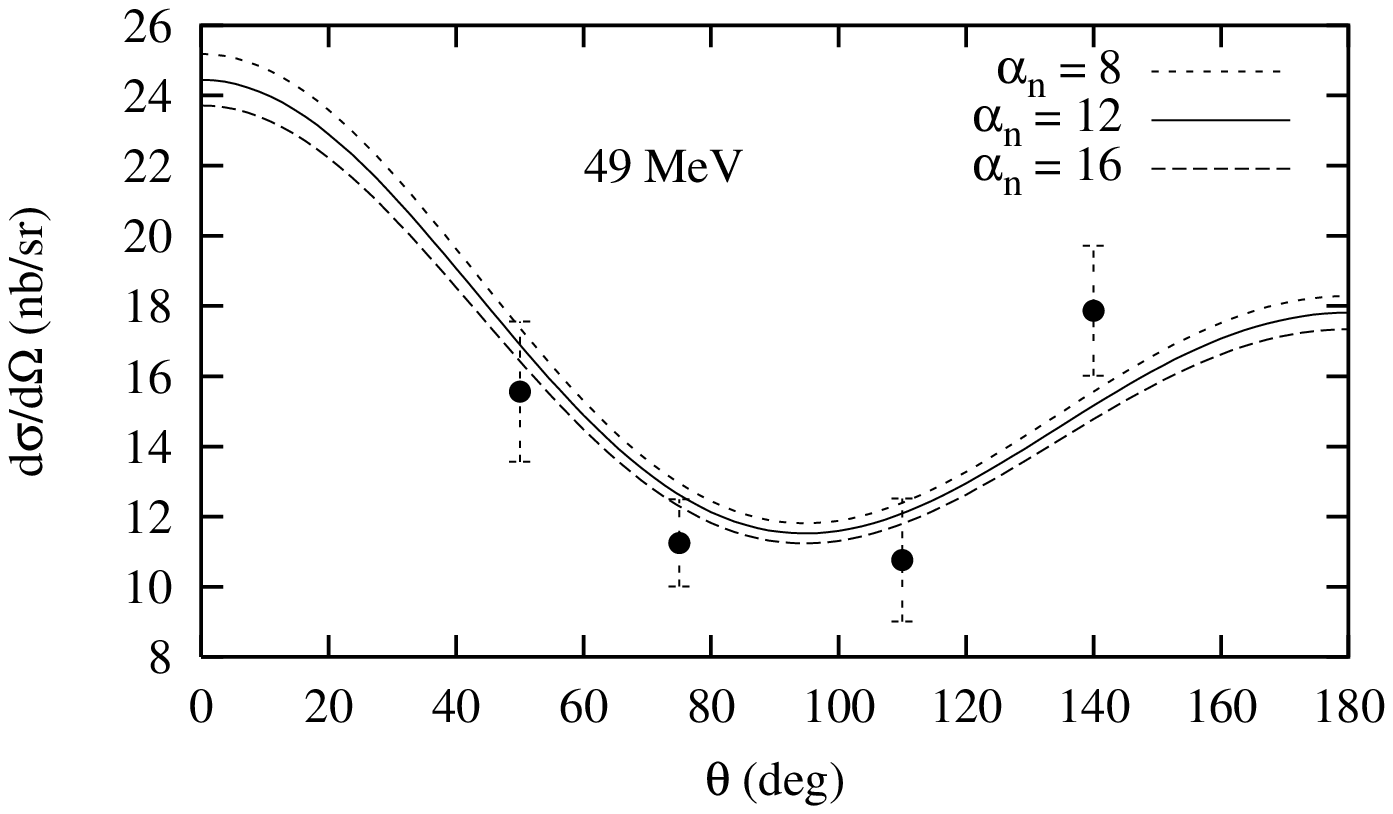}
\epsfig{file=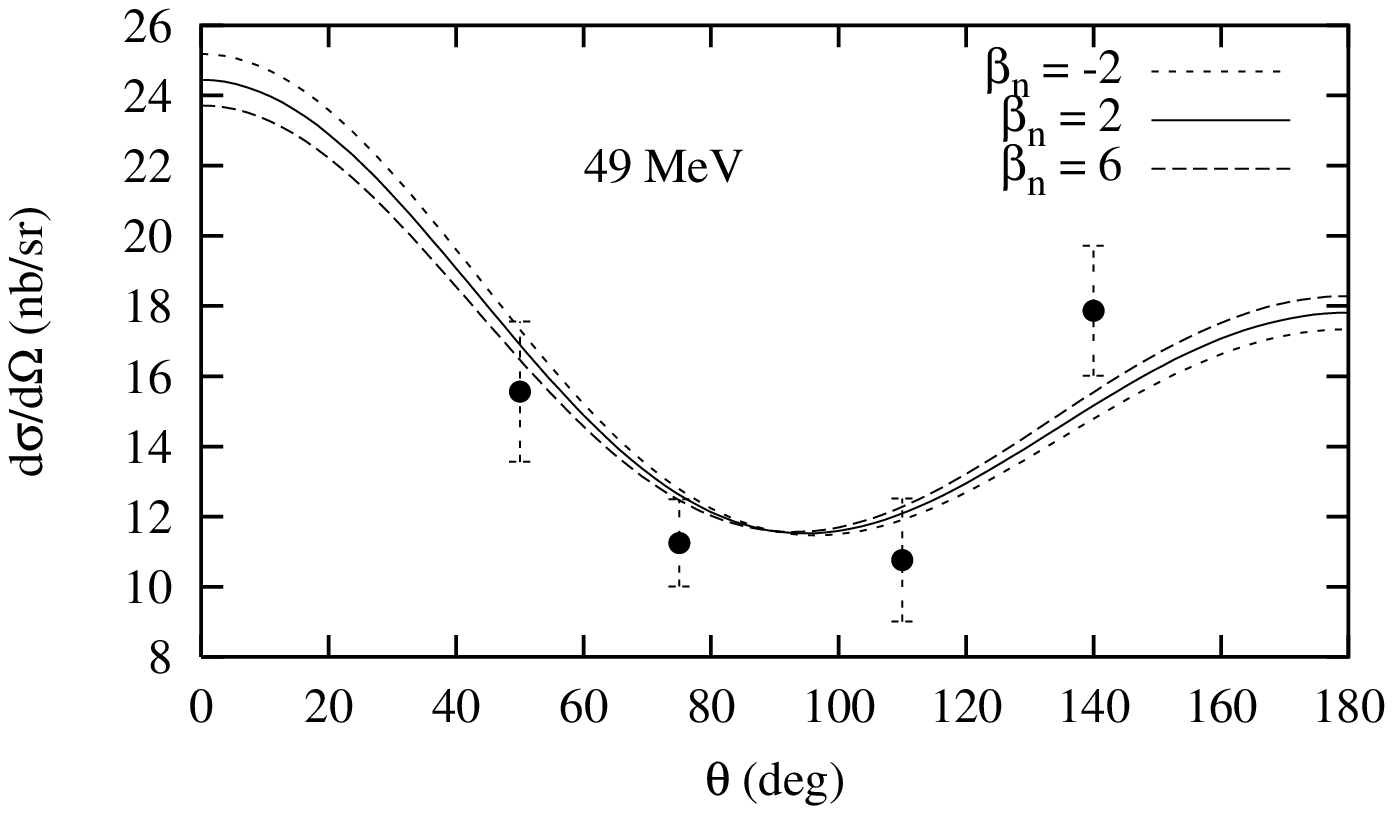}
\caption[Effect of varying $\alpha_n$ and $\beta_n$ separately in the Compton cross-sections at 49, 69, and 95 MeV]{ 
Effect of varying $\alpha_n$ and $\beta_n$ separately in the Compton cross-sections at 49, 69, and 95 MeV.  $\beta_n = 2.0$ in the top graph and
$\alpha_n = 12.0$ in the bottom.  \label{fig:e2-1} }
\end{figure}

\begin{figure}
\centering
\epsfig{file=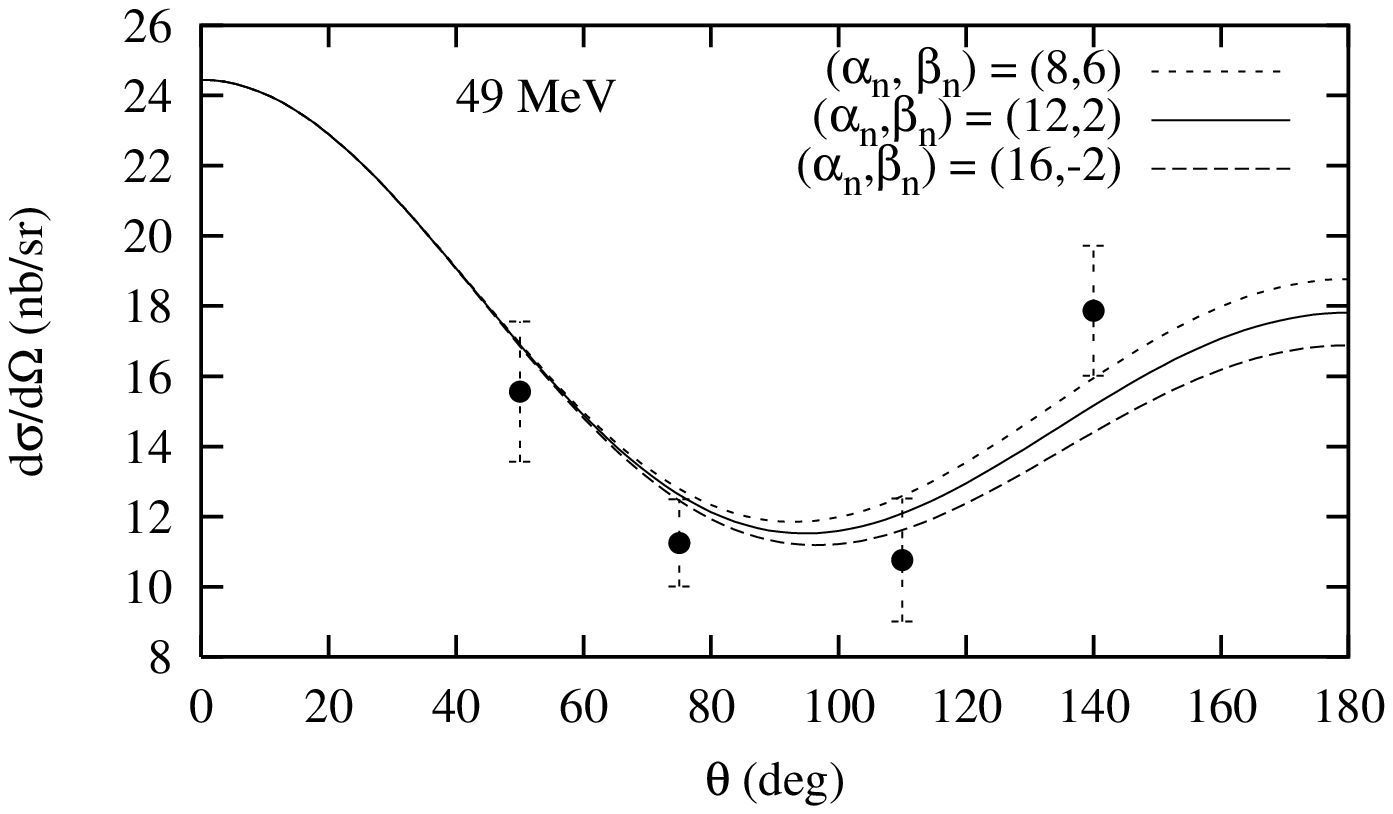}
\epsfig{file=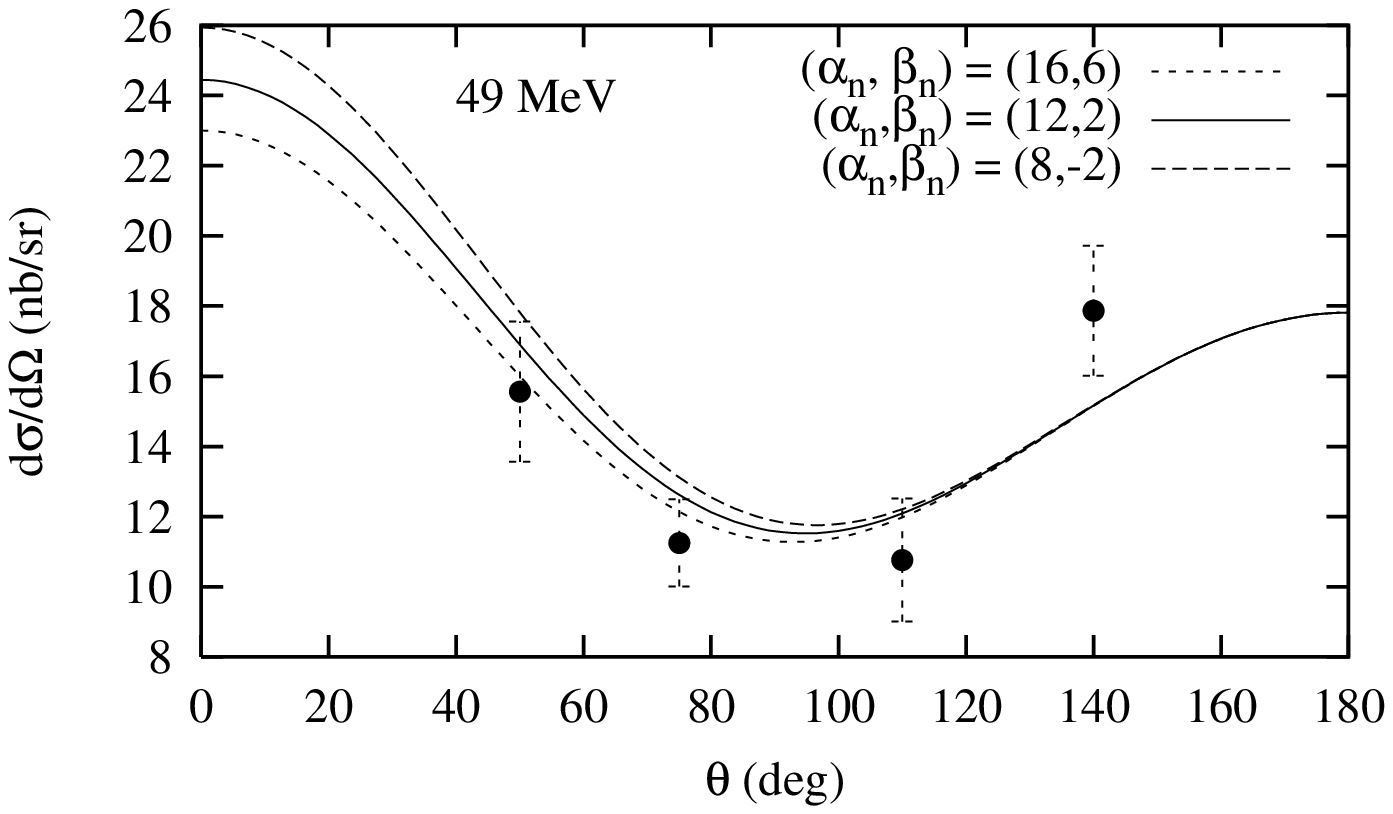}
\caption[Effect of varying both $\alpha_n$ and $\beta_n$  in the Compton cross-sections at 49, 69, and 95 MeV]{Effect 
of varying both $\alpha_n$ and $\beta_n$  in the Compton cross-sections at 49, 69, and 95 MeV.   \label{fig:e2-2} }
\end{figure}

Having seen the size of the major contributions, we now investigate
the dependence of the cross-section on the size of the neutron polarizabilities.  
Figures~\ref{fig:e2-1} and~\ref{fig:e2-2} show the effects of changing the magnitudes of $\alpha_n$ and $\beta_n$
at the three energies we have been looking at.  We first vary them one at a time,
and then together. The $\alpha_n$ term has the same $(1+ \cos^2{\theta})$ dependence
as the seagull term at low energies, while the $\beta_n$ term goes like $\cos{\theta}$.  
This means that the magnetic polarizability term gives no contribution at $90^{\circ}$.
Note that the largest effects for both terms are at extreme forward and backward angles, while
the experimental data points are in the middle.

Figure~\ref{fig:e2-1} shows the changes to the cross-section over the 
range of $\alpha_n = 12.0 \pm 4.0$ and $\beta_n = 2.0 \pm 4.0$.  These 
errors are slightly larger than the errors quoted in \cite{sh91} but are used to illustrate the sensitivity to
the size of the error bars.  The graph of 49 MeV shows that the experimental error bars are  too large to provide
significant distinction  between the values of $\alpha_n = 8.0 - 12.0$ and  $\beta_n = 2.0 - 6.0$.

There are only two data points at 69 MeV,  and one of these is not consistent with the calculated cross-section.  
The data point at the greatest angle is also difficult to fit at 49 MeV.  Other calculations show
similar inconsistencies with these two points \cite{ar95,ch98,lv98}.
Looking at the graphs where the polarizabilities both vary (Figure~\ref{fig:e2-2}), 
we see that having a lower $\alpha_n$ and a higher $\beta_n$ than the accepted values provide the best fit
to these points.  At 69 MeV there is not much else to say as the other point provides little information
except that the calculation fits the data using an acceptable range for the values of the polarizabilities.

The situation looks more promising at 95 MeV.  The effects of the polarizabilities 
are larger, and the range of $\pm 4.0$ probably will extend beyond the experimental error bars
at angles such as $50^{\circ}$ and $140^{\circ}$.    

\begin{figure}
\centering
\epsfig{file=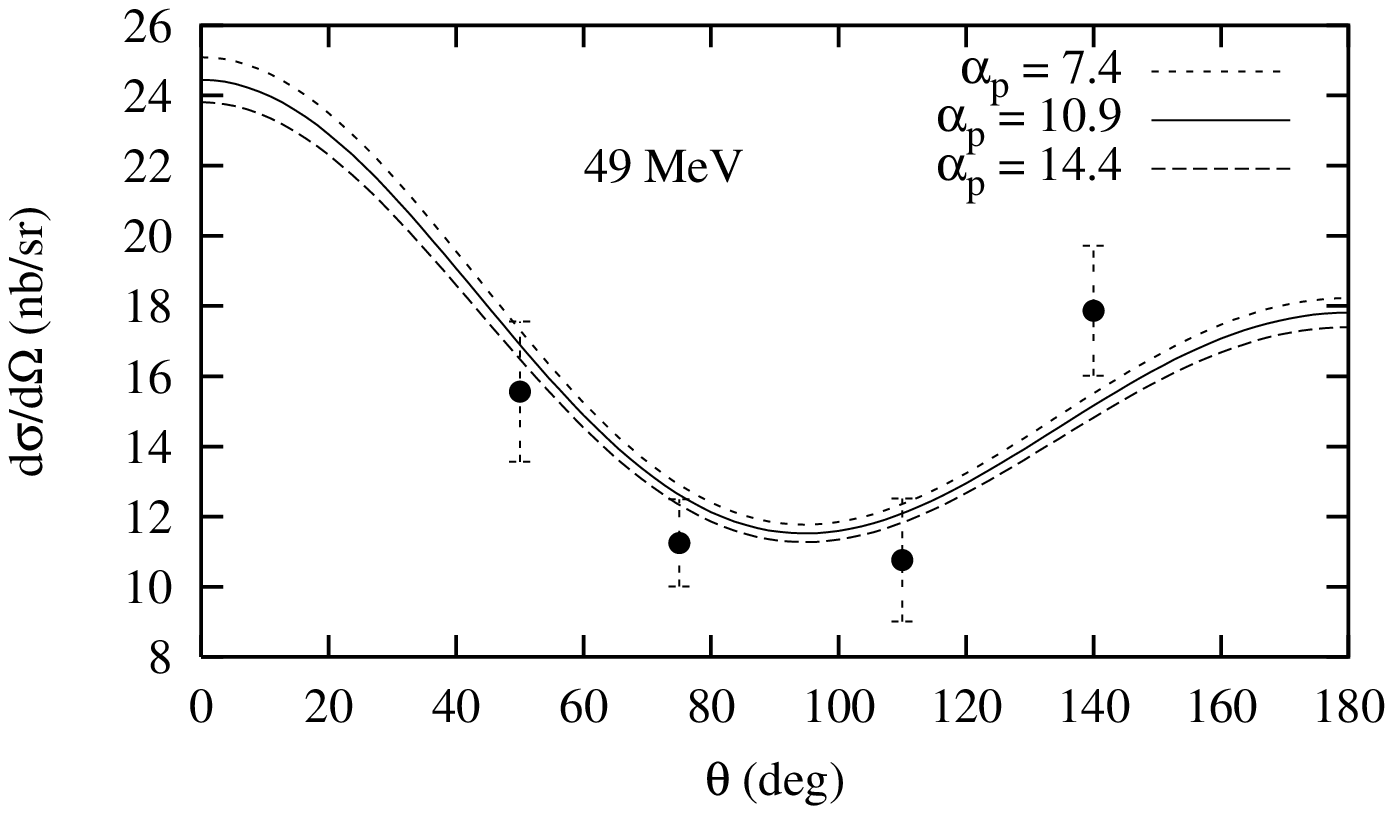}
\epsfig{file=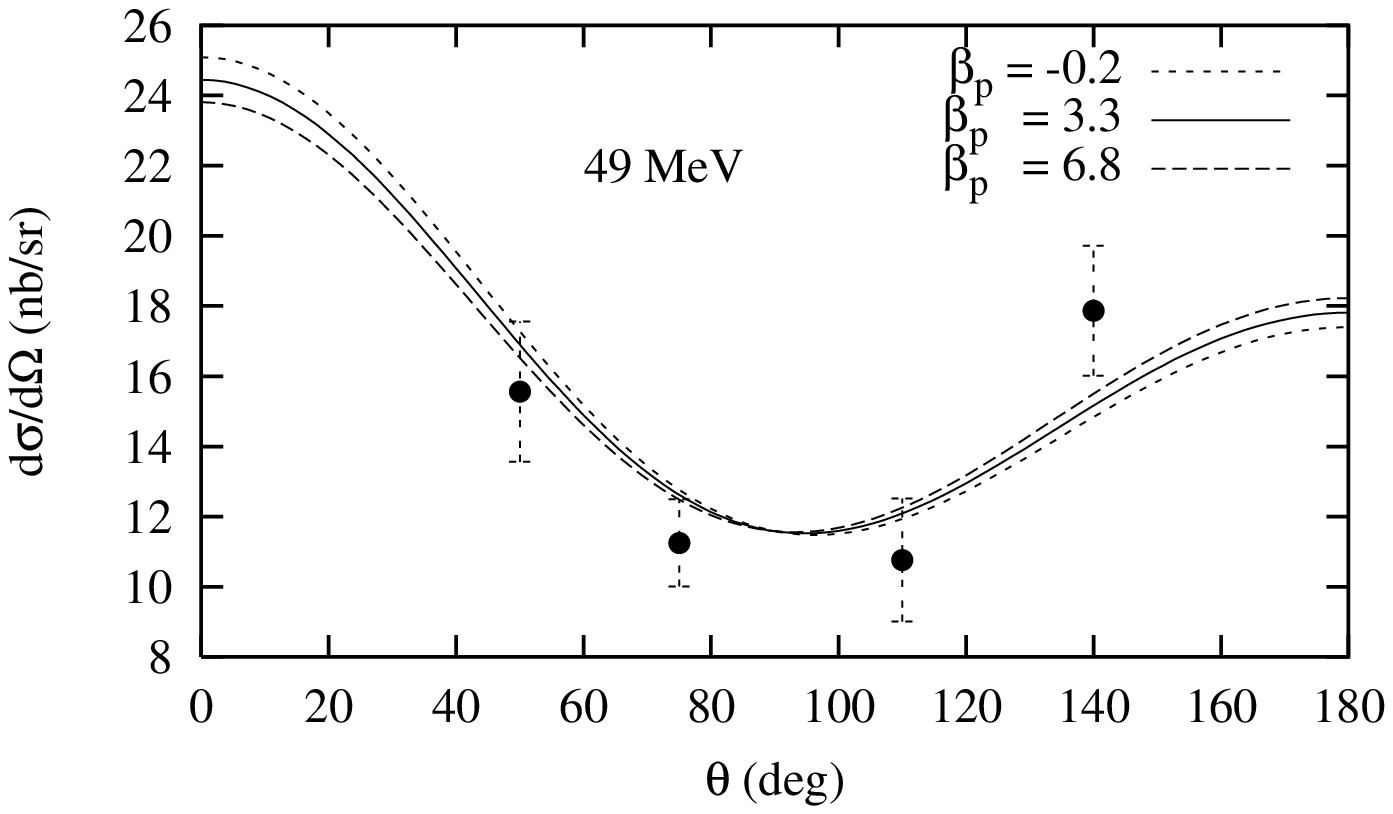}
\caption[Effect of varying $\alpha_p$ and $\beta_p$ separately in the Compton cross-sections at 49, 69, and 95 MeV]{ 
Effect of varying $\alpha_p$ and $\beta_p$ separately in the Compton cross-sections at 49, 69, and 95 MeV.  $\beta_n = 3.3$ in the top graph and
$\alpha_n = 10.9$ in the bottom.  \label{fig:e2-3} }
\end{figure}

\begin{figure}
\centering
\epsfig{file=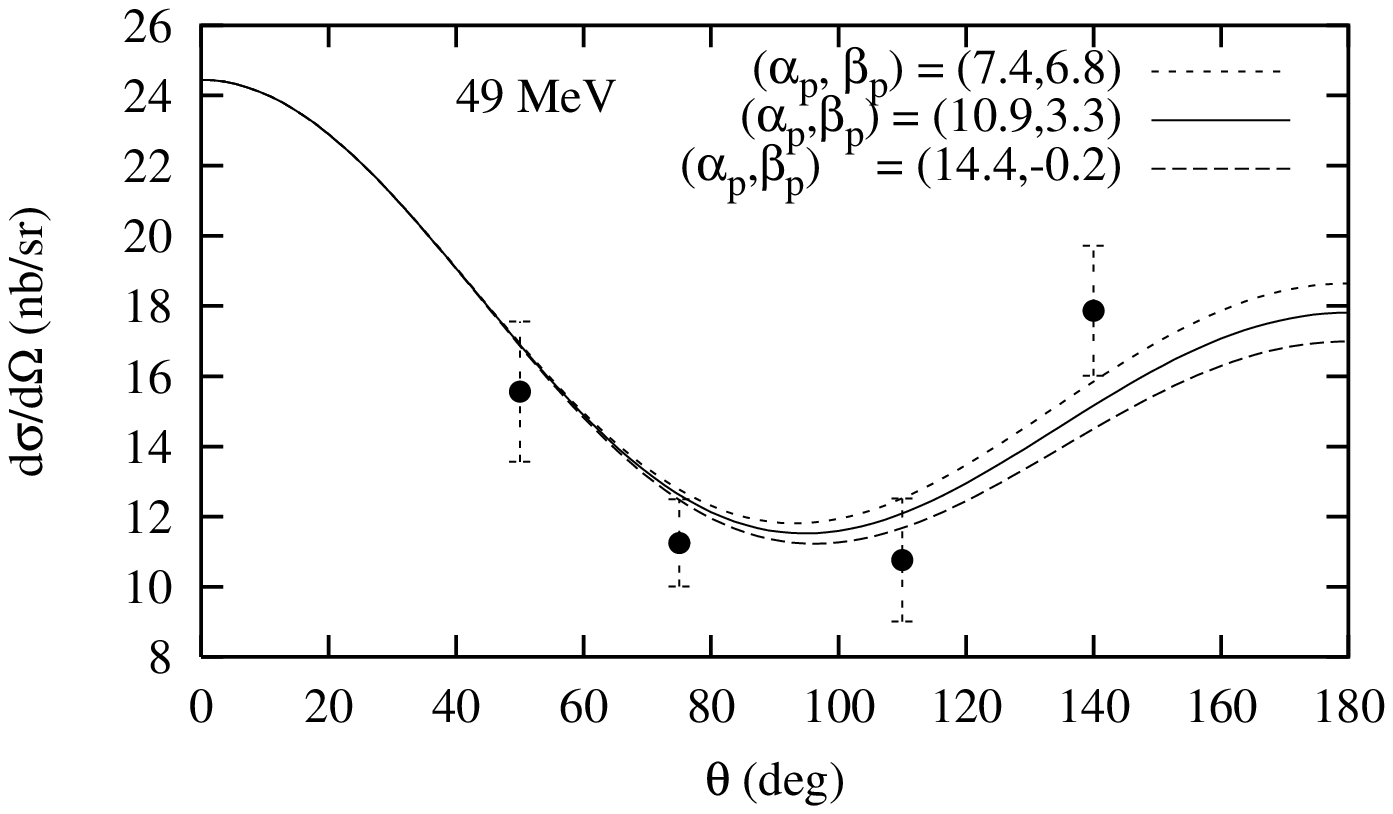}
\epsfig{file=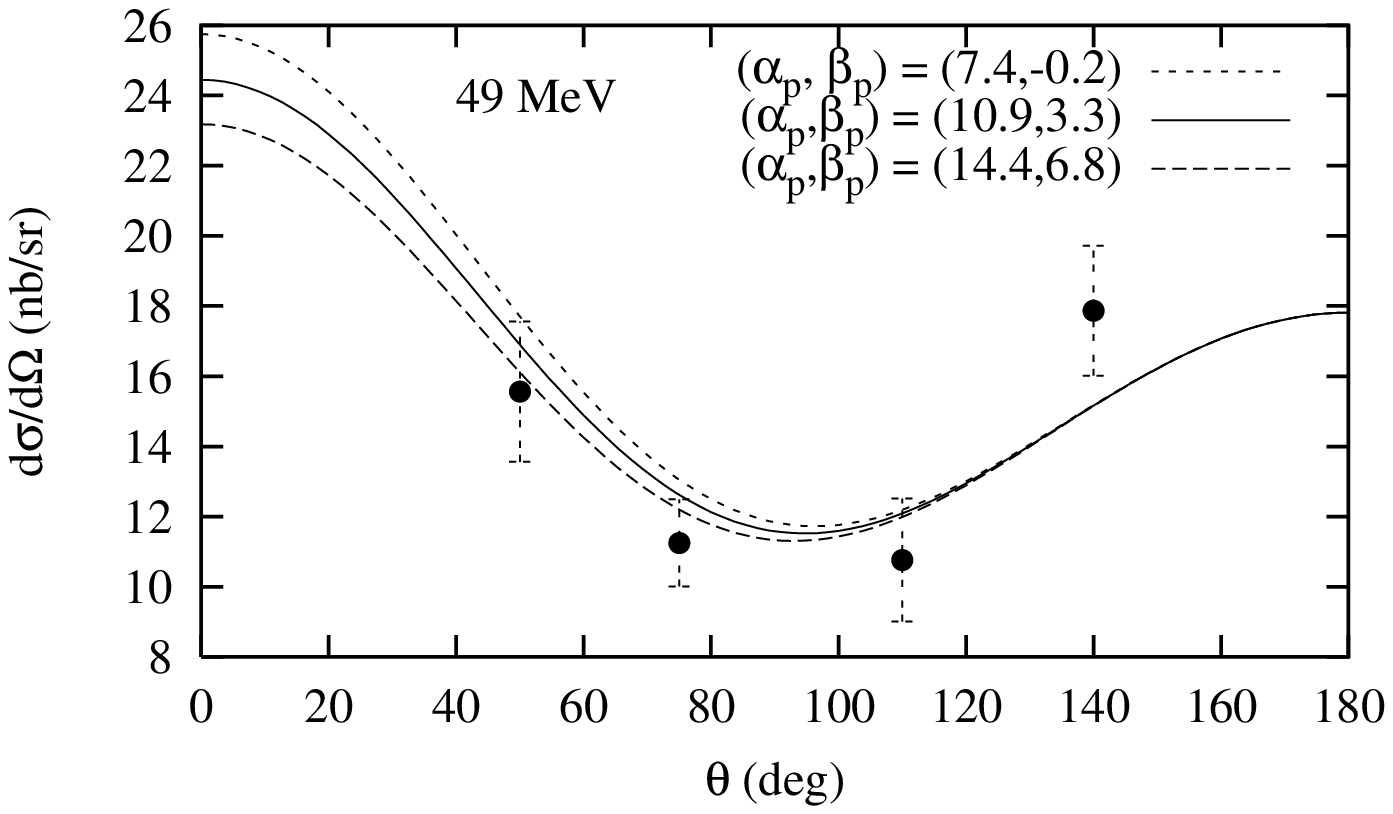}
\caption[Effect of varying both $\alpha_p$ and $\beta_p$  in the Compton cross-sections at 49, 69, and 95 MeV]{
Effect of varying both $\alpha_p$ and $\beta_p$  in the Compton cross-sections at 49, 69, and 95 MeV.   \label{fig:e2-4} }
\end{figure}

We must keep in mind that the proton polarizabilities also have uncertainties, and they contribute to the
deuteron cross-section just as much as the neutron.  Another set of graphs as described above is given for $\alpha_p$ and $\beta_p$
(Figures~\ref{fig:e2-3} and~\ref{fig:e2-4}),
but using only the ranges as the actual experimental error given in \cite{fe91}.   The same trends stand out here, most notably that
a smaller value of $\alpha$ and a larger value of $\beta$  would provide a better description of the data points at the back angles. 

\begin{figure}
\centering
\epsfig{file=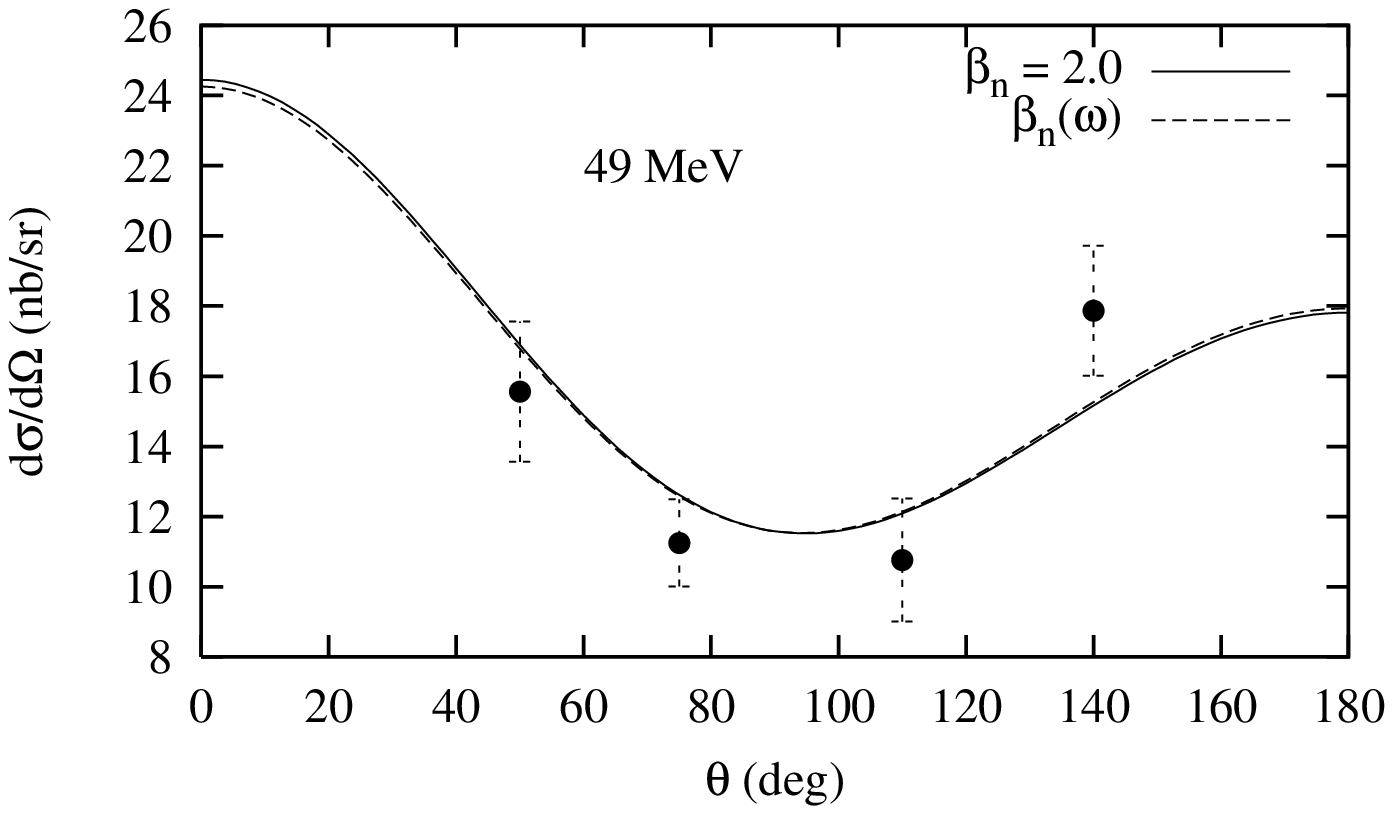}
\epsfig{file=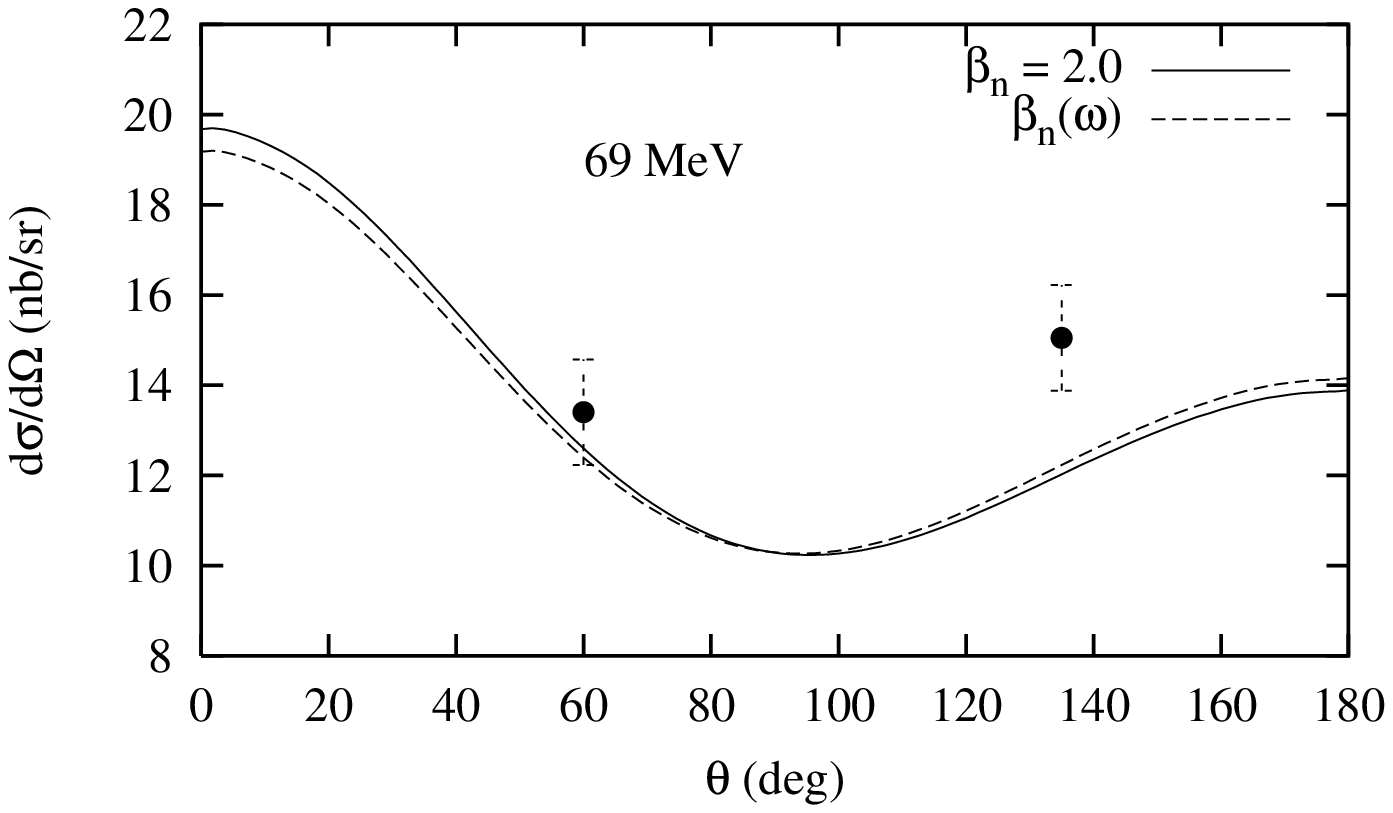}
\caption{Comparison of energy-dependent vs. static $\beta$ in the Compton cross-sections at 49, 69, and 95 MeV.\label{fig:e2-12}}
\end{figure}

An argument can be made that $\beta$ has an energy dependence which has been thus far neglected, causing 
its value to be underestimated.  We have already discussed in Section 2.2 that the $\Delta$ resonance gives a large
contribution to the magnetic polarizability.  For Compton scattering, this part of $\beta$ can be written 
in terms of matrix elements as
\begin{equation}
\beta^{\Delta}_N = 2\frac{ \left|\mx{\Delta}{\mu_z}{N\gamma}\right|^2 }{m_{\Delta}-m_N-\omega} \approx
2 \frac{ \left|\mx{\Delta}{\mu_z}{N\gamma}\right|^2 }{300-\omega}.
\end{equation}
The photon energy was omitted in the static case of Section 2.1.  In Figure~\ref{fig:e2-12}, we compare the
total cross-section with the static value of $\beta_n = 2.0$ to the one where $\beta_n$ is a function of $\omega$.
The energy-dependent polarizability is in fact more consistent with the data,  although the problem at the back angles
still persists.

\clearpage

\begin{figure}
\centering
\epsfig{file=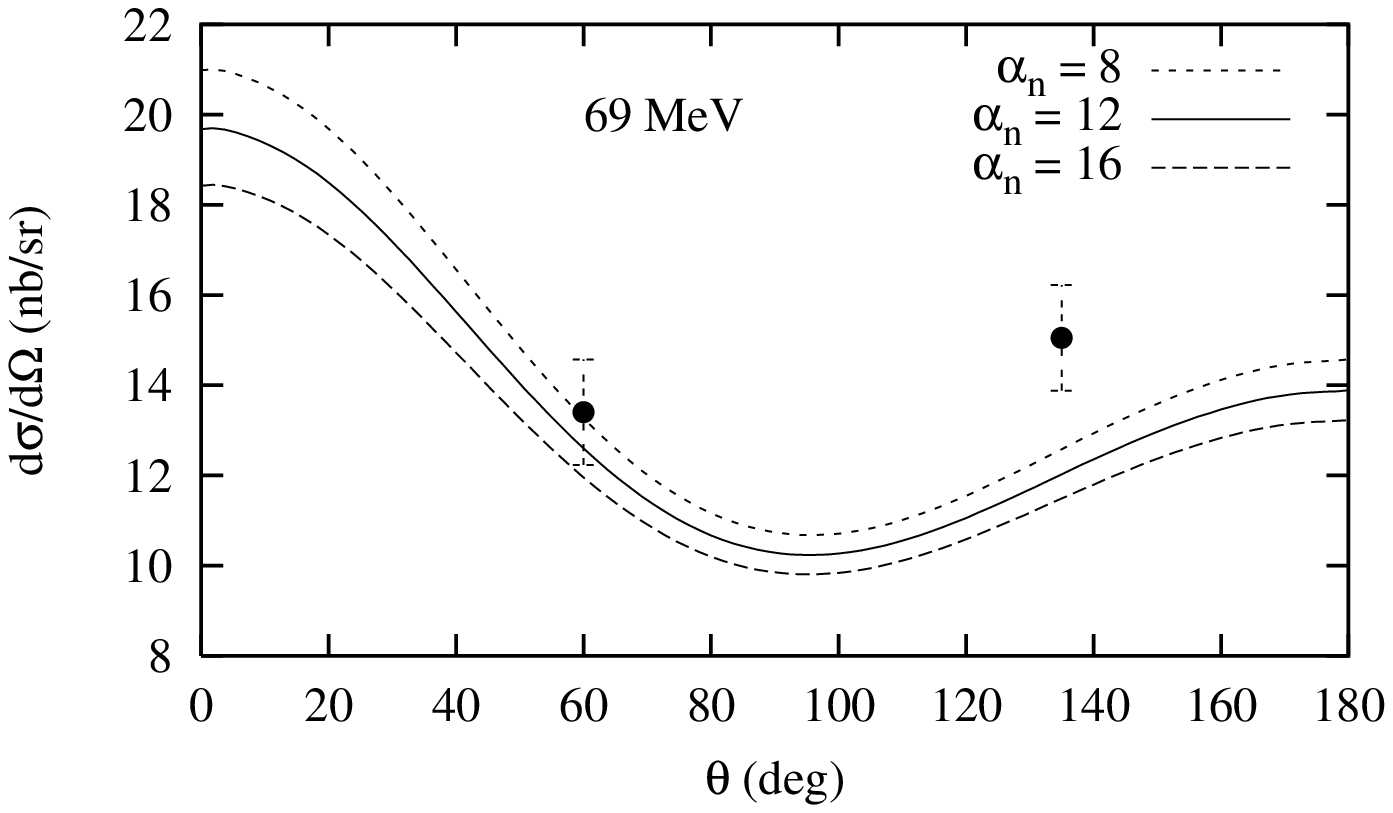}
\epsfig{file=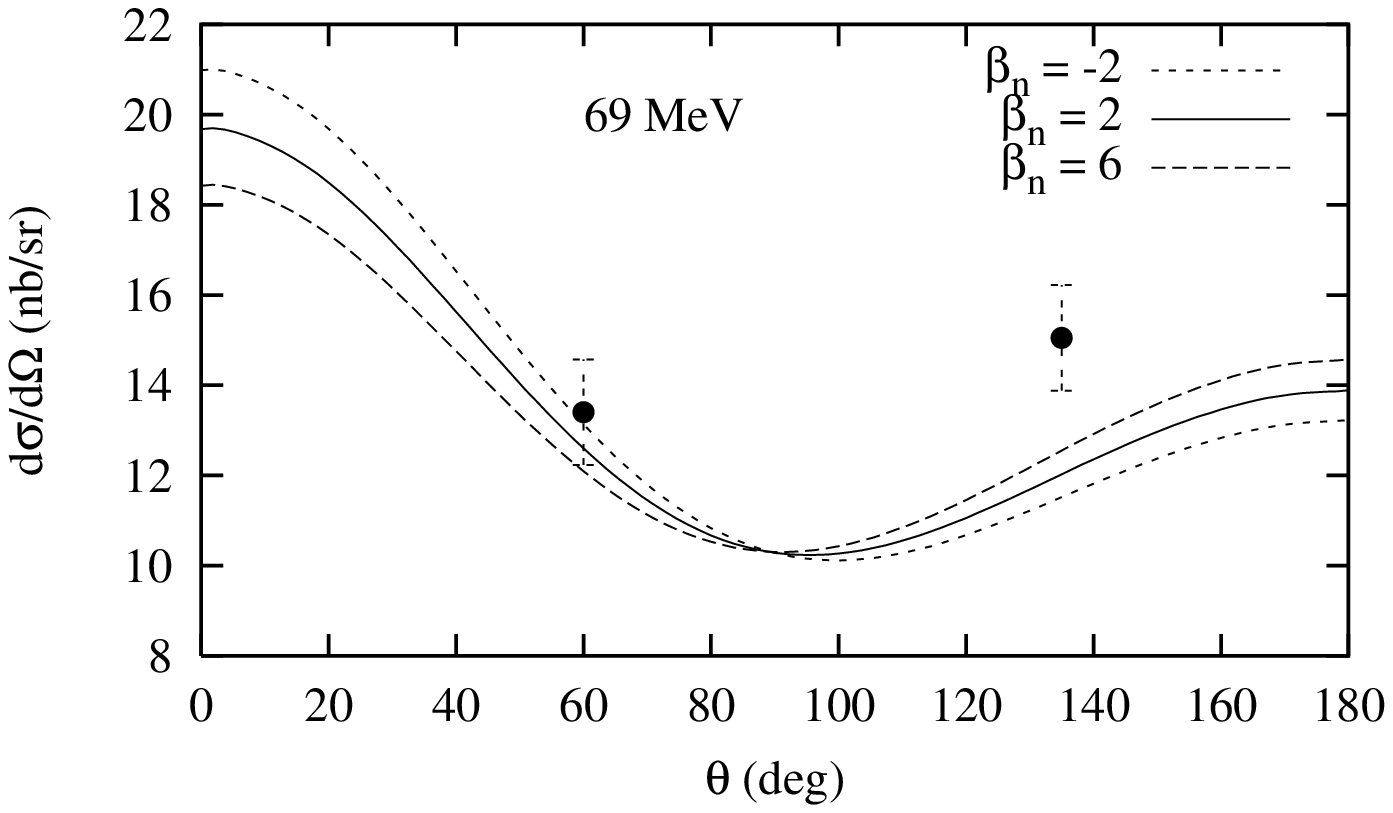}
\end{figure}
\clearpage

\begin{figure}
\centering
\epsfig{file=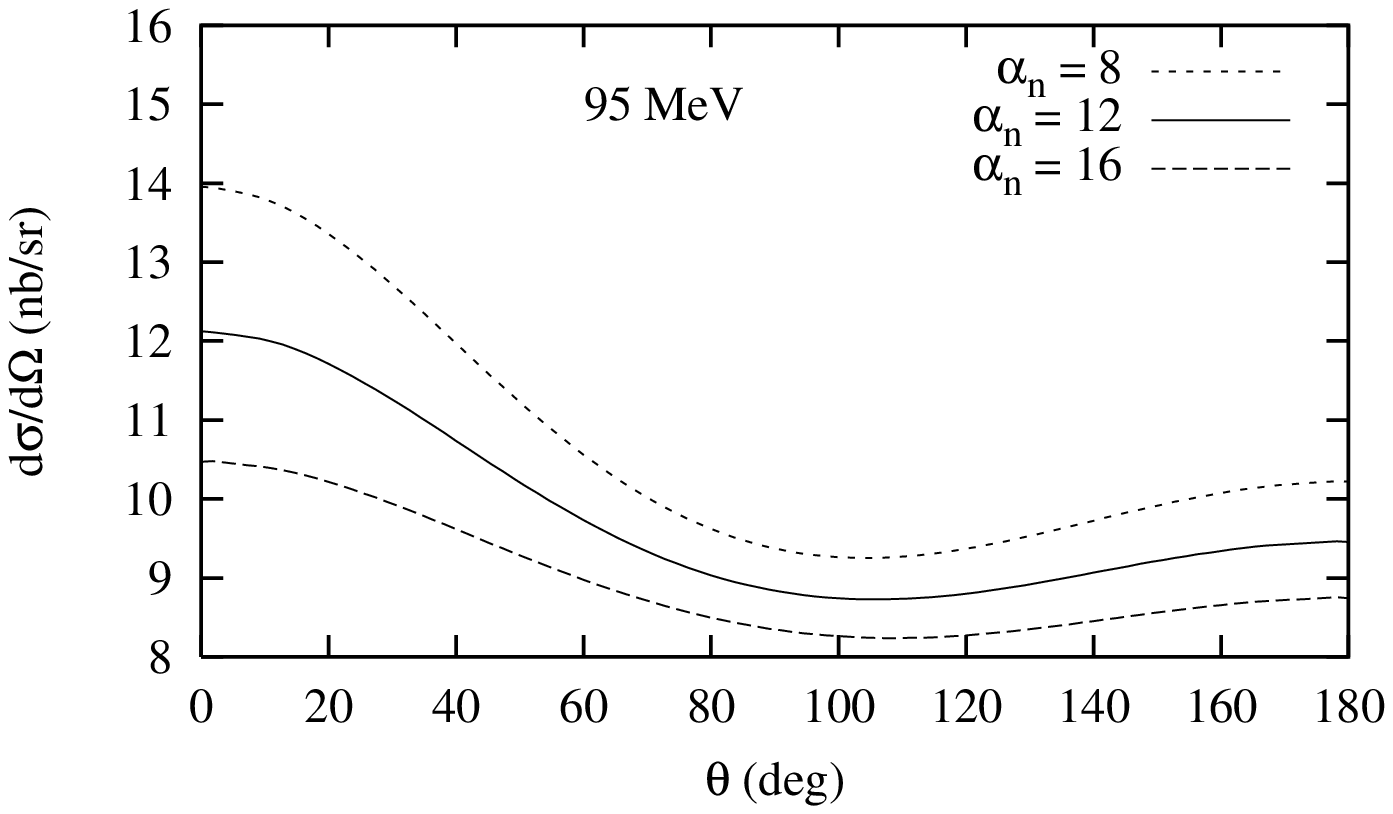}
\epsfig{file=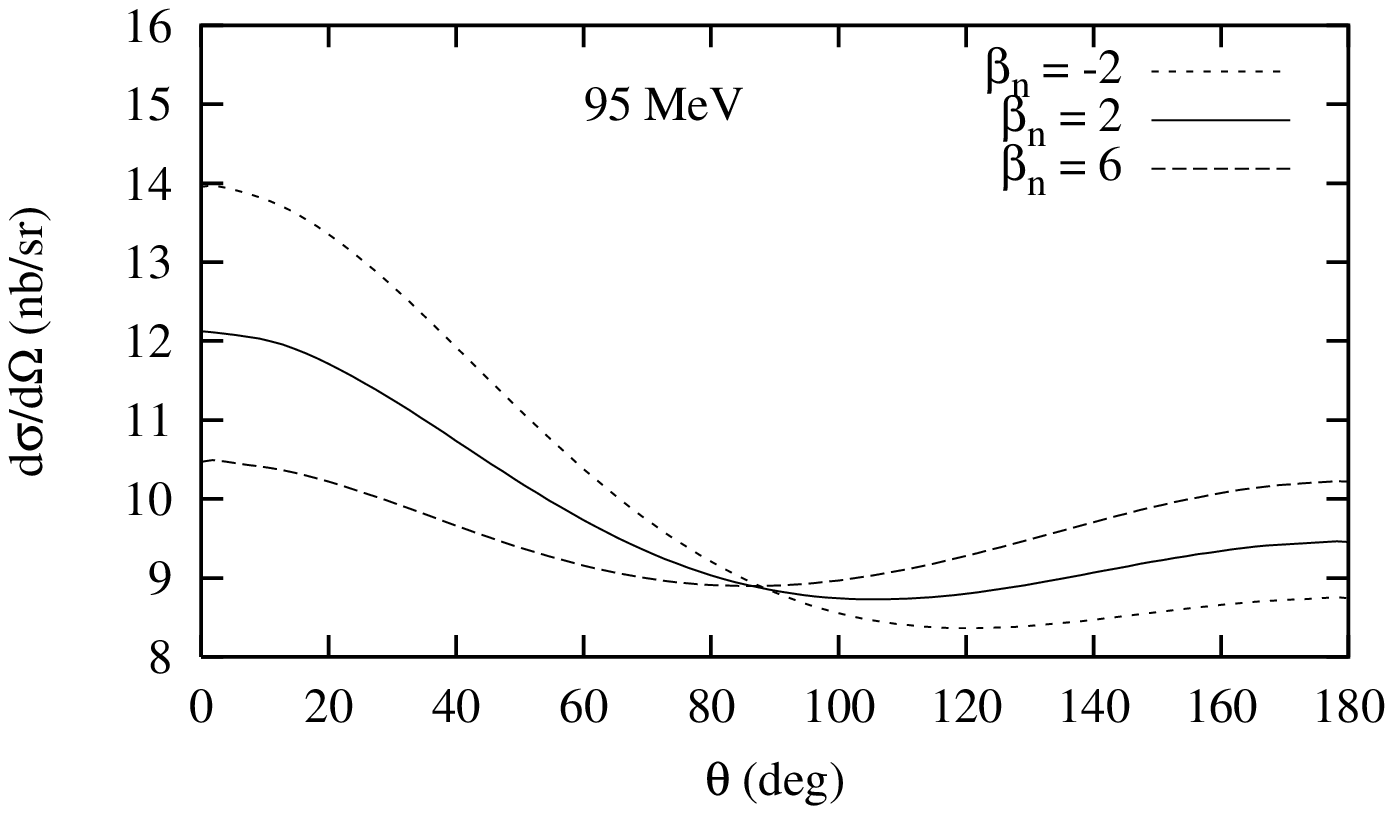}
\end{figure}
\clearpage

\clearpage

\begin{figure}
\centering
\epsfig{file=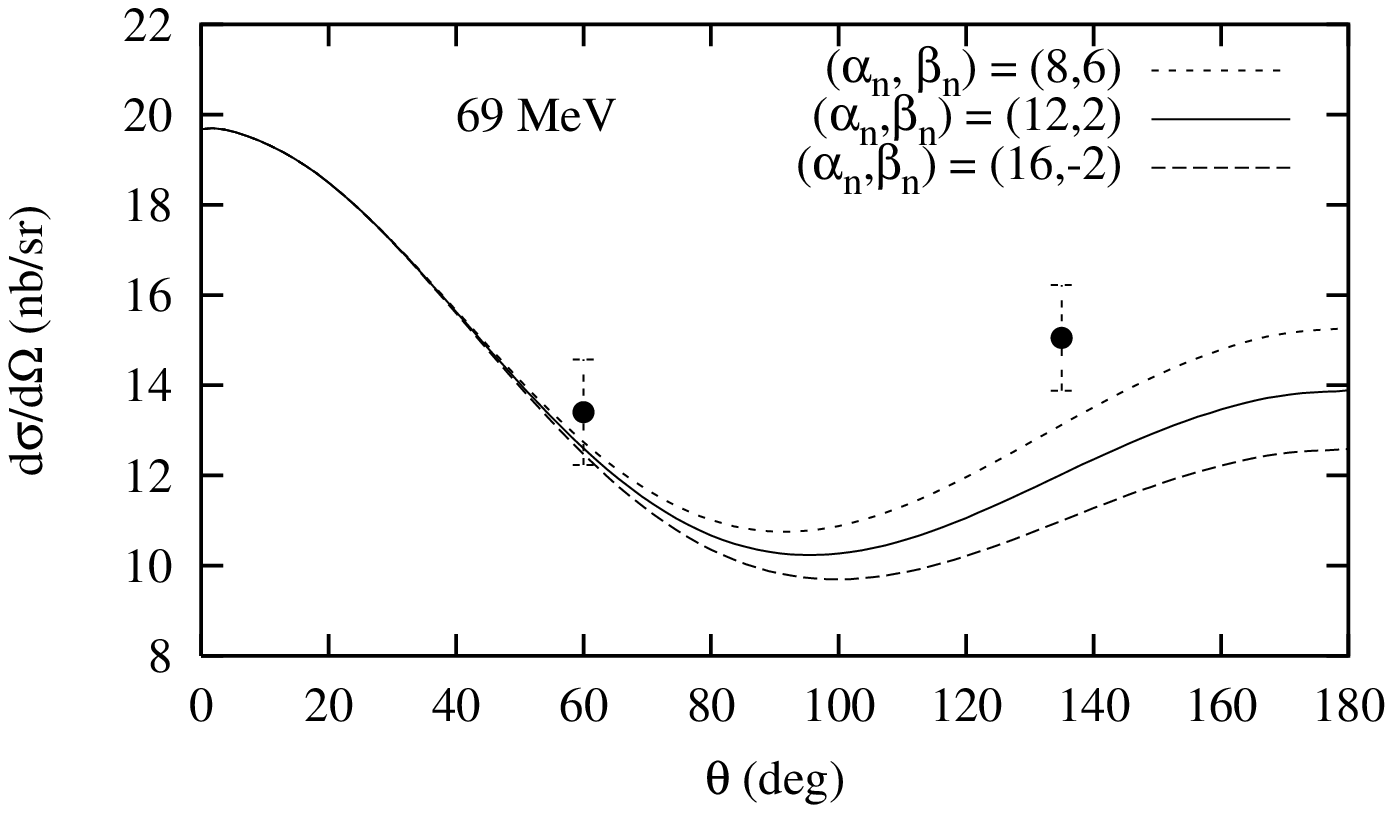}
\epsfig{file=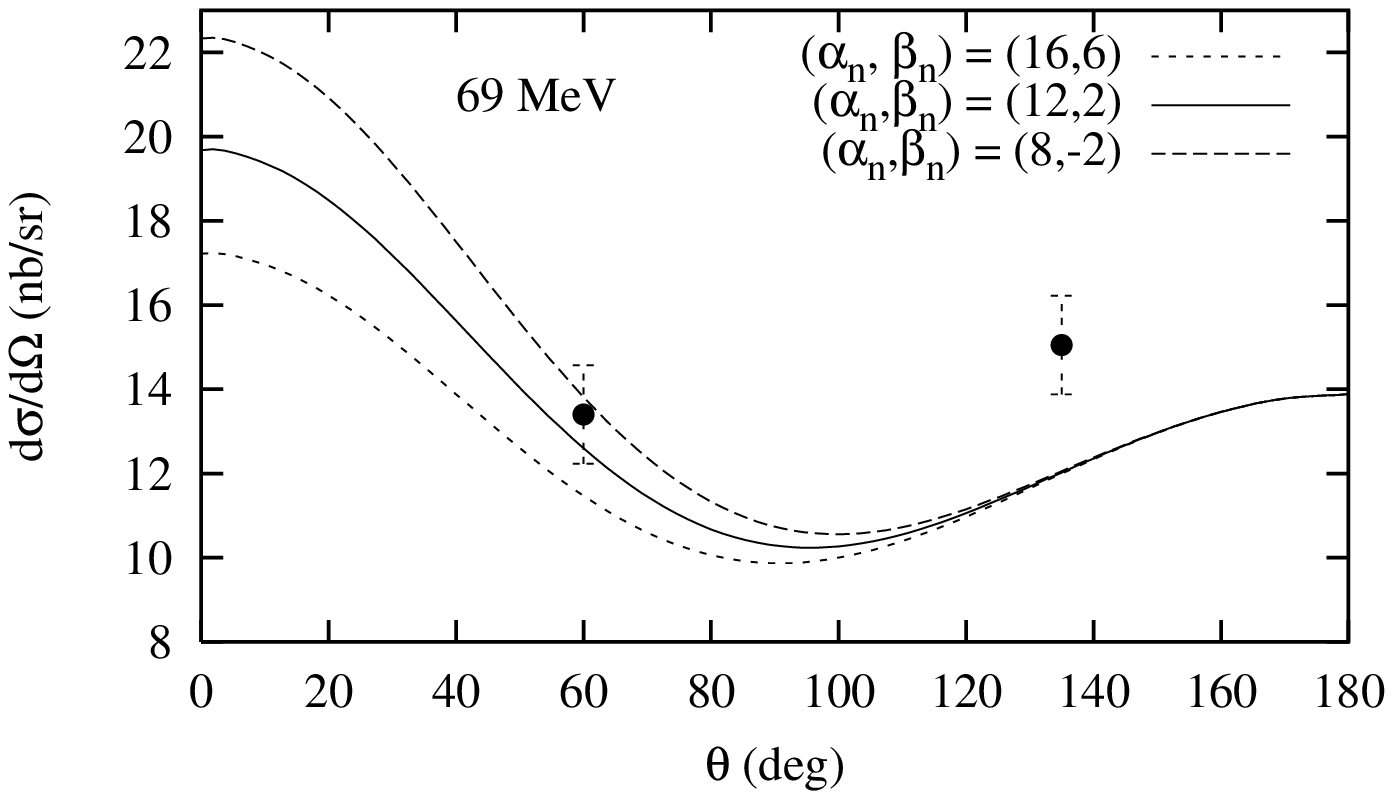}
\end{figure}
\clearpage

\begin{figure}
\centering
\epsfig{file=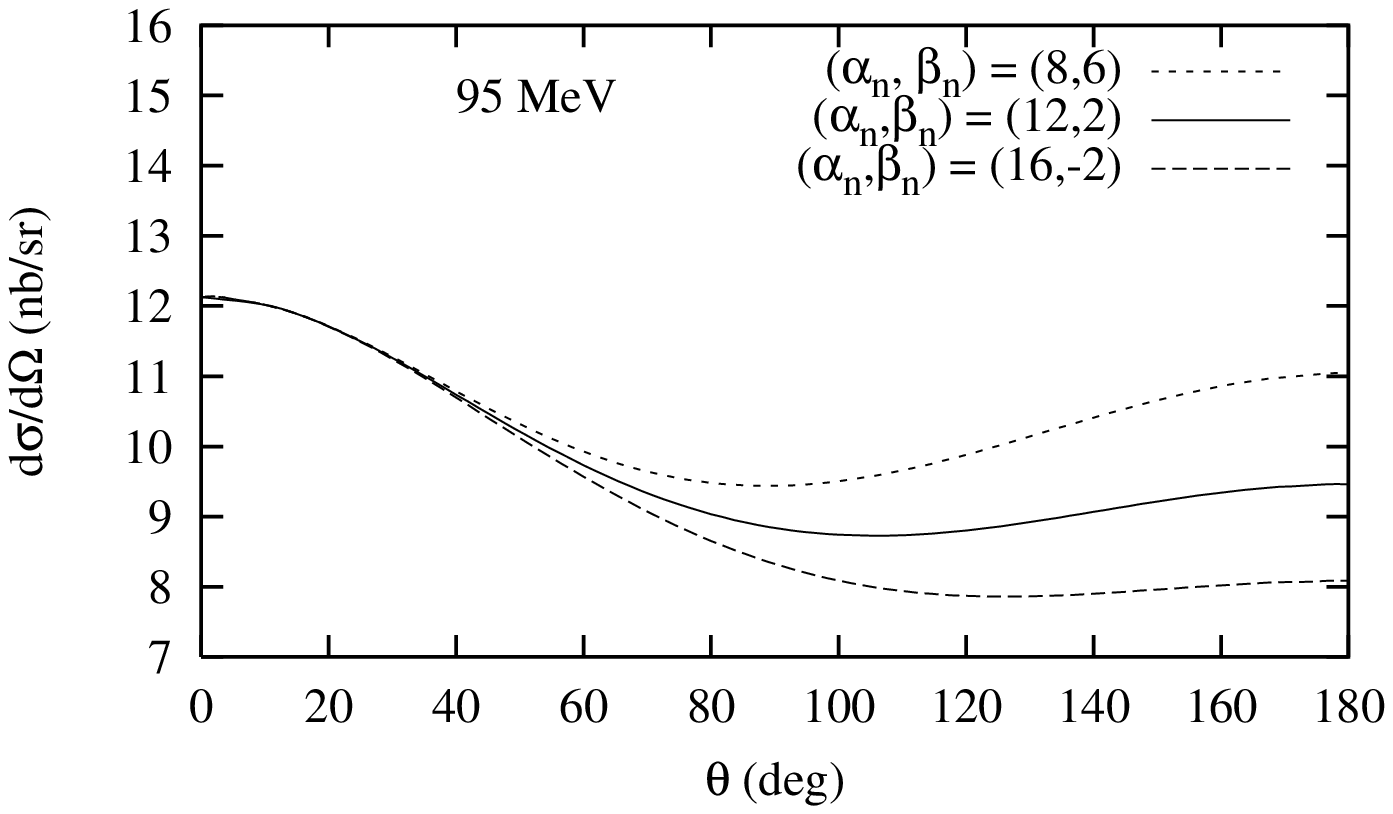}
\epsfig{file=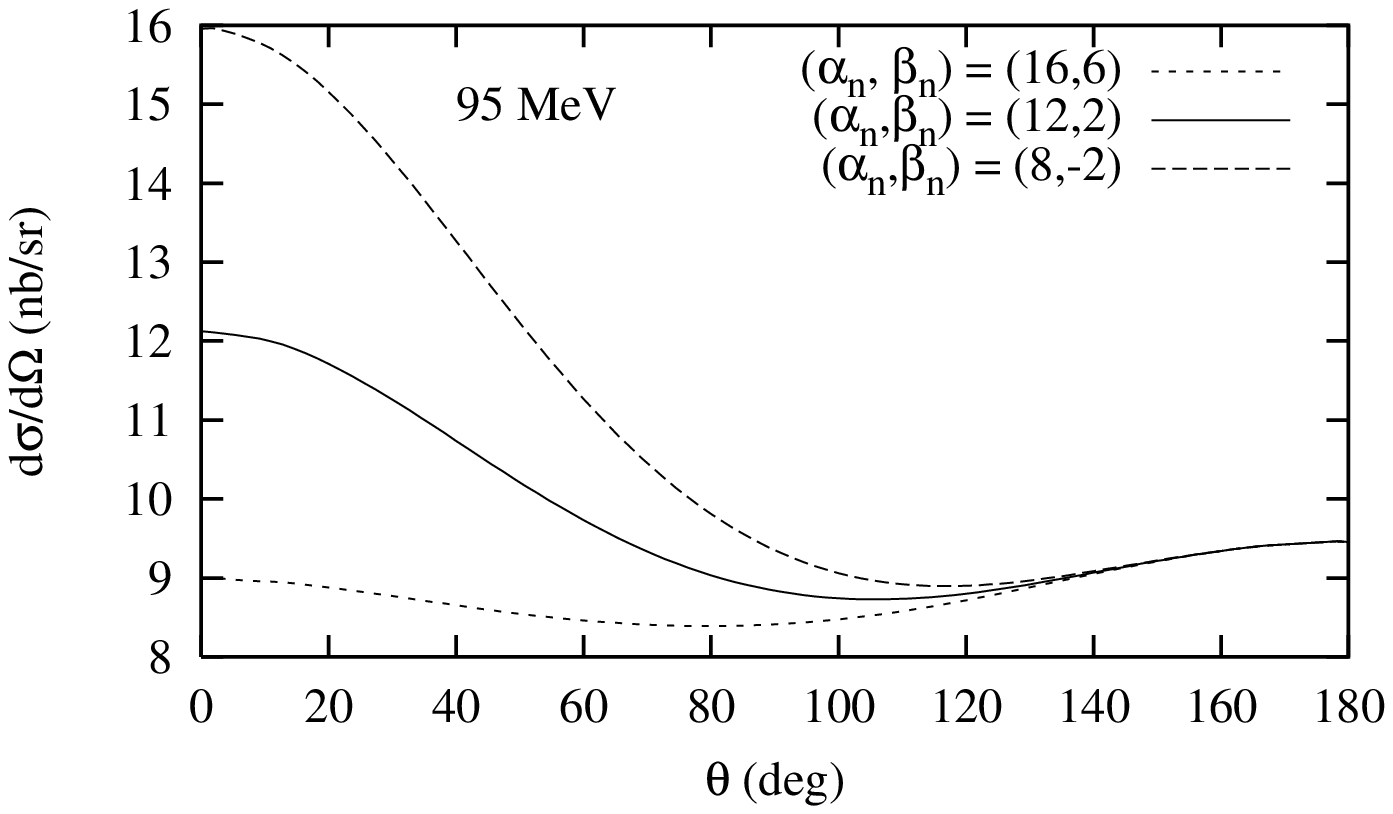}
\end{figure}

\clearpage

\begin{figure}
\centering
\epsfig{file=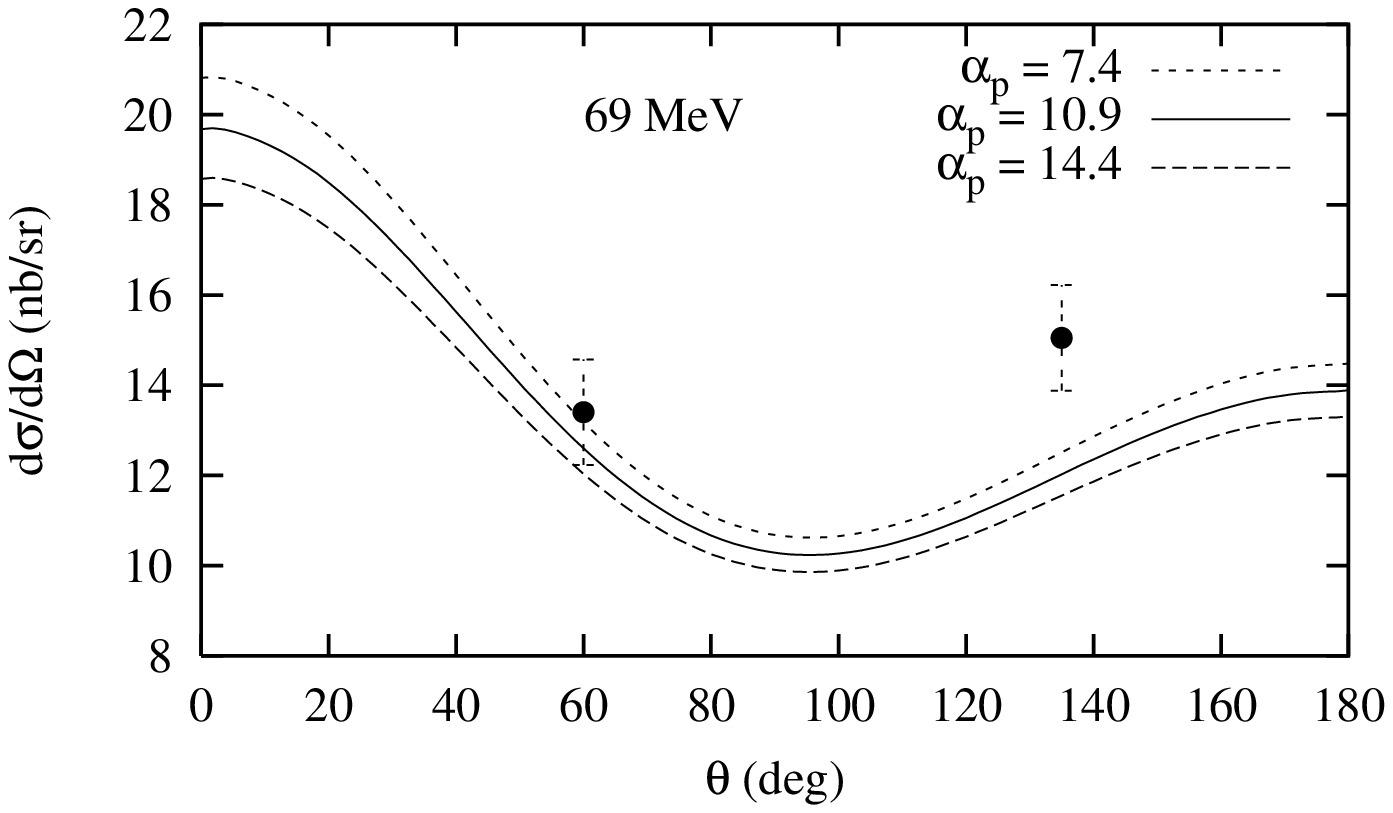}
\epsfig{file=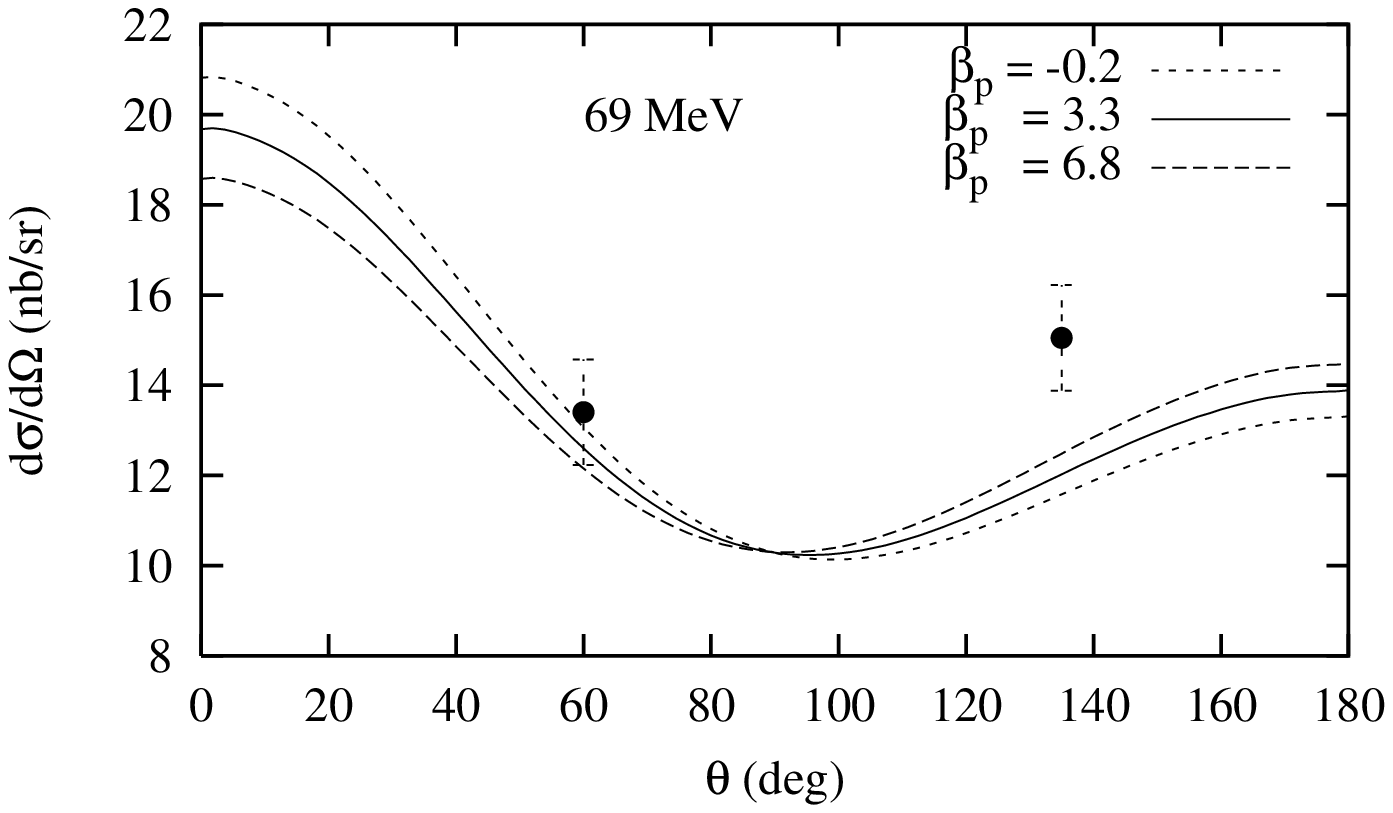}
\end{figure}
\clearpage

\begin{figure}
\centering
\epsfig{file=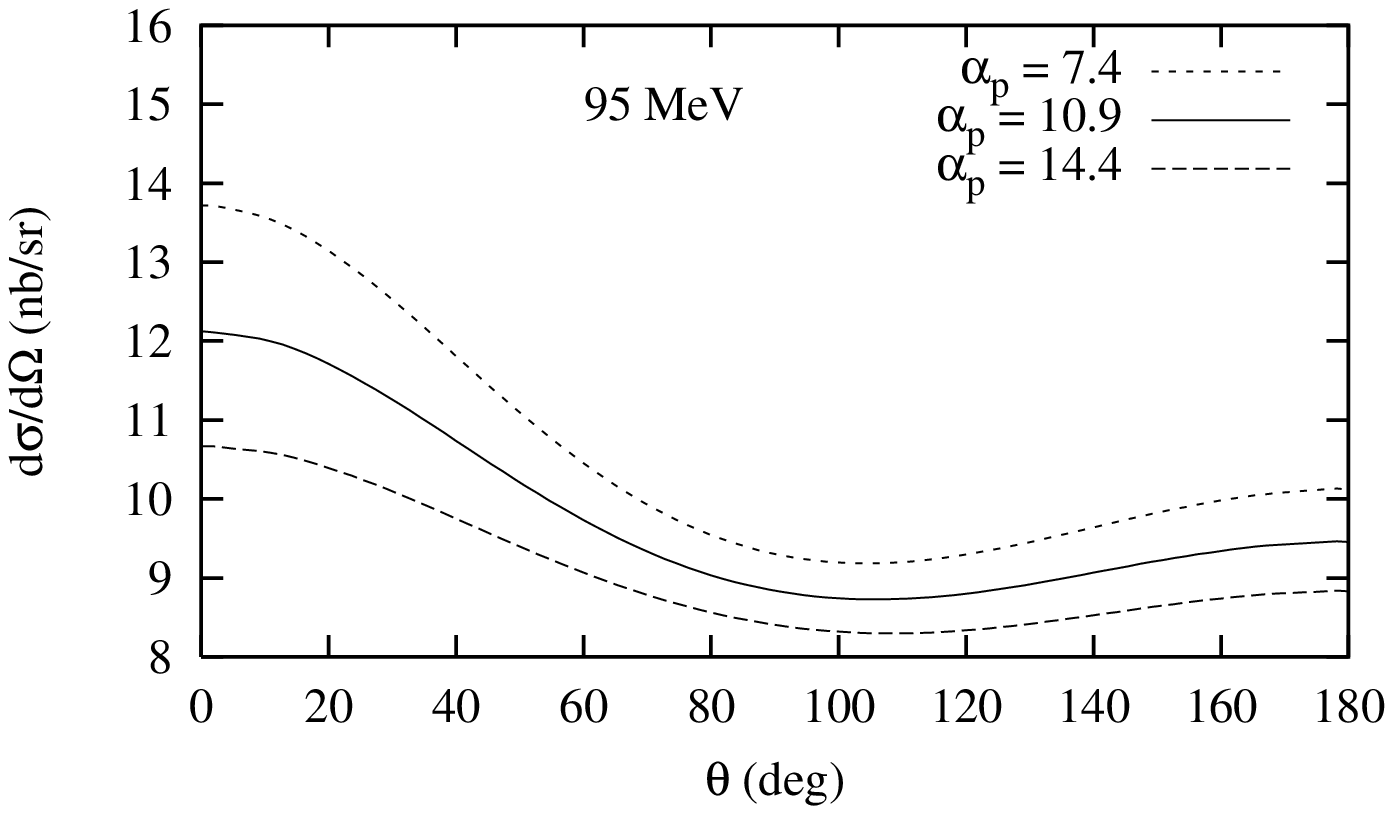}
\epsfig{file=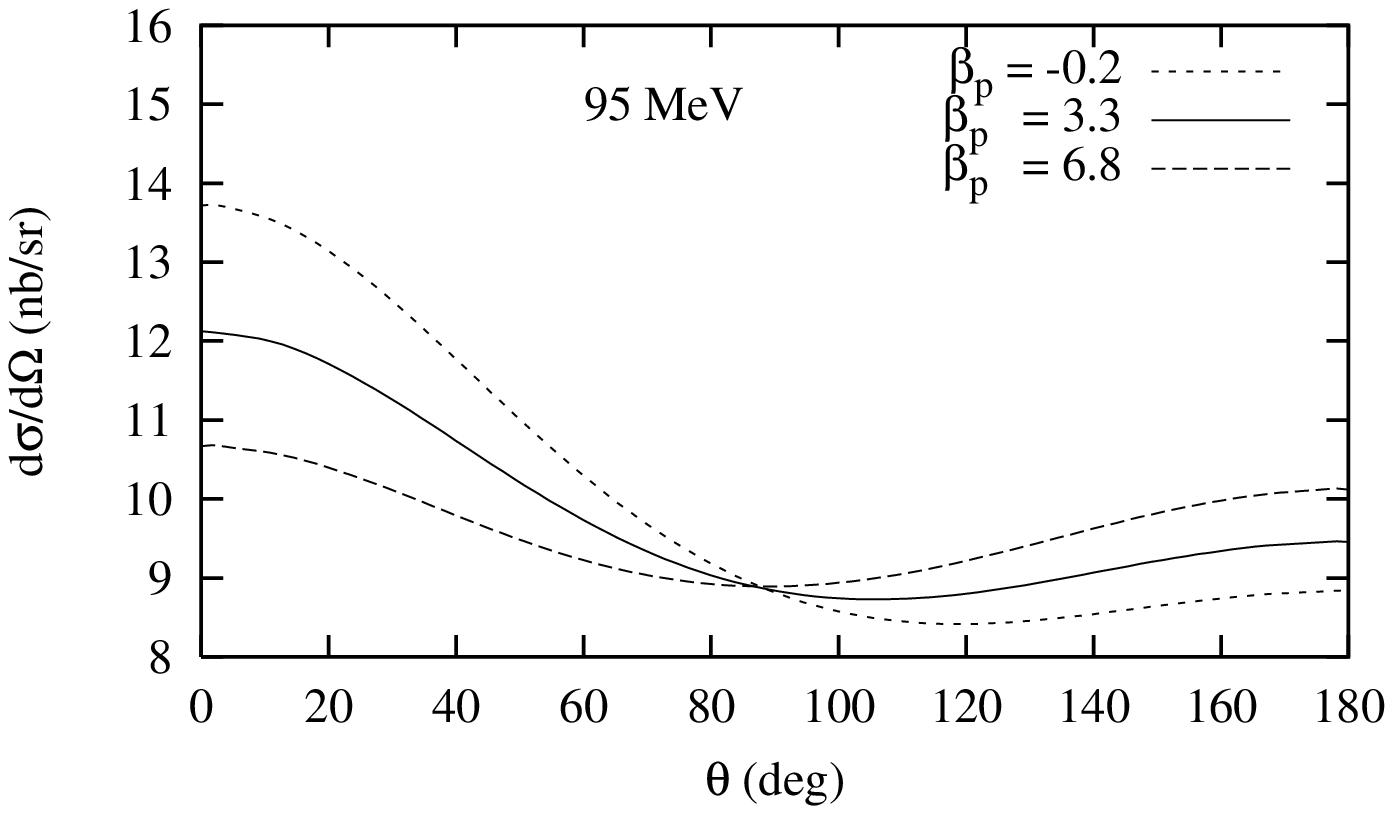}
\end{figure}
\clearpage

\clearpage

\begin{figure}
\centering
\epsfig{file=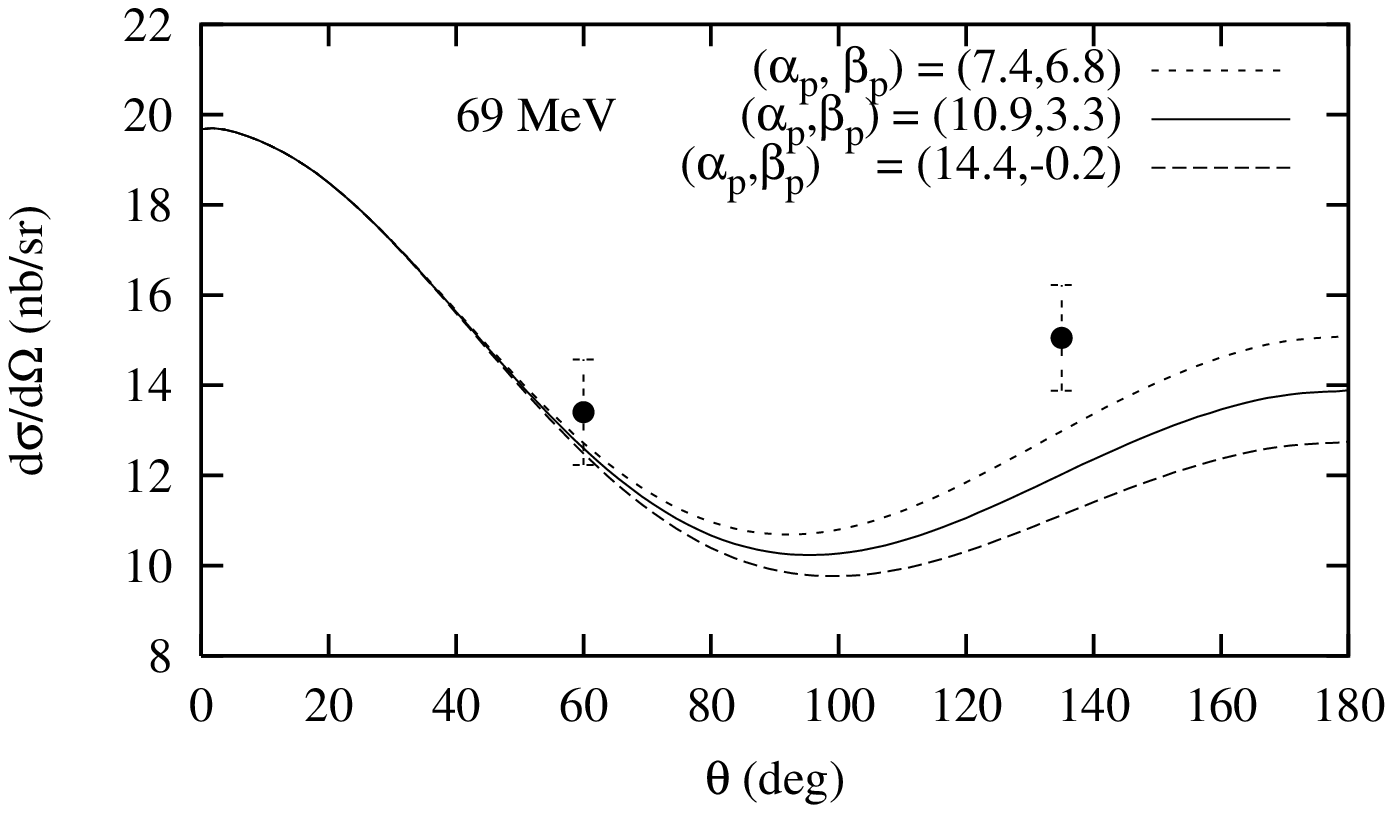}
\epsfig{file=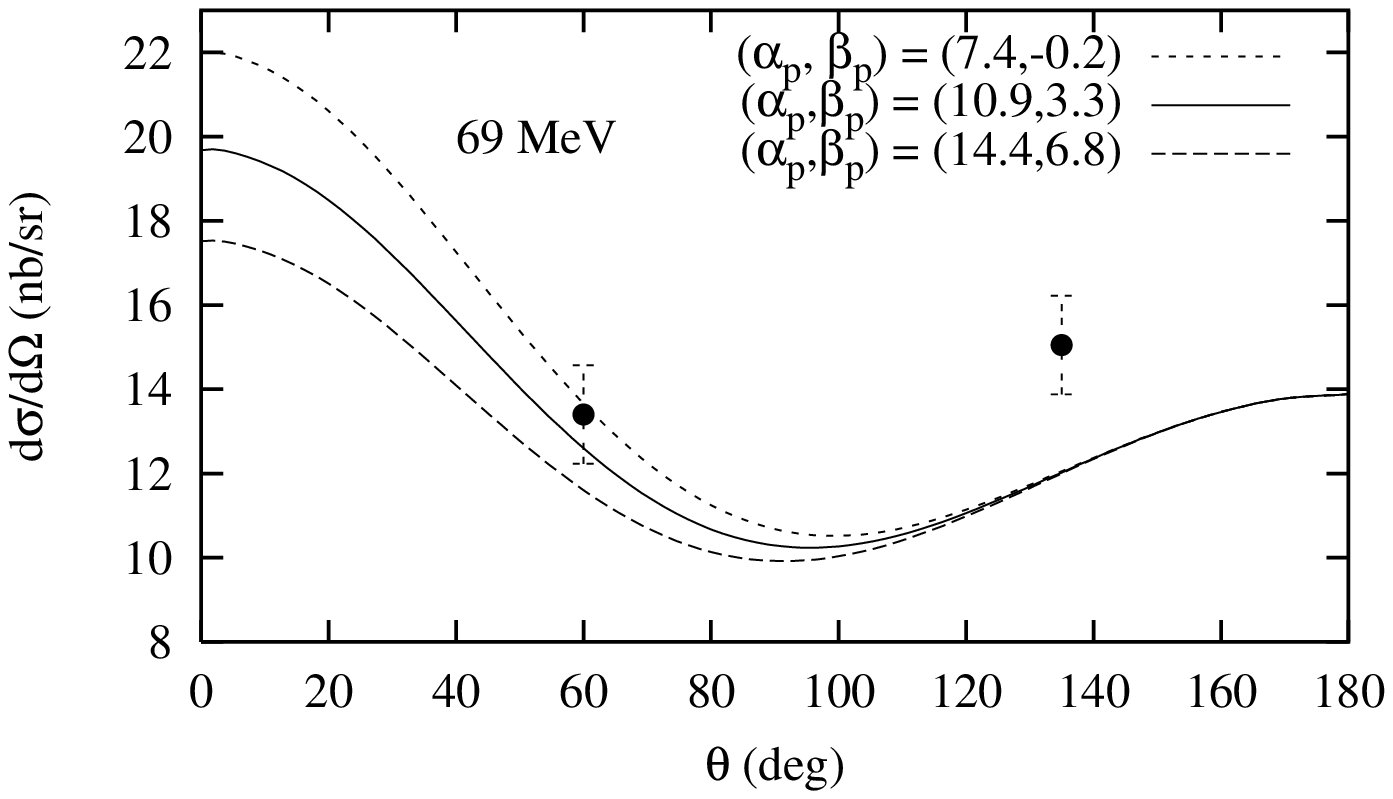}
\end{figure}
\clearpage

\begin{figure}
\centering
\epsfig{file=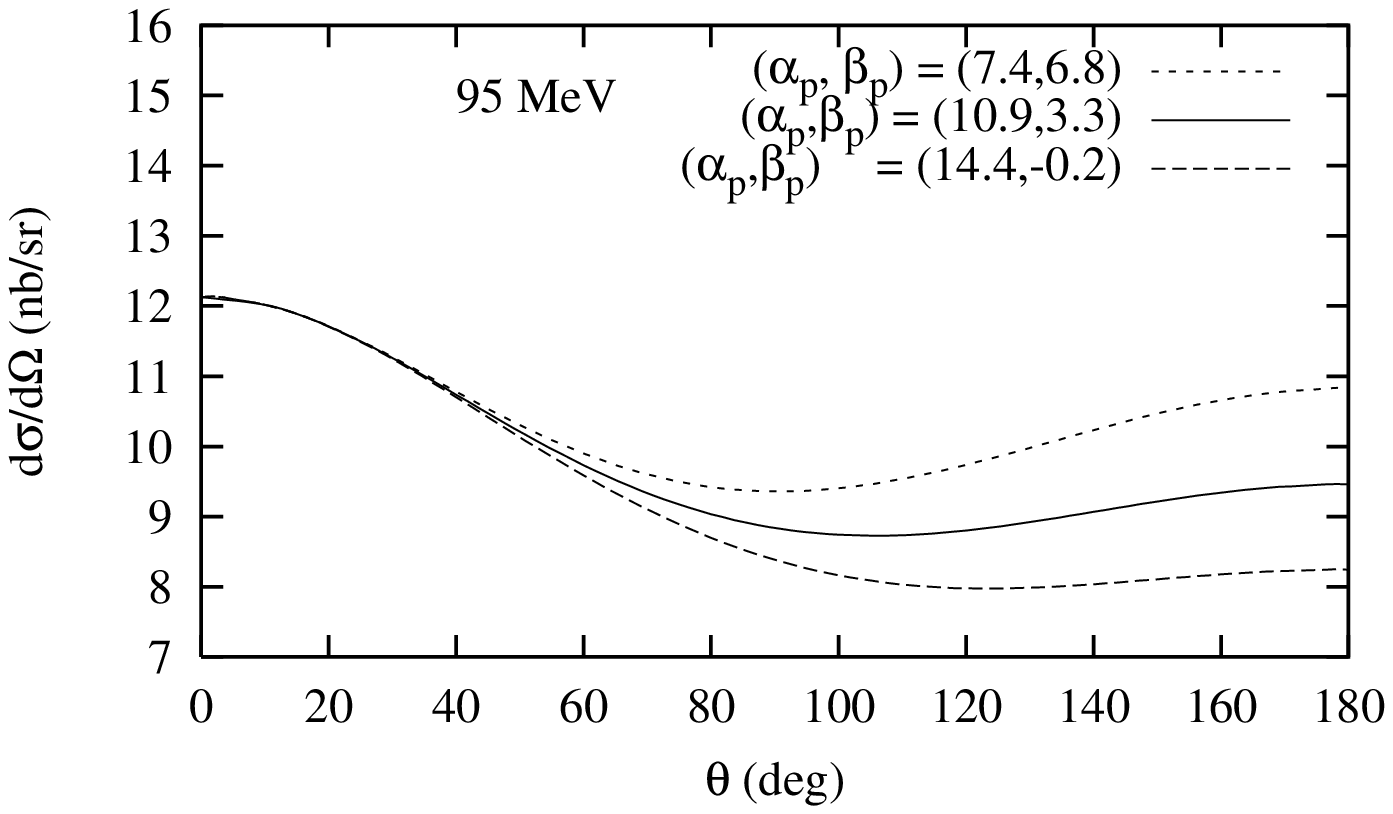}
\epsfig{file=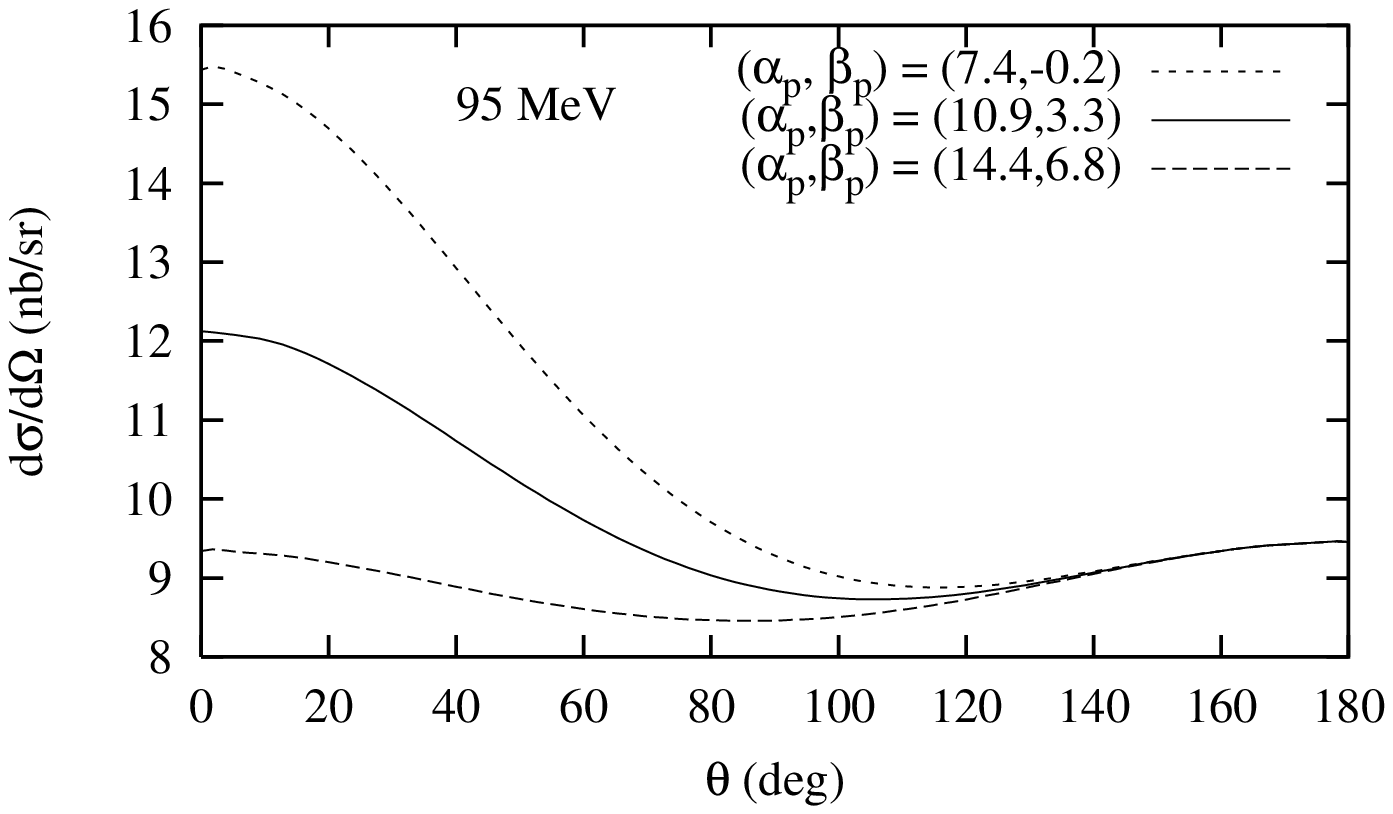}
\end{figure}

\clearpage

\begin{figure}
\centering
\epsfig{file=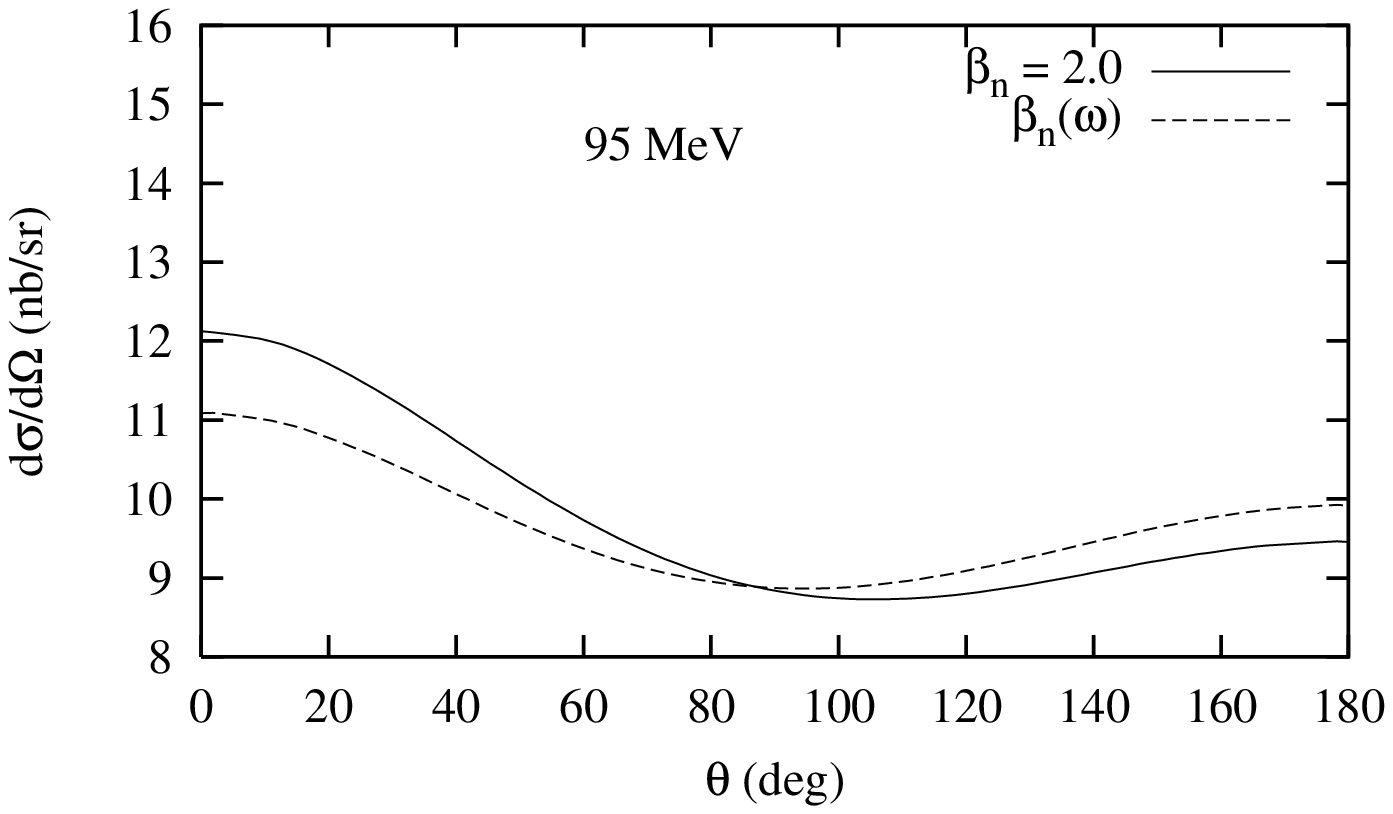}
\end{figure}
\clearpage

\section{Sources of Error}
We would like to study the effects of some of the smaller terms more closely.
The results presented here basically reinforce the general discussion of Section 4.1.
For each term, we show a graph comparing the full cross-section with a cross-section that has that effect subtracted out.
We can then estimate the absolute error in the value $\alpha_n$ (we fix $\beta_n = 2.0$ in this section) caused
by neglecting various effects.  These are listed in Table~\ref{t:e3-1} and shown in the figures.

\begin{figure}
\centering
\epsfig{file=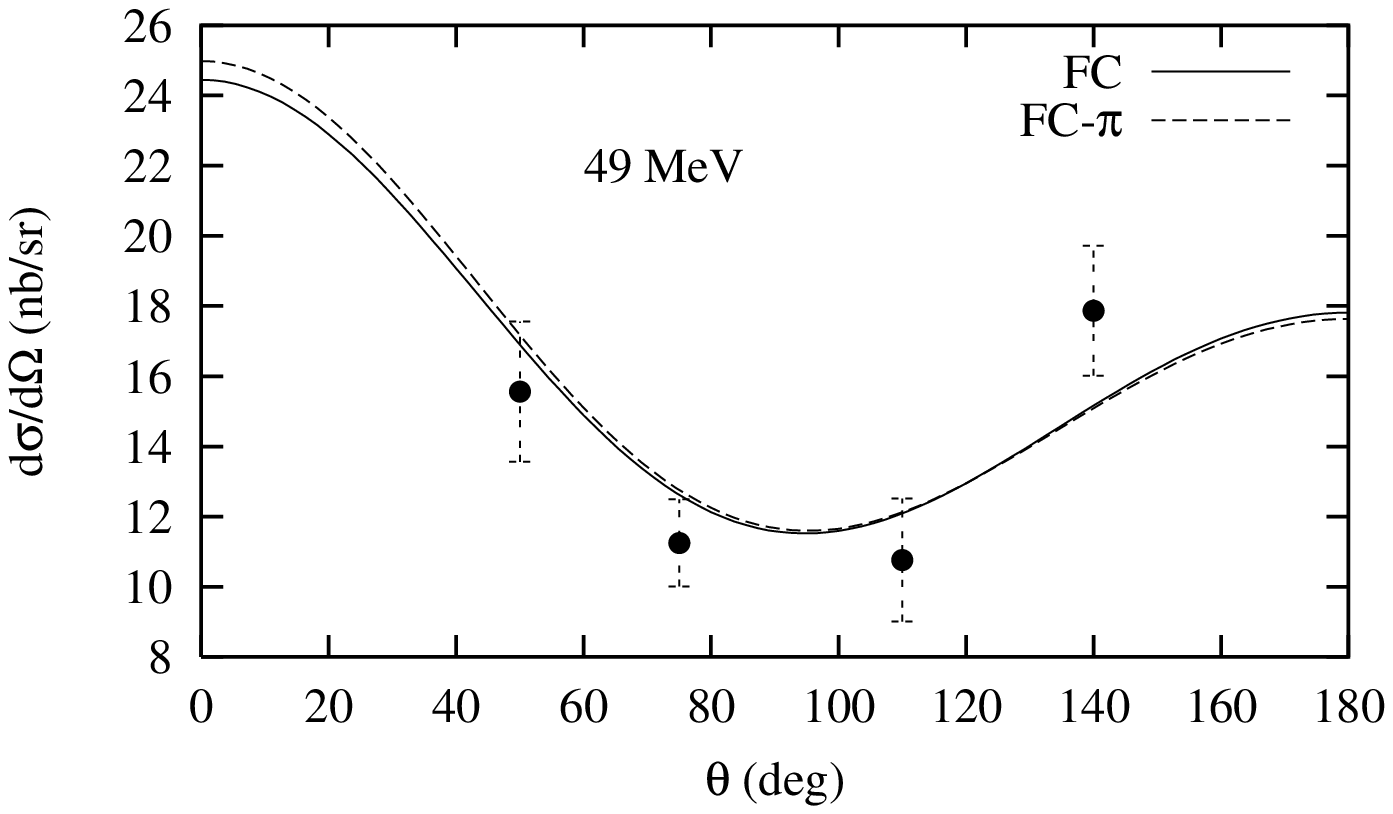}
\epsfig{file=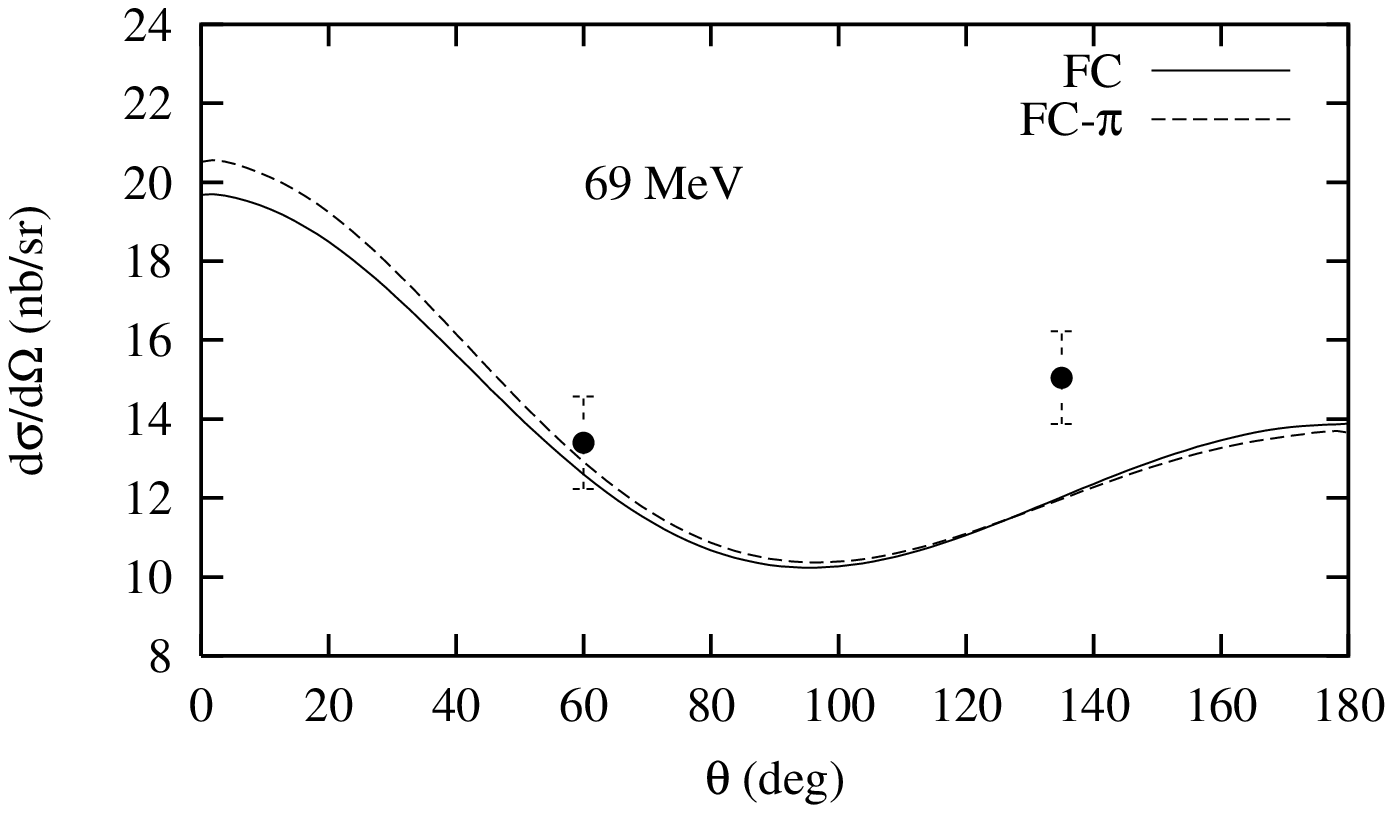}
\caption[Effect of neglecting pion terms on the full differential cross-section at 49, 69, and 95 MeV]{ 
\label{fig:e3-10} Effect of neglecting pion terms on the full differential cross-section (FC) at 49, 69, and 95 MeV.}
\end{figure}

First, we examine the gauge invariant set of pion terms.  
Examining Table~\ref{t:e3-1}, we see that the value $\alpha_n$ would be slightly underestimated  at the forward angles
and overestimated at backward angles, and would not be affected around $120-140^{\circ}$.   The magnitude of this error
is about the same as the experimental error in the polarizability, so these effects would certainly need to be taken into account
to make an accurate determination of  $\alpha_n$ in this type of experiment.  The effect is same at all energies since
both the $\pi$ terms and polarizability terms have an $\omega^2$ dependence. This is shown in Figure~\ref{fig:e3-10}.

\begin{figure}
\centering
\epsfig{file=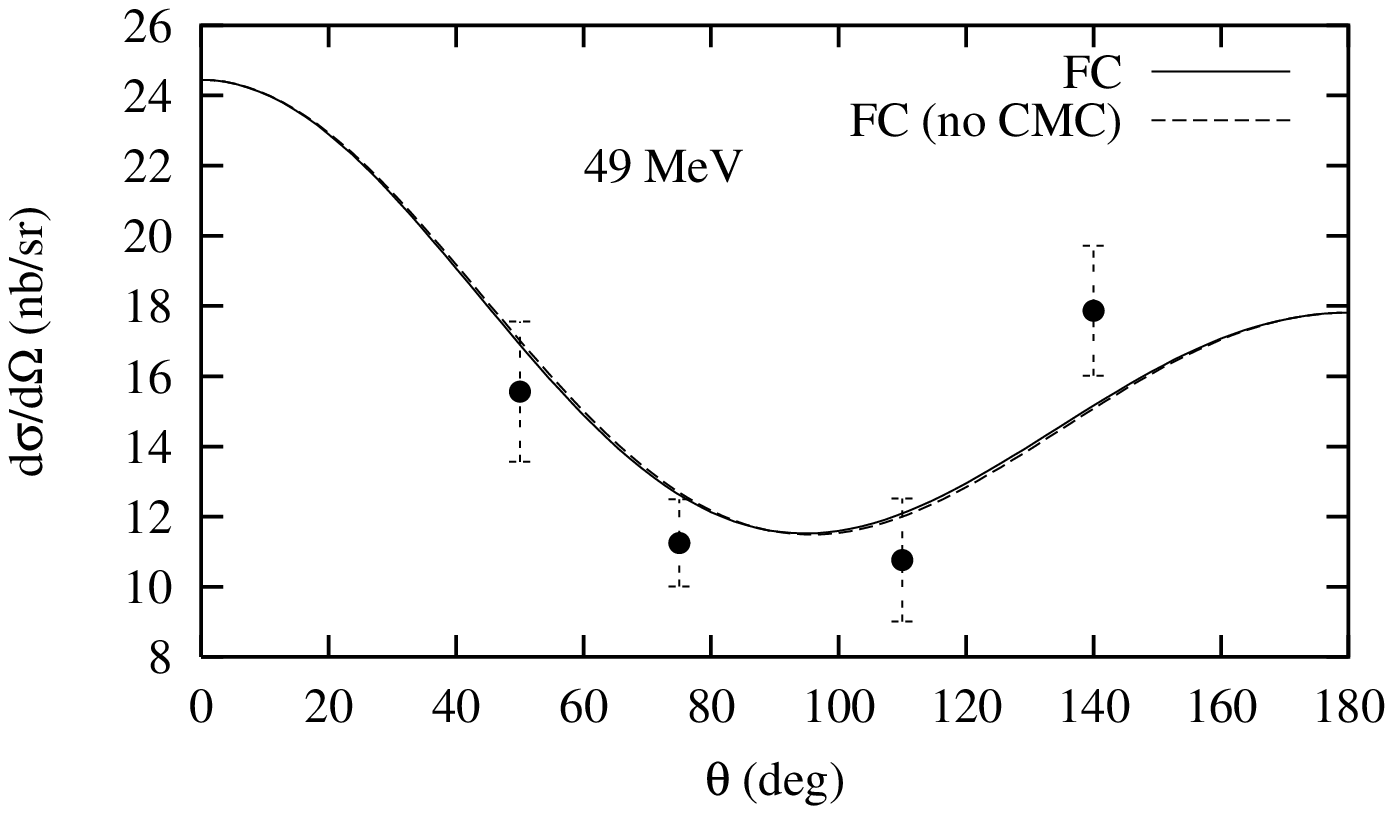}
\epsfig{file=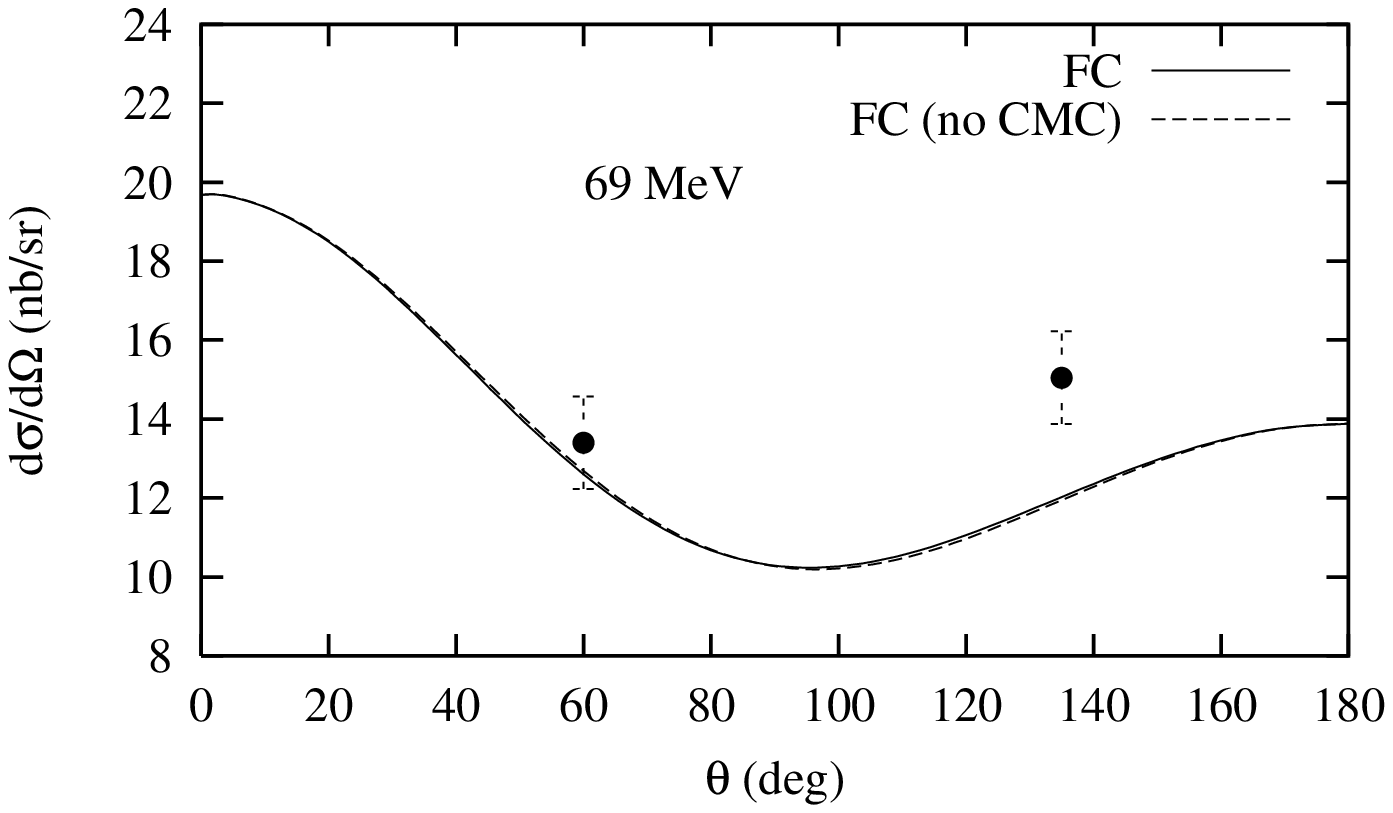}
\caption[Effect of neglecting recoil  terms on the full differential cross-section at 49, 69, and 95 MeV]{ 
\label{fig:e3-11} Effect of neglecting recoil  terms (CMC) on the full differential cross-section (FC) at 49, 69, and 95 MeV.}
\end{figure}

The recoil (CMC) corrections are difficult to see on the graphs (Figure~\ref{fig:e3-11}), and this is also shown in the table.  The largest errors
are actually at the lowest energies, indicating that these terms are of order $\omega^1$.  The   small values for $\Delta\alpha_n$
at 95 MeV (especially at forward angles) indicate that these terms can be neglected here.

\begin{figure}
\centering
\epsfig{file=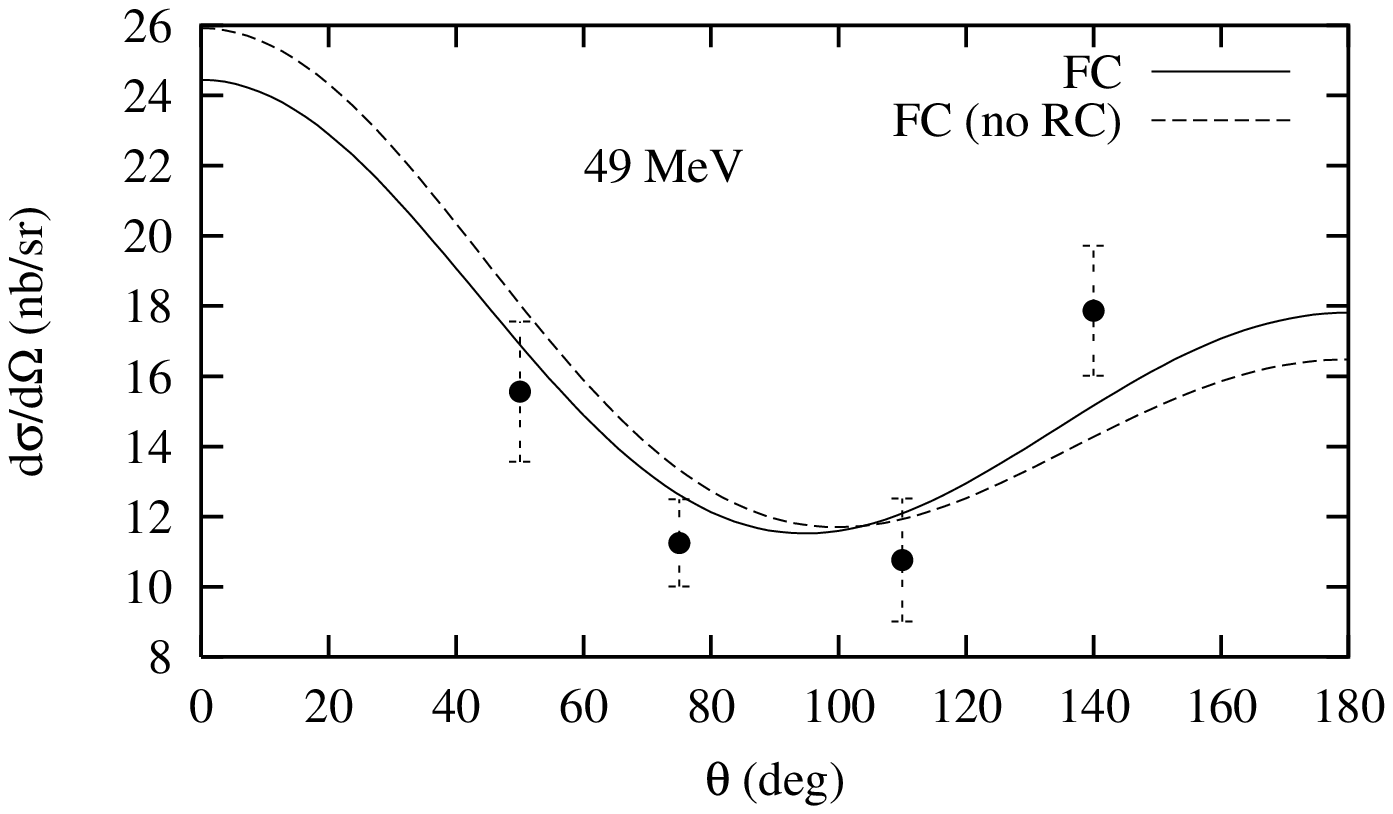}
\epsfig{file=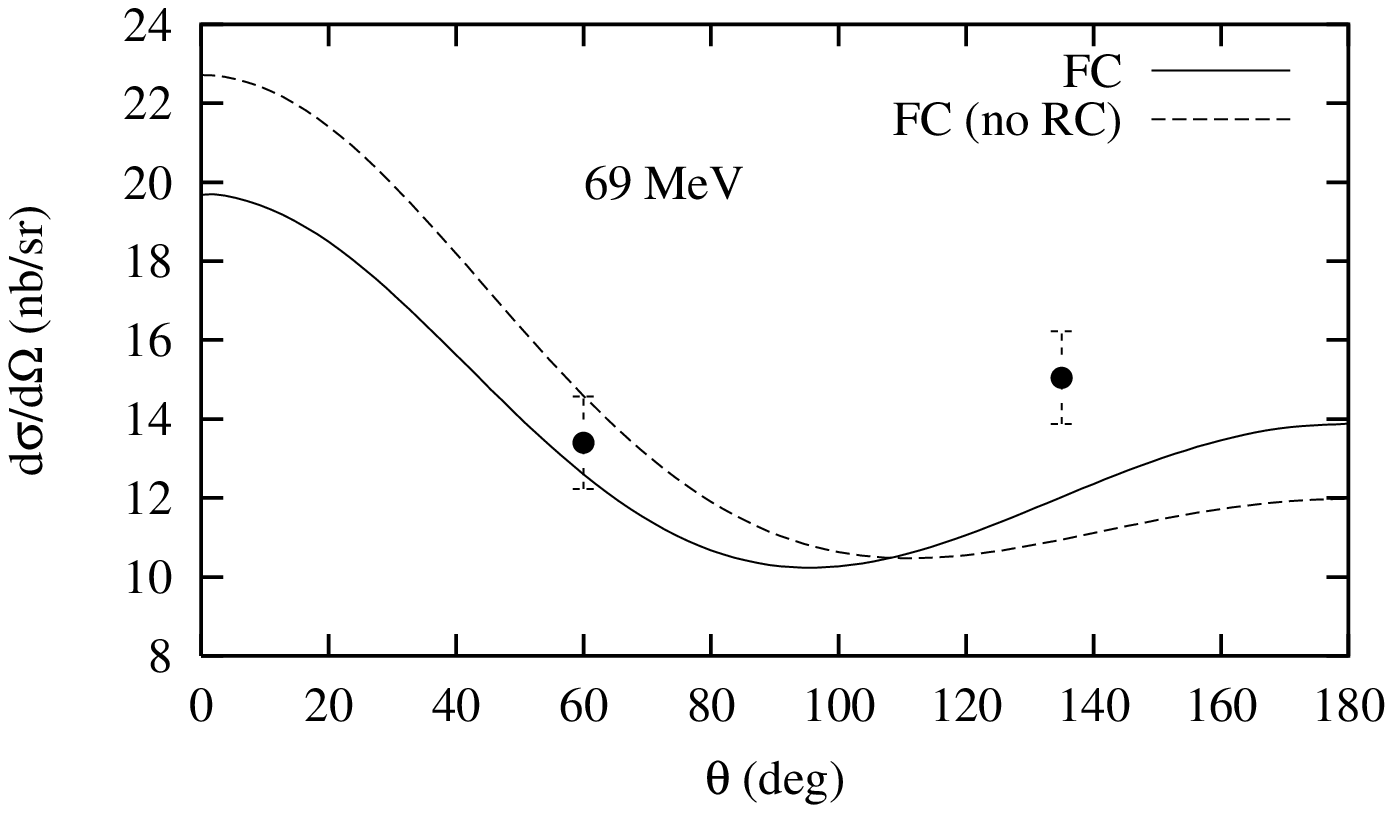}
\caption[Effect of neglecting relativistic terms on the full differential cross-section  at 49, 69, and 95 MeV]{ 
\label{fig:e3-12} Effect of neglecting relativistic  terms (RC) on the full differential cross-section (FC) at 49, 69, and 95 MeV.}
\end{figure}

The relativistic corrections (RC) (Figure~\ref{fig:e3-12}) are not a ``small'' correction: 
even at 49 MeV they would have to be included in a calculation
to make an accurate determination of  $\alpha_n$.  The amount of error that would be introduced by neglecting
these terms is comparable to $\alpha_n$ itself at some angles. They also scale like $\omega^2$ so the energy dependence should not be very strong.
This is the case at backward angles, but at forward angles this effect increases with energy, perhaps due to interference
with another $\omega^2$ term.  Again, this effect is smallest around $110^{\circ}$  at each energy.

\begin{figure}
\centering
\epsfig{file=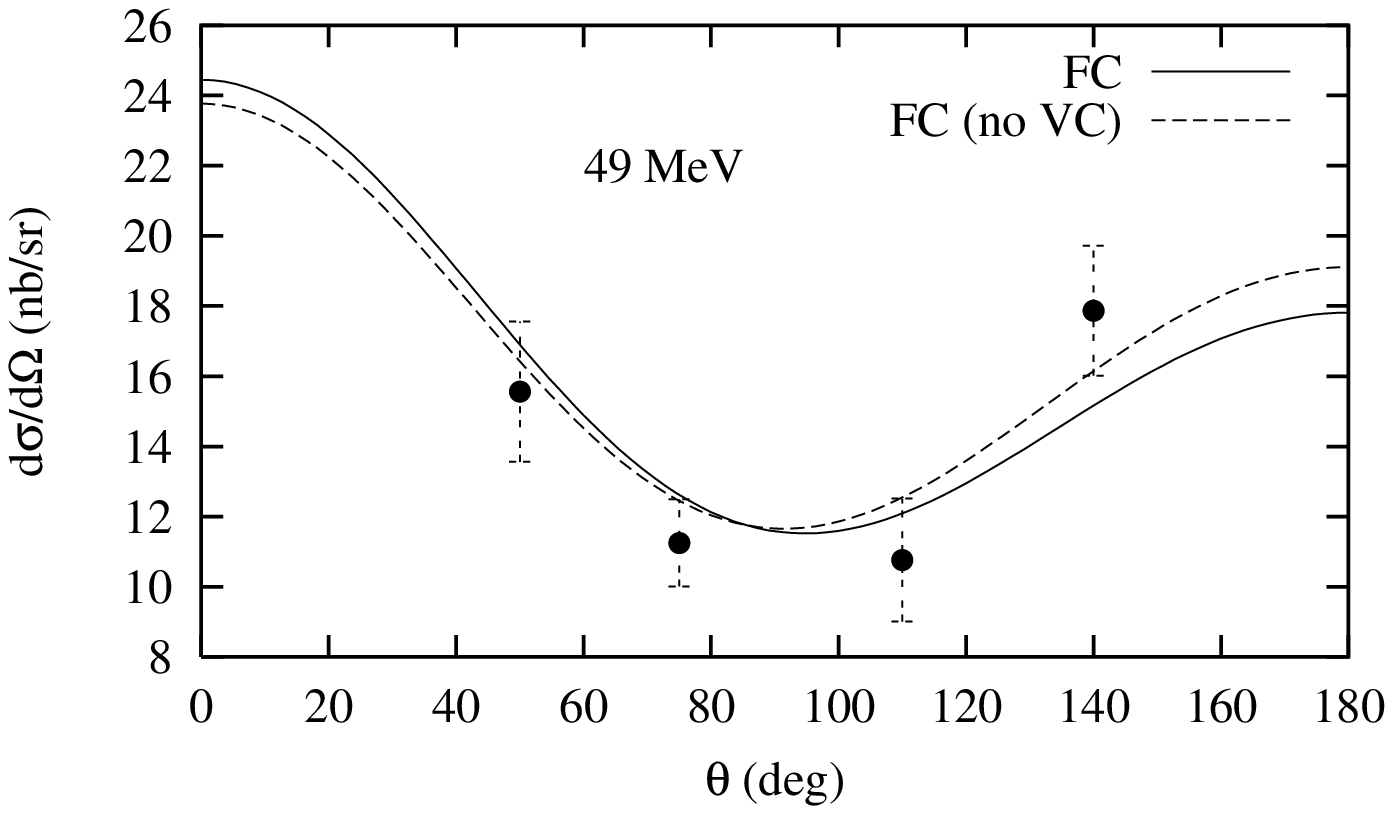}
\epsfig{file=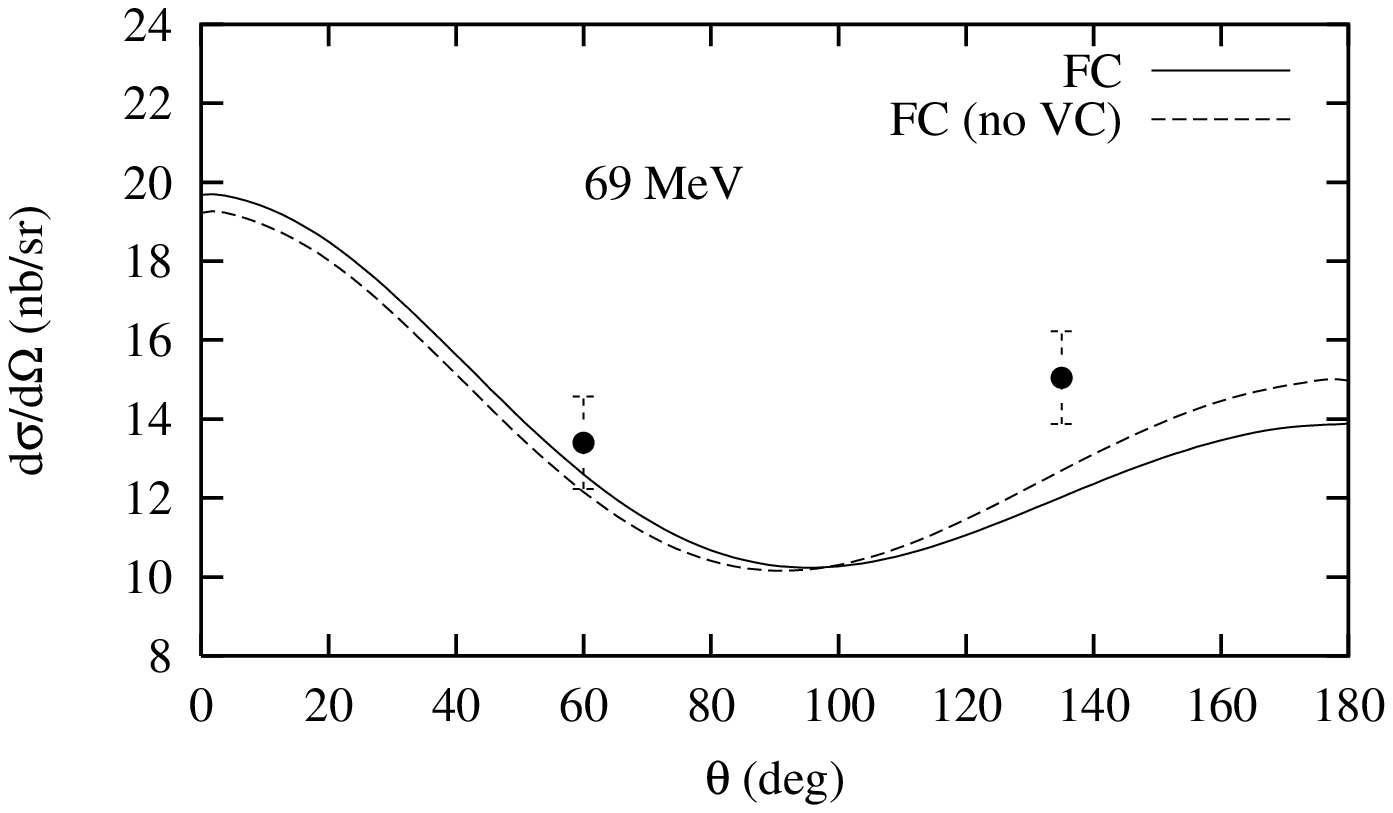}
\caption[ Effect of neglecting vertex corrections  on the full differential cross-section at 49, 69, and 95 MeV.]{ 
\label{fig:e3-13} Effect of neglecting vertex corrections (VC) on the full differential cross-section (FC) at 49, 69, and 95 MeV.}
\end{figure}

We turn to the vertex corrections.  These terms would also need to be included for an accurate estimate of the polarizability.
Their effect is actually greater than the effects of all of the other pion terms combined.   The errors decrease as energy increases, but even
at 95 MeV the errors, particularly at middle angles, are at least as large as previous errors in polarizability determinations.  
At backward angles the effect seems to be independent of energy. The graphs are shown in Figure~\ref{fig:e3-13}.

\begin{figure}
\centering
\epsfig{file=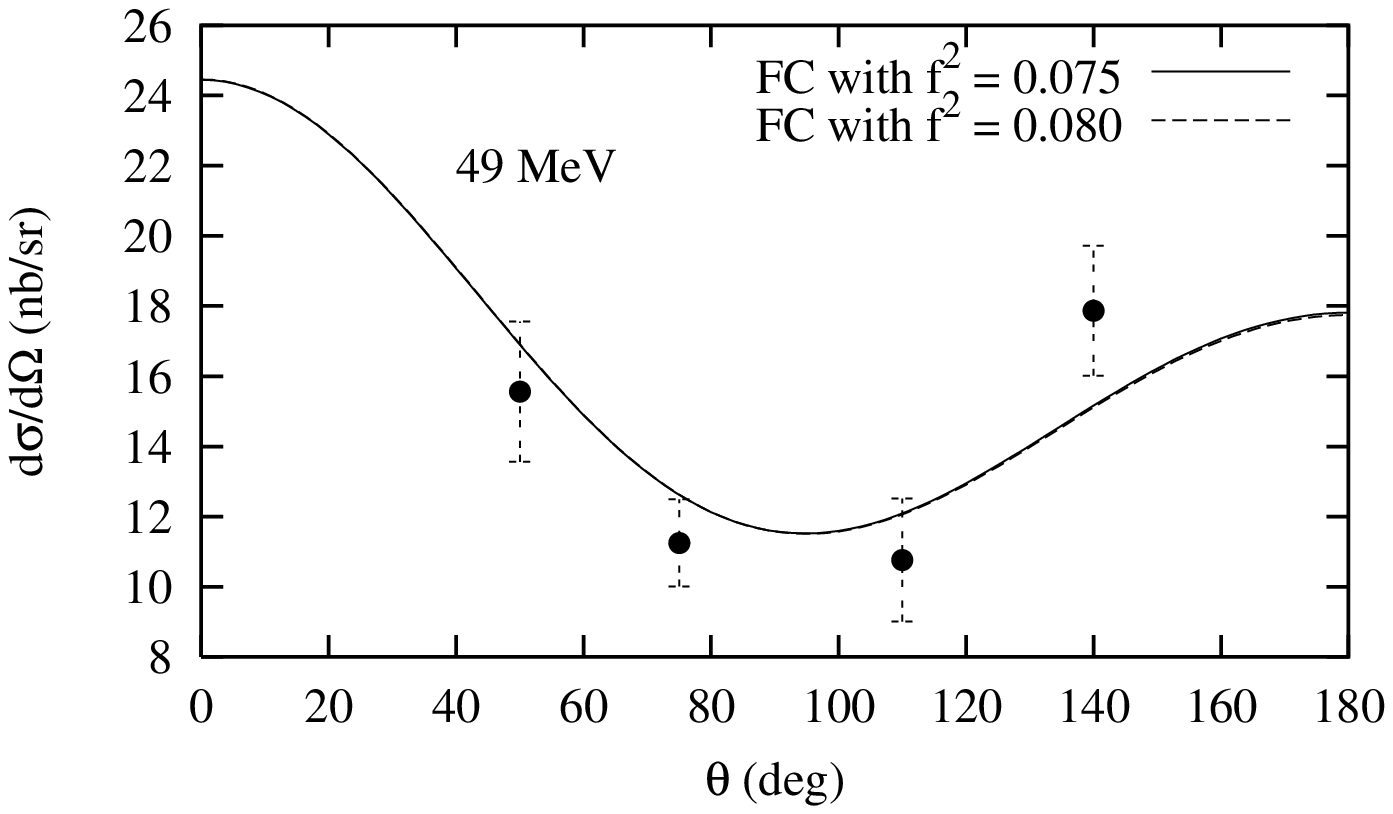}
\epsfig{file=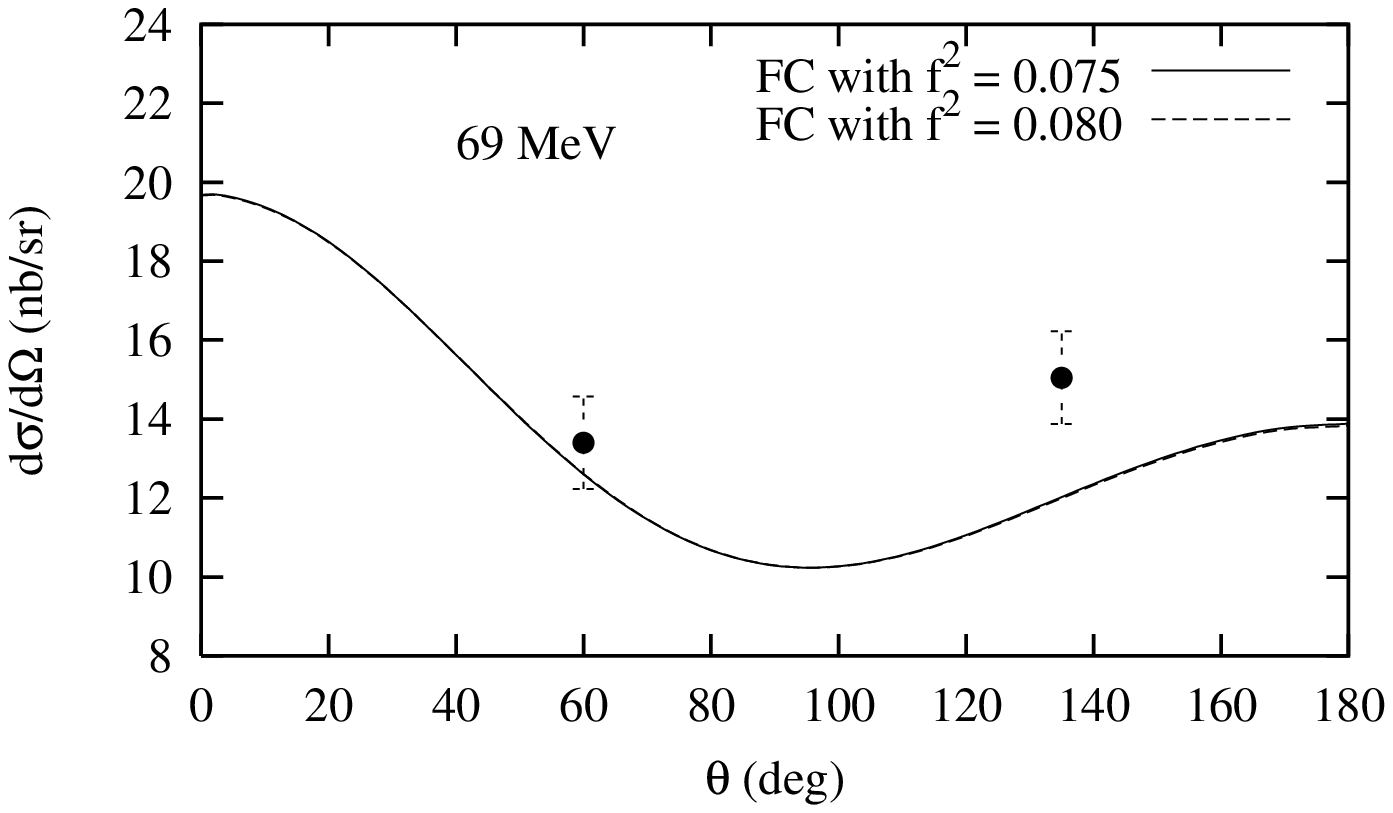}
\caption{ \label{fig:e3-14} Effect of changing $f^2$ on the full differential cross-section  at 49, 69, and 95 MeV.}
\end{figure}

Finally, we make an estimate of the effect of the uncertainty in the $\pi N$ coupling constant $f$.  We have been using 
the value $f^2/(\hbar c) = 0.075 $, but the uncertainty in this makes a value of 0.08 also reasonable.
The effect of this is not large, as seen in Figure~\ref{fig:e3-14}),
but it can be seen to have a minor effect at higher energies in the middle angles. 

\begin{table}
\caption[ Absolute error in electric polarizability generated by omitting various terms at 49, 69, and 95 MeV as 
a function of angle.]{ \label{t:e3-1} Absolute error in electric polarizability generated by omitting various terms - pion ($\pi$),
recoil (CMC),  relativistic (RC), and vertex (VC) corrections - at 49, 69, and 95 MeV as 
a function of angle. The dependence of the cross-section on the value of $f^2$ is also examined.} 
\begin{tabular}{c|ccc|ccc}
	\multicolumn{1}{r}{}&
	\multicolumn{3}{c}{$\pi$}&
	\multicolumn{3}{c}{CMC}\\
\underline{$\theta$(deg)} & \underline{49 MeV} & \underline{69 MeV} & \underline{95 MeV} &
\underline{49 MeV} & \underline{69 MeV} & \underline{95 MeV} \\ 
&&&&&&\\
0 & -2.9 & -2.6 & -2.0 & 0.0 & 0.0 & 0.0  \\
10 & -2.9 & -2.6 & -2.0 & -0.1 & -0.0 & -0.0 \\
20 & -2.8 & -2.5 & -2.0 & -0.2 & -0.1 & -0.1 \\
30 & -2.7 & -2.4 & -1.9  & -0.5 & -0.2 & -0.1  \\
40 & -2.5 & -2.4 & -1.8 & -0.8 & -0.4 & -0.2 \\
50 & -2.3 & -2.1 & -1.7& -1.0 & -0.5 & -0.2 \\
60 & -2.1 & -2.0 & -1.6 & -1.1 & -0.5 & -0.2  \\
70 & -1.8 & -1.8 & -1.6& -1.0 & -0.4 & -0.1 \\
80 & -1.6 & -1.6 & -1.7 & -0.6 & -0.2 & 0.0  \\
90 & -1.2 & -1.4 & -1.7 & 0.1 & 0.2 & 0.3   \\
100 & -0.9 & -1.1 & -1.6 & 0.8 & 0.5 & 0.4 \\
110 & -0.4 & -0.7 & -1.3 & 1.2 & 0.7 & 0.5 \\
120 & 0.0 & -0.3 & -0.9 & 1.4 & 0.8 & 0.5  \\
130 & 0.4 & 0.1 & -0.4 & 1.2 & 0.7 & 0.4 \\
140 & 0.8 & 0.6 & 0.1 & 0.9 & 0.5 & 0.3   \\
150 & 1.1 & 0.9 & 0.5 & 0.6 & 0.3 & 0.2 \\
160 & 1.3 & 1.2 & 0.9 & 0.2 & 0.1 & 0.1 \\
170 & 1.4 & 1.3 & 1.1 & 0.1 & 0.0 & 0.0 \\
180 & 1.5 & 1.4 & 1.1  & 0.0 & 0.0 & 0.0 \\
\end{tabular}

\end{table}
\clearpage

\begin{table}
\begin{tabular}{c|ccc|ccc}
        \multicolumn{1}{r}{}&
        \multicolumn{3}{c}{RC}&
        \multicolumn{3}{c}{VC} \\
\underline{$\theta$(deg)} & \underline{49 MeV} & \underline{69 MeV} & \underline{95 MeV} &
\underline{49 MeV} & \underline{69 MeV} & \underline{95 MeV} \\
&&&&&&\\
0 & -7.9 & -9.2 & -11.9 & 3.7 & 1.4 & 0.9  \\
10 & -8.0 & -9.3 & -12.0 & 3.7 & 1.5 & 1.0  \\
20 & -8.3 & -9.7 & -12.5 & 3.8 & 1.6 & 1.2 \\
30 & -8.7 & -10.2 & -13.2 & 3.9 & 1.9 & 1.7  \\
40  & -9.2 & -10.8 & -14.2 & 4.0 & 2.2 & 2.3  \\
50  & -9.6 & -11.5 & -15.2 & 4.0 & 2.5 & 3.2  \\
60  & -9.7 & -11.8 & -15.8 & 3.7 & 2.7 & 4.1  \\
70  & -9.2 & -11.5 & -15.5 & 2.8 & 2.7 & 5.0   \\
80  & -7.7 & -10.0 & -13.7 & 1.2 & 2.2 & 5.4  \\
90   & -4.9 & -7.1 & -10.1 & -1.1 & 1.2 & 5.0   \\
100  & -1.4 & -3.3 & -5.4 & -3.7 & -0.3 & 3.9  \\
110  & 2.1 & 0.7 & -0.7 & -6.1 & -1.9 & 2.4 \\
120   & 5.1 & 4.1 & 3.4 & -8.0 & -3.4 & 0.9  \\
130  & 7.4 & 6.7 & 6.6 & -9.2 & -4.5 & -0.5  \\
140  & 9.0 & 8.5 & 8.8 & -10.1 & -5.3 & -1.6  \\
150  & 10.0 & 9.8 & 10.3 & -10.6 & -5.9 & -2.4  \\
160  & 10.7 & 10.6 & 11.3 & -10.9 & -6.3 & -2.9  \\
170  & 11.0 & 11.0 & 11.8 & -11.0 & -6.5 & -3.2  \\
180  & 11.1 & 11.1 & 12.0 & -11.1 & -6.6 & -3.3  \\
\end{tabular}

\end{table}
\clearpage

\begin{table}
\begin{tabular}{c|ccc}
        \multicolumn{1}{r}{}&
        \multicolumn{3}{c}{$f^2$} \\
\underline{$\theta$(deg)} & \underline{49 MeV} & \underline{69 MeV} & \underline{95 MeV}\\
&&&\\
0 & -0.1 & 0.1 & 0.0 \\
10  & -0.1 & 0.1 & 0.0 \\
20  & -0.1 & 0.1 & 0.0 \\
30  & -0.1 & 0.0 & 0.0 \\
40  & -0.1 & 0.0 & 0.0  \\
50 & -0.1 & 0.0 & -0.1 \\
60  & -0.1 & -0.1 & -0.2 \\
70  & -0.1 & -0.1 &  -0.3 \\
80 & 0.0 & -0.1 & -0.3 \\
90  & 0.1 & 0.0 & -0.3 \\
100  & 0.2 & 0.0 & -0.2  \\
110 & 0.4 & 0.1 & -0.2 \\
120  & 0.5 & 0.2 & -0.1  \\
130 & 0.5 & 0.2 & 0.0  \\
140  & 0.5 & 0.2 & 0.0  \\
150  & 0.6 & 0.2 & 0.0 \\
160 & 0.6 & 0.3 & 0.0  \\
170  & 0.6 & 0.3 & 0.0   \\
180  & 0.6 & 0.3 & 0.0 \\
\end{tabular}

\end{table}
\clearpage

\clearpage

\begin{figure}
\centering
\epsfig{file=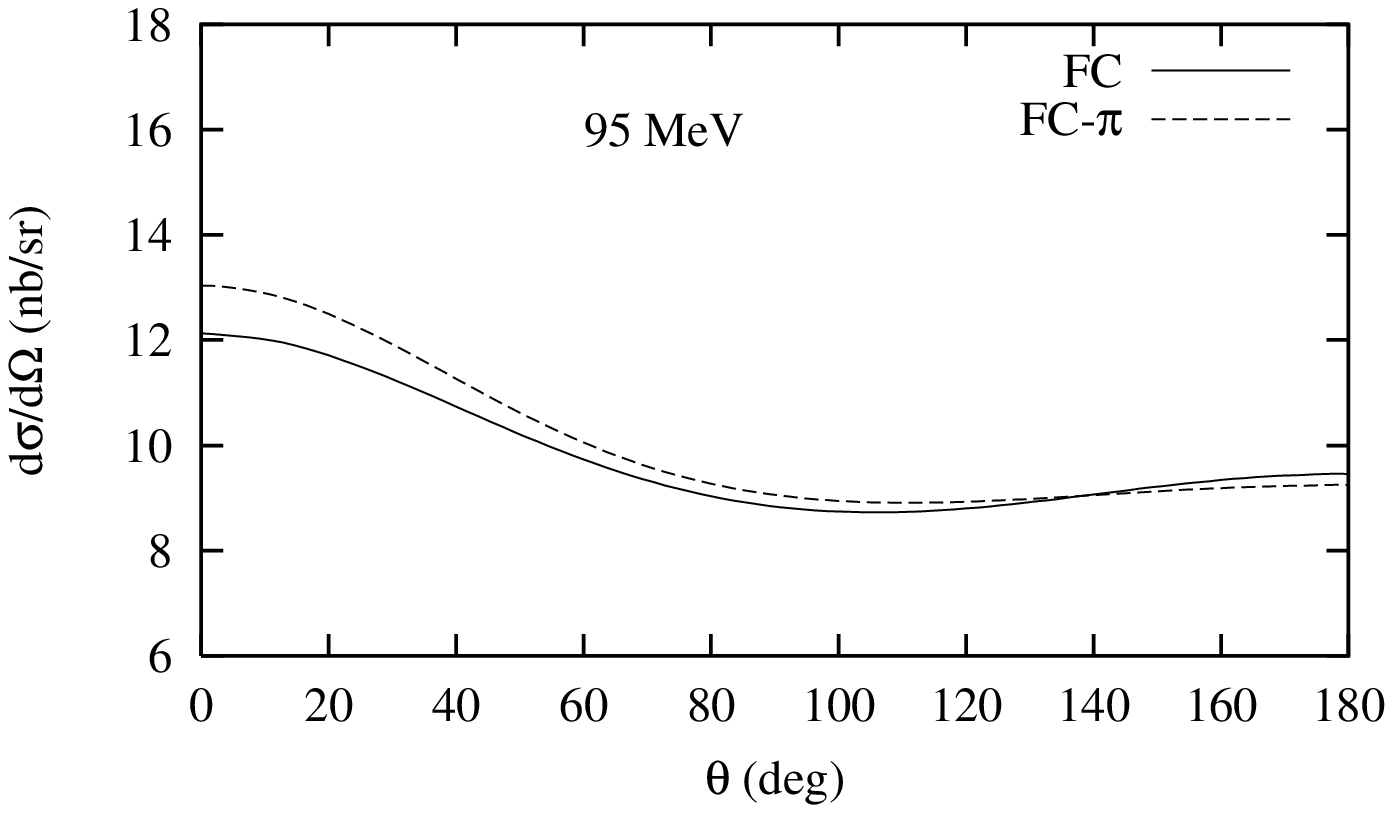}
\end{figure}
\clearpage

\clearpage

\begin{figure}
\centering
\epsfig{file=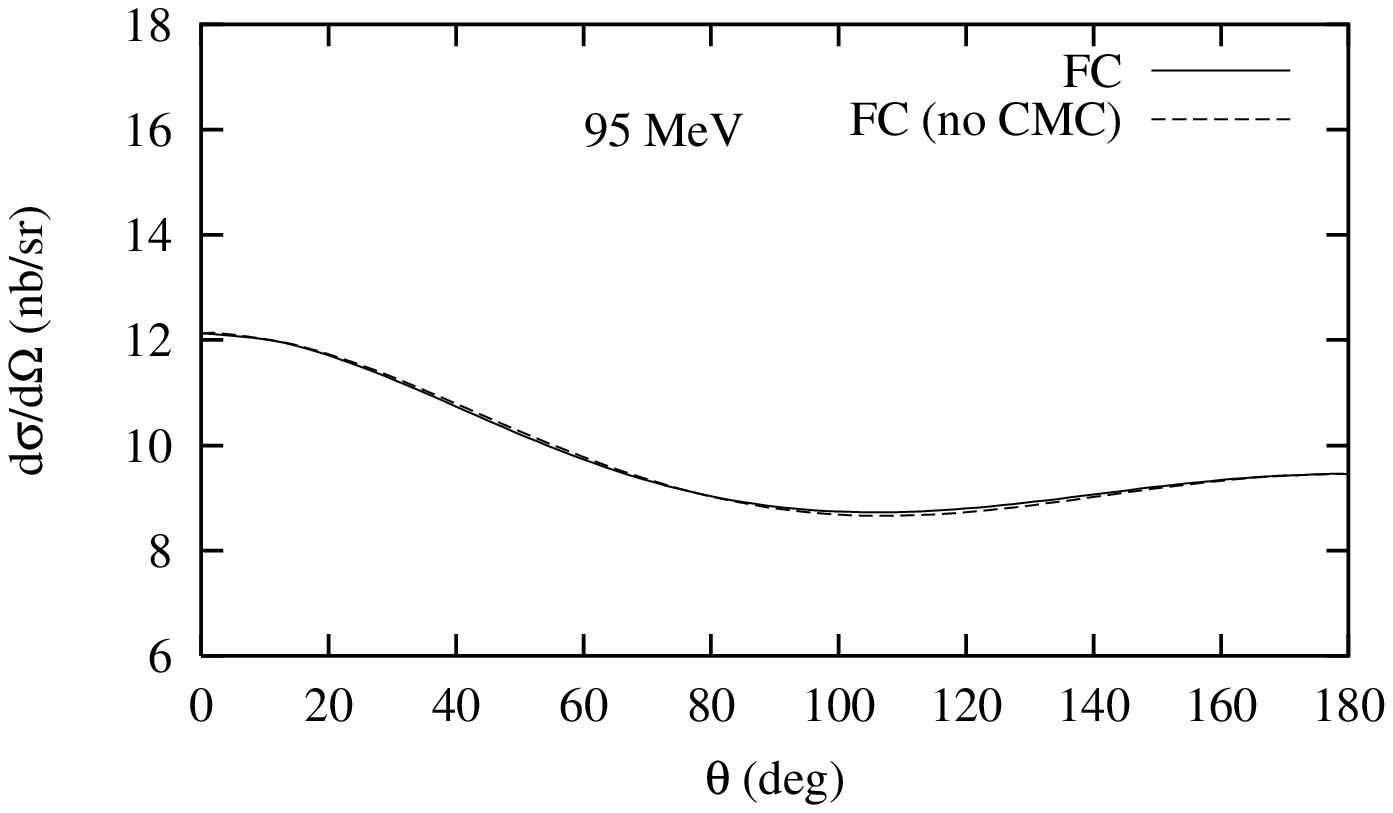}
\end{figure}
\clearpage

\clearpage

\begin{figure}
\centering
\epsfig{file=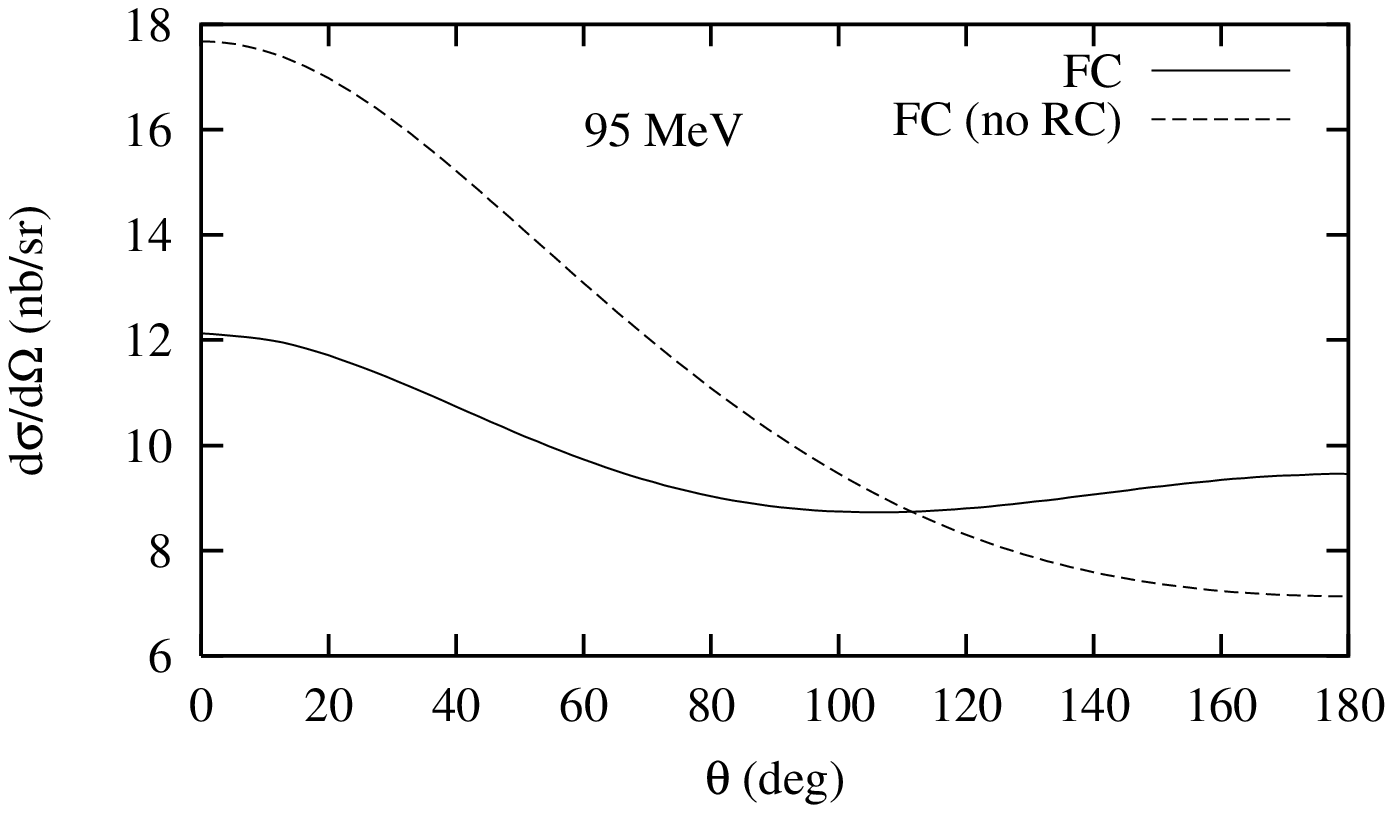}
\end{figure}
\clearpage

\clearpage

\begin{figure}
\centering
\epsfig{file=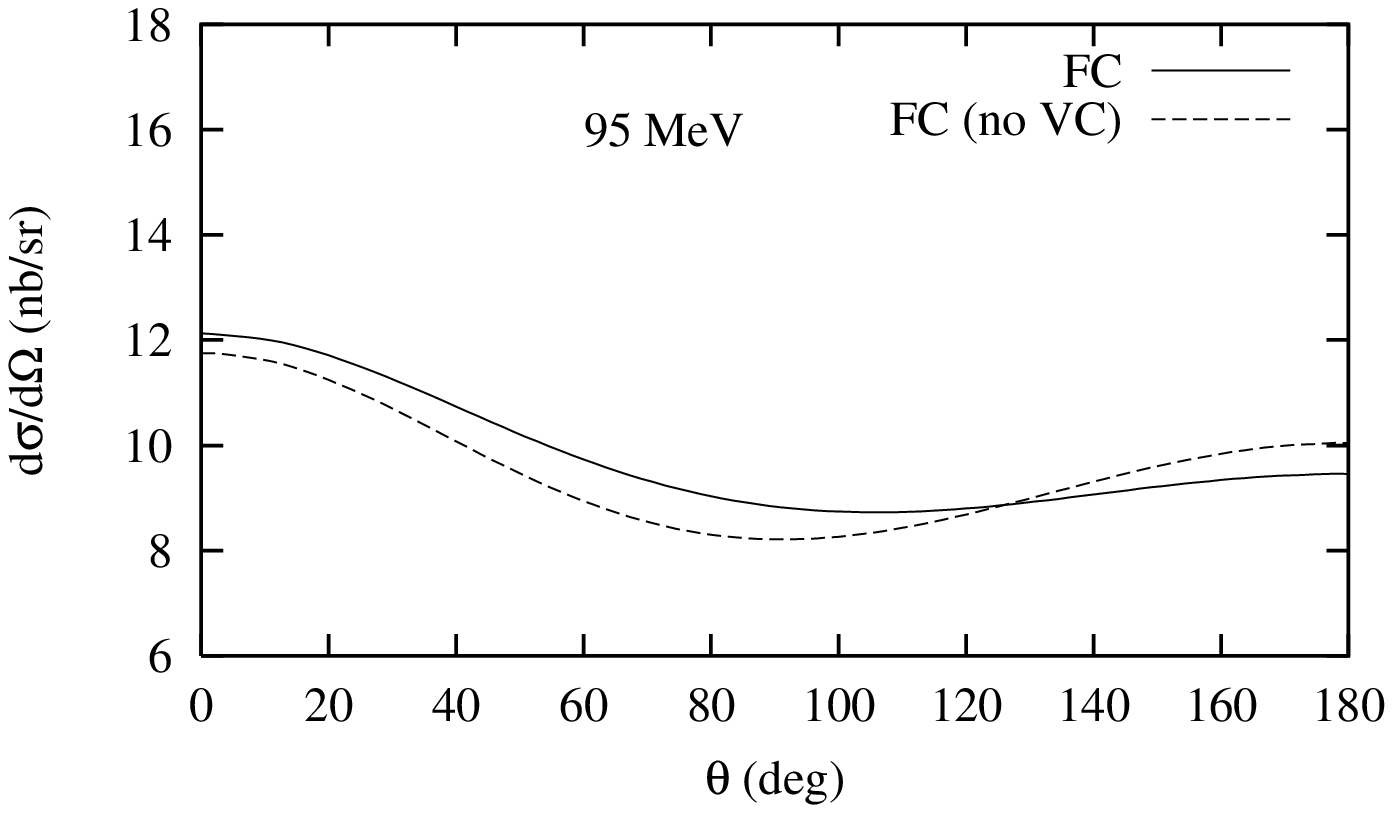}
\end{figure}
\clearpage

\clearpage

\begin{figure}
\centering
\epsfig{file=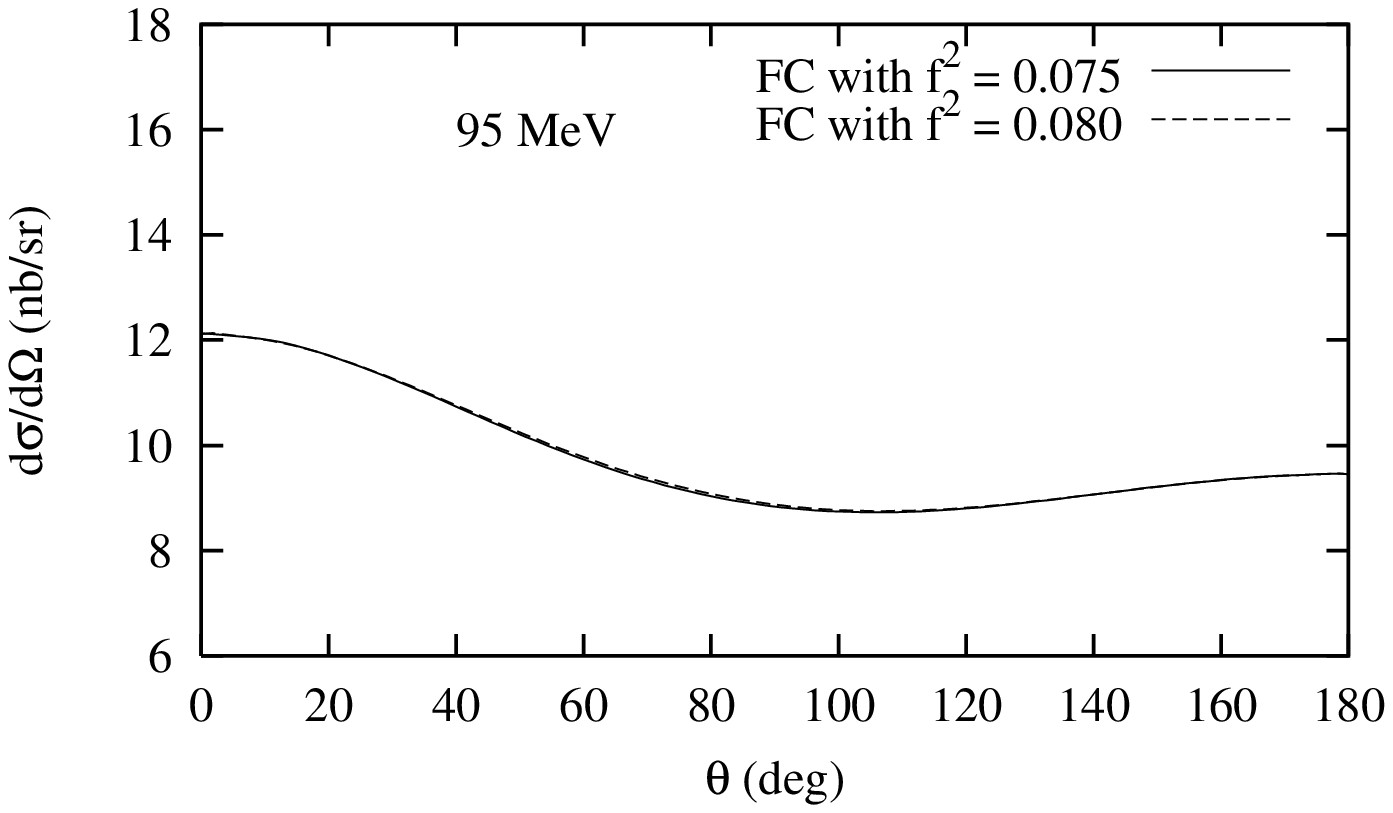}
\end{figure}

\section{Tensor-Polarized Deuteron Compton Scattering}
Another interesting observable is the tensor-polarized deuteron Compton scattering cross-section. It is
defined by  \cite{chen98}
\begin{equation}
\left( \frac{d\sigma}{d\Omega} \right)_T = \frac{1}{4} \left[ 2 \frac{d\sigma}{d\Omega} (M_i = 0) - 
\frac{d\sigma}{d\Omega} (M_i = 1) - \frac{d\sigma}{d\Omega} (M_i = -1) \right] .
\end{equation} 
This observable has not been well-studied; a recent calculation in effective field theory can be found in \cite{chen98}.
Although this cross-section does indeed have a slight dependence on the polarizabilities,  we do not expect this
to give a measurement of $\alpha$ and $\beta$.   However, it is easy to calculate within our 
formalism, and  we hope that our predictions will be of use  
to others interested in this quantity.

\begin{figure}
\centering
\epsfig{file=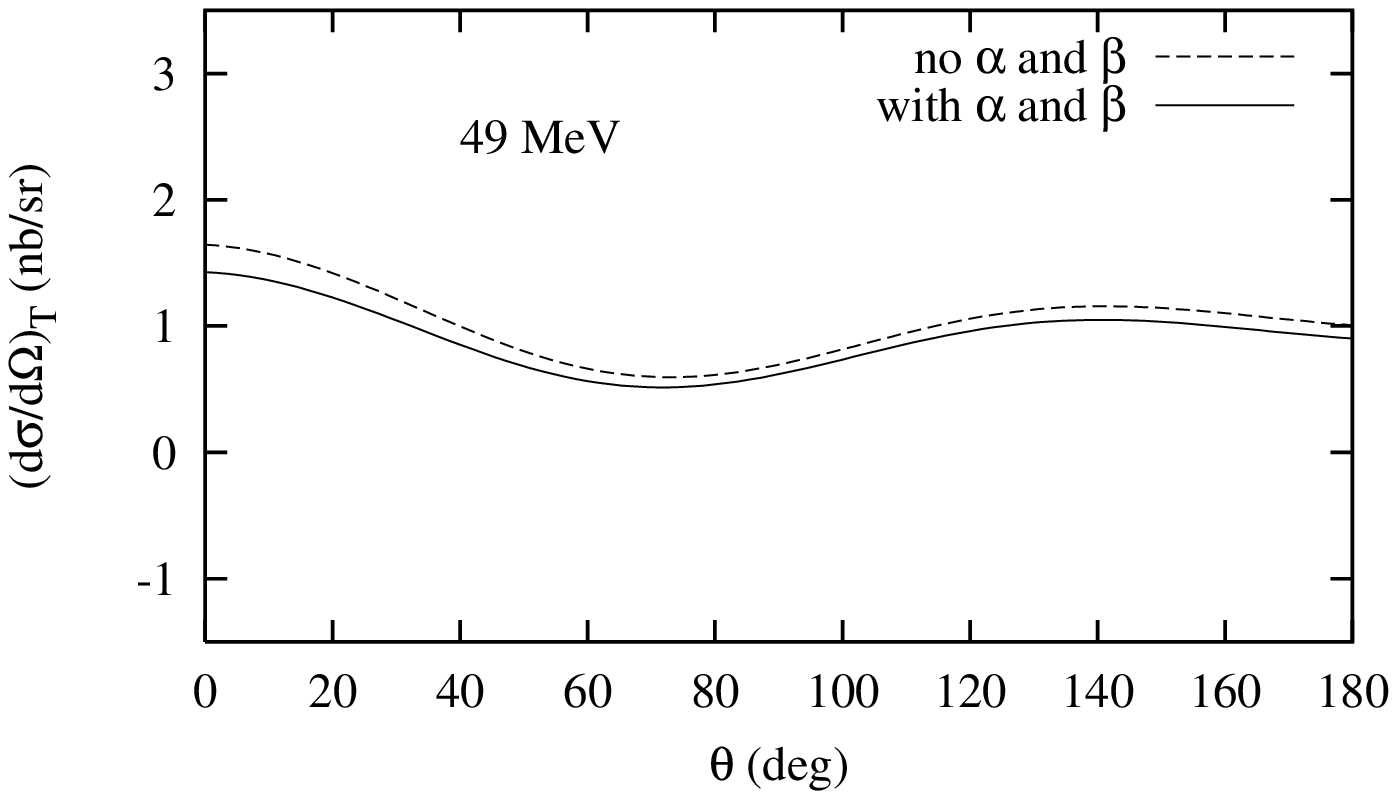}
\epsfig{file=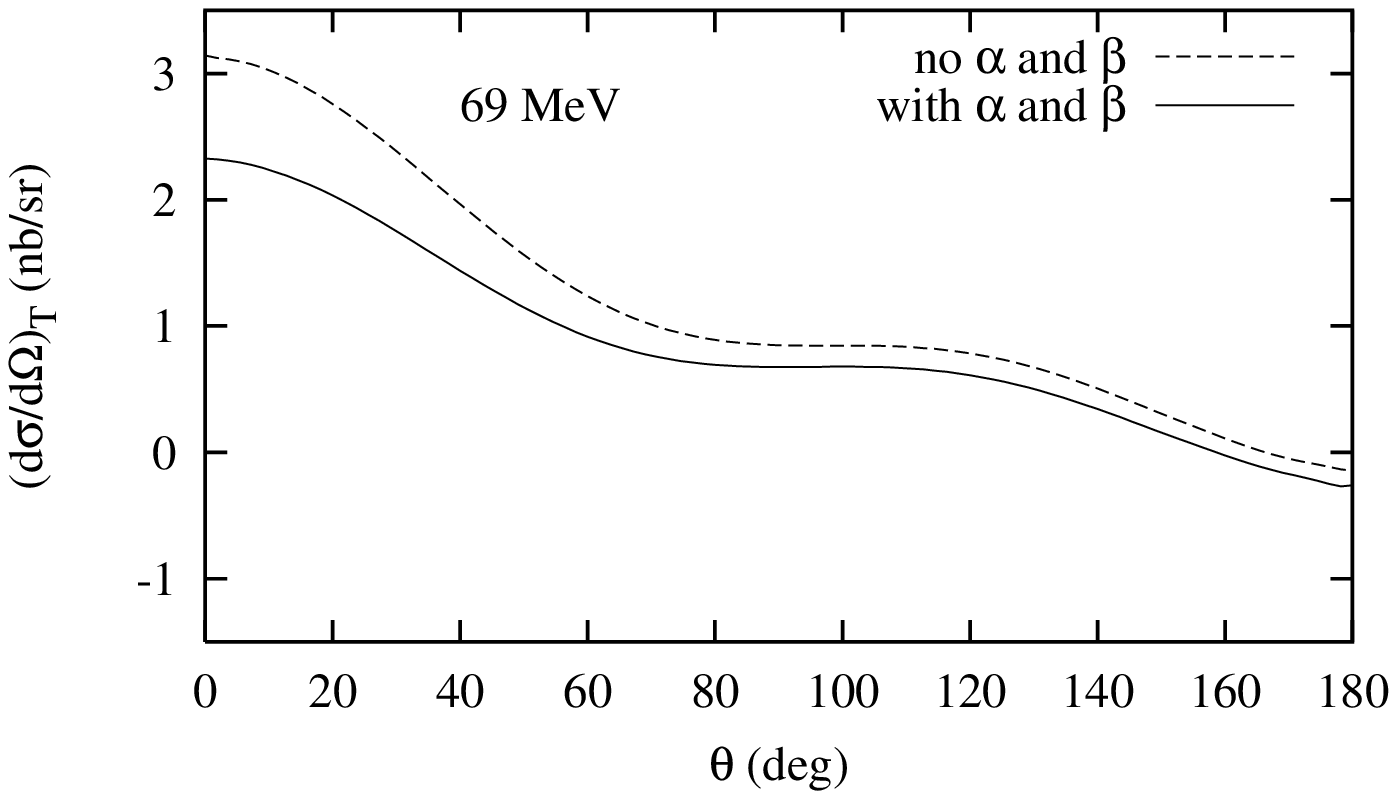}
\caption[Tensor-polarized deuteron Compton scattering cross-section, with and without polarizabilities]{Tensor-polarized deuteron 
Compton scattering cross-section, with and without polarizabilities.  The values of $\alpha_n$ = 12.0, $\beta_n$ = 2.0, $\alpha_p$ = 10.9,
and $\beta_p$ = 3.3 are used in the solid curve.  Both curves include all other interactions described in the Appendices  \label{fig:e4-1} }
\end{figure}

Figure~\ref{fig:e4-1} shows the total polarized cross-section for two different energies, both with and without the polarizabilities.
These results are similar to the unpolarized case in that the effect of $\alpha$ and $\beta$ is greater at forward angles and at higher
energies.  However, the absolute magnitude of this effect is at most 10\% of its magnitude in the unpolarized case.   Moreover, the 
polarized cross-section is much more sensitive to individual terms, as we will show shortly.   It does not seem that this quantity could produce
an accurate measurement of the polarizabilities, but since no experiment has been performed yet, we must wait for this to
be done to say anything for certain.   

In an effort to be more consistent with the calculation of \cite{chen98}, we now try to isolate the contributions of the magnetic interactions
and the pion-exchange terms at one of these energies.  The pion-exchange terms include any pions that give rise to the tensor force (and 
the $l=2$ state)  in  the deuteron.    However, we will include only one pion exchange in the pion terms and no pion exchanges
in the magnetic terms.  Therefore, the $D$-state deuteron wavefunction $u_2(r)$ will be set to zero in all terms except for the Thomson term;
the $D$-state contributions to this term will be grouped with the pion-exchange terms.    

The results for the magnetic interactions (MI)  are shown in Figure~\ref{fig:e4-2}. These interactions include all of the terms that are calculated
in Appendices F and G, as well as the seagull term (as defined in Appendix A) without the deuteron $D$-state.    
In addition, we show the effect of including the deuteron $D$-state in the calculation of the magnetic terms (but not the
seagull term).  This is a non-negligible effect.  Finally, 
the relativistic correction (RC) is added (including the $D$-state).  Since this is a spin-orbit effect,
we group this with the magnetic interactions.  Both the relativistic term and the deuteron $D$-states have a large
effect in the calculation;  this observable is very sensitive to these individual effects.

Figure~\ref{fig:e4-3} shows the pion terms.  This includes all terms in  Appendix J, both the gauge invariant set and
the vertex corrections, along with the $D$-state contributions of the seagull term.  We also show the effect of including
the deuteron $D$-state in the pion terms of Appendix J.  This effect is not a large as for the magnetic interaction case.

\begin{figure}
\centering
\epsfig{file=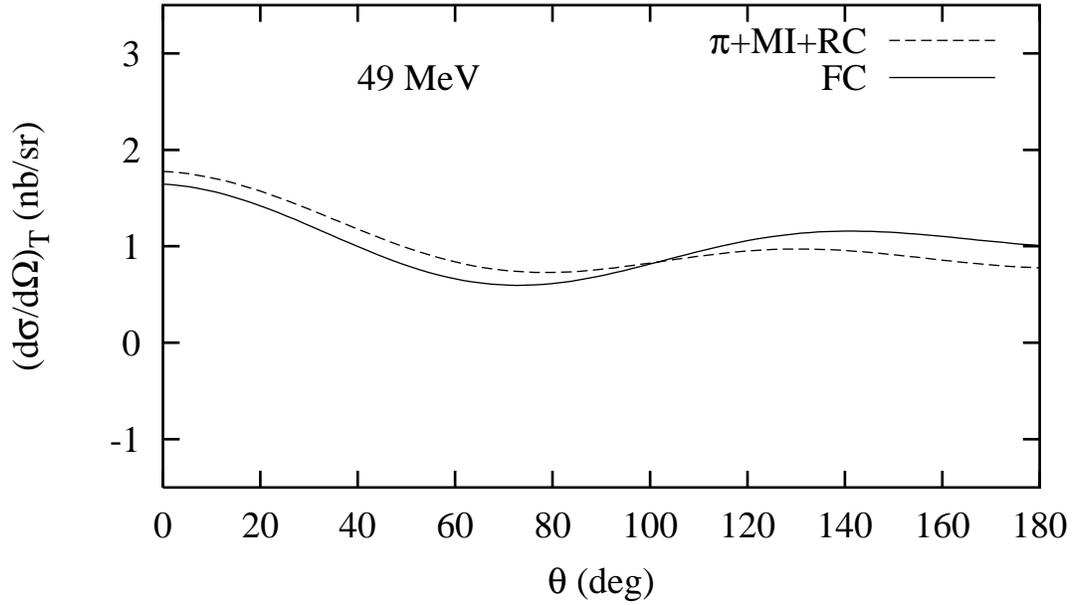}
\caption[Effect of interference terms in tensor-polarized deuteron Compton scattering cross-section]{
Effect of interference terms between the pion and magnetic terms in tensor-polarized deuteron Compton scattering cross-section.
The dashed curve is the sum of the total cross-sections from the pion and magnetic terms individually, while
the solid curve is the total cross-section including both the pion and magnetic terms (but not the polarizabilities).
All deuteron $D$-states are included.  The difference between the curves is the interference between the pion and magnetic
interactions.  \label{fig:e4-4} }
\end{figure}

Finally, we would like to estimate the magnitude of the interference of the pion and magnetic terms with each other.  
The solid curves of Figures~\ref{fig:e4-2} and \ref{fig:e4-3} show the interference of the magnetic and pion
terms with the $S$-state seagull term (as well as with themselves), since the $S$-state seagull term alone gives no contribution to the tensor-polarized
cross-section.   The dashed curve in Figure~\ref{fig:e4-4} shows the sum of these two curves.  In addition, the full calculation (not
including polarizabilities) is shown as the solid curve.  The difference between these curves  is the interference of the
magnetic and pion interactions with each other.  This is seen to be a minor effect, but is largest at forward and backward angles.

\clearpage

\begin{figure}
\centering
\epsfig{file=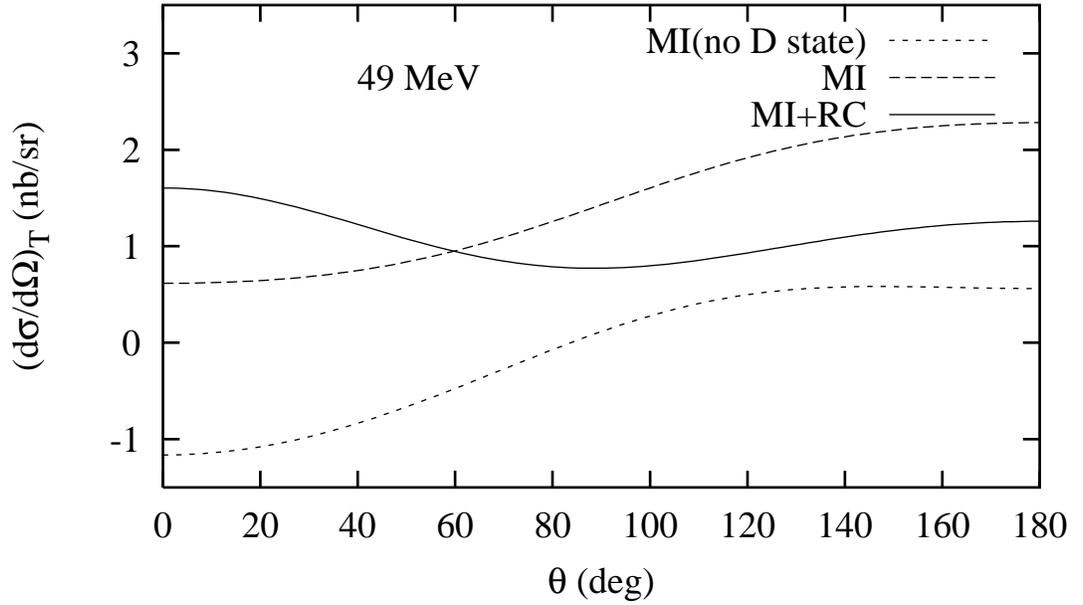}
\caption[Tensor-polarized deuteron Compton scattering cross-section, including only magnetic interactions]{ Tensor-polarized deuteron 
Compton scattering cross-section, including only magnetic interactions(MI).  The dotted curve contains all interactions in Appendices F and G,
as well as the $S$-state seagull term, but only the contributions from the deuteron $S$-state are included.  These contributions
are added in the dashed curve (but still only the $S$-state is included in the seagull term).  Relativistic corrections (RC) are added
in the solid curve. \label{fig:e4-2} }
\end{figure}
\clearpage

\begin{figure}
\centering
\epsfig{file=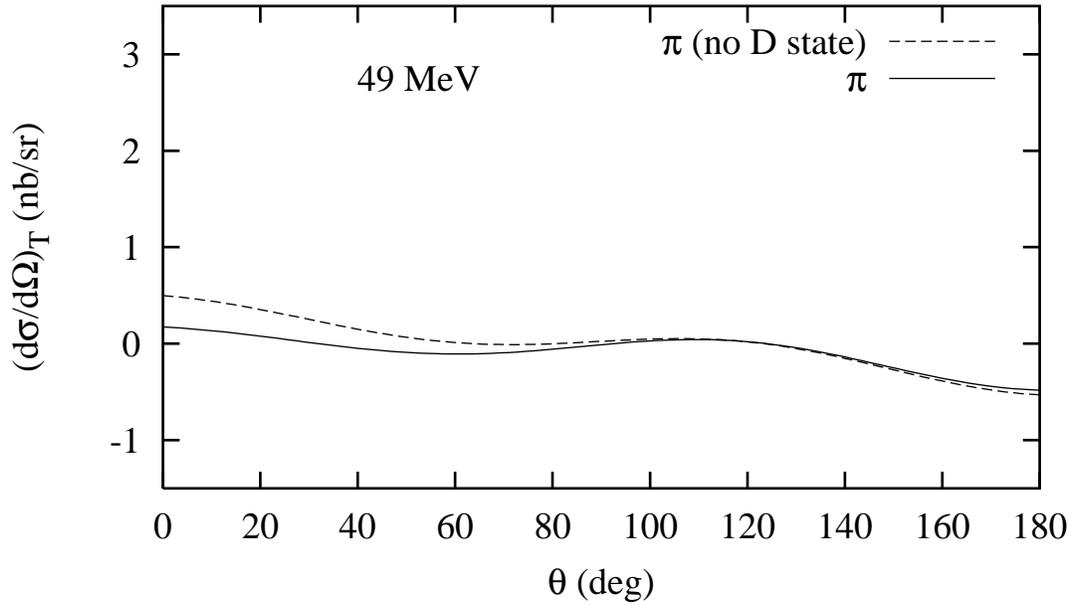}
\caption[Tensor-polarized deuteron Compton scattering cross-section, including only pion interactions]{ 
Tensor-polarized deuteron Compton scattering cross-section, including only pion interactions ($\pi$).  These consist of 
the gauge-invariant set of terms described in Section 3.6, as well as the vertex corrections and contributions to the
seagull term from the deuteron $D$-state.  The dashed curve includes only the deuteron $S$-state in the
gauge-invariant pion terms and the vertex corrections, while the solid curve includes both the $S$- and $D$-states.
\label{fig:e4-3} }
\end{figure} 
\clearpage

\chapter{Conclusion \label{ch:conc} }
We have presented a calculation of deuteron Compton scattering valid for energies less than 100 MeV.
All terms that we believe to be important are included in this formulation.  The seagull, lowest-order
multipoles, and polarizability effects dominate the cross-section.  Relativistic corrections must also 
be incorporated in the calculation.  Although smaller than the preceding terms, pion-exchange and 
vertex corrections are also needed to accurately determine the polarizabilities.  We conclude that
recoil effects and multipoles of second-order and higher are negligible in this energy range.

This is the most extensive study so far of the effects of various terms in the cross-section on the value of $\alpha_n$.
However, certain corrections which are believed to be small  have been omitted.  No contributions from 
the $\Delta$ resonance have been explicitly included, although the magnetic polarizability itself is thought
to contain a large paramagnetic contribution from the $\Delta$.   The $\Delta$ resonance has been shown to
have a small effect on deuteron photodisintegration at energies less than 100 MeV \cite{yi88}.   In addition, considering the rather small contributions
from $\pi$-exchange currents, heavier mesons such as the $\rho$ have also been neglected.  
Higher-order relativistic corrections have not been included, but may need to be investigated further
at energies near 100 MeV because of the large effects of the leading-order terms there.

The only full calculation of this type that has been published to date was by Weyrauch \cite{we88, we90}.
There are several important differences between his calculation and the present one.  
One of these is the $NN$ potential used.  He works with a separable $NN$ T matrix, while here
we have found the Green's function for an intermediate $NN$ state interacting via the Reid93 potential.
He also assumes meson-exchange currents  and relativistic corrections to be negligible.  Probably the most important
difference, however,  is in gauge invariance.  Gauge invariance is broken in the separable potential model and 
needs to be artificially restored.  We have made no such assumptions here.  Gauge invariance is a natural consequence
of including all of the appropriate terms.

The cross-section predicted by Weyrauch is higher than the experimental data.  However,
results published recently Arenhovel \cite{ar95} and Levchuk and L'vov \cite{lv95,lv98} are in agreement with the data.  
They have also used a diagrammatic approach and have included realistic potentials and
meson-exchange currents, but the full details of their calculations have not yet been made available.
Lastly, a new calculation using effective field theories also agrees with the data \cite{ch98}.

We return to the basic question of the feasibility of making an accurate determination 
of the neutron polarizability from a deuteron Compton scattering experiment. 
The available data at 49 and 69 MeV show that the ranges $\alpha_n = 12.0 \pm 4.0$ and $\beta_n = 2.0 \pm 4.0$
are reasonable, but a more accurate estimate is difficult since the polarizability terms are
small relative to the experimental error bars.  The uncertainties in the proton 
polarizabilities must also be taken into account.  These  problems are compounded by the data point at the greatest angle at each 
energy, which we find difficult to describe (other calculations also find disagreement with this point). 
The closest fit to the all of the experimental data comes from lowering 
the value of $\alpha_n$ and
raising $\beta_n$.  Throwing out the data point in question 
allows a closer fit, but there would still be a wide range of reasonable values for the polarizability.

Perhaps the most promising estimates of the polarizability will come from data at higher energies.  Assuming that
the experimental error bars are not significantly larger,  we should find that the range of polarizabilities that fit the
data is smaller.  Sensitivity to both $\alpha$ and $\beta$ is largest at the forward and backward angles, 
while the cross-section at $90^{\circ}$ is the least sensitive to the polarizabilities.
Many of the ``small'' corrections, however,  are particularly small at $120-140^{\circ}$,
and the cross-section is still very sensitive
to the polarizabilities at these angles.   The proton polarizabilities, unfortunately, will also make a larger contribution
to the cross-section, and there is no way to determine the neutron polarizabilities more accurately than those of the 
proton in this experiment.   This energy is at the limit of the validity of the low-energy expansion, which
may introduce errors as well.  In any case, studies at higher energies should provide additional information 
on the polarizabilities.

\bibliographystyle{phreport}
\bibliography{karakow-th}

\appendix
\chapter{Seagull Term \label{app:seagull} }
\noindent The interaction Hamiltonian is 
\begin{equation}
 H^{SG} = \sum_{j=n,p} \frac{e_j^2}{2 m_j} A^2(\vec{x}_j) =  \frac{e^2}{2 m_p} A^2(\vec{x}_p) ,
\end{equation}
since the neutron has no charge.  Using
\begin{equation}
 \vec{A}(\vec{x_j}) = \frac{1}{\sqrt{V} } \sum_{\vec{k},\lambda = \pm 1 } \sqrt{\frac{2\pi\hbar }{\omega } } [ a_k 
{\hat{\epsilon}}_{\lambda} e^{i\vec{k}
\cdot{\vec{x}}_j } + a_k^{\dagger} {\hat{\epsilon}}_{\lambda}^{\ast} e^{-i\vec{k}\cdot{\vec{x}}_j } ] ,
\label{eq:seagull2}
\end{equation}
\noindent we obtain the transition matrix
\begin{eqnarray}
\mathcal{T}_{fi}^{SG}  & \equiv & \mx{d_f, \vec{P}_f, {\gamma}_f}{ H^{SG} }{ d_i, \vec{P}_i, {\gamma}_i }  \\
	    & = & \frac{e^2}{2m_p } \frac{1}{V} \frac{2\pi\hbar}{\sqrt{\winit\wfin}} 
\mx{d_f,\vec{P}_f}{2 \eps e^{i\kinit\cdot{\vec{x}}_p } \cdot \epspr e^{-i\kfin\cdot{\vec{x}}_p} }{d_i,\vec{P}_i} \\
	    & = & \int d^3r d^3R \frac{e^2}{m_p}  \frac{1}{V} \frac{2\pi\hbar}{\sqrt{\winit\wfin}}(\epsdp) \times\nonumber\\
& & \ \ \ \ \ \ \mx{d_f, \vec{P}_f}{ e^{-\frac{i}{2}\vec{q}\cdot\vec{r}} e^{-i\vec{q}\cdot\vec{R} } }{ \vec{r},\vec{R} } 
\mxemp{ \vec{r},\vec{R} }{d_i, \vec{P}_i}, \label{eq:seagull3}
\end{eqnarray} 
where $ \vec{q} \equiv \kfin - \kinit $, $ \vec{r} \equiv {\vec{x}}_p - {\vec{x}}_n $ ,  $\vec{R} \equiv 
({\vec{x}}_p + {\vec{x}}_n)/2$, $\vec{p} \equiv \vec{p}_p - \vec{p}_n $, and 
$\vec{P} \equiv \vec{p}_p + \vec{p}_n$. 
The notation ${\mathcal{T}}_{fi}$ means that $\mathcal{T}$ is 
a function of the final kinematic variables ${\vec{k}}_f, M_f, {\lambda}_f$ as well as of the 
corresponding initial ones.  Assuming the center-of-mass wavefunctions for the deuteron states to be plane waves, the integral
over $R$ can be performed to obtain
\begin{equation}
 \matrixt{SG} = \int d^3r \frac{e^2}{m_p} \frac{1}{V} \frac{2\pi\hbar}{\sqrt{\winit\wfin}} (\epsdp) 
\frac{{\delta}({\vec{P}}_f + \kfin - {\vec{P}}_i - \kinit)}{V} \mx{d_f}{e^{-\frac{i}{2}\vec{q}\cdot\vec{r}} }
{\vec{r}} \mxemp{\vec{r} }{d_i}. 
\end{equation} 
\noindent We now define the scattering amplitude $\matrixm{}$ by
\begin{equation}
 \matrixt{} = \ \frac{1}{V} \frac{2\pi\hbar}{\sqrt{\winit\wfin}} 
\frac{ {\delta}({\vec{P}}_f + \kfin - {\vec{P}}_i - \kinit)}{V} \matrixm{}  \label{eq:seagull10},
\end{equation}
\noindent and so  
\begin{equation}
 \matrixm{SG} = \sum_{l=0,2} \sum_{l'=0,2} \int  d^3r \frac{e^2}{m_p}  (\epsdp)  
\mx{l' S_f J_f M_f r_f}{ e^{-\frac{i}{2}\vec{q}\cdot\vec{r}} }{\vec{r}}\mxemp{\vec{r} }{l S_i J_i M_i r_i} .
\end{equation}
\noindent The deuteron wavefunction can be written as
\begin{equation}
 \mxemp{\vec{r}}{d_i} =  \frac{u_0(r)}{r}{\mathcal{Y}}_M^{01J}(\hat{r}) + 
\frac{u_2(r)}{r}{\mathcal{Y}}_M^{21J}(\hat{r})  \label{eq:seagull7} ,
\end{equation}
\noindent where $u_0$, $u_2$ are the radial wavefunctions for the $S$, $D$ states, respectively, and
${\mathcal{Y}}_M^{LSJ}(\hat{r}) = \mxemp{\hat{r}}{LSJM}$ is the combination of the angular and spin
wavefunctions.
Inserting the partial wave decomposition of the exponential,
\begin{equation}
 e^{-\frac{i}{2}\vec{q}\cdot\vec{r}} = \sum_{L=0}^{\infty} \sum_{M = -L}^{L} 4\pi (-i)^L 
j_L(\frac{qr}{2}) Y_{LM}^{\ast}(\hat{q})Y_{LM}(\hat{r}) \label{eq:seagull1} ,
\end{equation} 
\noindent and the wavefunctions into the matrix element gives
\begin{eqnarray}
\matrixm{SG} & = & 
 \frac{e^2}{m_p}(\epsdp )\sum_{l l' L M}\int dr\ 4\pi (-i)^L j_L(\frac{qr}{2}) Y_{LM}^{\ast}(\hat{q}) 
\nonumber\times\\
& & \ \ \ \ \ u_{\sss l}(r) u_{\sss l'}(r) \mxba{ Y_{LM} } .\label{eq:seagull9}
\end{eqnarray}
\noindent We can use the Wigner-Eckart theorem on the $Y_{LM}$ matrix element to get
\begin{equation}
 \mxba{ Y_{LM} } = (-1)^{1-M_f} \threej{1}{l}{1}{-M_f}{M}{M_i} \mxredba{ Y_L }. 
\end{equation}
\noindent This reduced matrix element can be further simplified.  First,
we rewrite $Y_{LM}$ as a tensor product $[Y_L\otimes 1]_{LM}$, where 1 is an operator in spin space and
where the tensor product is defined by 
\begin{equation} 
 [T_{k_1} \otimes T_{k_2}]_{kn} \equiv \sum_{n_1 n_2} (-1)^{-k_1+k_2-n} \sqrt{2k+1} 
\threej{k_1}{k_2}{k}{n_1}{n_2}{-n} T_{k_1 n_1} T_{k_2 n_2}. \label{eq:seagull4}
\end{equation}
\noindent We then separate
the spin and orbital angular momentum parts using the relationship
\begin{eqnarray}
\lefteqn{ \mxred{ j_1 j_2 j  }{ [T_{k_1} \otimes T_{k_2}]_{k} }{ {j_1}' {j_2}' j' }  =}  \label{eq:seagull5}\\
& & \sqrt{(2j+1)(2j'+1)(2k+1)}  \ninej{j_1}{j_2}{j}{{j_1}'}{{j_2}'}{j'}{k_1}{k_2}{k} 
\mxred{ j_1 }{T_{k_1}}{ {j_1}' }\mxred{ j_2 }{ T_{k_2} }{ {j_2}' }.  \nonumber
\end{eqnarray}
\noindent This gives 
\begin{eqnarray}
\mxba{ Y_{LM} }   
& =  & (-1)^{1-M_f} \threej{1}{l}{1}{-M_f}{M}{M_i} \ninej{l'}{1}{1}{l}{1}{1}{L}{0}{L} \times\nonumber\\
& & \ \ \ \ \ \ \ 3 \sqrt{2L+1} \mxred{l'}{Y_L}{l} \mxred{1}{1}{1} \\
& = & (-1)^{1-M_f} \threej{1}{L}{1}{-M_f}{M}{M_i} \sixj{1}{l'}{1}{l}{1}{l} \times\nonumber\\
& & \ \ \ \ \ \ \ \ 3 \mxred{l'}{Y_L}{l},
\end{eqnarray}
\noindent where the $9j$ symbol has been simplified using 
\begin{equation}
 \ninej{a}{b}{e}{c}{d}{e}{f}{f}{0} = 
 \frac{(-1)^{b+c+e+f}}{\sqrt{(2e+1)(2f+1)}} \sixj{a}{b}{e}{d}{c}{f} ,
\end{equation}
\noindent and $\mxred{1}{1}{1} = \sqrt{3}$ has been found from the Wigner-Eckart theorem.
The formulas for several useful reduced matrix elements, including $\mxred{l'}{Y_L}{l}$,
can be be found in Appendix K.  Putting everything
together into $\matrixm{SG}$, and then removing the formal sum over $M$ by substituting $M = M_f-M_i$, we obtain
\begin{eqnarray}
\lefteqn{ \matrixm{SG}  =  \frac{3e^2}{m_p} (\epsdp) \sum_{l' l L} (-1)^{1-M_f+l'} (-i)^L  \times} \\
& &  Y_{L,M_f-M_i}^{\ast}(\hat{q}) \sqrt{4\pi(2l'+1)(2L+1)(2l+1)} \sixj{1}{l'}{1}{l}{1}{L} \times\nonumber\\
& & \threej{l}{L}{l'}{0}{0}{0} \threej{1}{L}{1}{-M_f}{M_f-M_i}{M_i} 
\int_0^{\infty} dr u_{\sss l'}(r) j_{\sss L}(\frac{qr}{2}) u_{\sss l}(r). \nonumber
\end{eqnarray}
\noindent These sums can be evaluated explicitly since there are only 5 triplets $(l', L, l)$ which give a non-zero 
contribution.  Defining 
\begin{equation}
 I_L^{l'l} \equiv \int_0^{\infty} dr\ u_{\sss l'}(r) j_{\sss L}(\frac{qr}{2}) u_{\sss l}(r) ,
\end{equation}
\noindent we can write the final result as
\begin{eqnarray}
 \matrixm{SG} & = & \frac{\sqrt{12\pi}e^2}{m_p} (\epsdp) (-1)^{1-M_f}  \times\label{eq:seagull8}\\
& & \ \ \ \ \ \ \left[ \threej{1}{0}{1}{-M_f}{0}{M_f} Y_{00}^{\ast}(\hat{q}) \right.  (I_0^{00}+I_0^{22}) {\delta}_{M_f, M_i} + 
\nonumber \\
& & \ \ \ \ \ \ \ \ \left. \threej{1}{2}{1}{-M_f}{M_f-M_i}{M_i} Y_{2, M_f-M_i}^{\ast}(\hat{q}) (I_2^{02} +
I_2^{20} - \frac{I_2^{22}}{\sqrt{2}}) \right]. \nonumber
\end{eqnarray}


\chapter{Electric Polarizability Term \label{app:alpha} }


\noindent The interaction Hamiltonian is 
\begin{equation}
 H^{\alpha} = \sum_{j=n,p} -\frac{{\alpha}_j}{2} E^2({\vec{x}}_j) =  \sum_{j} -\frac{{\alpha}_j}{2} 
\frac{\partial \vec{A}({\vec{x}}_j )}{\partial t}\cdot\frac{\partial \vec{A}({\vec{x}}_j )}{\partial t} .
\end{equation}
\noindent We insert equation~(\ref{eq:seagull2}) for the vector potential into $\matrixt{}$ to get
\begin{eqnarray}
 \matrixt{\alpha} & = & -\frac{\wfin\winit}{2V} \frac{2\pi\hbar}{\sqrt{\wfin\winit}}
2\bra{d_f, \vec{P}_f} \left[ {\alpha}_p \left( \eps  e^{i\kinit\cdot{\vec{x}}_p} \cdot
 \epspr e^{-i\kfin\cdot{\vec{x}}_p} \right) \right. + \\  
& & \left. {\alpha}_n \left( \eps e^{i\kinit\cdot{\vec{x}}_n}  \cdot
 \epspr {\alpha}_n e^{-i\kfin\cdot\vec{x}_n}  \right) \right] \ket{d_i, \vec{P}_i} \nonumber\\
 & = & -\frac{\wfin\winit}{V} \frac{2\pi\hbar}{\sqrt{\wfin\winit}} (\epsdp) \mx{d_f,\vec{P}_f}{ \left( {\alpha}_p 
 e^{-i\vec{q}\cdot{\vec{x}}_p} +{\alpha}_n e^{-i\vec{q}\cdot{\vec{x}}_n} \right) }{d_i, \vec{P}_i} 
\end{eqnarray}
\noindent After integrating out the center-of-mass states, this becomes
\begin{eqnarray}
\matrixm{\alpha} & = & -\wfin\winit (\epsdp) \sum_{l=0,2} \sum_{l'=0,2} \label{eq:alpha1} \times\\
& & \ \ \ \ \ \ \int d^3r \mx{l' 1 1 M_f r_f}{ \left( {\alpha}_p  e^{-\frac{i}{2}\vec{q}\cdot\vec{r}} +
{\alpha}_n e^{\frac{i}{2}\vec{q}\cdot\vec{r}} \right) }{ \vec{r} } \mxemp{\vec{r}}{l 1 1 M_i r_i}.\nonumber
\end{eqnarray}
\noindent We can expand the exponentials into partial waves using equation~(\ref{eq:seagull1}) and insert the 
deuteron wavefunctions (equation \ref{eq:seagull7}) to get
\begin{eqnarray}
\matrixm{\alpha} & = &  -\wfin\winit (\epsdp) \sum_{l' l} \sum_{L=0}^{\infty} \sum_{M=-L}^{L} \int_0^{\infty}  dr\ u_{\sss l'}(r) 
j_{\sss L}(\frac{qr}{2}) u_{\sss l}(r)\times\\
& &\ \ \ \ \ 4\pi (-i)^L \left[ {\alpha}_p + (-1)^L {\alpha}_n \right] Y_{LM}^{\ast}(\hat{q}) \nonumber
\mxba{ Y_{LM} } .
\end{eqnarray}  
\noindent This is identical to equation~(\ref{eq:seagull9}) if we replace $\frac{e^2}{m_p}$ by $-\wfin\winit
 \left[ {\alpha}_p + (-1)^L {\alpha}_n \right]$, so we can immediately write down the final answer by making 
this substitution in equation~(\ref{eq:seagull8}).  Thus,
\begin{eqnarray}
\matrixm{\alpha} & = & -\wfin\winit\sqrt{12\pi} (\epsdp) (-1)^{1-M_f} ( {\alpha}_p + {\alpha}_n )  
\label{eq:alpha2}\times \\
 & & \ \ \ \left[ \threej{1}{0}{1}{-M_f}{0}{M_f} Y_{00}^{\ast}(\hat{q}) (I_0^{00} + I_0^{22})\delta_{M_f,M_i} 
\right. + \nonumber\\  
 & & \ \ \ \ \ \ \left. \threej{1}{2}{1}{-M_f}{M_f-M_i}{M_i} Y_{2,M_f-M_i}^{\ast}(\hat{q})(
I_2^{02} + I_2^{20} - \frac{I_2^{22}}{\sqrt{2}} ) \right]. \nonumber
\end{eqnarray}
\noindent The factor $(-1)^L$ has been omitted since $L$ must be even.

\chapter{Magnetic Polarizability Term \label{app:beta} }
\noindent The interaction Hamiltonian is 
\begin{equation}
 H^{\beta} = \sum_{j=n,p} -\frac{{\beta}_j}{2} B^2({\vec{x}}_j) =  \sum_{j} -\frac{{\beta}_j}{2} 
\left[\vec{\nabla}_j\times\vec{A}({\vec{x}}_j )\right]\cdot\left[\vec{\nabla}_j\times\vec{A}({\vec{x}}_j )\right].
\end{equation}
\noindent Using this to calculate $\matrixt{\beta}$ gives
\begin{eqnarray}
\matrixt{\beta} & = & \frac{1}{2V} \frac{2\pi\hbar}{\sqrt{\wfin\winit}} 
 2 \langle d_f , \vec{P}_f \mid \left[ {\beta}_p \left(
i\kinit\times\eps e^{i\kinit\cdot{\vec{x}}_p} \right) \cdot \left( -i\kfin\times\epspr 
e^{-i\kfin\cdot{\vec{x}}_p} \right)  + \right. \nonumber\\
& & \ \ \ \ \left. {\beta}_n \left( i\kinit\times\eps e^{i\kinit\cdot{\vec{x}}_n} \right) \cdot 
\left( -i\kfin\times\epspr e^{-i\kfin\cdot{\vec{x}}_n} \right) \right] \mid d_i , \vec{P}_i\rangle  ,
\end{eqnarray}
\noindent and so
\begin{eqnarray}
\matrixm{\beta} & = & -\wfin\winit \left[ \left( \khatinit\times\eps \right) \cdot \left( \khatfin\times\epspr
\right) \right] \times\nonumber \\
& &\ \ \ \int d^3r \mx{d_f}{ \left( {\beta}_p e^{-\frac{i}{2}\vec{q}\cdot\vec{r} } +
{\beta}_n e^{\frac{i}{2}\vec{q}\cdot\vec{r} } \right) }{ \vec{r} }
\mxemp{ \vec{r} }{ d_i}.
\end{eqnarray}
\noindent This is the same as equation~(\ref{eq:alpha1}), but with $\alpha$ replaced by $\beta$,
and with a different dot product.  Therefore,  we can make these substitutions in equation~(\ref{eq:alpha2}) to get
\begin{eqnarray}
\matrixm{\beta} & = & -\wfin\winit \sqrt{12\pi}  \left[ \left( \khatinit\times\eps \right) \cdot \left(\khatfin 
\times\epspr\right) \right] (-1)^{1-M_f} ({\beta}_p+{\beta}_n)\times\nonumber \\
& &\ \ \ \left[ \threej{1}{0}{1}{-M_f}{0}{M_f} Y_{00}^{\ast}(\hat{q})(I_0^{00} + I_0^{22}) 
\delta_{M_f,M_i} + \right. \nonumber\\
& &\ \ \ \ \ \ \left. \threej{1}{2}{1}{-M_f}{M_f-M_i}{M_i} Y_{2,M_f-M_i}^{\ast}(\hat{q}) 
(I_2^{02} + I_2^{20} - \frac{I_2^{22}}{\sqrt{2}}) \right].
\end{eqnarray} 

\chapter{Phase Space Factors \label{app:phase} }
\noindent To calculate the differential cross section, we use Fermi's golden rule:
\begin{eqnarray}
d\sigma_{fi} & = & \frac{2\pi}{\hbar} \frac{\delta(\Delta E)}{\underbrace{1/V}_{\mathrm{initial }
\;\gamma\;\mathrm{flux}}}
\overbrace{ \frac{V d^3k_f}{(2\pi)^3} }^{\mathrm{final}\;\gamma} 
\overbrace{ \frac{V d^3p_f}{(2\pi)^3} }^{\mathrm{final}\; d}
\left| \matrixt{ } \right|^2  \\
& = &  \frac{2\pi}{\hbar} \frac{\delta(\Delta E)}{1/V}\frac{V d^3k_f}{(2\pi)^3}\frac{V d^3p_f}{(2\pi)^3}
\left|  \frac{2\pi\hbar}{\sqrt{\wfin\winit}}\frac{1}{V^2} \delta (\Delta\vec{p}) \matrixm{} \right|^2 \\
& = & \frac{\hbar}{\wfin\winit} \delta(\Delta E)  d^3k_f d^3p_f \delta (\Delta\vec{p}) 
\left| \matrixm{} \right|^2  ,
\end{eqnarray}
\noindent where we have used the definition of $\matrixm{}$ (equation \ref{eq:seagull10}),
as well as $| \delta(\Delta\vec{P})|^2 = V(2\pi)^3\delta(\Delta\vec{P})$.
Performing the integration over $p_f$ removes one of the $\delta$-functions;  
$d\sigma$ becomes
\begin{eqnarray}
d{\sigma}_{fi} & = & \hbar\frac{\delta(E'-E)}{\wfin\winit} {\wfin}^2 d\wfin d{\Omega}_f
\left| \matrixm{} \right|^2 ,
\end{eqnarray}
\noindent and so 
\begin{eqnarray}
\left(\frac{d\sigma}{d{\Omega}}\right)_{fi} & = & \hbar\frac{\wfin}{\winit} \frac{1}{\partial E'/ \partial\wfin } 
\left| \matrixm{} \right|^2 \label{eq:phase1},
\end{eqnarray}
\noindent where $E(E')$ is the total energy of the initial(final) deuteron and photon combined.
The partial derivative is calculated in the lab frame.  We define 4-vectors $\tilde{p}_i =
(m_d, 0)$, $\tilde{p}_f = (E_f, {\vec{p}}_f) $,  $\tilde{k}_i = (\winit, \kinit)$, and
$\tilde{k}_f = (\wfin, \kfin)$. By energy-momentum conservation, $\tilde{p}_i + \tilde{k}_i=
 \tilde{p}_f + \tilde{k}_f$.  Moving $\tilde{k}_f$ to the left-hand side and then squaring both sides gives
\begin{eqnarray}
\tilde{k}_i\cdot\tilde{p}_i - \tilde{k}_f\cdot\tilde{p}_i - \tilde{k}_f\cdot\tilde{k}_i & = & 0.
\end{eqnarray}
\noindent This simplifies to
\begin{equation}
\wfin  =  \frac{m_d\winit}{m_d+\winit-\winit\cos\theta} \label{eq:phase2},
\end{equation}
\noindent where $\theta$ is the angle between $\kfin$ and $\kinit$.
Next,  we use energy conservation ($E_f=\winit + m_d - \wfin $) to write
\begin{eqnarray}
E_f    & = & \sqrt{ {\winit}^2 + {\wfin}^2 + m_d^2 -  2\wfin\winit\cos\theta} \label{eq:phase3},
\end{eqnarray}
\noindent where we have again used equation~(\ref{eq:phase2}) .
Now we take the derivative:
\begin{eqnarray}
\frac{\partial E'}{\partial\wfin} & = & \frac{\partial (E_f + \wfin)}{\partial\wfin} \\
& = & \frac{\wfin - \winit\cos\theta}{E_f} + 1 \\
& = & \frac{m_d \winit}{E_f \wfin}.
\end{eqnarray} 
\noindent This has been simplified using equations (\ref{eq:phase2}) and (\ref{eq:phase3}) . 
Plugging this back into equation (\ref{eq:phase1}),  we see that 
\begin{eqnarray}
\left(\frac{d\sigma}{d\Omega}\right)_{fi} & = & 
\left( \frac{\wfin}{\winit} \right)^2 \frac{E_f}{m_d} \mid\matrixm{} \mid^2 , \label{eq:phase4}
\end{eqnarray}
\noindent where $\matrixm{} = \matrixm{SG} + \matrixm{\alpha} + \matrixm{\beta} + \cdots$. 
If we are interested in computing the spin-averaged cross section, we sum over the final 
 and average over the initial polarizations to get
\begin{eqnarray}
\frac{d\sigma}{d\Omega} & = & \frac{1}{6} \sum_{M_f M_i {\lambda}_f {\lambda}_i} 
 \left( \frac{\wfin}{\winit} \right)^2 \frac{E_f}{m_d} \mid\matrixm{} \mid^2.
\end{eqnarray} 

\chapter{Multipole Expansion Derivation \label{app:mult} }
\noindent It is convenient to obtain the multipole expansion for $\epsexp,$
where ${\hat{\epsilon}}_{\lambda}$ is a spherical vector.  This derivation
is similar to the one found in \cite{rose55}; some additional details are included here.

We set $\vec{k}$ to be in the direction of the $\hat{z}$-axis for now,
which means that $ {\hat{\epsilon}}_{\lambda} $ becomes  $ {\hat{r}}_{\lambda} $, the unit vector
in the spherical basis, with $\lambda = \pm 1$. 
Defining $\omega \equiv |\vec{k}|$,
\begin{eqnarray}
\left.\epsexp\right|_{\hat{k} = \hat{z} } & = &  {\hat{r}}_{\lambda} \sum_{l=0}^{\infty} 
\sum_{m=-l}^l 4\pi i^l j_{\sss l}(\omega r)
Y_{lm}^{\ast}(\hat{k}) Y_{lm}(\hat{r}) \\
& = & {\hat{r}}_{\lambda} \sum_{l} i^l \sqrt{4\pi(2l+1)} j_{\sss l}(\omega r) Y_{l0}(\hat{r}) \label{eq:multexp1} ,
\end{eqnarray}
\noindent where we have used equation~(\ref{eq:seagull1}) and
$Y_{lm}(\hat{z}) = \delta_{m0}\sqrt{\frac{2l+1}{4\pi}}$.  
We now define the vector spherical harmonic $\vsh{J}{L}{M}$ by
\begin{equation}
\vsh{J}{L}{M}  = \sum_{\nu=-1,0,1}  (-1)^{L-M-1} \sqrt{2J+1} \threej{1}{L}{J}{-\nu}{M+\nu}{-M} 
{\hat{r}}_{-\nu} Y_{L,M+\nu}(\hat{r}). \label{eq:multexp11}
\end{equation}
\noindent By rearranging this to get 
\begin{equation}
Y_{lm}(\hat{r}) {\hat{r}}_{-\lambda} = \sum_{JM} (-1)^{-l+M+1} \sqrt{2J+1} \threej{1}{l}{J}{-\lambda}{m}{-M} 
\vsh{J}{l}{M},  \label{eq:multexp6}
\end{equation}
\noindent we can simplify equation~(\ref{eq:multexp1}):
\begin{eqnarray}
\lefteqn{\left. \epsexp\right|_{\hat{k} = \hat{z} }   = } \\
& & \sum_{J=| l-1 | }^{l+1} \sum_{l=0}^{\infty} (-1)^{l+\lambda -1} i^l \sqrt{4\pi(2l+1)(2J+1)} 
 \threej{1}{l}{J}{\lambda}{0}{-\lambda} j_{\sss l}(\omega r) \vsh{J}{l}{\lambda} \nonumber.
\end{eqnarray}
\noindent Explicitly writing out the $J$ sum yields
\begin{eqnarray}
\lefteqn{\left. \epsexp\right|_{\hat{k} = \hat{z} } =   }  \\
&  & \sum_{l=0}^{\infty} (-1)^{l+\lambda -1} i^l \sqrt{4\pi(2l+1)} j_{\sss l}(\omega r)
 \left\{ \sqrt{2l-1} \threej{1}{l}{l-1}{\lambda}{0}{-\lambda} \vsh{l-1}{l}{\lambda} + \right. \nonumber \\
& &\left. \sqrt{2l+1} \threej{1}{l}{l}{\lambda}{0}{-\lambda} \vsh{l}{l}{\lambda} +  
 \sqrt{2l+3} \threej{1}{l}{l+1}{\lambda}{0}{-\lambda} \vsh{l+1}{l}{\lambda} \right\} \nonumber \\
& = & \sum_{l=0}^{\infty} i^l \sqrt{4\pi} j_l(\omega r) \left\{ \sqrt{\frac{l-1}{2}} \vsh{l-1}{l}{\lambda} - \right. \label{eq:multexp2}\\  
& & \ \ \ \ \ \ \ \ \ \ \ \ \ \ \ \ \ \ \ \ \ \ \ \ \ \ \ \ \ \ \left.\lambda 
\sqrt{\frac{2l+1}{2}} \vsh{l}{l}{\lambda} + \sqrt{\frac{l+2}{2}} \vsh{l+1}{l}{\lambda}\!\right\} ,
\nonumber
\end{eqnarray}
\noindent after evaluating the $3j$ symbols.  Next, we manipulate the summation index 
in order to rearrange this into a form where all of the
vector spherical harmonics have the same total angular momentum.  
We call this new index $L$, and let $L=l-1$ in the first term of equation~(\ref{eq:multexp2}),
$L=l$ in the second term, and $L=l+1$ in the third term.  Thus,
\begin{eqnarray}
\left.  \epsexp\right|_{\hat{k} = \hat{z} } 
&  = & \sum_{L=1}^{\infty} i^L \sqrt{2\pi(2L+1)}  \left\{ i \sqrt{\frac{L}{2L+1}} j_{\sss L+1}(\omega r)
\vsh{L}{L+1}{\lambda} - \right.  \\
& & \ \ \ \ \ \ \left. \lambda j_{\sss L}(\omega r) \vsh{L}{L}{\lambda} - i \sqrt{\frac{L+1}{2L+1}} j_{\sss L-1}(\omega r) 
\vsh{L}{L-1}{\lambda}  \right\}. \nonumber
\nonumber
\end{eqnarray}
\noindent Now we would like to return $\vec{k}$ to its original orientation, where it has spherical
 coordinates~$(\omega,\vartheta,\varphi)$.  This can be attained by rotating  $\vec{r}$ by the Euler angles
$(0,-\vartheta,-\varphi)$ by means of the Wigner $d$-functions $\wignerd$, so that
\begin{eqnarray} 
\epsexp &  = & \sum_{L=1}^{\infty} \sum_{M=-L}^L i^L \sqrt{2\pi(2L+1)} \label{eq:multexp17}
\left\{  i \sqrt{\frac{L}{2L+1}} j_{\sss L+1}(\omega r)\vsh{L}{L+1}{M} - \right.  \\
& & \left.  \lambda j_{\sss L}(\omega r) 
\vsh{L}{L}{M} - i \sqrt{\frac{L+1}{2L+1}} j_{\sss L-1}(\omega r) \vsh{L}{L-1}{M}  
\right\} \wignerd\nonumber .
\end{eqnarray}
\noindent Before simplifying this further, we need to show that 
\begin{eqnarray}
\lefteqn{\frac{1}{\sqrt{L(L+1)}} \left\{ \omega\vec{r} j_{\sss L}(\omega r) Y_{LM}(\hat{r}) + 
\frac{1}{\omega}\vec{\nabla}_r\left(1+r\frac{d}{dr}\right) 
j_{\sss L}(\omega r) Y_{LM}(\hat{r}) \right\} = } \nonumber \\
& &  \sqrt{\frac{L+1}{2L+1}} j_{\sss L-1}(\omega r) \vsh{L}{L-1}{M} - 
\sqrt{\frac{L}{2L+1}} j_{\sss L+1}(\omega r) \vsh{L}{L+1}{M} \label{eq:multexp3} 
\end{eqnarray}
\noindent and
\begin{equation}
\frac{1}{\sqrt{L(L+1)} } \vec{L} Y_{LM}(\hat{r}) j_{\sss L}(\omega r) 
 = j_{\sss L}(\omega r) \vsh{L}{L}{M}.  \label{eq:multexp4}
\end{equation}
\noindent We start with the first term on the left-hand side ($LHS1$) of equation~(\ref{eq:multexp3}):
\begin{eqnarray}
LHS1 & = & \sum_{\nu=-1,0,1}\frac{\omega}{\sqrt{L(L+1)}} j_{\sss L}(\omega r) 
\sqrt{\frac{4\pi}{3}} (-1)^{\nu} Y_{1\nu}(\hat{r})
{\hat{r}}_{-\nu} Y_{LM}(\hat{r}),
\end{eqnarray}
\noindent where we have used the decomposition of an arbitrary vector $\vec{V}$ into 
spherical components,    
\begin{equation}
\vec{V}  =  \sum_{\nu=-1,0,1} (-1)^{\nu} V_{\nu} {\hat{r}}_{-\nu} \label{eq:multexp10} ,
\end{equation}
along with 
\begin{equation}
r_{\nu}  =  \sqrt{\frac{4\pi}{3}} Y_{1\nu}(\hat{r}).
\end{equation}
\noindent Next, the spherical harmonics are combined using  
\begin{eqnarray}
 Y_{jm} Y_{j'm'} & = & \sum_{J=|j-j'|}^{j+j'} \sum_{M=-J}^{J} (-1)^M \sqrt{\frac{(2j+1)(2j'+1)(2J+1)}{4\pi}}  \\
& & \ \ \ \ \threej{j}{j'}{J}{0}{0}{0}
\threej{j}{j'}{J}{m}{m'}{-M} Y_{JM} \label{eq:multexp15} 
\end{eqnarray}
\noindent to get
\begin{eqnarray}
LHS1 & = & \sum_{\nu=-1,0,1} \sum_{\tilde{J}=|1-L|}^{1+L} \sum_{\tilde{M}=-\tilde{J} }^{\tilde{J} }
\frac{\omega}{\sqrt{L(L+1)}} j_{\sss L}(kr) (-1)^{\nu+\tilde{M}}
\sqrt{(2L+1)(2\tilde{J}+1)} \times \nonumber \\
& & \ \ \ \ \ \threej{1}{L}{\tilde{J}}{0}{0}{0} \threej{1}{L}{\tilde{J}}{\nu}{M}{-\tilde{M}} 
Y_{\tilde{J}\tilde{M}}(\hat{r}) {\hat{r}}_{-\nu}.
\end{eqnarray}
\noindent Looking back at equation~(\ref{eq:multexp6}), we see that the spherical harmonic can be replaced
by a vector spherical harmonic:
\begin{eqnarray}
LHS1 & = & \sum_{\nu=-1,0,1} \sum_{\tilde{J}=|1-L|}^{1+L} \sum_{ J'= |1-\tilde{J}|}^{1+\tilde{J} } 
(-1)^{1-\tilde{J}}  \omega j_{\sss L}(\omega r) 
\sqrt{\frac{(2L+1)(2\tilde{J}+1)(2J'+1)}{L(L+1)}} \times  \nonumber \\
& & \threej{1}{L}{\tilde{J}}{0}{0}{0} \vsh{J'}{\tilde{J}}{M'} \sum_{\tilde{M}M'} 
\threej{1}{L}{\tilde{J}}{\nu}{M}{-\tilde{M}} \threej{1}{\tilde{J}}{J'}{-\nu}{\tilde{M}}{-M'}.
\end{eqnarray}
\noindent This can be simplified because of the orthogonality of the 3j symbols,
\begin{equation}
\sum_{m_1m_2} \threej{j_1}{j_2}{j_3}{m_1}{m_2}{m_3} \threej{j_1}{j_2}{j_3'}{m_1}{m_2}{m_3'} =
\frac{\delta_{j_3{j_3}'}\delta_{m_3{m_3}'} }{2j_3+1},
\end{equation}
\noindent to yield
\begin{eqnarray}
LHS1 & = & \sum_{\tilde{J}}  (-1)^{1-\tilde{J}}  \omega j_{\sss L}(\omega r) \sqrt{\frac{2\tilde{J}+1}{L(L+1)}} 
\threej{1}{L}{\tilde{J}}{0}{0}{0} \vsh{L}{\tilde{J}}{M} \label{eq:multexp7}.
\end{eqnarray}
\noindent $\tilde{J}$ can only be equal to $L+1$ or $L-1$.   Inserting this into equation~(\ref{eq:multexp7}) 
and then writing out the $3j$ symbols algebraically, we obtain
\begin{eqnarray}
LHS1  & = & 
(-1)^L \omega j_{\sss L}(\omega r) \left[ \sqrt{\frac{2L+3}{L(L+1)}} \threej{1}{L}{L+1}{0}{0}{0} 
\vsh{L}{L+1}{M} + \right. \nonumber \\
& &\ \ \ \ \ \left. \sqrt{\frac{2L-1}{L(L+1)}} \threej{1}{L}{L-1}{0}{0}{0} \vsh{L}{L-1}{M} \right] \\
&  = & \omega j_{\sss L}(\omega r) \left[ -\frac{\vsh{L}{L+1}{M}}{\sqrt{L(2L+1)}} +
\frac{\vsh{L}{L-1}{M}}{\sqrt{(L+1)(2L+1)}} \right] \label{eq:multexp8}.
\end{eqnarray}
\noindent The second half of the left side of equation~(\ref{eq:multexp3}) can be simplified with repeated 
applications of the gradient formula (equation \ref{eq:formula13})
and the recursion relations for spherical Bessel functions (equations \ref{eq:formula11} and \ref{eq:formula12})
to get
\begin{eqnarray}
\lefteqn{ LHS2  =  
 \frac{\vsh{L}{L+1}{M}}{\sqrt{L(2L+1)}} \left\{ \left[ \frac{L^2}{2L+1}j_{\sss L-1}(\omega r) - 
\frac{L(L+1)}{2L+1}j_{\sss L+1}(\omega r) + \right. \right.} \\
& & \left. \left. \left( \omega r - \frac{L^2}{\omega r} \right) j_{\sss L}(\omega r) \right]  + 
\frac{\vsh{L}{L-1}{M}}{\sqrt{(L+1)(2L+1)}} \left[ \frac{L(L+1)}{2L+1} j_{\sss L-1}(\omega r) - \right. \right. \nonumber\\
& & \left. \left. \frac{(L+1)^2}{2L+1} j_{\sss L+1}(\omega r) + \left( \frac{(L+1)^2}{\omega r} - 
\omega r \right) j_{\sss L}(\omega r) \right] \right\} \nonumber.
\end{eqnarray}
\noindent Combining this with equation~(\ref{eq:multexp8}) and then using 
the Bessel function recursion relations again gives
\begin{eqnarray}
\lefteqn{LHS1 + LHS2 = } \\
& &  \frac{\vsh{L}{L+1}{M}}{\sqrt{L(2L+1)}} \left\{ \left[ \frac{L^2}{2L+1}j_{\sss L-1}(\omega r) -
\frac{L(L+1)}{2L+1}j_{\sss L+1}(\omega r) -  \right.\right. \nonumber\\
& & \left. \frac{L^2}{\omega r} j_{\sss L}(\omega r) \right]  + 
\frac{\vsh{L}{L-1}{M}}{\sqrt{(L+1)(2L+1)}} \left[ \frac{L(L+1)}{2L+1} j_{\sss L-1}(\omega r) - \right.\nonumber\\
& & \left. \left. \frac{(L+1)^2}{2L+1} j_{\sss L+1}(\omega r) +  
\frac{(L+1)^2}{\omega r} j_{\sss L}(\omega r)\right] \right\} \nonumber \\
& = & -\sqrt{\frac{L}{2L+1}} j_{\sss L+1}(\omega r) \vsh{L}{L+1}{M} + 
\sqrt{\frac{L+1}{2L+1}} j_{\sss L-1}(\omega r) \vsh{L}{L-1}{M} \label{eq:multexp18}.
\end{eqnarray}
\noindent This is identical to equation~(\ref{eq:multexp3}).
To prove equation~(\ref{eq:multexp4}), we first write out the left-hand side in terms of its 
spherical components using equation~(\ref{eq:multexp10}):
\begin{eqnarray}
\frac{1}{\sqrt{L(L+1)}} \vec{L} Y_{LM}(\hat{r})  
& = & \sum_{\nu} \frac{(-1)^{\nu} }{\sqrt{L(L+1)}} L_{\nu} {\hat{r}}_{-\nu} Y_{LM}(\hat{r}) .
\end{eqnarray}
\noindent Using the fact that \cite[p. 122]{rose57}
\begin{equation} 
L_{\nu} Y_{LM} = (-1)^{-M-\nu+1+L} \sqrt{L(L+1)(2L+1)} \threej{L}{1}{L}{M}{\nu}{-M-\nu} Y_{L,M+\nu},
\end{equation}
\noindent this becomes
\begin{equation}
\frac{1}{\sqrt{L(L+1)}} \vec{L} Y_{LM}(\hat{r})
 =  \sum_{\nu} (-1)^{1+L-M} \sqrt{2L+1} \threej{L}{1}{L}{M}{\nu}{-M-\nu} Y_{L,M+\nu}(\hat{r}) 
{\hat{r}}_{-\nu}.
\end{equation} 
\noindent With the help of equation~(\ref{eq:multexp11}) we can immediately write
\begin{equation}
\frac{1}{\sqrt{L(L+1)}} \vec{L} Y_{LM}(\hat{r}) = \vsh{L}{L}{M} \label{eq:multexp12},
\end{equation} 
\noindent which is what we were trying to establish.  After substituting equation~(\ref{eq:multexp12}) and
equation~(\ref{eq:multexp18}) in  equation~(\ref{eq:multexp17}), we get the final result
for the multipole expansion:
\begin{eqnarray}
\lefteqn{ \epsexp = 
 \sum_{L = 1}^{\infty} \sum_{M=-L}^{L}\wignerd i^L\sqrt{\frac{2\pi(2L+1)}{L(L+1)}}\times}  \label{eq:multexp99} \\
& &  \left\{ -\frac{i}{\omega}\vec{\nabla_r} \left( 1+r\frac{d}{dr} \right) j_{\sss L}(\omega r) Y_{LM}(\hat{r}) -
i\omega \vec{r} j_{\sss L}(\omega r) Y_{LM}(\hat{r}) - 
\lambda \vec{L}  Y_{LM}(\hat{r})  j_{\sss L}(\omega r) \right\}. \nonumber
\end{eqnarray}

\chapter{Largest Dispersive Terms \label{app:main} }
\noindent The remaining terms will be calculated using second-order perturbation theory:
\begin{eqnarray}
\matrixt{} & = & \sum_{C, \vec{P}_C} \left\{ \frac{ \mx{d_f, \vec{P}_f ,{\gamma}_f}{ 
H^{\mathrm{int}} }{C, \vec{P}_C} 
\mx{C, \vec{P}_C}{H^{\mathrm{int}}}{d_i, \vec{P}_i, {\gamma}_i} }{\hbar{\omega}_i+E_{d_i}-E_C-
P_C^2/2m_d +i\varepsilon} 
+ \right.  \label{eq:mainterm1}\\
& & \left. \frac{ \mx{d_f, \vec{P}_f, {\gamma}_f}{H^{\mathrm{int}}}{C, \vec{P}_C, {\gamma}_i, {\gamma}_f }
\mx{C, \vec{P}_C, {\gamma}_i, {\gamma}_f }{H^{\mathrm{int}}}{d_i, \vec{P}_i, {\gamma}_i} } 
{-\hbar{\omega}_f+E_{d_i}-E_C-P_C^2/2m_d+i\varepsilon} \right\} \nonumber .
\end{eqnarray}
\noindent Here $C$ denotes the internal quantum numbers of the intermediate $np$ state, 
and $\sum_C$ is shorthand for all sums and integrals
which are needed to describe this complete set of states except for the center-of-mass states. These  are 
written separately as $\vec{P}_C$.  At this point, it is most convenient to 
use $H^{\mathrm{int}} = -\int \vec{J}(\vec{\dum})\cdot \vec{A}(\vec{\dum}) d^3\dum$,
and then expand $\vec{A}$ into multipoles according to equation~(\ref{eq:multexp99}).  There are
3 terms in this expansion; therefore, equation~(\ref{eq:mainterm1}) contains a total of 18 terms. 
The largest contributions at low energies (which we will call $\matrixt{a}$) arise from the gradient operator in 
equation~(\ref{eq:multexp99}).  These are the leading-order contributions; the other
terms are of order $\omega/m_N$ smaller.  This gradient term includes the case where there is an $E1$ interaction
at both $\gamma N$ vertices.  

With this in mind, 
we define $\Phi_i(\vec{r})$ and ${\Phi}_f(\vec{r})$ by 
\begin{eqnarray}
\Phi_i(\vec{r}) & \equiv & -\frac{1}{\sqrt{V}} \sqrt{\frac{2\pi\hbar}{\winit} } \sum_{L=1}^{\infty} \sum_{M=-L}^{L} \wignerdi 
\frac{i^{L+1}}{\winit} \sqrt{ \frac{2\pi(2L+1)}{L(L+1)} }  \times \label{eq:mainterm10} \\
& & \ \ \ \ \ \ \left( 1+r\frac{d}{dr} \right) j_{\sss L}(\winit r) Y_{LM}(\hat{r}) \, \nonumber \\
\Phi_f(\vec{r}) & \equiv & \frac{1}{\sqrt{V}} \sqrt{\frac{2\pi\hbar}{\wfin} } \sum_{L'=1}^{\infty} \sum_{M'=-L'}^{L'}
 (-1)^{L'-\mufin} \wignerdf
\frac{i^{L'+1}}{\wfin} \times \nonumber \\
& & \ \ \ \ \ \  \sqrt{ \frac{2\pi(2L'+1)}{L'(L'+1)} }
\left( 1+r\frac{d}{dr} \right) j_{\sss L'}(\wfin r) Y_{L'M'}(\hat{r}), \label{eq:mainterm11}
\end{eqnarray}
\noindent which means that 
\begin{eqnarray}
\frac{1}{\sqrt{V}} \sqrt{\frac{2\pi\hbar}{\winit} } \eps 
e^{i\kinit\cdot\vec{r}} \label{eq:mainterm2} & = &
\vec{\nabla}_r\Phi_i(\vec{r}) + \cdots, \\
\frac{1}{\sqrt{V}} \sqrt{\frac{2\pi\hbar}{\wfin} } \epspr 
e^{-i\kfin\cdot\vec{r}} \label{eq:mainterm3}
& = & \vec{\nabla}_r\Phi_f(\vec{r}) + \cdots.
\end{eqnarray}
\noindent We have chosen 
a coordinate system where $\kinit$ lies on the $\hat{z}$-axis and where $\kfin$
has spherical coordinates $(\wfin,\theta,\phi)$.  Using equations (\ref{eq:mainterm2}) and
(\ref{eq:mainterm3}) in the uncrossed term of equation~(\ref{eq:mainterm1}) gives  
\begin{eqnarray}
\lefteqn{\matrixt{a,\mathrm{uncr}}  = \sum_{C,\vec{P}_C} \mx{d_f, \vec{P}_f}{\int 
\vec{J}(\vec{\dum}')\cdot\vec{\nabla}_{\dum '}\Phi_f({\vec{\dum}}') d^3\dum'}{C, \vec{P}_C} \times}  \\
& & \frac{1}{\hbar\winit + E_{d_i}-E_C-P_C^2/2m_d+i\varepsilon}
\mx{C,\vec{P}_C }{ \int \vec{J}(\vec{\dum})\cdot\vec{\nabla}_{\dum}\Phi_i(\vec{\dum}) d^3\dum}{d_i, \vec{P}_i}. \nonumber
\end{eqnarray}
\noindent Integrating by parts and then using current conservation,
\begin{eqnarray}
\vec{\nabla}_{\dum}\cdot\vec{J}(\vec{\dum}) & = & -\frac{i}{\hbar}\left[ H,\rho(\vec{\dum}) \right], \\
\rho(\vec{\dum}) & = & \sum_{j=n,p} e_j \delta(\vec{\dum} - {\vec{x}}_j),
\end{eqnarray}
\noindent gives
\begin{eqnarray}
\lefteqn{ \matrixt{a,\mathrm{uncr}}  =  - \sum_{C,\vec{P}_C} \mx{d_f, \vec{P}_f}{\int  
[H,\rho(\vec{\dum}')]\Phi_f({\vec{\dum}}') d^3\dum'}{C,\vec{P}_C} \times}\label{eq:mainterm13} \\ 
& & \frac{1}{\hbar^2(\hbar\winit + E_{d_i}-E_C-P_C^2/2m_d+i\varepsilon)}
\mx{C,\vec{P}_C}{\int  [H,\rho(\vec{\dum})]\Phi_i(\vec{\dum}) d^3\dum}{d_i, \vec{P}_i}. \nonumber
\end{eqnarray}
\noindent Performing the integrations yields
\begin{eqnarray}
\lefteqn{ \matrixt{a, \mathrm{uncr}}  =  -\sum_{C,\vec{P}_C} \mx{d_f, 
\vec{P}_f}{[H,(e/\hbar)\Phi_f(\vec{x}_p)]}{C,\vec{P}_C} \times} \\ 
& & \frac{1}{\hbar\winit+E_{d_i}-E_C-P_C^2/2m_d+i\varepsilon}
\mx{C,\vec{P}_C}{[H, (e/\hbar)\Phi_i(\vec{x}_p)]}{d_i, \vec{P}_i}.  \nonumber
\end{eqnarray}
\noindent It is easiest to evaluate $\matrixt{}$ by writing the commutators in terms of $\vec{r}$ and $\vec{R}$:
\begin{eqnarray}
[H,(e/\hbar)\Phi_i(\vec{x}_p)] & = & \left[ \frac{p_p^2}{2m_p} +\frac{p_n^2}{2m_n} +V,(e/\hbar)\Phi_i(\vec{x}_p)\right] \\
& = & -\frac{ie}{m_p} \vec{\nabla}_r\Phi_i(\vec{r}/2) e^{i \vec{k}_i\cdot\vec{R}} \cdot \vec{P} +\nonumber\\
& & \ \ \ \ \ \left[ \frac{p^2}{m_p} + V , (e/\hbar)\Phi_i(\vec{r}/2) \right]  e^{i \vec{k}_i\cdot\vec{R}},\label{eq:mainterm14}
\end{eqnarray}
\noindent and
\begin{eqnarray}
[H,(e/\hbar)\Phi_f(\vec{x}_p)] & = &
  -\frac{ie}{m_p} \vec{\nabla}_r\Phi_f(\vec{r}/2) e^{-i\vec{k}_f\cdot\vec{R}} \cdot \vec{P} +\nonumber\\
& & \ \ \ \ \ \ \left[ \frac{p^2}{m_p} + V, (e/\hbar)\Phi_f(\vec{r}/2) \right]  e^{-i\vec{k}_f\cdot\vec{R}}, \label{eq:mainterm15}
\end{eqnarray}
\noindent where equations (\ref{eq:mainterm2}) and  (\ref{eq:mainterm3}) have been used, along with  
 $\vec{p} \equiv (\vec{p}_p - \vec{p}_n)/2$ and $\vec{P} \equiv \vec{p}_p + \vec{p}_n$. The terms containing
$\vec{P}$ are recoil corrections and will be calculated in Appendix H. Defining  
the dimensionless functions $\hat{\Phi}_i \equiv (e/\hbar)\Phi_i(\vec{r}/2)$
and  $\hat{\Phi}_f \equiv (e/\hbar)\Phi_f(\vec{r}/2)$, along with $H^{np} \equiv \frac{p^2}{m_p} + V$, we get
\begin{equation}
\matrixt{a, \mathrm{uncr}}  =  -\sum_{C,\vec{P}_C} \frac { \mx{d_f, \vec{P}_f}{[H^{np}, 
\hat{\Phi}_f] e^{-i\vec{k}_f\cdot\vec{R}}}{C,\vec{P}_C}
\mx{C,\vec{P}_C}{[H^{np}, \hat{\Phi}_i] e^{i \vec{k}_i\cdot\vec{R}}}{d_i, \vec{P}_i} }{\hbar\winit+E_{d_i}-E_C-
P_C^2/2m_d+i\varepsilon}.
\end{equation}
\noindent After evaluating the commutators and integrating out the center-of-mass motion, this becomes
\begin{eqnarray}
\matrixt{a, \mathrm{uncr}} & = &  
\delta(\vec{P}_i + \vec{k}_i - \vec{P}_f - \vec{k}_f) \left\{
 \frac{1}{2} \mx{d_f}{[H^{np}, \hat{\Phi}_f] \hat{\Phi}_i}{d_i} - \nonumber\right. \\
& & \frac{1}{2}\mx{d_f}{\hat{\Phi}_f [H^{np}, \hat{\Phi}_i]}{d_i} -
\left[ \hbar\winit - \frac{(\hbar\winit)^2}{2m_d}  \right] \mx{d_f}{\hat{\Phi}_f \hat{\Phi}_i}{d_i} + \nonumber\\
& & \left.\left[ \hbar\winit - \frac{(\hbar\winit)^2}{2m_d}  \right]^2 \sum_C \frac{\mx{d_f}{\hat{\Phi}_f}{C}
\mx{C}{ \hat{\Phi}_i}{d_i} }{\hbar\winit-\frac{(\hbar\winit)^2}{2m_d}+E_{d_i}-E_C+i\varepsilon} 
\right\} \label{eq:mainterm4}.
\end{eqnarray}
\noindent Similar treatment on the crossed term of equation~(\ref{eq:mainterm1}) yields
\begin{eqnarray}
\matrixt{a,\mathrm{cr}} & = & \delta(\vec{P}_i + \vec{k}_i - \vec{P}_f - \vec{k}_f) \left\{
\frac{1}{2}\mx{d_f}{[H^{np},\hat{\Phi}_i]\hat{\Phi}_f}{d_i} - \nonumber \right. \\
& & \frac{1}{2}\mx{d_f}{\hat{\Phi}_i[H^{np},\hat{\Phi}_f]}{d_i} + 
\left[ \hbar\wfin + \frac{(\hbar\wfin)^2}{2m_d} \right] \mx{d_f}{\hat{\Phi}_i\hat{\Phi}_f}{d_i}+\nonumber\\
& & \left. \left[ \hbar\wfin + \frac{(\hbar\wfin)^2}{2m_d} \right]^2  \sum_C \label{eq:mainterm5}
\frac{ \mx{d_f}{\hat{\Phi}_i}{C} 
\mx{C}{ \hat{\Phi}_f}{d_i} }{-\hbar\wfin-  \frac{(\hbar\wfin)^2}{2m_d} +E_{d_i}-E_C+i\varepsilon} \right\}.
\end{eqnarray}
\noindent We now add equations (\ref{eq:mainterm4}) and (\ref{eq:mainterm5}) to get 
\begin{equation}
\matrixt{a}  =  \delta(\vec{P}_i + \vec{k}_i - \vec{P}_f - \vec{k}_f) 
[\matrixt{a1} + \matrixt{a2} + \matrixt{a3} + \matrixt{a4}],  
\end{equation}
\noindent where 
\begin{eqnarray}
\matrixt{a1} & \equiv & \left[ \hbar\winit - \frac{(\hbar\winit)^2}{2m_d}  \right]^2 \sum_C 
\frac{ \mx{d_f}{\hat{\Phi}_f}{C}
\mx{C}{ \hat{\Phi}_i}{d_i} }{\hbar\winit- \frac{(\hbar\winit)^2}{2m_d}+E_{d_i}-E_C+i\varepsilon}
, \label{eq:mainterm6}  \\
\matrixt{a2} & \equiv & \left[ \hbar\wfin + \frac{(\hbar\wfin)^2}{2m_d} \right]^2  \sum_C \label{eq:mainterm7}
\frac{ \mx{d_f}{\hat{\Phi}_i}{C} \mx{C}{ \hat{\Phi}_f}{d_i}}{-\hbar\wfin- \frac{(\hbar\wfin)^2}{2m_d}+
E_{d_i}-E_C+i\varepsilon},  \\
\matrixt{a3} & \equiv & \left[\hbar\wfin + \frac{(\hbar\wfin)^2}{2m_d} - \hbar\winit +  \frac{(\hbar\winit)^2}{2m_d} \right] 
\mx{d_f}{\hat{\Phi}_f\hat{\Phi}_i}{d_i} \label{eq:mainterm8},  \\
\matrixt{a4} & \equiv & \frac{1}{2} \mx{d_f}{\left[ \ [H^{np},\hat{\Phi}_i],  \hat{\Phi}_f \right] +
\left[\ [ H^{np},\hat{\Phi}_f  ] ,  \hat{\Phi}_i \right] }{d_i} \label{eq:mainterm9}.
\end{eqnarray}
\noindent $\matrixt{a1}$ will be calculated first.  We write out  $\hat{\Phi}_i$ and $\hat{\Phi}_f$ explicitly (see 
equations~\ref{eq:mainterm10} and \ref{eq:mainterm11}), 
and then convert $\matrixt{}$ to  $\matrixm{}$ using equation (\ref{eq:seagull10}) to get  
\begin{eqnarray}
\lefteqn{ \matrixm{a1}  =  -e^2  \left[ \hbar\winit - \frac{(\hbar\winit)^2}{2m_d}  \right]^2 
\sum_C \sum_{LM} \sum_{L'M'} 
\int d^3r d^3r' \sqrt{\frac{2\pi(2L'+1)}{L'(L'+1)} }  (-1)^{L'-\mufin} \times  } \nonumber \\
& & \mx{d_f}{ \wignerdf \frac{i^{L'+1}}{\hbar\wfin}  
\left( 1 + r'\frac{d}{dr'} \right) j_{\sss L'}(\frac{\wfin r'}{2})Y_{L'M'}(\hat{r}')}{\vec{r}'} 
\nonumber \times \\
& & \mx{\vec{r}'}{\frac{1}{\hbar\winit- \frac{(\hbar\winit)^2}{2m_d}+
E_{d_i}-E_C+i\varepsilon}}{C} \mxemp{C}{\vec{r}} \bra{\vec{r}} 
\wignerdi \frac{i^{L+1}}{\hbar\winit} \times  \nonumber \\
& & \sqrt{\frac{2\pi(2L+1)}{L(L+1)} }
\left( 1 + r\frac{d}{dr} \right) j_{\sss L}(\frac{\winit r}{2}) Y_{LM}(\hat{r}) \ket{d_i} .
\end{eqnarray}
\noindent The sum over $C$ can be  explicitly divided into radial, angular, and spin sums;
the radial sum can be collapsed. Removing the complete sets of angular and spin states 
(which reduces $d^3r$ to $dr$) yields
\begin{eqnarray}
\lefteqn{ \matrixm{a1}  =  \frac{2\pi e^2}{(\hbar\winit)(\hbar\wfin)} 
\left[ \hbar\winit - \frac{(\hbar\winit)^2}{2m_d}  \right]^2 \sum_{l=0,2} \sum_{l'=0,2}
\sum_{L_CS_CJ_CM_C} \sum_{LM} \sum_{L'M'} 
(-1)^{L'-\mufin} i^{L+L'} \times } \nonumber \\
& & \int r^2 dr\ r^{\prime 2} dr' \wignerdi \wignerdf \sqrt{\frac{(2L+1)(2L'+1)}{LL'(L+1)(L'+1)}} \times\nonumber \\
& &  \frac{u_{\sss l'}(r')}{r'}  \left( 1 + r'\frac{d}{dr'} \right) j_{\sss L'}(\frac{\wfin r'}{2})
\mx{r'}{\frac{1}{\hbar\winit -  \frac{(\hbar\winit)^2}{2m_d} - E_b - H^{np} +i\varepsilon}}{r} \times\nonumber\\
& & \frac{u_{\sss l}(r)}{r} 
\left( 1 + r\frac{d}{dr} \right) j_{\sss L}(\frac{\winit r}{2}) \mxbc{Y_{L'M'}} \times\nonumber\\
& & \mxca{ Y_{LM} } .
\end{eqnarray}
\noindent Defining $E_0 \equiv \hbar\winit -  \frac{(\hbar\winit)^2}{2m_d} - E_b$, 
where $E_b$ is the deuteron binding energy, we can 
rearrange $\matrixm{a1}$ to get
\begin{eqnarray}
\lefteqn{ \matrixm{a1} = \frac{2\pi e^2}{(\hbar\winit)(\hbar\wfin)} 
\left[ \hbar\winit - \frac{(\hbar\winit)^2}{2m_d}  \right]^2 
\sum_{ll'} \sum_{L_CS_CJ_CM_C} \sum_{LM} \sum_{L'M'}
(-1)^{L'-\mufin} i^{L+L'} \times }\nonumber \\
& & \wignerdi \wignerdf \sqrt{\frac{(2L+1)(2L'+1)}{LL'(L+1)(L'+1)}}  \times\nonumber\\
& & \int dr\ dr'\ r\ r'u_{\sss l}(r) u_{\sss l'}(r') 
\psi_{\sss L'}(\frac{\wfin r'}{2})  \green \psi_{\sss LL}(\frac{\winit r}{2}) \times \nonumber \\
& &  \mxbc{Y_{L'M'}} \mxca{ Y_{LM} }, 
\end{eqnarray}
\noindent where $\green \equiv \mx{r'}{\frac{1}{E_0-H^{np}}}{r}$. 
We have also defined a function $\psi_L$ by
\begin{equation}
\psi_{\sss L}(x) \equiv j_{\sss L}(x) + x \frac{d}{dx} j_{\sss L}(x). \label{eq:mainterm16}
\end{equation}
The $M$ sums are formal and are removed to produce the final result:
\begin{eqnarray}
\lefteqn{ \matrixm{a1}   =   \frac{-2\pi e^2}{(\hbar\winit)(\hbar\wfin)} 
\left[ \frac{(\hbar\winit)^2}{2m_d} - \hbar\winit \right]^2
\sum_{l=0,2} \sum_{l'=0,2} \sum_{L,L'=1}^{\infty} \sum_{S_C=0,1} 
\sum_{J_C=|1-L|}^{1+L} \sum_{L_C=|J_C-S_C|}^{J_C+S_C}\times }\nonumber\\
& & \ \ \ \ i^{L+L'}(-1)^{L'-\mufin-M_f+J_C-M_i-\muinit} \wignerdfnom \times\nonumber \\
& & \ \ \ \ \sqrt{\frac{(2L+1)(2L'+1)}{LL'(L+1)(L'+1)}} \threej{1}{L'}{J_C}{-M_f}{M_f-M_i-\muinit}{\muinit+M_i} \times \nonumber \\
& & \ \ \ \ \threej{J_C}{L}{1}{-\muinit-M_i}{\muinit}{M_i} \mxredbc{Y_{L'}} \mxredca{ Y_{L} } \times\nonumber\\ 
& & \ \ \ \ \int dr\ dr'\ r\ r' u_{\sss l}(r) u_{\sss l'}(r') 
\psi_{\sss L'}(\frac{\wfin r'}{2})  \green \psi_{\sss L}(\frac{\winit r}{2}).  
\end{eqnarray}
\noindent Equation~(\ref{eq:mainterm7}) can be handled similarly.  The result is
\begin{eqnarray}
\matrixm{a2} & = & \frac{-2\pi e^2}{(\hbar\winit)(\hbar\wfin)} 
\left[ \hbar\wfin + \frac{ (\hbar\wfin)^2 }{2m_d} \right]^2 
\sum_{l=0,2} \sum_{l'=0,2} \sum_{L,L'=1}^{\infty}  \sum_{S_C=0,1}
\sum_{J_C=|1-L|}^{1+L} \sum_{L_C=|J_C-S_C|}^{J_C+S_C}\times \nonumber\\ 
& & i^{L+L'}(-1)^{L'-\mufin+J_C-\muinit} 
\wignerdfnom  \sqrt{\frac{(2L+1)(2L'+1)}{LL'(L+1)(L'+1)}} \times\nonumber\\
& & \threej{1}{L}{J_C}{-M_f}{\muinit}{-\muinit+M_f} \threej{J_C}{L'}{1}{\muinit-M_f}{M_f-M_i-\muinit}{M_i} 
\times \nonumber \\
& & \mxredbc{Y_{L}} \mxredca{ Y_{L'} } \times\nonumber\\
& & \int  dr\ dr'\ r\ r' u_{\sss l'}(r) u_{\sss l}(r') 
\psi_{\sss L'}(\frac{\wfin r'}{2})  \greenpr \psi_{\sss L}(\frac{\winit r}{2}),
\end{eqnarray}
\noindent where $E_0' \equiv -\hbar\wfin -  \frac{ (\hbar\wfin)^2 }{2m_d} - E_b$.

Equation~(\ref{eq:mainterm8}) contains no intermediate states and therefore no Green's function
is needed.  Using the definitions of $\hat{\Phi}_i$ and $\hat{\Phi}_f$ from equations (\ref{eq:mainterm10}) and (\ref{eq:mainterm11}), we get
\begin{eqnarray}
\matrixm{a3} & = & \frac{2\pi e^2}{(\hbar\wfin)(\hbar\winit)}
\left[ \hbar\wfin  + \frac{ (\hbar\wfin)^2 }{2m_d} - \hbar\winit +  \frac{ (\hbar\winit)^2 }{2m_d} \right] 
\sum_{ll'} \sum_{LM} \sum_{L'M'} (-1)^{L'-\mufin} i^{L+L'} \times \nonumber \\
& & \ \ \ \ \wignerdi \wignerdf \sqrt{\frac{(2L+1)(2L'+1)}{LL'(L+1)(L'+1)}} \times \nonumber \\
& & \ \ \ \ \mxba{Y_{LM}Y_{L'M'}} \times \nonumber \\
& & \ \ \ \ \int dr u_{\sss l}(r) u_{\sss l'}(r) \psi_{\sss L'}(\frac{\wfin r}{2}) \psi_{\sss L}(\frac{\winit r}{2}) .
\end{eqnarray}
\noindent The spherical harmonics are combined according to equation~(\ref{eq:formula10}) to produce:
\begin{eqnarray}
\matrixm{a3}
& = & \frac{e^2 \sqrt{\pi}}{(\hbar\wfin)(\hbar\winit)}  
\left[ \hbar\wfin  + \frac{ (\hbar\wfin)^2 }{2m_d} - \hbar\winit +  \frac{ (\hbar\winit)^2 }{2m_d} \right]\sum_{ll'} 
\sum_{LM} \sum_{L'M'} \sum_{\tilde{L}\tilde{M}}   i^{L+L'} \times \nonumber \\
& &  (-1)^{L'-\mufin + \tilde{M}} 
\wignerdi \wignerdf  \sqrt{\frac{2\tilde{L}+1}{LL'(L+1)(L'+1)}} \times \nonumber \\
& & (2L+1)(2L'+1) \threej{L}{L'}{\tilde{L}}{0}{0}{0} 
\threej{L}{L'}{\tilde{L}}{M}{M'}{-\tilde{M} } \times \nonumber \\
& & \mxba{Y_{\tilde{L}\tilde{M}} } \int dr\ u_{\sss l}(r) u_{\sss l'}(r)
\psi_{\sss L'}(\frac{\wfin r}{2}) \psi_{\sss L}(\frac{\winit r}{2}) .
\end{eqnarray}
\noindent Finally, the $M$ sums are removed to yield the answer
\begin{eqnarray}
\lefteqn{ \matrixm{a3}
 =  -\frac{e^2 \sqrt{\pi}}{(\hbar\wfin)(\hbar\winit)} 
\left[ \hbar\wfin  + \frac{ (\hbar\wfin)^2 }{2m_d} - \hbar\winit +  \frac{ (\hbar\winit)^2 }{2m_d} \right] 
\sum_{l=0,2} \sum_{l'=0,2} \sum_{\tilde{L}=|L-L'|}^{L+L'} \sum_{L=1}^{\infty} \sum_{L'=1}^{\infty} \times} \nonumber\\
& & \wignerdfnom (2L+1)(2L'+1) \sqrt{\frac{2\tilde{L}+1}{LL'(L+1)(L'+1)}} \times \nonumber \\
& &  \threej{L}{L'}{\tilde{L}}{0}{0}{0} 
\threej{L}{L'}{\tilde{L}}{\muinit}{M_f-M_i-\muinit}{M_i-M_f} \threej{1}{\tilde{L}}{1}{-M_f}{M_f-M_i}{M_i} 
\times \nonumber \\
& &  i^{L+L'} (-1)^{L'-\mufin -M_i}  \mxredba{Y_{\tilde{L} } }
\int dr u_{\sss l}(r) u_{\sss l'}(r) \psi_{\sss L'}(\frac{\wfin r}{2}) \psi_{\sss L}(\frac{\winit r}{2}).
\end{eqnarray}
\noindent The Hamiltonians in the double commutators in the $\matrixt{a4}$ term (equation~\ref{eq:mainterm9}) 
will be split into kinetic and potential energy parts.  Inserting a complete set of $\vec{r}$ states into the
kinetic energy term, we see that
\begin{eqnarray}
\matrixm{a4, \mathrm{KE} } & = & \frac{e^2\pi}{(\hbar\wfin)(\hbar\winit)} \sum_{LM} \sum_{L'M'} i^{L+L'} (-1)^{L'-\mufin} 
\sqrt{ \frac{(2L+1)(2L'+1)}{LL'(L+1)(L'+1)} } \nonumber \times \\
& & \int d^3r \bra{d_f}\left[ \ [\frac{p^2}{m_p}, \psi_{\sss L}(\frac{\winit r}{2}) Y_{LM}(\hat{r})], 
\psi_{\sss L'}(\frac{\wfin r}{2}) Y_{L'M'}(\hat{r}) \right] + \nonumber\\
& & \left[ \ [\frac{p^2}{m_p}, \psi_{\sss L'}(\frac{\wfin r}{2}) Y_{L'M'}(\hat{r})], 
\psi_{\sss L}(\frac{\winit r}{2}) Y_{LM}(\hat{r}) \right] \ket{\vec{r}} \times \nonumber \\
& & \mxemp{\vec{r}}{d_i} \wignerdi \wignerdf .
\end{eqnarray}
\noindent Evaluating the double commutators yields
\begin{eqnarray}
\matrixm{a4, \mathrm{KE}} & = & -\frac{4e^2\pi}{m_p \wfin\winit} \sum_{LM} \sum_{L'M'} i^{L+L'} (-1)^{L'-\mufin}
\sqrt{ \frac{(2L+1)(2L'+1)}{LL'(L+1)(L'+1)} } \int d^3r \nonumber \times \\
& & \mx{d_f}{ \vec{\nabla}_r\left[ \psi_{\sss L}(\frac{\winit r}{2}) Y_{LM}(\hat{r}) \right] \cdot
\vec{\nabla}_r\left[ \psi_{\sss L'}(\frac{\wfin r}{2}) Y_{L'M'}(\hat{r}) \right] }{\vec{r} } 
\mxemp{\vec{r}}{d_i} \times \nonumber \\ 
& & \wignerdi \wignerdf,
\end{eqnarray}
\noindent Next, using  the identity $\vec{\nabla}f\cdot\vec{\nabla}g = \frac{1}{2} [ \nabla^2(fg) -
f \nabla^2g - g \nabla^2 f] $, we get
\begin{eqnarray}
\matrixm{a4, \mathrm{KE}} 
& = & -\frac{2e^2\pi}{m_p \wfin\winit} \sum_{LM} \sum_{L'M'} i^{L+L'} (-1)^{L'-\mufin}
\sqrt{ \frac{(2L+1)(2L'+1)}{LL'(L+1)(L'+1)} } \int d^3r \nonumber \times \\
& &  \ \ \ \ \ \ \bra{d_f} \nabla_r^2\left[  \psi_{\sss L}(\frac{\winit r}{2}) Y_{LM}(\hat{r}) 
\psi_{\sss L'}(\frac{\wfin r}{2}) Y_{L'M'}(\hat{r}) \right] - \nonumber \\
& & \ \ \ \ \ \ \psi_{\sss L'}(\frac{\wfin r}{2}) Y_{L'M'}(\hat{r}) 
\nabla_r^2\left[  \psi_{\sss L}(\frac{\winit r}{2}) Y_{LM}(\hat{r}) \right] - \nonumber \\
& & \ \ \ \ \ \ \psi_{\sss L}(\frac{\winit r}{2}) Y_{LM}(\hat{r})\nabla_r^2\left[ \psi_{\sss L'}(\frac{\wfin r}{2}) 
Y_{L'M'}(\hat{r}) \right] \ket{d_f} \times \nonumber \\ 
& & \ \ \ \ \ \ \wignerdi \wignerdf \label{eq:mainterm12}.
\end{eqnarray}
\noindent Since 
\begin{equation}
 \nabla^2\left[Y_{LM}f(r)\right] = Y_{LM}\left[ \frac{1}{r} \frac{d^2}{dr^2} r - \frac{L(L+1)}{r^2}\right] f(r),
\end{equation}
\noindent each of the three terms inside the 
bracket in equation~(\ref{eq:mainterm12}) can be simplified. Thus,
\begin{eqnarray}
\matrixm{a4, \mathrm{KE}} & = & -\frac{e^2\sqrt{\pi}}{m_p \wfin\winit} \sum_{LM} \sum_{L'M'} \sum_{\tilde{L}\tilde{M}}
i^{L+L'} (-1)^{L'-\mufin + \tilde{M}} (2L+1)(2L'+1 ) \nonumber \times \\
& & \ \ \ \ \sqrt{ \frac{2\tilde{L}+1}{LL'(L+1)(L'+1)} } \threej{L}{L'}{\tilde{L}}{0}{0}{0}
\threej{L}{L'}{\tilde{L}}{M}{M'}{-\tilde{M}} \nonumber \times \\
& & \ \ \ \ \wignerdi \wignerdf \times \nonumber \\
& & \ \ \ \ \int d^3r \bra{d_f} Y_{\tilde{L}\tilde{M}} \left\{ \left[ \frac{1}{r} \frac{d^2}{dr^2} r - 
\frac{\tilde{L}(\tilde{L}+1)}{r^2} \right] \psi_{\sss L}(\frac{\winit r}{2}) \psi_{\sss L'}(\frac{\wfin r}{2})  -
\right. \nonumber \\
& &  \ \ \ \ \psi_{\sss L}(\frac{\winit r}{2}) \left[ \frac{1}{r} \frac{d^2}{dr^2} r -
\frac{L'(L'+1)}{r^2} \right] \psi_{\sss L'}(\frac{\wfin r}{2}) - \nonumber \\
& & \ \ \ \ \left. \psi_{\sss L'}(\frac{\wfin r}{2}) \left[ \frac{1}{r} \frac{d^2}{dr^2} r -
\frac{L(L+1)}{r^2} \right] \psi_{\sss L}(\frac{\winit r}{2}) \right\} \ket{\vec{r}} \mxemp{ \vec{r} }{d_i}, 
\end{eqnarray}
\noindent where equation (\ref{eq:formula10}) has been used to combine the spherical harmonics.
After some algebra this becomes
\begin{eqnarray}
\matrixm{a4, \mathrm{KE}}
& = & -\frac{e^2\sqrt{\pi}}{m_p \wfin\winit} \sum_{ll'} \sum_{LM} \sum_{L'M'} \sum_{\tilde{L}\tilde{M}}
i^{L+L'} (-1)^{L'-\mufin + \tilde{M}} (2L+1)(2L'+1 ) \nonumber \times \\
& & \ \ \ \ \sqrt{ \frac{2\tilde{L}+1}{LL'(L+1)(L'+1)} } \threej{L}{L'}{\tilde{L}}{0}{0}{0}
\threej{L}{L'}{\tilde{L}}{M}{M'}{-\tilde{M}} \nonumber \times \\
& & \ \ \ \ \wignerdi \wignerdf \mxba{Y_{\tilde{L}\tilde{M} } } \times\nonumber \\
& & \ \ \ \ \int dr\ u_{\sss l}(r) u_{\sss l'}(r)
\left\{ 2 \left[ \frac{d}{dr}\psi_{\sss L}(\frac{\winit r}{2}) \right]
\left[ \frac{d}{dr}\psi_{\sss L'}(\frac{\wfin r}{2}) \right]  + \right. \nonumber\\
& & \ \ \ \ \left. \frac{ L(L+1) + L'(L'+1) - \tilde{L}(\tilde{L}+1) }{r^2} \psi_{\sss L}(\frac{\winit r}{2})
\psi_{\sss L'}(\frac{\wfin r}{2}) \right\} .
\end{eqnarray}
\noindent We now remove the formal sums over $M$ to obtain the final result:
\begin{eqnarray}
\matrixm{a4, \mathrm{KE}} & = & \frac{e^2\sqrt{\pi}}{m_p \wfin\winit} 
\sum_{l=0,2} \sum_{l'=0,2} \sum_{\tilde{L}=|L-L'|}^{L+L'} \sum_{L=1}^{\infty} \sum_{L'=1}^{\infty} i^{L+L'} (-1)^{L'-\mufin - M_i}
\times\nonumber\\
& &  \ \ \ \ (2L+1)(2L'+1 ) \sqrt{ \frac{2\tilde{L}+1}{LL'(L+1)(L'+1)} }
\threej{L}{L'}{\tilde{L}}{0}{0}{0} \nonumber \times \\
& & \ \ \ \ \threej{L}{L'}{\tilde{L}}{\muinit}{M_f-M_i-\muinit}{M_i-M_f} 
\threej{1}{\tilde{L}}{1}{-M_f}{M_f-M_i}{M_i} \nonumber \times \\
& & \ \ \ \ \wignerdfnom \mxredba{  Y_{\tilde{L}} }  \times \nonumber \\
& & \ \ \ \ \int dr\ u_{\sss l}(r) u_{\sss l'}(r) 
\left\{ 2 \left[ \frac{d}{dr}\psi_{\sss L}(\frac{\winit r}{2}) \right]
\left[ \frac{d}{dr}\psi_{\sss L'}(\frac{\wfin r}{2}) \right] + \nonumber \right. \\
& & \ \ \ \ \left. \frac{ L(L+1) + L'(L'+1) - \tilde{L}(\tilde{L}+1) }{r^2} \psi_{\sss L}(\frac{\winit r}{2})
\psi_{\sss L'}(\frac{\wfin r}{2}) \right\}.
\end{eqnarray}
The potential energy part of the double commutator term, $\matrixt{a4, \mathrm{PE} }$,
will be calculated in Appendix J  using the one pion exchange potential.

\chapter{Other Dispersive Terms \label{app:next} }
\noindent The next set of terms (that we call $\matrixt{b}$) to be investigated are those
 in equation~(\ref{eq:mainterm1}) where the substitution $\vec{A} 
= \vec{\nabla}\Phi$  is made only once. This includes the 
situation where there is an $E1$ interaction at one $\gamma N$ vertex, and a different
type of interaction at the other.  There are four types of terms:
\begin{eqnarray}
\lefteqn{ \matrixt{b}  =  \sum_{C, \vec{P}_C} \left\{  \mx{d_f, \vec{P}_f}{\int \vec{J}(\vec{\dum}')\cdot\vec{\nabla}_{\dum '}\Phi_f(
\vec{\dum}') d^3{\dum}'}{C, \vec{P}_C}  \right. \times } \nonumber\\
& & \ \ \ \ \ \ \frac{1}{ \hbar\winit + E_{d_i} - E_C -P_C^2/2m_d + i\varepsilon }
\mx{C, \vec{P}_C}{\int \vec{J}(\vec{\dum})\cdot\vec{A}(\vec{\dum})d^3\dum}{d_i, \vec{P}_i, \gamma_i}  + \nonumber \\
& & \mx{d_f,  \vec{P}_f, \gamma_f}{\int \vec{J}(\vec{\dum}')\cdot\vec{A}(\vec{\dum}')
d^3{\dum}'}{C, \vec{P}_C} \frac{1}{ \hbar\winit + E_{d_i} - E_C  -P_C^2/2m_d+ i\varepsilon } \times\nonumber\\
& & \ \ \ \ \ \ \mx{C, \vec{P}_C}{\int \vec{J}(\vec{\dum})\cdot\vec{\nabla}_{\dum }\Phi_i(\vec{\dum})
d^3\dum}{d_i, \vec{P}_i} + \nonumber \\
& & \mx{d_f, \vec{P}_f}{\int \vec{J}(\vec{\dum})\cdot\vec{\nabla}_{\dum }\Phi_i(\vec{\dum}) d^3\dum}{C, \vec{P}_C}
\frac{1}{ -\hbar\wfin + E_{d_i} - E_C  -P_C^2/2m_d+ i\varepsilon } \times\nonumber\\
& & \ \ \ \ \ \ \mx{C, \vec{P}_C, \gamma_i, \gamma_f}{\int \vec{J}(\vec{\dum}')\cdot\vec{A}(\vec{\dum}')d^3{\dum}'}{d_i, \vec{P}_i, \gamma_i}
+ \nonumber \\
& & \mx{d_f,  \vec{P}_f, \gamma_f}{\int \vec{J}(\vec{\dum})\cdot\vec{A}(\vec{\dum})
d^3\dum}{C, \vec{P}_C, \gamma_i, \gamma_f} \times\nonumber\\
& & \ \ \ \ \ \ \frac{1}{ -\hbar\wfin+ E_{d_i} - E_C  -P_C^2/2m_d+ i\varepsilon } \times\nonumber\\
& & \ \ \ \ \ \ \left. 
 \mx{C, \vec{P}_C}{\int \vec{J}(\vec{\dum}')\cdot\vec{\nabla}_{\dum '}\Phi_f(\vec{\dum}') d^3{\dum}'}{d_i, \vec{P}_i}\right\}   .
\end{eqnarray}
\noindent Using the same steps that were used to derive  equation~(\ref{eq:mainterm4}) (again ignoring
 the $\vec{P}$ terms in equations \ref{eq:mainterm14} and \ref{eq:mainterm15}), this becomes
\begin{eqnarray}
\lefteqn{ \matrixt{b}  =  i\sum_{C, \vec{P}_C}  \left\{ \mx{d_f, \vec{P}_f}{\hat{\Phi}_f e^{-i\vec{k}_f\cdot\vec{R}} }{C, 
\vec{P}_C} \mx{C, \vec{P}_C}{\int \vec{J}(\vec{\dum})\cdot\vec{A}(\vec{\dum})
d^3\dum}{d_i, \vec{P}_i, \gamma_i} + \right. } \nonumber \\
& &\mx{d_f, \vec{P}_f}{\left[\frac{(\hbar\winit)^2}{2m_d} -\hbar\winit\right]\hat{\Phi}_f 
e^{-i\vec{k}_f\cdot\vec{R}}}{C, \vec{P}_C} 
\frac{1}{ \hbar\winit + E_{d_i} - E_C  -P_C^2/2m_d+ i\varepsilon } \times \nonumber \\
& & \ \ \ \ \ \ \mx{C, \vec{P}_C}{\int \vec{J}(\vec{\dum})\cdot\vec{A}(\vec{\dum})d^3\dum}{d_i, \vec{P}_i, \gamma_i} -\nonumber\\
& & \mx{d_f,  \vec{P}_f, \gamma_f}{\int \vec{J}(\vec{\dum}')\cdot\vec{A}(\vec{\dum}')d^3{\dum}'}{C, \vec{P}_C} 
\mx{C, \vec{P}_C}{ \hat{\Phi}_i e^{i\vec{k}_i\cdot\vec{R}}}{d_i, \vec{P}_i} + \nonumber \\
& & \mx{d_f,  \vec{P}_f, \gamma_f}{\int \vec{J}(\vec{\dum}')\cdot\vec{A}(\vec{\dum}')
d^3{\dum}'}{C, \vec{P}_C} 
\frac{1}{ \hbar\winit + E_{d_i} - E_C  -P_C^2/2m_d+ i\varepsilon } \times\nonumber \\
& & \ \ \ \ \ \ \mx{C, \vec{P}_C}{\left[\hbar\winit - \frac{(\hbar\winit)^2}{2m_d} \right]
\hat{\Phi}_i  e^{i\vec{k}_i\cdot\vec{R}}}{d_i, \vec{P}_i} + \nonumber\\
& & \mx{d_f, \vec{P}_f}{\hat{\Phi}_i e^{i\vec{k}_i\cdot\vec{R}}}{C, \vec{P}_C} 
\mx{C, \vec{P}_C, \gamma_i, \gamma_f}{\int \vec{J}(\vec{\dum}')\cdot\vec{A}
(\vec{\dum}')d^3{\dum}' }{d_i, \vec{P}_i, \gamma_i} + \nonumber \\
& & \mx{d_f, \vec{P}_f}{\left[ \frac{(\hbar\wfin)^2}{2m_d}+\hbar\wfin \right] 
\hat{\Phi}_i  e^{i\vec{k}_i\cdot\vec{R}}}{C, \vec{P}_C} \times\nonumber\\
& & \ \ \ \ \ \ \frac{1}{ -\hbar\wfin + E_{d_i} - E_C  -P_C^2/2m_d+ i\varepsilon }\times\nonumber \\
& & \ \ \ \ \ \ \mx{C, \vec{P}_C, \gamma_i, \gamma_f}{\int \vec{J}(\vec{\dum}')\cdot\vec{A}(\vec{\dum}')d^3{\dum}'}{d_i, \vec{P}_i, \gamma_i}
-\nonumber\\
& & \mx{d_f,  \vec{P}_f, \gamma_f}{\int \vec{J}(\vec{\dum})\cdot\vec{A}(\vec{\dum})
d^3\dum}{C, \vec{P}_C, \gamma_i, \gamma_f} \mx{C, \vec{P}_C}{\hat{\Phi}_f e^{-i\vec{k}_f\cdot\vec{R}}}{d_i, 
\vec{P}_i} -\nonumber \\
& & \mx{d_f,  \vec{P}_f, \gamma_f}{\int \vec{J}(\vec{\dum})\cdot\vec{A}(\vec{\dum})
d^3\dum}{C, \vec{P}_C, \gamma_i, \gamma_f}  \times\nonumber\\
& & \ \ \ \ \ \ \frac{1}{ -\hbar\wfin+ E_{d_i} - E_C  -P_C^2/2m_d+ i\varepsilon } \times\nonumber\\
& & \ \ \ \ \ \ \left. \mx{C, \vec{P}_C}{\left[ \frac{(\hbar\wfin)^2}{2m_d} + \hbar\wfin \right] \hat{\Phi}_f e^{-i\vec{k}_f\cdot\vec{R}}}{d_i,
\vec{P}_i} \right\} .
\end{eqnarray}
\noindent The first and seventh terms above cancel each other, as do the third and fifth.  
Therefore, we only need to consider terms with energy denominators. We replace   
$\vec{A}$ by
$\vec{A}^{(1)} + \vec{A}^{(2)}$, where $\vec{A}^{(1)}$ and $\vec{A}^{(2)}$ are the non-gradient terms in
the multipole expansion of equation~(\ref{eq:multexp99}): 
\begin{eqnarray}
\vec{A}^{(1)}(\vec{\dum}) & = & -\frac{1}{\sqrt{V}} \sum_{\vec{k}, \lambda = \pm 1} \sum_{L=1}^{\infty} \sum_{M=-L}^{L} 
\sqrt{ \frac{2\pi\hbar}{\omega} } 
\lambda \sqrt{ \frac{2\pi(2L+1)}{L(L+1)} } i^L j_{\sss L}(\omega\dum) \vec{L} 
Y_{LM}(\hat{\dum}) \times \nonumber  \\
& & \left\{ a_{k,\lambda} \wignerd - a_{k,\lambda}^{\dagger}  
(-1)^{L-\lambda} \wignerdneg \right\}, \label{eq:nextterm2} \\
\vec{A}^{(2)}(\vec{\dum}) & = & -\frac{1}{\sqrt{V}} \sum_{\vec{k}, \lambda = \pm 1} \sum_{L=1}^{\infty} \sum_{M=-L}^{L} 
\sqrt{ \frac{2\pi\hbar}{\omega} }
\sqrt{ \frac{2\pi(2L+1)}{L(L+1)} } i^{L+1} \omega\vec{\dum} j_{\sss L}(\omega\dum) Y_{LM}(\hat{\dum}) \times \nonumber  \\
& & \left\{ a_{k,\lambda} \wignerd - a_{k,\lambda}^{\dagger}  
(-1)^{L-\lambda} \wignerdneg \right\} \label{eq:nextterm3} .
\end{eqnarray}
\noindent An explicit form for $\vec{J}$ is also needed, so we write $\vec{J} =
\vec{J}^{(\sigma)} + \vec{J}^{(p)}$, where
\begin{eqnarray}
\vec{J}^{(\sigma)}(\vec{\dum}) & = & \frac{e\hbar}{2m_p} \sum_{j=n,p} \left[ \vec{\nabla}_{\dum} \times \mu_j
\vec{\sigma}_j \delta(\vec{\dum} - \vec{x}_j) \right], \label{eq:nextterm4} \\
\vec{J}^{(p)}(\vec{\dum}) & = &  \frac{1}{2m_p} \sum_{j=n,p} \left\{ e_j \delta(\vec{\dum}-\vec{x}_j), 
\vec{p}_j \right\}.
\label{eq:nextterm5}
\end{eqnarray}
\noindent Here $\mu_j$ is the magnetic moment and $\vec{p}_j$  the momentum operator for the $j$th nucleon. To 
keep track of the sixteen different terms in $\matrixt{b}$, we write
\begin{equation}
\matrixt{b} = \matrixt{b1} + \matrixt{b2} + \matrixt{b3} + \matrixt{b4},
\end{equation}
\noindent where
\begin{eqnarray}
 \matrixt{b1} &  \equiv &   -i \sum_{C,\vec{P}_C} \mx{d_f, \vec{P}_f}{\left[ \hbar\winit -\frac{(\hbar\winit)^2}{2m_d} 
 \right] \hat{\Phi}_f e^{-i\vec{k}_f\cdot\vec{R} }}{C,\vec{P}_C}
 \times  \label{eq:nextterm6} \\
& & \ \ \ \ \ \ \frac{1}{ \hbar\winit + E_{d_i} - E_C -P_C^2/2m_d + i\varepsilon } \times\nonumber\\
& & \ \ \ \ \ \ \mx{C,\vec{P}_C}{\int \vec{J}(\vec{\dum})\cdot\vec{A}(\vec{\dum})d^3\dum}{d_i, \vec{P}_i, \gamma_i}  , \nonumber \\
\matrixt{b2} & \equiv  & i\sum_{C,\vec{P}_C} \mx{d_f,  \vec{P}_f, \gamma_f}{\int \vec{J}(\vec{\dum}')\cdot\vec{A}(\vec{\dum}')
d^3{\dum}'}{C,\vec{P}_C} \times \label{eq:nextterm7} \\
& & \ \ \ \ \ \ \frac{1}{ \hbar\winit + E_{d_i} - E_C  -P_C^2/2m_d+ i\varepsilon } \times\nonumber\\
& & \ \ \ \ \ \ \mx{C,\vec{P}_C}{\left[\hbar\winit-  
\frac{(\hbar\winit)^2}{2m_d}\right]\hat{\Phi}_i  e^{i\vec{k}_i\cdot\vec{R}}}{d_i, \vec{P}_i} , \nonumber\\
\matrixt{b3}  & \equiv &  i\sum_{C,\vec{P}_C} \mx{d_f, \vec{P}_f}{\left[ \hbar\wfin +
\frac{(\hbar\wfin)^2}{2m_d}\right] \hat{\Phi}_i e^{i\vec{k}_i\cdot\vec{R}}}{C,\vec{P}_C}
\times \label{eq:nextterm8}\\
& & \ \ \ \ \ \ \frac{1}{ -\hbar \wfin + E_{d_i} - E_C  -P_C^2/2m_d+ i\varepsilon } \times\nonumber\\
& & \ \ \ \ \ \ \mx{C,\vec{P}_C, \gamma_i, \gamma_f}{\int \vec{J}(\vec{\dum}')\cdot\vec{A}(\vec{\dum}')d^3{\dum}'}{d_i,
\vec{P}_i, \gamma_i}, \nonumber\\
\matrixt{b4} &  \equiv &  -i\sum_{C,\vec{P}_C}  \mx{d_f,  \vec{P}_f, \gamma_f}{\int \vec{J}(\vec{\dum})\cdot\vec{A}(\vec{\dum})
d^3\dum}{C,\vec{P}_C , \gamma_i, \gamma_f}
\times \label{eq:nextterm9}\\
& & \ \ \ \ \ \ \frac{1}{ -\hbar\wfin + E_{d_i}  - E_C  -P_C^2/2m_d+ i\varepsilon } \times\nonumber\\
& & \ \ \ \ \ \ \mx{C,\vec{P}_C}{\left[ \hbar\wfin +
\frac{(\hbar\wfin)^2}{2m_d} \right]\hat{\Phi}_f e^{-i\vec{k}_f\cdot\vec{R}}}{d_i, \vec{P}_i} ,\nonumber
\end{eqnarray}
\noindent and
\begin{equation}
\matrixt{b1} = \matrixt{b1,\sigma 1}+\matrixt{b1,\sigma 2}+\matrixt{b1,p1}+\matrixt{b1,p2} 
\ \ \ \ \ \ \ \ \ \ \mathrm{etc.}
\end{equation}
\noindent The superscript $\sigma 2$, for example,  indicates that this is the term with $\vec{J}^{(\sigma)}\cdot
\vec{A}^{(2)}.$   As a reminder, $\hat{\Phi} \equiv (e/h) \Phi(\vec{x}_p).$ 

We start by examining the integral
$\int \vec{J}^{(\sigma)}(\vec{\dum})\cdot\vec{A}^{(1)}(\vec{\dum})d^3\dum $.  Including only the $\vec{\dum}$-dependent
terms for now, this integral becomes
\begin{eqnarray}
\lefteqn{\int \vec{J}^{(\sigma)}(\vec{\dum})\cdot\vec{A}^{(1)}(\vec{\dum})d^3\dum } \nonumber \\  
& \sim &\int \sum_{j=n,p} \left[ \vec{\nabla}_{\dum} \times \mu_j \vec{\sigma}_j \delta(\vec{\dum} - \vec{x}_j) \right]
\cdot j_{\sss L}(\omega\dum) \vec{L} Y_{LM}(\hat{\dum}) d^3\dum \\
& = & \int \left[ \vec{\nabla}_{\dum} \times  j_{\sss L}(\omega\dum) \vec{L} Y_{LM}(\hat{\dum}) \right] \cdot
\left[\mu_p \vec{\sigma}_p \delta(\vec{\dum} - \vec{x}_p) +
\mu_n \vec{\sigma}_n \delta(\vec{\dum} - \vec{x}_n) \right] d^3\dum  \nonumber
\end{eqnarray}
\noindent After performing the integrals, we substitute $\vec{r} \equiv \vec{x}_p - \vec{x}_n$
and neglect recoil terms to get
\begin{eqnarray}
\lefteqn{ \int \vec{J}^{(\sigma)}(\vec{\dum})\cdot\vec{A}^{(1)}(\vec{\dum})d^3\dum } \nonumber\\
& \sim & \left[ \vec{\nabla}_r \times  j_{\sss L}(\omega r/2) \vec{L} Y_{LM}(\hat{r}) \right] \cdot
\left[\mu_p \vec{\sigma}_p - (-1)^L \mu_n \vec{\sigma}_n \right] \label{eq:nextterm1}.
\end{eqnarray}  
Using equation~(\ref{eq:multexp12}),  
the spherical Bessel recursion relations (equations~\ref{eq:formula11} and
\ref{eq:formula12}), and  the curl version of the gradient formula (equations~\ref{eq:formula14}-\ref{eq:formula16}),
equation~(\ref{eq:nextterm1}) becomes 
\begin{eqnarray}
\lefteqn{\int \vec{J}^{(\sigma)}(\vec{\dum})\cdot\vec{A}^{(1)}(\vec{\dum})d^3\dum } \nonumber \\
&  \sim & \left[ -i\omega \sqrt{ \frac{L^2(L+1)}{2L+1} } j_{\sss L+1}(\frac{\omega r}{2})
\vsh{L}{L+1}{M} + \right.  \\
& & \ \ \ \ \ \ \ \ \left. i\omega \sqrt{ \frac{L(L+1)^2}{2L+1} } j_{\sss L-1}(\frac{\omega r}{2})
\vsh{L}{L-1}{M} \right] \cdot 
\left[\mu_p \vec{\sigma}_p - (-1)^L \mu_n \vec{\sigma}_n \right] \nonumber \\
& = & \left[ -i\omega \sqrt{ \frac{L^2(L+1)}{2L+1} } j_{\sss L+1}(\frac{\omega r}{2})\vsh{L}{L+1}{M} + \right. \nonumber\\
& & \ \ \ \ \ \ \left. i\omega \sqrt{ \frac{L(L+1)^2}{2L+1} } j_{\sss L-1}(\frac{\omega r}{2})\vsh{L}{L-1}{M} \right] \cdot \nonumber \\
& & \ \ \ \ \ \ \ \ \ \ \ \ \ \left[\left( \mu_p - (-1)^L \mu_n \right) \vec{S} + 
\left( \mu_p + (-1)^L \mu_n \right) \vec{t} \right] \label{eq:nextterm10},
\end{eqnarray}
\noindent where we have defined $\vec{S} \equiv (\vec{\sigma}_p + \vec{\sigma}_n)/2$ and 
$\vec{t} \equiv (\vec{\sigma}_p - \vec{\sigma}_n)/2$.

We are now ready to calculate the $\matrixt{b1,\sigma 1}$ term.  Removing the center-of-mass
motion immediately and inserting equations (\ref{eq:mainterm10}),
(\ref{eq:nextterm2}), (\ref{eq:nextterm4}), and (\ref{eq:nextterm10}) into (\ref{eq:nextterm6}), we get
\begin{eqnarray}
\lefteqn{ \matrixm{b1,m1}  =    -i\sum_C \sum_{LM} \sum_{L'M'} \int d^3r\ d^3r' \bra{d_f} \frac{e}{\hbar}
\left[ \hbar\winit - \frac{(\hbar\winit)^2}{2m_d}  \right] (-1)^{L'-\mufin} \times} \label{eq:nextterm11} \\
& & \wignerdf \frac{i^{L'+1}}{\wfin} \sqrt{ \frac{2\pi(2L'+1)}{L'(L'+1)} } \left( 1 + r'\frac{d}{dr'} \right)
j_{\sss L'}(\frac{\wfin r'}{2}) \times \nonumber \\
& & Y_{L'M'}(\hat{r}') \ket{\vec{r}'} \mxemp{\vec{r}'}{C} \mx{C}{ \frac{1}{\hbar\winit -  \frac{(\hbar\winit)^2}{2m_d} 
+E_{d_i}-E_C+ i\varepsilon} }{\vec{r} } \wignerdi \times \nonumber \\
& & \bra{\vec{r}} -\frac{e\hbar\muinit}{2m_p} \sqrt{ \frac{2\pi(2L+1)}{L(L+1)} } i^L  
\left[ -i\winit \sqrt{ \frac{L^2(L+1)}{2L+1} } j_{\sss L+1}(\frac{\winit r}{2})\vsh{L}{L+1}{M} + \right. \nonumber \\
& & \left. i\winit \sqrt{ \frac{L(L+1)^2}{2L+1} } j_{\sss L-1}(\frac{\winit r}{2})\vsh{L}{L-1}{M} \right] \cdot 
\left[\left( \mu_p - (-1)^L \mu_n \right) \vec{S} + \right. \nonumber\\
& & \left. \left( \mu_p + (-1)^L \mu_n \right) \vec{t} \right] \ket{d_i} \nonumber .
\end{eqnarray}
\noindent  The $\vec{t}$ term  gives zero contribution since this operator changes the spin,
but $S_i=S_C=1$. The $L+1$
term in the fourth line  will also be  dropped since it is of smaller order than the
$L-1$ term,  and the $L-1$ term turns out to be small itself. We now use the fact that
\begin{eqnarray}
\vsh{J}{L}{M} \cdot \vec{V} & = & \left[ Y_{L} \otimes V \right]_{JM}, \label{eq:nextterm16}
\end{eqnarray}
\noindent where $\vec{V}$ is any vector operator,  to get
\begin{eqnarray}
\lefteqn{ \matrixm{b1,\sigma 1}  =  \sum_{L_C S_C J_C M_C} \sum_{ll'} \sum_{LM} \sum_{L'M'} \frac{i\pi e^2\muinit}{m_p}
\left[  \frac{(\hbar\winit)^2}{2m_d}-\hbar\winit \right] 
i^{L+L'} (-1)^{L'-\mufin} \frac{\winit}{\wfin} \times}  \nonumber \\
& & \wignerdi \wignerdf \sqrt{ \frac{(L+1)(2L'+1)}{L'(L'+1)} } \left[ \mu_p - (-1)^L \mu_n \right] 
\times \nonumber \\
& & \mxbc{Y_{L'M'} } \mxca{ \left[ Y_{L-1} \otimes S \right]_{LM} } \times \nonumber \\
& & \int dr\ dr'\ r\ r'\ u_{\sss l}(r)\ u_{\sss l'}(r') 
\psi_{\sss L'}(\frac{\wfin r'}{2})
\green j_{\sss L-1}(\frac{\winit r}{2}). 
\end{eqnarray}
Again, $\psi_{\sss L}(x) \equiv j_{\sss L}(x) + x \frac{d}{dx} j_{\sss L}(x).$
The $M$ sums can be removed to produce the final result:  
\begin{eqnarray}
\lefteqn{ \matrixm{b1,\sigma 1}  =    \sum_{l=0,2} \sum_{l'=0,2} \sum_{L=1}^{\infty} \sum_{L'=1}^{\infty}
\sum_{S_C=0,1} \sum_{J_C = |1-L|}^{1+L} \sum_{L_C = |J_C-S_C|}^{J_C+S_C} \times} \\
& & \frac{i\pi e^2\muinit}{m_p}\frac{\winit}{\wfin}
\left[ \hbar\winit - \frac{ (\hbar\winit)^2 }{2m_d}  \right] i^{L+L'} 
(-1)^{L'+J_C-\mufin-\muinit-M_f-M_i} \times \nonumber \\
& & \wignerdfnom \sqrt{ \frac{(L+1)(2L'+1)}{L'(L'+1)} } 
\left[ \mu_p - (-1)^L \mu_n \right] \times \nonumber \\
& &\threej{1}{L'}{J_C}{-M_f}{M_f-M_i-\muinit}{\muinit + M_i} \threej{J_C}{L}{1}{-\muinit - M_i}{\muinit}{M_i}
\times \nonumber \\
& & \mxredbc{Y_{L'} } \mxredca{ \left[ Y_{L-1} \otimes S \right]_{L} } \times \nonumber \\
& & \int_{0}^{\infty} \int_{0}^{\infty} dr\ dr'\ r\ r'\ u_{\sss l}(r) \ u_{\sss l'}(r')
\psi_{\sss L'}(\frac{\wfin r'}{2}) \green j_{\sss L-1}(\frac{\winit r}{2}).\nonumber
\end{eqnarray}
\noindent The other $\sigma 1$ terms are very similar so we just write the results here:
\begin{eqnarray}
\lefteqn{ \matrixm{b2,m1}  =  \sum_{l=0,2} \sum_{l'=0,2} \sum_{L=1}^{\infty} \sum_{L'=1}^{\infty}
\sum_{S_C=0,1} \sum_{J_C = |1-L|}^{1+L} \sum_{L_C = |J_C-S_C|}^{J_C+S_C} \times} \\ 
& & \frac{-i\pi e^2\mufin}{m_p} \left[\hbar\wfin +
 \frac{ (\hbar\wfin)^2 }{2m_d} \right] i^{L+L'} (-1)^{L'+J_C-\mufin-\muinit-M_f-M_i} \times \nonumber \\ 
& & \wignerdfnom \sqrt{ \frac{(L'+1)(2L+1)}{L(L+1)} }
\left[ \mu_p - (-1)^{L'} \mu_n \right] \times \nonumber \\
& &\threej{1}{L'}{J_C}{-M_f}{M_f-M_i-\muinit}{\muinit + M_i} \threej{J_C}{L}{1}{-\muinit - M_i}{\muinit}{M_i}
\times \nonumber \\
& & \mxredbc{\left[ Y_{L'-1} \otimes S \right]_{L'} } \mxredca{Y_{L} } \times \nonumber \\
& & \int_{0}^{\infty} \int_{0}^{\infty} dr\ dr'\ r\ r'\ u_{\sss l}(r) \ u_{\sss l'}(r')
\psi_{\sss L}(\frac{\winit r}{2}) 
\green j_{\sss L'-1}(\frac{\wfin r}{2}),\nonumber 
\end{eqnarray}
\begin{eqnarray}
\lefteqn{ \matrixm{b3,m1}  =  \sum_{l=0,2} \sum_{l'=0,2} \sum_{L=1}^{\infty} \sum_{L'=1}^{\infty}
\sum_{S_C=0,1} \sum_{J_C = |1-L|}^{1+L} \sum_{L_C = |J_C-S_C|}^{J_C+S_C} \times} \\ 
& & \frac{-i\pi e^2\mufin}{m_p}\frac{\wfin}{\winit}
\left[ \hbar\wfin + \frac{ (\hbar\wfin)^2 }{2m_d} \right] 
i^{L+L'} (-1)^{L'+J_C-\mufin-\muinit} \times \nonumber \\
& & \wignerdfnom \sqrt{ \frac{(L'+1)(2L+1)}{L(L+1)} }
\left[ \mu_p - (-1)^{L'} \mu_n \right] \times \nonumber \\
& &\threej{1}{L}{J_C}{-M_f}{\muinit}{-\muinit + M_f} \threej{J_C}{L'}{1}{\muinit - M_f}{M_f-M_i-\muinit}{M_i}
\times \nonumber \\
& & \mxredbc{Y_{L}} \mxredca{\left[ Y_{L'-1} \otimes S \right]_{L'} }\times \nonumber \\
& & \int_{0}^{\infty} \int_{0}^{\infty} dr\ dr'\ r\ r'\ u_{\sss l'}(r) \ u_{\sss l}(r')
\psi_{\sss L}(\frac{\winit r}{2}) 
\greenpr j_{\sss L'-1}(\frac{\wfin r}{2}),  \nonumber
\end{eqnarray}
\begin{eqnarray}
\lefteqn{ \matrixm{b4,m1}  =  \sum_{l=0,2} \sum_{l'=0,2} \sum_{L=1}^{\infty} \sum_{L'=1}^{\infty}
\sum_{S_C=0,1} \sum_{J_C = |1-L|}^{1+L} \sum_{L_C = |J_C-S_C|}^{J_C+S_C} \times} \\ 
& & \frac{i\pi e^2\muinit}{m_p}
\left[ \hbar\winit - \frac{ (\hbar\winit)^2 }{2m_d} \right] 
i^{L+L'} (-1)^{L'+J_C-\mufin-\muinit} \times \nonumber \\
& & \wignerdfnom \sqrt{ \frac{(L+1)(2L'+1)}{L'(L'+1)} }
\left[ \mu_p - (-1)^L \mu_n \right] \times \nonumber \\
& &\threej{1}{L}{J_C}{-M_f}{\muinit}{-\muinit + M_f} \threej{J_C}{L'}{1}{\muinit - M_f}{M_f-M_i-\muinit}{M_i}
\times \nonumber \\
& & \mxredbc{\left[ Y_{L-1} \otimes S \right]_{L} }\mxredca{Y_{L'} }  \times \nonumber \\
& & \int_{0}^{\infty} \int_{0}^{\infty} dr\ dr'\ r\ r'\  u_{\sss l'}(r)\ u_{\sss l}(r')
\psi_{\sss L'}(\frac{\wfin r'}{2}) 
\greenpr j_{\sss L-1}(\frac{\winit r}{2}). \nonumber
\end{eqnarray}
\noindent We turn now to the $\sigma 2$ terms, following the same steps used in the derivation of
equation~(\ref{eq:nextterm10}):
\begin{eqnarray}
\lefteqn{\int \vec{J}^{(\sigma)}(\vec{\dum}) \cdot\vec{A}^{(2)}(\vec{\dum}) d^3\dum} \nonumber \\
& \sim & \int \sum_{j=n,p} \left[ \vec{\nabla}_{\dum} 
\times \mu_j \vec{\sigma}_j \delta(\vec{\dum} - \vec{x}_j)
\right] \cdot \vec{\dum} j_{\sss L}(\omega \dum) Y_{LM}(\hat{\dum}) d^3\dum \\
& = & \left[ \vec{\nabla} \times \frac{\vec{r}}{2} j_{\sss L}(\frac{\omega r}{2}) Y_{LM}(\hat{r}) \right]
\cdot \left[ \left( \mu_p + (-1)^L \mu_n \right) \vec{S} + \left( \mu_p - (-1)^L \mu_n \right) \vec{t} \right]\\
& = & \left[ -i \sqrt{L(L+1)} j_{\sss L}(\frac{\omega r}{2}) \vsh{L}{L}{M} \right] \cdot
\left[ \left( \mu_p + (-1)^L \mu_n \right) \vec{S} + \right. \nonumber\\
& & \ \ \ \ \ \ \ \ \left. \left( \mu_p - (-1)^L \mu_n \right) \vec{t} \right].
\end{eqnarray}
\noindent Putting this into equation~(\ref{eq:nextterm6}) gives  (again neglecting the $\vec{t}$ 
 terms):
\begin{eqnarray}
\lefteqn{ \matrixm{b1,\sigma 2}  = -i \int d^3r\ d^3r' \sum_C \sum_{LM} \sum_{L'M'} \bra{d_f} (e/\hbar)\left[ 
\hbar\winit - \frac{ (\hbar\winit)^2}{2m_d}  \right]  (-1)^{L'-\mufin}  \times } \\
& & \wignerdf\frac{i^{L'+1}}{\wfin} \sqrt{\frac{2\pi(2L'+1)}{L'(L'+1)}} \left( 1+r'\frac{d}{dr'} \right) 
j_{\sss L'}(\frac{\wfin r'}{2}) Y_{L'M'}(\hat{r}') \ket{\vec{r}' } \times \nonumber \\
& & \mxemp{\vec{r}'}{C} \frac{1}{\hbar\winit -\frac{ (\hbar\winit)^2}{2m_d} + E_{d_i} - 
E_C + i\varepsilon} \mxemp{C}{\vec{r}} 
\bra{ \vec{r} } -\frac{e\hbar\winit}{2m_p} \sqrt{\frac{2\pi(2L+1)}{L(L+1)}} i^{L+1} \times \nonumber \\
& & \wignerdi \left[ -i \sqrt{L(L+1)}
j_{\sss L}(\frac{\winit r}{2}) \vsh{L}{L}{M} \right] \cdot \left[ \left( \mu_p + (-1)^L \mu_n \right) \vec{S}
\right]  \ket{d_i}. \nonumber
\end{eqnarray}
\noindent Using equation~(\ref{eq:nextterm16}) to convert the dot products into tensor products, we see that this becomes
\begin{eqnarray}
\lefteqn{ \matrixm{b1,\sigma 2}  =  \sum_{L_C S_C J_C M_C} \sum_{ll'} \sum_{LM} \sum_{L'M'} \frac{-\pi e^2 }{m_p} \frac{\winit}{\wfin} 
\left[ \hbar\winit - \frac{ (\hbar\winit)^2 }{2m_d}   \right]  i^{L+L'} (-1)^{L'-\mufin}  \times} \nonumber  \\
& & \wignerdi \wignerdf \sqrt { \frac{ (2L+1)(2L'+1) }{L'(L'+1)} }  \left[ \mu_p + (-1)^L \mu_n \right]
\times \nonumber \\
& & \mxbc{Y_{L'M'}} \mxca{ \left[ Y_L \otimes S \right]_{LM} } \times \nonumber \\
& & \int_{0}^{\infty} \int_{0}^{\infty} dr\ dr'\ r\ r'\ u_{\sss l}(r) \ u_{\sss l'}(r') 
\psi_{\sss L'}(\frac{\wfin r'}{2}) 
\green j_{\sss L}(\frac{\winit r}{2}). 
\end{eqnarray}
\noindent The final expression for $\matrixm{b1,\sigma 2}$ after removing the $M$ sums is
\begin{eqnarray}
\matrixm{b1,\sigma 2} & = & \sum_{l=0,2} \sum_{l'=0,2} \sum_{L=1}^{\infty} \sum_{L'=1}^{\infty}
\sum_{S_C=0,1} \sum_{J_C = |1-L|}^{1+L} \sum_{L_C = |J_C-S_C|}^{J_C+S_C} \times\\
& & \frac{\pi e^2 }{m_p} \frac{\winit}{\wfin}
\left[ \hbar\winit - \frac{ (\hbar\winit)^2}{2m_d}  \right] i^{L+L'} 
(-1)^{L'+J_C-M_f-M_i-\muinit-\mufin} \nonumber \times \\
& &  \wignerdfnom \sqrt { \frac{ (2L+1)(2L'+1) }{L'(L'+1)} }  \left[ \mu_p + (-1)^L \mu_n \right]
\times \nonumber \\
& &\threej{1}{L'}{J_C}{-M_f}{M_f-M_i-\muinit}{\muinit + M_i} \threej{J_C}{L}{1}{-\muinit - M_i}{\muinit}{M_i}
\times \nonumber \\
& & \mxredbc{Y_{L'}} \mxredca{ \left[ Y_L \otimes S \right]_{L} } \times \nonumber \\
& & \int_{0}^{\infty} \int_{0}^{\infty} dr\ dr'\ r\ r'\  u_{\sss l}(r) \ u_{\sss l'}(r')
\psi_{\sss L'}(\frac{\wfin r'}{2}) 
\green j_{\sss L}(\frac{\winit r}{2}).\nonumber
\end{eqnarray}
\noindent The expressions for the other $\sigma 2$ terms are:
\begin{eqnarray}
\matrixm{b2,\sigma 2} & = & \sum_{l=0,2} \sum_{l'=0,2} \sum_{L=1}^{\infty} \sum_{L'=1}^{\infty}
\sum_{S_C=0,1} \sum_{J_C = |1-L|}^{1+L} \sum_{L_C = |J_C-S_C|}^{J_C+S_C} \times\\
& &  \frac{-\pi e^2 }{m_p} \left[ \hbar\wfin +
\frac{ (\hbar\wfin)^2 }{2m_d} \right]
i^{L+L'} (-1)^{L'+J_C-M_f-M_i-\muinit-\mufin} \nonumber \times \\
& &  \wignerdfnom \sqrt { \frac{ (2L+1)(2L'+1) }{L(L+1)} }  \left[ \mu_p + (-1)^{L'}\mu_n \right]
\times \nonumber \\
& &\threej{1}{L'}{J_C}{-M_f}{M_f-M_i-\muinit}{\muinit + M_i} \threej{J_C}{L}{1}{-\muinit - M_i}{\muinit}{M_i}
\times \nonumber \\
& & \mxredbc{\left[ Y_{L'} \otimes S \right]_{L'} } \mxredca{  Y_L } \times \nonumber \\
& & \int_{0}^{\infty} \int_{0}^{\infty} dr\ dr'\ r\ r'\ u_{\sss l}(r) \ u_{\sss l'}(r')
\psi_{\sss L}(\frac{\winit r}{2}) 
\green j_{\sss L'}(\frac{\wfin r'}{2}), \nonumber \\
\matrixm{b3,\sigma 2} & = & \sum_{l=0,2} \sum_{l'=0,2} \sum_{L=1}^{\infty} \sum_{L'=1}^{\infty}
\sum_{S_C=0,1} \sum_{J_C = |1-L|}^{1+L} \sum_{L_C = |J_C-S_C|}^{J_C+S_C} \times\\
& &  \frac{-\pi e^2 }{m_p}
\frac{\wfin}{\winit} \left[ \hbar\wfin + \frac{ (\hbar\wfin)^2 }{2m_d} \right]   i^{L+L'}
 (-1)^{L'+J_C-\muinit-\mufin} \nonumber \times \\
& &  \wignerdfnom \sqrt { \frac{ (2L+1)(2L'+1) }{L(L+1)} }  \left[ \mu_p + (-1)^{L'} \mu_n \right]
\times \nonumber \\
& &\threej{1}{L}{J_C}{-M_f}{\muinit}{-\muinit + M_f} \threej{J_C}{L'}{1}{\muinit - M_f}{M_f-M_i-\muinit}{M_i}
\times \nonumber \\
& & \mxredbc{ Y_L } \mxredbc{\left[ Y_{L'} \otimes S \right]_{L'} }\times \nonumber \\
& & \int_{0}^{\infty} \int_{0}^{\infty} dr\ dr'\ r\ r'\ u_{\sss l'}(r) \ u_{\sss l}(r')
\psi_{\sss L}(\frac{\winit r}{2}) 
\greenpr  j_{\sss L'}(\frac{\wfin r'}{2}), \nonumber \\
\matrixm{b4,\sigma 2} & = & \sum_{l=0,2} \sum_{l'=0,2} \sum_{L=1}^{\infty} \sum_{L'=1}^{\infty}
\sum_{S_C=0,1} \sum_{J_C = |1-L|}^{1+L} \sum_{L_C = |J_C-S_C|}^{J_C+S_C} \times\\
& &  \frac{\pi e^2 }{m_p} 
\left[ \hbar\winit - \frac{ (\hbar\winit) }{2m_d} \right]
 i^{L+L'}  (-1)^{L'+J_C-\muinit-\mufin} \nonumber \times \\
& &  \wignerdfnom \sqrt { \frac{ (2L+1)(2L'+1) }{L'(L'+1)} }  \left[ \mu_p + (-1)^{L} \mu_n \right]
\times \nonumber \\
& &\threej{1}{L}{J_C}{-M_f}{\muinit}{-\muinit + M_f} \threej{J_C}{L'}{1}{\muinit - M_f}{M_f-M_i-\muinit}{M_i}
\times \nonumber \\
& & \mxredbc{ \left[ Y_L \otimes S \right]_{L} } \mxredca{Y_{L'}} \times \nonumber \\
& & \int_{0}^{\infty} \int_{0}^{\infty} dr\ dr'\ r\ r'\ u_{\sss l'}(r) \ u_{\sss l}(r')
\psi_{\sss L'}(\frac{\wfin r'}{2}) 
\greenpr j_{\sss L}(\frac{\winit r}{2}) \nonumber .
\end{eqnarray}
Next, we will look at the $p1$ terms.  Inserting equations~(\ref{eq:nextterm2}) and (\ref{eq:nextterm5})
into the $\vec{\dum}$-integral  gives 
\begin{eqnarray}
\lefteqn{ \int \vec{J}^{(p)}(\vec{\dum})\cdot\vec{A}^{(1)}(\vec{\dum}) d^3\dum } \nonumber\\
& \sim & \int \sum_{j=n,p} \left\{ \vec{p}_j, e_j \delta(\vec{\dum} - \vec{x}_j) \right\} \cdot
\vec{L} Y_{LM}(\hat{\dum}) j_{\sss L}(\omega\dum) d^3\dum \nonumber\\
& = & i e\hbar \vec{\nabla}_{x_p} \left[ Y_{LM}(\hat{x}_p) 
j_{\sss L}(\omega x_p) \right] \cdot \vec{L}. 
\end{eqnarray}
\noindent A second term, which would involve the gradient of the wavefunction,
gives no contribution and has been dropped.  The gradient in the above equation can be evaluated
with the help of equation (\ref{eq:formula13}).  We find
\begin{eqnarray}
\lefteqn{ \int \vec{J}^{(p)}(\vec{\dum})\cdot\vec{A}^{(2)}(\vec{\dum}) d^3\dum } \\
& \sim & i e\hbar\omega \left\{ \sqrt{ \frac{L}{2L+1} } j_{\sss L-1}(\frac{\omega r}{2}) 
[ Y_{L-1} \otimes L ]_{LM} + \sqrt{ \frac{L+1}{2L+1} } j_{\sss L+1}(\frac{\omega r}{2})
[ Y_{L+1} \otimes L ]_{LM} \right\}. \nonumber
\end{eqnarray} 
\noindent We are now ready to calculate $\matrixm{b1,p1}$.  Including only the main ($L-1$) term
from above, we insert two complete sets of $\vec{r}$ states into equation (\ref{eq:nextterm6}) to get
\begin{eqnarray}
\lefteqn{ \matrixm{b1,p1}  =   -i \sum_C \sum_{LM} \sum_{L'M'} \int d^3r \int d^3r' \bra{d_f} \frac{e}{\hbar}
\left[ \hbar\winit - \frac{(\hbar\winit)^2}{2m_d}  \right]  (-1)^{L'-\mufin} \times} \\
& &  \frac{i^{L'+1}}{\wfin} \wignerdf
\sqrt{\frac{2\pi(2L'+1)}{L'(L'+1)}}
 \left( 1 + r'\frac{d}{dr'} \right) j_{\sss L'}(\frac{\wfin r'}{2}) Y_{L'M'}(\hat{r}') \ket{\vec{r}'} \times\nonumber\\
& & \bra{ \vec{r}' } \frac{1}{\hbar\winit + E_{d_i} - E_C + i\varepsilon} \ket{C} 
\mxemp{C}{\vec{r}} \bra{\vec{r}} -\frac{\muinit}{2m_p} \sqrt{ \frac{2\pi(2L+1)}{L(L+1)}} i^L \times\nonumber\\
& & \wignerdi ie\hbar\winit\sqrt{ \frac{L}{2L+1} } j_{\sss L-1}(\frac{\winit r}{2})[ Y_{L-1} \otimes L ]_{LM} \ket{d_i}. \nonumber
\end{eqnarray}
\noindent After some simplification, this becomes
\begin{eqnarray}
\lefteqn{ \matrixm{b1,p1}  =  \sum_{L_C S_C J_C M_C} \sum_{ll'} \sum_{LM} \sum_{L'M'}  \frac{\pi ie^2\muinit}{m_p} \frac{\winit}{\wfin}
\left[ \frac{(\hbar\winit)^2}{2m_d} - \hbar\winit \right] i^{L+L'} (-1)^{L'-\mufin} \times} \\
& & \wignerdf\wignerdi \sqrt{ \frac{2L'+1}{L'(L+1)(L'+1)} }  \nonumber\times\\
& &\mxbc{ Y_{L'M'} } \mxca{ [Y_{L-1} \otimes L]_{LM} } \times\nonumber\\
& &\int_{0}^{\infty} \int_{0}^{\infty}  dr\ dr'\ r\ r'\ u_{\sss l'}(r') \psi_{\sss L'}(\frac{\wfin r'}{2}) \green
j_{\sss L-1}(\frac{\winit r}{2}) u_{\sss l}(r) . \nonumber 
\end{eqnarray}
\noindent The final expression without the $M$ sums is
\begin{eqnarray}
\matrixm{b1,p1} & = & \sum_{l=0,2} \sum_{l'=0,2} \sum_{L=1}^{\infty} \sum_{L'=1}^{\infty}
\sum_{S_C=0,1} \sum_{J_C = |1-L|}^{1+L} \sum_{L_C = |J_C-S_C|}^{J_C+S_C} \times\\
& &  \frac{\pi ie^2\muinit}{m_p} \frac{\winit}{\wfin}
\left[ \hbar\winit - \frac{ (\hbar\winit)^2 }{2m_d} \right] i^{L+L'} (-1)^{L'+J_C-\muinit-\mufin-M_f-M_i} \times\nonumber\\
& & \wignerdfnom  \sqrt{ \frac{2L'+1}{L'(L+1)(L'+1)} }  \nonumber\times\\
& & \threej{1}{L'}{J_C}{-M_f}{M_f-M_i-\muinit}{\muinit+M_i}
\threej{J_C}{L}{1}{-\muinit-M_i}{\muinit}{M_i} \nonumber\times\\
& &\mxredbc{ Y_{L'} } \mxredca{ [Y_{L-1} \otimes L]_{L} } \times\nonumber\\
& &\int_{0}^{\infty} \int_{0}^{\infty}  dr\ dr'\ r\ r'\ u_{\sss l'}(r') \psi_{\sss L'}(\frac{\wfin r'}{2}) \green
j_{\sss L-1}(\frac{\winit r}{2}) u_{\sss l}(r). \nonumber 
\end{eqnarray}
\noindent We simply list the other $p1$ terms:
\begin{eqnarray}
\matrixm{b2,p1} & = & \sum_{l=0,2} \sum_{l'=0,2} \sum_{L=1}^{\infty} \sum_{L'=1}^{\infty}
\sum_{S_C=0,1} \sum_{J_C = |1-L|}^{1+L} \sum_{L_C = |J_C-S_C|}^{J_C+S_C} \times\\
& &  \frac{-i\pi e^2\mufin}{m_p} 
\left[ \hbar\wfin + \frac{ (\hbar\wfin)^2 }{2m_d} \right]
i^{L+L'} (-1)^{L'+J_C-\mufin-\muinit-M_f-M_i} \times \nonumber \\
& & \wignerdfnom \sqrt{ \frac{(2L+1)}{L(L+1)(L'+1)} } \times \nonumber \\
& &\threej{1}{L'}{J_C}{-M_f}{M_f-M_i-\muinit}{\muinit + M_i} \threej{J_C}{L}{1}{-\muinit - M_i}{\muinit}{M_i}
\times \nonumber \\
& & \mxredbc{\left[ Y_{L'-1} \otimes L \right]_{L'} } \mxredca{Y_{L} } \times \nonumber \\
& & \int_{0}^{\infty} \int_{0}^{\infty}  dr\ dr'\ r\ r'\ u_{\sss l}(r) \ u_{\sss l'}(r')
\psi_{\sss L}(\frac{\winit r}{2}) 
\green j_{\sss L'-1}(\frac{\wfin r}{2}), \nonumber \\
\matrixm{b3,p1} & = & \sum_{l=0,2} \sum_{l'=0,2} \sum_{L=1}^{\infty} \sum_{L'=1}^{\infty}
\sum_{S_C=0,1} \sum_{J_C = |1-L|}^{1+L} \sum_{L_C = |J_C-S_C|}^{J_C+S_C} \times\\ 
& &  \frac{-i\pi e^2\mufin}{m_p}\frac{\wfin}{\winit}
\left[ \hbar\wfin + \frac{ (\hbar\wfin)^2 }{2m_d} \right] 
i^{L+L'} (-1)^{L'+J_C-\mufin-\muinit} \times \nonumber \\
& & \wignerdfnom \sqrt{ \frac{(2L+1)}{L(L+1)(L'+1)} } \times \nonumber \\
& &\threej{1}{L}{J_C}{-M_f}{\muinit}{-\muinit + M_f} \threej{J_C}{L'}{1}{\muinit - M_f}{M_f-M_i-\muinit}{M_i}
\times \nonumber \\
& & \mxredbc{Y_{L}} \mxredca{\left[ Y_{L'-1} \otimes L \right]_{L'} }\times \nonumber \\
& & \int_{0}^{\infty} \int_{0}^{\infty}  dr\ dr'\ r\ r'\ u_{\sss l'}(r) \ u_{\sss l}(r')
\psi_{\sss L}(\frac{\winit r}{2}) 
\greenpr j_{\sss L'-1}(\frac{\wfin r}{2}), \nonumber \\
\matrixm{b4,p1} & = &  \sum_{l=0,2} \sum_{l'=0,2} \sum_{L=1}^{\infty} \sum_{L'=1}^{\infty}
\sum_{S_C=0,1} \sum_{J_C = |1-L|}^{1+L} \sum_{L_C = |J_C-S_C|}^{J_C+S_C} \times\\
& & \frac{i\pi e^2\muinit}{m_p}
\left[ \hbar\winit - \frac{ (\hbar\winit)^2 }{2m_d} \right] 
i^{L+L'} (-1)^{L'+J_C-\mufin-\muinit} \times \nonumber \\
& & \wignerdfnom \sqrt{ \frac{(2L'+1)}{L'(L+1)(L'+1)} }\times \nonumber \\
& &\threej{1}{L}{J_C}{-M_f}{\muinit}{-\muinit + M_f} \threej{J_C}{L'}{1}{\muinit - M_f}{M_f-M_i-\muinit}{M_i}
\times \nonumber \\
& & \mxredbc{\left[ Y_{L-1} \otimes L \right]_{L} }\mxredca{Y_{L'} }  \times \nonumber \\
& & \int_{0}^{\infty} \int_{0}^{\infty}  dr\ dr'\ r\ r'\ u_{\sss l'}(r) \ u_{\sss l}(r')
\psi_{\sss L'}(\frac{\wfin r'}{2}) 
\greenpr j_{\sss L-1}(\frac{\winit r}{2}). \nonumber 
\end{eqnarray}
\noindent Lastly, we calculate the $p2$ terms. Again, we simplify the $\vec{\dum}$-dependent terms first:
\begin{eqnarray}
\lefteqn{\int \vec{J}^{(p)}(\vec{\dum}) \cdot \vec{A}^{(2)}(\vec{\dum}) d^3\dum} \\
& \sim & \sum_{j=n,p} \int \left\{ e_j \delta(\vec{\dum} - \vec{x}_j) , \vec{p}_j \right\} \cdot
\vec{\dum} j_{\sss L}(\omega\dum) Y_{LM}(\hat{\dum}) d^3\dum \\
& = & \frac{e\hbar}{2i} \left\{ \vec{\nabla}_r \cdot \left[ \vec{r} 
j_{\sss L}(\frac{\omega r}{2}) Y_{LM}(\hat{r})
\right] + 2 j_{\sss L}(\frac{\omega r}{2}) Y_{LM}(\hat{r}) \vec{r} \cdot \vec{\nabla}_r \right\} .
\end{eqnarray}
\noindent Using the gradient formula (equation \ref{eq:formula13}) and the form of the deuteron wavefunction 
(equation \ref{eq:seagull7}), we can show that
\begin{eqnarray}
\vec{\nabla} \cdot \left[ \vec{r} j_{\sss L}(\frac{\winit r}{2}) Y_{LM}(\hat{r}) \right] & = &
\left[ 3j_{\sss L}(\frac{\winit r}{2}) + \frac{\winit r}{2} j_{\sss L}'(\frac{\winit r}{2}) \right] Y_{LM}(\hat{r})
\end{eqnarray}
\noindent and
\begin{eqnarray}
\vec{r}\cdot \vec{\nabla} \mxemp{\vec{r}}{d_i} & = &
\sum_l r \frac{d}{dr} \frac{u_{\sss l}(r)}{r} \mathcal{Y}_{M_i}^{l11}(\hat{r}).
\end{eqnarray}
\noindent Therefore,
\begin{eqnarray}
\lefteqn{\int \vec{J}^{(p)}(\vec{\dum}) \cdot \vec{A}^{(2)}(\vec{\dum}) d^3\dum} \\
& \sim & \frac{e\hbar}{2i} Y_{LM}(\hat{r}) \left\{ 3j_{\sss L}(\frac{\omega r}{2}) + 
\frac{\omega r}{2} j_{\sss L}'(\frac{\omega r}{2}) 
+ 2 j_{\sss L}(\frac{\omega r}{2}) r \frac{d}{dr} \right\},
\end{eqnarray}
\noindent where the derivative in the last term acts on a wavefunction.
\noindent Inserting this into equation (\ref{eq:nextterm6}), we get
\begin{eqnarray}
\lefteqn{ \matrixm{b1,p2} = -i \sum_{L_C S_C J_C M_C} \sum_{ll'} \sum_{LM} \sum_{L'M'} \int d^3r d^3r'
\bra{d_f} \left[  \hbar\winit - \frac{ (\hbar\winit)^2}{2m_d}   \right] \times} \\
& & (-1)^{L'-\mufin} \frac{ei^{L'+1}}{\hbar\wfin} \wignerdf \sqrt{ \frac{2\pi (2L'+1)}{L'(L'+1)} }
\left( 1 + r' \frac{d}{dr'} \right) j_{\sss L'}(\frac{\wfin r'}{2}) \times\nonumber\\
& & Y_{L'M'}(\hat{r}) \ket{\vec{r}}
\mxemp{\vec{r}}{C}  \frac{1}{\hbar\winit   -\frac{ (\hbar\winit)^2}{2m_d} +
E_{d_i} - E_C + i\varepsilon} \mxemp{C}{\vec{r}} \times\nonumber\\
& & \bra{\vec{r}} -\sqrt{ \frac{2\pi (2L+1)}{L(L+1)} } i^{L+1} \frac{\winit}{2m_p} \wignerdi \frac{e\hbar}{2i}
Y_{LM}(\hat{r}) \times\nonumber\\
& & \left\{ 3j_{\sss L}(\frac{\winit r}{2}) +
\frac{\winit r}{2} j_{\sss L}'(\frac{\winit r}{2})
+ 2 j_{\sss L}(\frac{\winit r}{2}) r \frac{d}{dr} \right\} \ket{d_i} .\nonumber
\end{eqnarray}
\noindent This simplifies to
\begin{eqnarray}
\matrixm{b1,p2} & = & \sum_{L_C S_C J_C M_C} \sum_{ll'} \sum_{LM} \sum_{L'M'} \frac{-e^2\pi}{2m_p}
\frac{\winit}{\wfin} \left[  \hbar\winit - \frac{ (\hbar\winit)^2}{2m_d}   \right]
(-1)^{L'-\mufin} \times \\
& &  i^{L'+L} \wignerdf \wignerdi \sqrt{ \frac{(2L+1)(2L'+1)}{LL'(L+1)(L'+1)} }\times\nonumber\\
& & \mxbc{Y_{L'M'} } \mxca{Y_{LM} } \times\nonumber \\
& & \int_{0}^{\infty} \int_{0}^{\infty}  r'\ dr'\ r\ dr\ u_{\sss l'}(r') \psi_{\sss L'}(\frac{\wfin r'}{2}) \green\times\nonumber\\ 
& & \left[ j_{\sss L}(\frac{\winit r}{2}) u_{\sss l}(r) + \frac{\winit r}{2} j_{\sss L}'(\frac{\winit r}{2})  u_{\sss l}(r)+
2rj_{\sss L}(\frac{\winit r}{2}) u_{\sss l}'(r) \right] . \nonumber 
\end{eqnarray}
\noindent After removing the $M$ sums, this becomes
\begin{eqnarray}
\lefteqn{ \matrixm{b1,p2}  =   \sum_{l=0,2} \sum_{l'=0,2} \sum_{L=1}^{\infty} \sum_{L'=1}^{\infty}
\sum_{S_C=0,1} \sum_{J_C = |1-L|}^{1+L} \sum_{L_C = |J_C-S_C|}^{J_C+S_C} \frac{ e^2\pi}{2m_p} \frac{\winit}{\wfin} \times}\\
& & \left[  \hbar\winit - \frac{ (\hbar\winit)^2}{2m_d}  \right]
 i^{L'+L} (-1)^{L'+ J_C - \mufin - \muinit - M_f - M_i} \wignerdfnom \times\nonumber\\
& & \sqrt{ \frac{(2L+1)(2L'+1)}{LL'(L+1)(L'+1)} } \threej{1}{L'}{J_C}{-M_f}{M_f-M_i-\muinit}{M_i+\muinit}
\times\nonumber\\ 
& & \threej{J_C}{L}{1}{-M_i-\muinit}{\muinit}{M_i} \mxredbc{Y_{L'} } \mxredca{Y_{L} } \times\nonumber \\
& & \int_{0}^{\infty} \int_{0}^{\infty}  r'\ dr'\ r\ dr\ u_{\sss l'}(r') \psi_{\sss L'}(\frac{\wfin r'}{2}) \green\times\nonumber\\ 
& & \left[ j_{\sss L}(\frac{\winit r}{2}) u_{\sss l}(r) + \frac{\winit r}{2} j_{\sss L}'(\frac{\winit r}{2})  u_{\sss l}(r)+
2rj_{\sss L}(\frac{\winit r}{2}) u_{\sss l}'(r) \right]. \nonumber
\end{eqnarray}
\noindent The other $p2$ terms are:
\begin{eqnarray}
\lefteqn{ \matrixm{b2,p2}  =  - \sum_{l=0,2} \sum_{l'=0,2} \sum_{L=1}^{\infty} \sum_{L'=1}^{\infty}
\sum_{S_C=0,1} \sum_{J_C = |1-L|}^{1+L} \sum_{L_C = |J_C-S_C|}^{J_C+S_C} \frac{ \pi e^2}{2m_p} \times}\\
& & \left[ \hbar\wfin + \frac{ (\hbar\wfin)^2 }{2m_d} \right]
i^{L'+L} (-1)^{L'+ J_C - \mufin - \muinit - M_f - M_i} \wignerdfnom\times\nonumber\\
& & \sqrt{ \frac{(2L+1)(2L'+1)}{LL'(L+1)(L'+1)} } \threej{1}{L'}{J_C}{-M_f}{M_f-M_i-\muinit}{M_i+\muinit}
\times\nonumber\\
& & \threej{J_C}{L}{1}{-M_i-\muinit}{\muinit}{M_i} \mxredbc{Y_{L'} }  \times\nonumber \\
& &  \mxredca{Y_{L} } \int_{0}^{\infty} \int_{0}^{\infty}  r'\ dr'\ r\ dr\ u_{\sss l}(r) \psi_{\sss L}(\frac{\winit r}{2}) \green\times\nonumber\\ 
& & \left[ j_{\sss L'}(\frac{\wfin  r'}{2}) u_{\sss l'}(r') + \frac{\wfin r'}{2} j_{\sss L'}'(\frac{\wfin r'}{2}) 
u_{\sss l'}(r')+ 2r'j_{\sss L'}(\frac{\wfin r'}{2}) u_{\sss l'}'(r') \right], \nonumber \\
\lefteqn{ \matrixm{b3,p2}  =  - \sum_{l=0,2} \sum_{l'=0,2} \sum_{L=1}^{\infty} \sum_{L'=1}^{\infty}
\sum_{S_C=0,1} \sum_{J_C = |1-L|}^{1+L} \sum_{L_C = |J_C-S_C|}^{J_C+S_C}  \frac{ e^2\pi}{2m_p} \frac{\wfin}{\winit} \times} \nonumber\\
& & \left[ \hbar\wfin + \frac{ (\hbar\wfin)^2 }{2m_d} \right]
i^{L'+L} (-1)^{L'+ J_C - \mufin - \muinit } \wignerdfnom \times\nonumber\\
& & \sqrt{ \frac{(2L+1)(2L'+1)}{LL'(L+1)(L'+1)} } \threej{1}{L}{J_C}{-M_f}{\muinit}{M_f-\muinit}
\times\nonumber\\
& & \threej{J_C}{L'}{1}{-M_f+\muinit}{M_f-M_i-\muinit}{M_i} \mxredbc{Y_{L} }  \times\nonumber \\
& & \mxredca{Y_{L'} } \int_{0}^{\infty} \int_{0}^{\infty}  r'\ dr'\ r\ dr\ u_{\sss l'}(r) 
\psi_{\sss L}(\frac{\winit r}{2}) \greenpr\times\nonumber\\ 
& & \left[ j_{\sss L'}(\frac{\wfin  r'}{2}) u_{\sss l}(r') + \frac{\wfin r'}{2} j_{\sss L'}'(\frac{\wfin r'}{2})
u_{\sss l}(r')+ 2r'j_{\sss L'}(\frac{\wfin r'}{2}) u_{\sss l}'(r') \right], \nonumber \\
\lefteqn{ \matrixm{b4,p2} =   \sum_{l=0,2} \sum_{l'=0,2} \sum_{L=1}^{\infty} \sum_{L'=1}^{\infty}
\sum_{S_C=0,1} \sum_{J_C = |1-L|}^{1+L} \sum_{L_C = |J_C-S_C|}^{J_C+S_C} \frac{ e^2\pi}{2m_p} \times} \nonumber\\
& & \left[ \hbar\winit - \frac{ (\hbar\winit)^2 }{2m_d} \right]
i^{L'+L} (-1)^{L'+ J_C - \mufin - \muinit} \wignerdfnom \times\nonumber\\
& & \sqrt{ \frac{(2L+1)(2L'+1)}{LL'(L+1)(L'+1)} } \threej{1}{L}{J_C}{-M_f}{\muinit}{M_f-\muinit}
\times\nonumber\\
& & \threej{J_C}{L'}{1}{-M_f+\muinit}{M_f-M_i-\muinit}{M_i} \mxredbc{Y_{L} } \times\nonumber \\
& & \mxredca{Y_{L'} } \int_{0}^{\infty} \int_{0}^{\infty}  r'\ dr'\ r\ dr\ u_{\sss l}(r') 
\psi_{\sss L'}(\frac{\wfin r'}{2}) \greenpr\times\nonumber \\
& & \left[ j_{\sss L}(\frac{\winit r}{2}) u_{\sss l'}(r) + 
\frac{\winit r}{2} j_{\sss L}'(\frac{\winit r}{2})  u_{\sss l'}(r)+
2rj_{\sss L}(\frac{\winit r}{2}) u_{\sss l'}'(r) \right]. \nonumber
\end{eqnarray} 
\noindent Finally, there are terms in $\matrixt{}$ where the substitution $\vec{A} = \vec{\nabla}\Phi$
for the $\vec{A}$ in the Hamiltonian is not made at all.  We can instead substitute one of the other two 
terms in the multipole expansion.  
These terms are generally small, so only the ones with non-negligible amplitudes are listed below.  
To keep track of the various terms, the names will indicate which combinations of $\vec{J}$ and $\vec{A}$ were used
in its derivation.
For example, $\matrixm{p2,\sigma 1, \mathrm{cr} }$ is the term derived from
\begin{equation}
\sum_C \frac{ \mx{d_i}{\int \vec{J}^{(p)}\cdot\vec{A}^{(2)} }{C} \mx{C}{ \int \vec{J}^{(\sigma)}\cdot\vec{A}^{(1)} }{C} }{
-\hbar\wfin - \frac{(\hbar\wfin)^2}{2m_d} + E_{d_i} - E_C +i\varepsilon }.
\end{equation}
\noindent  The most important term below is the $\matrixm{\sigma 1, \sigma 1}$ term;
this is the double $M1$ interaction which dominates the photodisintegration cross-section
at threshold.  Here are the results:
\begin{eqnarray}
\lefteqn{ \matrixm{\sigma 1,\sigma 1, \mathrm{uncr} }  =  \sum_{l=0,2} \sum_{l'=0,2} \sum_{L=1}^{\infty} \sum_{L'=1}^{\infty}
\sum_{S_C=0,1} \sum_{J_C = |1-L|}^{1+L} \sum_{L_C = |J_C-S_C|}^{J_C+S_C}  -\frac{\pi e^2 (\hbar\wfin)(\hbar\winit) }{2m_p^2} \times}\\
& & i^{L+L'} (-1)^{L'+J_C-M_f-M_i-\muinit-\mufin} 
\muinit\mufin \wignerdfnom  \sqrt{(L+1)(L'+1)} \times\nonumber\\
& & \threej{1}{L'}{J_C}{-M_f}{M_f-M_i-\muinit}{M_i+\muinit}
\threej{J_C}{L}{1}{-M_i-\muinit}{\muinit}{M_i} \times\nonumber\\
& & \left\{ \mxredbc{ \left[\mu_p - (-1)^{L'} \mu_n \right] \left[ Y_{L'-1} \otimes S \right]_{L'} }\right.\nonumber\\
& & \ \ \ \ \ \ \ \mxredca{ \left[\mu_p - (-1)^{L} \mu_n \right] \left[ Y_{L-1} \otimes S \right]_{L} } + \nonumber\\
& & \mxredbc{ \left[\mu_p +  (-1)^{L'} \mu_n \right] \left[ Y_{L'-1} \otimes t  \right]_{L'} } \nonumber\\
& &  \ \ \ \ \ \ \ \ \left.\mxredca{ \left[\mu_p + 
(-1)^{L} \mu_n \right] \left[ Y_{L-1} \otimes t \right]_{L} } \right\}\times\nonumber\\
& & \int_{0}^{\infty} \int_{0}^{\infty}  r\ dr\ r'\ dr'\ u_{\sss l'}(r') j_{\sss L'-1}(\frac{\wfin r'}{2}) \green
j_{\sss L-1}(\frac{\winit r}{2}) u_{\sss l}(r),\nonumber \\
\lefteqn{ \matrixm{\sigma 1,\sigma 1, \mathrm{cr} }  =  \sum_{l=0,2} \sum_{l'=0,2} \sum_{L=1}^{\infty} \sum_{L'=1}^{\infty}
\sum_{S_C=0,1} \sum_{J_C = |1-L|}^{1+L} \sum_{L_C = |J_C-S_C|}^{J_C+S_C}  -\frac{\pi e^2 (\hbar\wfin)(\hbar\winit) }{2m_p^2}\times}\\ 
& & i^{L+L'} (-1)^{L'+J_C-\muinit-\mufin} 
\muinit\mufin \wignerdfnom  \sqrt{(L+1)(L'+1)} \times\nonumber\\
& & \threej{1}{L}{J_C}{-M_f}{\muinit}{M_f-\muinit}
\threej{J_C}{L'}{1}{\muinit-M_f}{M_f-M_i-\muinit}{M_i} \times\nonumber\\ 
& & \left\{ \mxredbc{ \left[\mu_p - (-1)^{L} \mu_n \right] \left[ Y_{L-1} \otimes S \right]_{L} }\right.\nonumber\\
& & \ \ \ \ \ \ \ \mxredca{ \left[\mu_p - (-1)^{L'} \mu_n \right] \left[ Y_{L'-1} \otimes S \right]_{L'} } + \nonumber\\
& & \mxredbc{ \left[\mu_p +  (-1)^{L} \mu_n \right] \left[ Y_{L-1} \otimes t  \right]_{L} } \nonumber\\
& &  \ \ \ \ \ \ \ \ \left.\mxredca{ \left[\mu_p +
(-1)^{L'} \mu_n \right] \left[ Y_{L'-1} \otimes t \right]_{L'} } \right\}\times\nonumber\\
& & \int_{0}^{\infty} \int_{0}^{\infty}  r\ dr\ r'\ dr'\ u_{\sss l}(r') j_{\sss L'-1}(\frac{\wfin r'}{2}) \greenpr
j_{\sss L-1}(\frac{\winit r}{2}) u_{\sss l'}(r),\nonumber \\
\lefteqn{ \matrixm{\sigma 2,\sigma 2, \mathrm{uncr} }  =  \sum_{l=0,2} \sum_{l'=0,2} \sum_{L=1}^{\infty} \sum_{L'=1}^{\infty}
\sum_{S_C=0,1} \sum_{J_C = |1-L|}^{1+L} \sum_{L_C = |J_C-S_C|}^{J_C+S_C} \frac{\pi e^2 (\hbar\wfin)(\hbar\winit) }{2m_p^2}\times}\\ 
& & i^{L+L'} (-1)^{L'+J_C-M_f-M_i-\muinit-\mufin} 
 \wignerdfnom  \sqrt{(2L+1)(2L'+1)} \times\nonumber\\
& & \threej{1}{L'}{J_C}{-M_f}{M_f-M_i-\muinit}{M_i+\muinit}
\threej{J_C}{L}{1}{-M_i-\muinit}{\muinit}{M_i} \times\nonumber\\  
& & \left\{ \mxredbc{ \left[\mu_p +  (-1)^{L'} \mu_n \right] \left[ Y_{L'} \otimes S \right]_{L'} }\right.\nonumber\\
& & \ \ \ \ \ \ \ \mxredca{ \left[\mu_p+ (-1)^{L} \mu_n \right] \left[ Y_{L} \otimes S \right]_{L} } + \nonumber\\
& & \mxredbc{ \left[\mu_p -  (-1)^{L'} \mu_n \right] \left[ Y_{L'} \otimes t  \right]_{L'} } \nonumber\\
& &  \ \ \ \ \ \ \ \ \left.\mxredca{ \left[\mu_p -
(-1)^{L} \mu_n \right] \left[ Y_{L} \otimes t \right]_{L} } \right\}\times\nonumber\\
& & \int_{0}^{\infty} \int_{0}^{\infty}  r\ dr\ r'\ dr'\ u_{\sss l'}(r') j_{\sss L'}(\frac{\wfin r'}{2}) \green
j_{\sss L}(\frac{\winit r}{2}) u_{\sss l}(r),\nonumber \\
\lefteqn{ \matrixm{\sigma 2,\sigma 2, \mathrm{cr} }  =  \sum_{l=0,2} \sum_{l'=0,2} \sum_{L=1}^{\infty} \sum_{L'=1}^{\infty}
\sum_{S_C=0,1} \sum_{J_C = |1-L|}^{1+L} \sum_{L_C = |J_C-S_C|}^{J_C+S_C} \frac{\pi e^2 (\hbar\wfin)(\hbar\winit) }{2m_p^2} \times}\\  
& & i^{L+L'} (-1)^{L'+J_C-\muinit-\mufin} 
\wignerdfnom  \sqrt{(2L+1)(2L'+1)} \times\nonumber\\ 
& & \threej{1}{L}{J_C}{-M_f}{\muinit}{M_f-\muinit}
\threej{J_C}{L'}{1}{\muinit-M_f}{M_f-M_i-\muinit}{M_i} \times\nonumber\\
& & \left\{ \mxredbc{ \left[\mu_p + (-1)^{L} \mu_n \right] \left[ Y_{L} \otimes S \right]_{L} }\right.\nonumber\\
& & \ \ \ \ \ \ \ \mxredca{ \left[\mu_p + (-1)^{L'} \mu_n \right] \left[ Y_{L'} \otimes S \right]_{L'} } + \nonumber\\
& & \mxredbc{ \left[\mu_p -  (-1)^{L} \mu_n \right] \left[ Y_{L} \otimes t  \right]_{L} } \nonumber\\
& &  \ \ \ \ \ \ \ \ \left.\mxredca{ \left[\mu_p -
(-1)^{L'} \mu_n \right] \left[ Y_{L'} \otimes t \right]_{L'} } \right\}\times\nonumber\\
& & \int_{0}^{\infty} \int_{0}^{\infty}  r\ dr\ r'\ dr'\ u_{\sss l}(r') j_{\sss L'}(\frac{\wfin r'}{2}) \greenpr
j_{\sss L}(\frac{\winit r}{2}) u_{\sss l'}(r),\nonumber 
\end{eqnarray}
\begin{eqnarray}
\lefteqn{ \matrixm{p1,p1, \mathrm{uncr} }  = \sum_{l=0,2} \sum_{l'=0,2} \sum_{L=1}^{\infty} \sum_{L'=1}^{\infty}
\sum_{S_C=0,1} \sum_{J_C = |1-L|}^{1+L} \sum_{L_C = |J_C-S_C|}^{J_C+S_C}  \times}\\ 
& &  -\frac{\pi e^2 (\hbar\wfin)(\hbar\winit) \muinit\mufin }{2m_p^2} i^{L+L'} (-1)^{L'+J_C-M_f-M_i-\muinit-\mufin}  \times\nonumber\\
& & \wignerdfnom \left[(L+1)(L'+1)\right]^{-\frac{1}{2}} \times\nonumber\\
& & \threej{1}{L'}{J_C}{-M_f}{M_f-M_i-\muinit}{M_i+\muinit}
\threej{J_C}{L}{1}{-M_i-\muinit}{\muinit}{M_i} \times\nonumber\\  
& & \mxredbc{ \left[ Y_{L'-1} \otimes L   \right]_{L'M'} }
\mxredca{   \left[ Y_{L-1} \otimes L \right]_{LM} } \times\nonumber\\
& & \int_{0}^{\infty} \int_{0}^{\infty}  r\ dr\ r'\ dr'\ u_{\sss l'}(r') j_{\sss L'-1}(\frac{\wfin r'}{2}) \green
j_{\sss L-1}(\frac{\winit r}{2}) u_{\sss l}(r),\nonumber \\
\lefteqn{ \matrixm{p1,p1, \mathrm{cr} } =  \sum_{l=0,2} \sum_{l'=0,2} \sum_{L=1}^{\infty} \sum_{L'=1}^{\infty}
\sum_{S_C=0,1} \sum_{J_C = |1-L|}^{1+L} \sum_{L_C = |J_C-S_C|}^{J_C+S_C} -\frac{\pi e^2 (\hbar\wfin)(\hbar\winit) \muinit\mufin }{2m_p^2}\times} \nonumber\\ 
& & i^{L+L'} (-1)^{L'+J_C-\muinit-\mufin}
\wignerdfnom \left[(L+1)(L'+1)\right]^{-\frac{1}{2}} \times\nonumber\\
& & \threej{1}{L}{J_C}{-M_f}{\muinit}{M_f-\muinit}
\threej{J_C}{L'}{1}{\muinit-M_f}{M_f-M_i-\muinit}{M_i} \times\nonumber\\
& & \mxredbc{ \left[ Y_{L-1} \otimes L   \right]_{LM} }                          
\mxredca{   \left[ Y_{L'-1} \otimes L \right]_{L'M'} } \times\nonumber\\                 
& & \int_{0}^{\infty} \int_{0}^{\infty}  r\ dr\ r'\ dr'\ u_{\sss l}(r') j_{\sss L'-1}(\frac{\wfin r'}{2}) \greenpr
j_{\sss L-1}(\frac{\winit r}{2}) u_{\sss l'}(r),\\
\lefteqn{ \matrixm{p2,p2, \mathrm{uncr} } = \sum_{l=0,2} \sum_{l'=0,2} \sum_{L=1}^{\infty} \sum_{L'=1}^{\infty}
\sum_{S_C=0,1} \sum_{J_C = |1-L|}^{1+L} \sum_{L_C = |J_C-S_C|}^{J_C+S_C}  \times}\\ 
& & \frac{\pi e^2 (\hbar\wfin)(\hbar\winit) }{2m_p^2} i^{L+L'} (-1)^{L'+J_C-M_f-M_i-\muinit-\mufin} \times\nonumber\\
& & \wignerdfnom \sqrt{ \frac{(2L+1)(2L'+1)}{LL'(L+1)(L'+1)}}  \times\nonumber\\
& & \threej{1}{L'}{J_C}{-M_f}{M_f-M_i-\muinit}{M_i+\muinit}
\threej{J_C}{L}{1}{-M_i-\muinit}{\muinit}{M_i} \times\nonumber\\
& & \mxredbc{ Y_{L'} } \mxredca{ Y_{L} } \times\nonumber\\
& & \int_{0}^{\infty} \int_{0}^{\infty}  r\ dr\ r'\ dr'\ \left[ j_{\sss L'}(\frac{\wfin r'}{2})  u_{\sss l'}(r') +
\frac{\wfin r'}{2} j_{\sss L'}'(\frac{\wfin r'}{2})  u_{\sss l'}(r') + \right. \nonumber\\
& & \ \ \ \ \ \ \left. 2r' j_{\sss L'}(\frac{\wfin r'}{2})  u_{\sss l'}'(r') \right] 
\green \left[ j_{\sss L}(\frac{\winit r}{2})  u_{\sss l}(r) + \right. \nonumber\\
& & \ \ \ \ \ \ \left. \frac{\winit r}{2} j_{\sss L}'(\frac{\winit r}{2})  u_{\sss l}(r) +
2r j_{\sss L}(\frac{\winit r}{2})  u_{\sss l}'(r) \right],\nonumber\\
\lefteqn{ \matrixm{p2,p2, \mathrm{cr} } =  \sum_{l=0,2} \sum_{l'=0,2} \sum_{L=1}^{\infty} \sum_{L'=1}^{\infty}
\sum_{S_C=0,1} \sum_{J_C = |1-L|}^{1+L} \sum_{L_C = |J_C-S_C|}^{J_C+S_C}  \frac{\pi e^2 (\hbar\wfin)(\hbar\winit) }{2m_p^2} \times} \\ 
& & i^{L+L'} (-1)^{L'+J_C-\muinit-\mufin} 
\wignerdfnom \sqrt{ \frac{(2L+1)(2L'+1)}{LL'(L+1)(L'+1)}}  \times\nonumber\\
& & \threej{1}{L}{J_C}{-M_f}{\muinit}{M_f-\muinit}
\threej{J_C}{L'}{1}{\muinit-M_f}{M_f-M_i-\muinit}{M_i} \times\nonumber\\
& & \mxredbc{ Y_{L}} \mxredca{ Y_{L'} } \times\nonumber\\
& & \int_{0}^{\infty} \int_{0}^{\infty}  r\ dr\ r'\ dr'\ \left[ j_{\sss L'}(\frac{\wfin r'}{2})  u_{\sss l}(r') +
\frac{\wfin r'}{2} j_{\sss L'}'(\frac{\wfin r'}{2})  u_{\sss l}(r') + \right. \nonumber\\
& & \ \ \ \ \ \ \left. 2r' j_{\sss L'}(\frac{\wfin r'}{2})  u_{\sss l}'(r') \right]        
\green \left[ j_{\sss L}(\frac{\winit r}{2})  u_{\sss l'}(r) + \right. \nonumber\\
& & \ \ \ \ \ \ \left. \frac{\winit r}{2} j_{\sss L}'(\frac{\winit r}{2})  u_{\sss l'}(r) +
2r j_{\sss L}(\frac{\winit r}{2})  u_{\sss l'}'(r) \right].\nonumber\\
\lefteqn{ \matrixm{\sigma 1,\sigma 2, \mathrm{uncr} }  =  \sum_{l=0,2} \sum_{l'=0,2} \sum_{L=1}^{\infty} \sum_{L'=1}^{\infty}
\sum_{S_C=0,1} \sum_{J_C = |1-L|}^{1+L} \sum_{L_C = |J_C-S_C|}^{J_C+S_C} \times}\\
& & \frac{i\pi e^2 (\hbar\wfin)(\hbar\winit) }{2m_p^2} i^{L+L'} (-1)^{L'+J_C-M_f-M_i-\muinit-\mufin}  \times\nonumber\\
& & \mufin \wignerdfnom  \sqrt{(2L+1)(L'+1)} \times\nonumber\\
& & \threej{1}{L'}{J_C}{-M_f}{M_f-M_i-\muinit}{M_i+\muinit}
\threej{J_C}{L}{1}{-M_i-\muinit}{\muinit}{M_i} \times\nonumber\\
& & \left\{ \mxredbc{ \left[\mu_p - (-1)^{L'} \mu_n \right] \left[ Y_{L'-1} \otimes S \right]_{L'} }\right.\nonumber\\
& & \ \ \ \ \ \ \ \mxredca{ \left[\mu_p + (-1)^{L} \mu_n \right] \left[ Y_{L} \otimes S \right]_{L} } + \nonumber\\
& & \mxredbc{ \left[\mu_p +  (-1)^{L'} \mu_n \right] \left[ Y_{L'-1} \otimes t  \right]_{L'} } \nonumber\\
& &  \ \ \ \ \ \ \ \ \left.\mxredca{ \left[\mu_p -
(-1)^{L} \mu_n \right] \left[ Y_{L} \otimes t \right]_{L} } \right\}\times\nonumber\\
& & \int_{0}^{\infty} \int_{0}^{\infty}  r\ dr\ r'\ dr'\ u_{\sss l'}(r') j_{\sss L'-1}(\frac{\wfin r'}{2}) \green
j_{\sss L}(\frac{\winit r}{2}) u_{\sss l}(r),\nonumber\\
\lefteqn{ \matrixm{\sigma 2,\sigma 1, \mathrm{uncr} }  = \sum_{l=0,2} \sum_{l'=0,2} \sum_{L=1}^{\infty} \sum_{L'=1}^{\infty}
\sum_{S_C=0,1} \sum_{J_C = |1-L|}^{1+L} \sum_{L_C = |J_C-S_C|}^{J_C+S_C} \times}\\
& &  \frac{i\pi e^2 (\hbar\wfin)(\hbar\winit) }{2m_p^2} i^{L+L'} (-1)^{L'+J_C-M_f-M_i-\muinit-\mufin} \times\nonumber\\
& & \muinit \wignerdfnom  \sqrt{(2L'+1)(L+1)} \times\nonumber\\
& & \threej{1}{L'}{J_C}{-M_f}{M_f-M_i-\muinit}{M_i+\muinit}
\threej{J_C}{L}{1}{-M_i-\muinit}{\muinit}{M_i} \times\nonumber\\
& & \left\{ \mxredbc{ \left[\mu_p + (-1)^{L'} \mu_n \right] \left[ Y_{L'} \otimes S \right]_{L'} }\right.\nonumber\\
& & \ \ \ \ \ \ \ \mxredca{ \left[\mu_p - (-1)^{L} \mu_n \right] \left[ Y_{L-1} \otimes S \right]_{L} } + \nonumber\\
& & \mxredbc{ \left[\mu_p -  (-1)^{L'} \mu_n \right] \left[ Y_{L'} \otimes t  \right]_{L'} } \nonumber\\
& &  \ \ \ \ \ \ \ \ \left.\mxredca{ \left[\mu_p +
(-1)^{L} \mu_n \right] \left[ Y_{L-1} \otimes t \right]_{L} } \right\}\times\nonumber\\
& & \int_{0}^{\infty} \int_{0}^{\infty}  r\ dr\ r'\ dr'\ u_{\sss l'}(r') j_{\sss L'}(\frac{\wfin r'}{2}) \green
j_{\sss L-1}(\frac{\winit r}{2}) u_{\sss l}(r),\nonumber\\
\lefteqn{ \matrixm{\sigma 1,p1, \mathrm{uncr} }  = \sum_{l=0,2} \sum_{l'=0,2} \sum_{L=1}^{\infty} \sum_{L'=1}^{\infty}
\sum_{S_C=0,1} \sum_{J_C = |1-L|}^{1+L} \sum_{L_C = |J_C-S_C|}^{J_C+S_C} -\frac{\pi e^2 (\hbar\wfin)(\hbar\winit) }{2m_p^2} \times} \nonumber\\
& & i^{L+L'} (-1)^{L'+J_C-M_f-M_i-\muinit-\mufin} 
\muinit\mufin \wignerdfnom  \sqrt{\frac{L'+1}{L+1} } \times\nonumber\\
& & \threej{1}{L'}{J_C}{-M_f}{M_f-M_i-\muinit}{M_i+\muinit}
\threej{J_C}{L}{1}{-M_i-\muinit}{\muinit}{M_i} \times\nonumber\\
& & \mxredbc{ \left[\mu_p - (-1)^{L'} \mu_n \right] \left[ Y_{L'-1} \otimes S \right]_{L'} }\nonumber\\
& & \ \ \ \ \ \ \ \mxredca{  \left[ Y_{L-1} \otimes L \right]_{L} } \times \nonumber\\
& & \int_{0}^{\infty} \int_{0}^{\infty}  r\ dr\ r'\ dr'\ u_{\sss l'}(r') j_{\sss L'-1}(\frac{\wfin r'}{2}) \green
j_{\sss L-1}(\frac{\winit r}{2}) u_{\sss l}(r), \\
\lefteqn{ \matrixm{p1,\sigma 1, \mathrm{uncr} }  =  \sum_{l=0,2} \sum_{l'=0,2} \sum_{L=1}^{\infty} \sum_{L'=1}^{\infty}
\sum_{S_C=0,1} \sum_{J_C = |1-L|}^{1+L} \sum_{L_C = |J_C-S_C|}^{J_C+S_C} -\frac{\pi e^2 (\hbar\wfin)(\hbar\winit) }{2m_p^2} \times}\nonumber \\
& & i^{L+L'} (-1)^{L'+J_C-M_f-M_i-\muinit-\mufin}
\muinit\mufin \wignerdfnom  \sqrt{\frac{L+1}{L'+1} } \times\nonumber\\
& & \threej{1}{L'}{J_C}{-M_f}{M_f-M_i-\muinit}{M_i+\muinit}
\threej{J_C}{L}{1}{-M_i-\muinit}{\muinit}{M_i} \times\nonumber\\
& &  \mxredbc{  \left[ Y_{L'-1} \otimes L \right]_{L'} }\nonumber\\
& & \ \ \ \ \ \ \ \mxredca{ \left[\mu_p - (-1)^{L} \mu_n \right] \left[ Y_{L-1} \otimes S \right]_{L} } \times \nonumber\\
& & \int_{0}^{\infty} \int_{0}^{\infty}  r\ dr\ r'\ dr'\ u_{\sss l'}(r') j_{\sss L'-1}(\frac{\wfin r'}{2}) \green
j_{\sss L-1}(\frac{\winit r}{2}) u_{\sss l}(r),\\
\lefteqn{ \matrixm{p 1,\sigma 2, \mathrm{uncr} }  =  \sum_{l=0,2} \sum_{l'=0,2} \sum_{L=1}^{\infty} \sum_{L'=1}^{\infty}
\sum_{S_C=0,1} \sum_{J_C = |1-L|}^{1+L} \sum_{L_C = |J_C-S_C|}^{J_C+S_C} -\frac{i\pi e^2 (\hbar\wfin)(\hbar\winit) }{2m_p^2} \times}\nonumber \\
& & i^{L+L'} (-1)^{L'+J_C-M_f-M_i-\muinit-\mufin}
\mufin \wignerdfnom  \sqrt{ \frac{2L+1}{L'+1} } \times\nonumber\\
& & \threej{1}{L'}{J_C}{-M_f}{M_f-M_i-\muinit}{M_i+\muinit}
\threej{J_C}{L}{1}{-M_i-\muinit}{\muinit}{M_i} \times\nonumber\\
& & \mxredbc{  \left[ Y_{L'-1} \otimes L \right]_{L'} }\nonumber\\
& & \ \ \ \ \ \ \ \mxredca{ \left[\mu_p + (-1)^{L} \mu_n \right] \left[ Y_{L} \otimes S \right]_{L} } \times \nonumber\\  
& & \int_{0}^{\infty} \int_{0}^{\infty}  r\ dr\ r'\ dr'\ u_{\sss l'}(r') j_{\sss L'-1}(\frac{\wfin r'}{2}) \green
j_{\sss L}(\frac{\winit r}{2}) u_{\sss l}(r),\\  
\lefteqn{ \matrixm{\sigma 2,p 1, \mathrm{uncr} }  = \sum_{l=0,2} \sum_{l'=0,2} \sum_{L=1}^{\infty} \sum_{L'=1}^{\infty}
\sum_{S_C=0,1} \sum_{J_C = |1-L|}^{1+L} \sum_{L_C = |J_C-S_C|}^{J_C+S_C} -\frac{i\pi e^2 (\hbar\wfin)(\hbar\winit) }{2m_p^2} \times}\nonumber \\
& & i^{L+L'} (-1)^{L'+J_C-M_f-M_i-\muinit-\mufin}
\muinit \wignerdfnom  \sqrt{ \frac{2L'+1}{L+1} } \times\nonumber\\
& & \threej{1}{L'}{J_C}{-M_f}{M_f-M_i-\muinit}{M_i+\muinit}
\threej{J_C}{L}{1}{-M_i-\muinit}{\muinit}{M_i} \times\nonumber\\
& & \mxredbc{ \left[\mu_p + (-1)^{L'} \mu_n \right] \left[ Y_{L'} \otimes S \right]_{L'} }\nonumber\\  
& & \ \ \ \ \ \ \ \mxredca{  \left[ Y_{L-1} \otimes L \right]_{L} } \times \nonumber\\
& & \int_{0}^{\infty} \int_{0}^{\infty}  r\ dr\ r'\ dr'\ u_{\sss l'}(r') j_{\sss L'}(\frac{\wfin r'}{2}) \green  
j_{\sss L-1}(\frac{\winit r}{2}) u_{\sss l}(r).
\end{eqnarray}

\chapter{Recoil Corrections \label{app:recoil} }
\noindent We will now return to the recoil corrections to the transition matrices which arise from the $\vec{P}$ operators in
equations (\ref{eq:mainterm14}) and (\ref{eq:mainterm15}).   The main contributions from these terms are 
\begin{eqnarray}
\lefteqn{ \matrixt{\mathrm{cm, uncr}}  \equiv } \\
& &  \sum_{C, \vec{P}_C} 
\frac{ \mx{d_f, \vec{P}_f }{ \frac{ie}{m_p} \vec{\nabla}\Phi_f(\frac{\vec{r}}{2})
e^{-i\vec{k}_f\cdot\vec{R} }\cdot\vec{P} }{C, \vec{P}_C} \mx{C, \vec{P}_C}{\left[ H^{np}, \hat{\Phi}_i \right]
e^{i\vec{k}_i\cdot\vec{R}}}{d_i,\vec{P}_i} }{
\hbar\winit + E_{d_i} -E_C - P_C^2/2m_d + i\varepsilon }, \nonumber\\ 
\lefteqn{ \matrixt{\mathrm{cm,cr} }   \equiv  } \\
& & \sum_{C, \vec{P}_C} 
\frac{ \mx{d_f, \vec{P}_f }{ \frac{ie}{m_p} \vec{\nabla}\Phi_i(\frac{\vec{r}}{2})
e^{i\vec{k}_i\cdot\vec{R} }\cdot\vec{P} }{C, \vec{P}_C} \mx{C, \vec{P}_C}{\left[ H^{np}, \hat{\Phi}_f \right]
e^{-i\vec{k}_f\cdot\vec{R}}}{d_i,\vec{P}_i} }{
-\hbar\wfin  + E_{d_i} -E_C - P_C^2/2m_d + i\varepsilon }. \nonumber
\end{eqnarray}
\noindent Since we are working in the lab frame, $\vec{P} \ket{\vec{P}_i} = 0.$
After integrating out the center-of-mass variables and using $\vec{A} = \vec{\nabla}\Phi$,
 $\matrixt{\mathrm{cm, uncr}}$ becomes
\begin{eqnarray} 
\lefteqn{\matrixt{\mathrm{cm, uncr}}  =   \sqrt{ \frac{2\pi\hbar}{V\wfin} }
\frac{ \delta(\vec{P}_i + \vec{k}_i - \vec{P}_f - \vec{k}_f) }{V} \times } \\
& &\sum_C  \frac{ \mx{d_f}{ \frac{ie\hbar}{m_p} (\vec{k}_i\cdot\epspr) e^{-\frac{i}{2}\vec{k}_f\cdot\vec{r} } }{C} 
\mx{C}{\left[ H^{np}, \hat{\Phi}_i \right] }{d_i} }{ \hbar\winit - \frac{(\hbar\winit)^2}{2m_d}+ E_{d_i} -E_C + i\varepsilon } \nonumber.
\end{eqnarray}
\noindent The commutator can now be evaluated; this produces two terms:
\begin{eqnarray}
\matrixt{\mathrm{cm, uncr}} & = &   \sqrt{ \frac{2\pi\hbar}{V\wfin} }
\frac{ \delta(\vec{P}_i + \vec{k}_i - \vec{P}_f - \vec{k}_f)  }{V} \left[
\matrixt{\mathrm{cm, uncr}1} + \matrixt{\mathrm{cm, uncr}2} \right],
\end{eqnarray}
\noindent where
\begin{eqnarray}
\matrixt{\mathrm{cm, uncr}1} & = & \sum_C  \frac{ \mx{d_f}{ \frac{ie\hbar}{m_p} (\vec{k}_i\cdot\epspr )
e^{-\frac{i}{2}\vec{k}_f\cdot\vec{r} } }{C}
\mx{C}{ \left[ \hbar\winit - \frac{(\hbar\winit)^2}{2m_d} \right] \hat{\Phi}_i }{d_i} }{ \hbar\winit -
\frac{(\hbar\winit)^2}{2m_d}+ E_{d_i} -E_C + i\varepsilon },\nonumber\\
\matrixt{\mathrm{cm, uncr}2} & = & -\mx{d_f}{ \frac{ie\hbar}{m_p} (\vec{k}_i\cdot\epspr)
e^{-\frac{i}{2}\vec{k}_f\cdot\vec{r} } \hat{\Phi}_i }{d_i}.
\end{eqnarray}
\noindent Starting with $\matrixt{\mathrm{cm, uncr}1}$, we insert equation~(\ref{eq:mainterm10}) 
for $\hat{\Phi}_i$ and expand the exponential into partial waves to get 
\begin{eqnarray}
\lefteqn{ \matrixm{\mathrm{cm, uncr}1}  = \sum_C \sum_{L=1}^{\infty} \sum_{L'=0}^{\infty} \sum_{MM'} \times}\\
& & \mx{d_f}{ \frac{ie\hbar}{m_p} (\vec{k}_i\cdot\epspr) 4\pi (-i)^{L'} j_{\sss L'}(\frac{\wfin r}{2}) Y_{L'M'}^{\ast}(\hat{k}_f)
Y_{L'M'}(\hat{r}) }{C} \times\nonumber \\
& & \frac{1}{ \hbar\winit - \frac{(\hbar\winit)^2}{2m_d}+ E_{d_i} -E_C + i\varepsilon } 
\bra{C} \left[ \hbar\winit - \frac{(\hbar\winit)^2}{2m_d} \right] \frac{-e}{\hbar} \times\nonumber\\
& & \wignerdi \frac{i^{L+1}}{\winit} \sqrt{ \frac{2\pi(2L+1)}{L(L+1)} }
\psi_{\sss L}(\frac{\winit r}{2}) 
Y_{LM}(\hat{r}) \ket{d_i},\nonumber
\end{eqnarray}
\noindent where the $\matrixt{}$ has been replaced with $\matrixm{}$, according to equation~(\ref{eq:seagull10}).
The function $\psi_{\sss L}$ is defined in equation~(\ref{eq:mainterm16}).
Next, we insert two complete sets of radial states:
\begin{eqnarray}
\lefteqn{ \matrixm{\mathrm{cm, uncr}1} = \sum_{ll'} \sum_{L_CS_CJ_CM_C} \sum_{L=1}^{\infty} \sum_{L'=0}^{\infty} \sum_{MM'} 
(\hat{k}_i\cdot\epspr) \frac{4\pi e^2}{m_p} \left[ \hbar\winit - \frac{(\hbar\winit)^2}{2m_d} \right]  \times} \\
& & i^{L-L'} Y_{L'M'}^{\ast}(\hat{k}_f) \wignerdi \sqrt{ \frac{2\pi(2L+1)}{L(L+1)} }\times\nonumber\\
& & \mx{l'11M_f}{Y_{L'M'}}{L_CS_CJ_CM_C} \mx{L_CS_CJ_CM_C}{Y_{LM}}{l11M_i} \times\nonumber\\
& & \int_{0}^{\infty} \int_{0}^{\infty}  r\ dr\ r'\ dr'\ u_{\sss l'}(r') j_{\sss L'}(\frac{\wfin r'}{2}) \green 
\psi_{\sss L}(\frac{\winit r}{2})  u_{\sss l}(r).\nonumber
\end{eqnarray}
\noindent The final answer is then obtained by removing the $M$ sums :
\begin{eqnarray}
\lefteqn{\matrixm{\mathrm{cm, uncr}1} = -\sum_{l=0,2} \sum_{l'=0,2} \sum_{S_C = 0,1} \sum_{J_C = |1-L|}^{1+L} \sum_{L_C=|J_C-S_C|}^{J_C+S_C}
 \sum_{L=1}^{\infty} \sum_{L'=0}^{\infty} \times}\\
& & (\hat{k}_i\cdot\epspr) \frac{4\pi e^2}{m_p} \left[ \hbar\winit- \frac{(\hbar\winit)^2}{2m_d} \right] (-1)^{J_C-M_f-M_i-\muinit}
\times\nonumber \\
& & i^{L-L'} Y_{L', M_f-M_i-\muinit}^{\ast}(\hat{k}_f) 
\sqrt{ \frac{2\pi(2L+1)}{L(L+1)} }\times\nonumber\\
& & \mxredbc{Y_{L'}} \mxredca{Y_L} \times\nonumber\\
& & \threej{1}{L'}{J_C}{-M_f}{M_f-M_i-\muinit}{M_i+\muinit} \threej{J_C}{L}{1}{-M_i-\muinit}{\muinit}{M_i} \times\nonumber\\
& & \int_{0}^{\infty} \int_{0}^{\infty} r\ dr\ r'\ dr'\ u_{\sss l'}(r') j_{\sss L'}(\frac{\wfin r'}{2}) \green
\psi_{\sss L}(\frac{\winit r}{2})  u_{\sss l}(r).\nonumber
\end{eqnarray}
\noindent We just list the results for the other $\matrixm{\mathrm{cm}}$ terms:
\begin{eqnarray}
\matrixm{\mathrm{cm, uncr}2} & = &  \sum_{l=0,2} \sum_{l'=0,2} \sum_{\tilde{L}=|L-L'|}^{L+L' }\sum_{L=1}^{\infty} \sum_{L'=0}^{\infty} 
(\hat{k}_i\cdot\epspr) \frac{4\pi e^2}{m_p} (-1)^{M_i} \times\\
& & i^{L-L'} Y_{L', M_f-M_i-\muinit}^{\ast}(\hat{k}_f)
(2L+1) \sqrt{ \frac{(2L'+1)(2\tilde{L}+1)}{2L(L+1)}} \times\nonumber\\
& & \mxredba{Y_{\tilde{L}}} \threej{L}{L'}{\tilde{L}}{0}{0}{0} \threej{L}{L'}{\tilde{L}}{\muinit}{M_f-M_i-\muinit}{M_i-M_f}
\times\nonumber\\
& & \threej{1}{\tilde{L}}{1}{-M_f}{M_f-M_i}{M_i} \int_{0}^{\infty}  dr\  u_{\sss l'}(r) j_{\sss L'}(\frac{\wfin r'}{2})
\psi_{\sss L}(\frac{\winit r}{2})  u_{\sss l}(r),\nonumber\\
\matrixm{\mathrm{cm, cr}1} & = &  \sum_{l=0,2} \sum_{l'=0,2} \sum_{S_C = 0,1} \sum_{J_C = |1-L|}^{1+L} \sum_{L_C=|J_C-S_C|}^{J_C+S_C}
\sum_{L'=1}^{\infty} \sum_{L=0}^{\infty}  \times\\
& &  (\hat{k}_f\cdot\eps) \frac{4\pi e^2}{m_p} \left[ \hbar\wfin+\frac{(\hbar\wfin)^2}{2m_d} \right] (-1)^{L'-\mufin-J_C}
\times\nonumber\\
& & i^{L+L'} 
\sqrt{ \frac{(2L+1)(2L'+1)}{2L'(L'+1)}}\wignerdf\times\nonumber\\
& & \mxredbc{Y_{L}} \mxredca{Y_{L'}} \times\nonumber\\
& & \threej{1}{L}{J_C}{-M_f}{0}{M_f} \threej{J_C}{L'}{1}{-M_f}{M_f-M_i}{M_i}  \times\nonumber\\
& & \int_{0}^{\infty} \int_{0}^{\infty} r\ dr\ r'\ dr'\ u_{\sss l}(r')
\psi_{\sss L'}(\frac{\wfin r'}{2})  
\greenpr  j_{\sss L}(\frac{\winit r}{2})  u_{\sss l'}(r),\nonumber\\
\matrixm{\mathrm{cm, cr}2} & = & \sum_{\tilde{L}=|L-L'|}^{L+L' } \sum_{l=0,2} \sum_{l'=0,2} \sum_{L=0}^{\infty} \sum_{L'=1}^{\infty} 
(\hat{k}_f\cdot\eps) \frac{2 e^2}{m_p}  (-1)^{L'-\mufin-M_i} \times\\
& & i^{L+L'} (2L'+1) (2L+1)
\sqrt{ \frac{\pi(2\tilde{L}+1)}{2L'(L'+1)} } \wignerdf 
\times\nonumber\\
& & \mxredba{Y_{\tilde{L}}} \threej{L}{L'}{\tilde{L}}{0}{0}{0} \threej{L}{L'}{\tilde{L}}{0}{M_f-M_i}{M_i-M_f}
\times\nonumber\\
& & \threej{1}{\tilde{L}}{1}{-M_f}{M_f-M_i}{M_i} \int_{0}^{\infty}  dr\  u_{\sss l'}(r)  j_{\sss L}(\frac{\winit r}{2})
\psi_{\sss L'}(\frac{\wfin r}{2})  u_{\sss l}(r). \nonumber
\end{eqnarray}

\chapter{Relativistic Corrections \label{app:rel} }
The relativistic correction to the charge density is given by
\begin{equation}
\rho^{R} = e\hbar \left[ \frac{2\kappa_p + 1}{4m_p^2} \vec{\nabla}_{\dum} \delta(\vec{\dum} - \vec{r}/2) \cdot
(\vec{\sigma}_p \times \vec{p}) - \frac{\kappa_n}{m_p^2} \vec{\nabla}_{\dum} \delta(\vec{\dum}+ \vec{r}/2) \cdot
(\vec{\sigma}_n \times \vec{p}) \right],
\end{equation}
where $\kappa$ is the anomalous magnetic moment.
We can calculate the effect of this correction by inserting 
\begin{equation}
\rho = \sum_j e_j \delta(\vec{\dum} - \vec{x}_j) + \rho^{R}
\end{equation}
into equation~(\ref{eq:mainterm13}).  Using the approximation that
$\vec{\nabla} \Phi_i \approx \vec{A} = \eps e^{i\kinit\cdot\vec{\dum} }$,
the integrals over the dummy variables can be performed immediately:
\begin{eqnarray}
\int [H, \rho^{R}] \Phi_i d^3\dum & = & [H^{np}, \hat{\Psi}_i] ,\\
\int [H, \rho^{R}] \Phi_f d^3\dum & = & [H^{np}, \hat{\Psi}_f] ,
\end{eqnarray}
where
\begin{eqnarray}
 \hat{\Psi}_i & \equiv & \frac{e}{4m_p^2} \left[
(2\kappa_p+1) e^{i\kinit\cdot\vec{r}/2} - 2\kappa_n e^{-i\kinit\cdot\vec{r}/2} \right] 
\eps\cdot(\vec{p}\times\vec{S}),\\
 \hat{\Psi}_f & \equiv &  \frac{e}{4m_p^2} \left[
(2\kappa_p+1) e^{-i\kfin\cdot\vec{r}/2} - 2\kappa_n e^{i\kfin\cdot\vec{r}/2} \right]
\epspr\cdot(\vec{p}\times\vec{S}) ,
\end{eqnarray}
and $\vec{S} \equiv (\vec{\sigma}_p + \vec{\sigma_n})/2.$ The term proportional to $\vec{t} \equiv (\vec{\sigma}_p - \vec{\sigma_n})/2$
is zero if we only work to first order in the relativistic correction, and therefore has not been included 
in the definition of $\hat{\Psi}$.  To this order, recoil corrections can also
be neglected.  Thus, the  correction to the scattering amplitude is given by
\begin{equation}
\matrixm{rel} = \matrixm{rel1} + \matrixm{rel2} + \matrixm{rel3} + \matrixm{rel4},
\end{equation}
where
\begin{eqnarray}
\matrixm{rel1} & \equiv & -\sum_{C} \frac { \mx{d_f}{[H^{np}, \hat{\Phi}_f] }{C}
\mx{C}{[H^{np}, \hat{\Psi}_i] }{d_i} }{\hbar\winit+E_{d_i}-E_C+ i\varepsilon}, \\
\matrixm{rel2}  & \equiv & -\sum_{C} \frac { \mx{d_f}{[H^{np}, \hat{\Psi}_f] }{C}
\mx{C}{[H^{np}, \hat{\Phi}_i] }{d_i} }{\hbar\winit+E_{d_i}-E_C +i\varepsilon}, \\
\matrixm{rel3} & \equiv & -\sum_{C} \frac { \mx{d_f}{[H^{np}, \hat{\Phi}_i] }{C}
\mx{C}{[H^{np}, \hat{\Psi}_f] }{d_i} }{-\hbar\wfin+E_{d_i}-E_C+i\varepsilon}, \\
\matrixm{rel4} & \equiv & -\sum_{C} \frac { \mx{d_f}{[H^{np}, \hat{\Psi}_i] }{C}
\mx{C}{[H^{np}, \hat{\Phi}_f] }{d_i} }{-\hbar\wfin+E_{d_i}-E_C+i\varepsilon}.
\end{eqnarray}
We now evaluate the commutators in the above terms, using, for example, the fact that
\begin{equation}
\mx{d_f}{[H^{np}, \hat{\Phi}_f] }{C} = \mx{d_f}{(E_{d_i} - E_C) \hat{\Phi}_f }{C}       
\end{equation}
to cancel some of the energy denominators.   This results in
\begin{eqnarray}
\matrixm{rel1} & = & -\frac{1}{2} \mx{d_f}{\hat{\Phi}_f[H^{np},\hat{\Psi}_i]}{d_i} + 
\frac{1}{2} \mx{d_f}{[H^{np},\hat{\Phi}_f]\hat{\Psi}_i}{d_i} - \label{eq:rc1}\\
& & (\hbar\winit) \mx{d_f}{\hat{\Phi}_f \hat{\Psi}_i}{d_i} + (\hbar\winit)^2 \sum_C \frac{\mx{d_f}{\hat{\Phi}_f}{C} 
\mx{C}{\hat{\Psi}_i}{d_i} }{\hbar\winit+E_{d_i}-E_C+ i\varepsilon}, \nonumber\\
\matrixm{rel2} & = &  -\frac{1}{2} \mx{d_f}{\hat{\Psi}_f[H^{np},\hat{\Phi}_i]}{d_i} +     
\frac{1}{2} \mx{d_f}{[H^{np},\hat{\Psi}_f] \hat{\Phi}_i}{d_i} - \label{eq:rc2}\\
& & (\hbar\winit) \mx{d_f}{\hat{\Psi}_f \hat{\Phi}_i}{d_i} + (\hbar\winit)^2 \sum_C \frac{\mx{d_f}{\hat{\Psi}_f}{C}    
\mx{C}{\hat{\Phi}_i}{d_i} }{\hbar\winit+E_{d_i}-E_C+ i\varepsilon}, \nonumber\\
\matrixm{rel3} & = & -\frac{1}{2} \mx{d_f}{\hat{\Phi}_i[H^{np},\hat{\Psi}_f]}{d_i} +     
\frac{1}{2} \mx{d_f}{[H^{np},\hat{\Phi}_i]\hat{\Psi}_f}{d_i} + \label{eq:rc3}\\
& & (\hbar\wfin) \mx{d_f}{\hat{\Phi}_i\hat{\Psi}_f}{d_i} + (\hbar\wfin)^2 \sum_C \frac{\mx{d_f}{\hat{\Phi}_i}{C}    
\mx{C}{\hat{\Psi}_f}{d_i} }{-\hbar\wfin+E_{d_i}-E_C+i\varepsilon}, \nonumber\\
\matrixm{rel4} & = &  -\frac{1}{2} \mx{d_f}{\hat{\Psi}_i[H^{np},\hat{\Phi}_f]}{d_i} +
\frac{1}{2} \mx{d_f}{[H^{np},\hat{\Psi}_i]\hat{\Phi}_f}{d_i} + \label{eq:rc4}\\
& & (\hbar\wfin) \mx{d_f}{\hat{\Psi}_i\hat{\Phi}_f}{d_i} + (\hbar\wfin)^2 \sum_C \frac{\mx{d_f}{\hat{\Psi}_i}{C}
\mx{C}{\hat{\Phi}_f}{d_i} }{-\hbar\wfin+E_{d_i}-E_C+ i\varepsilon}. \nonumber
\end{eqnarray}
The largest terms are the third terms of each group.  We call these the ``contact'' terms
and sum them to get
\begin{eqnarray}
\matrixm{rel, \mathrm{cont}}  & = & (\hbar\wfin - \hbar\winit) \mx{d_f}{\hat{\Phi}_i\hat{\Psi}_f + \hat{\Phi}_f\hat{\Psi}_i}{d_i} +
(\hbar\wfin) \mx{d_f}{[\hat{\Psi}_i,\hat{\Phi}_f]}{d_i} - \nonumber\\
& & \ \ \ \ \ \ \ \ \ \ (\hbar\winit) \mx{d_f}{[\hat{\Psi}_f,\hat{\Phi}_i]}{d_i}. \label{eq:rc6}
\end{eqnarray}
The first term is $\omega/m_d$ times smaller than the others, so we will neglect it for now.  
The commutator in the dominant terms is basically
\begin{equation}
[\vec{p}, \Phi_i] = -ei\vec{\nabla}\hat{\Phi}_i \approx -ei \eps e^{i\kinit\cdot\vec{r}/2},
\end{equation}
so that these terms are equal to
\begin{eqnarray}
\matrixm{rel, \mathrm{cont}} & = & \bra{d_f} \frac{ie^2}{4m_p^2} \vec{S}\cdot(\epspr\times\eps) 
\left[ (\hbar\wfin + \hbar\winit ) (2\kappa_p+1) e^{-i\vec{q}\cdot\vec{r}/2} - \right.\nonumber\\
& & \left. 2\kappa_n(\hbar\wfin e^{-i\vec{K}\cdot\vec{r}/2} - \hbar\winit e^{i\vec{K}\cdot\vec{r}/2} ) \right] \ket{d_i} ,
\end{eqnarray}
where $\vec{q} \equiv \kfin - \kinit$ and $\vec{K} \equiv (\kfin + \kinit)/2$.   
Note that the correction which depends on $\kappa_p$ is of order $\omega/m_d$ times larger than the $\kappa_n$ piece.

This term is easily evaluated; the result is
\begin{eqnarray}
\matrixm{rel, \mathrm{cont}, \kappa_p}  & = & -\frac{i\pi e^2}{m_p^2} 
\sum_{l=0,2} \sum_{l'=0,2} \sum_{j=-1,0,1} \sum_{\tilde{J} = 0}^{2} \sum_{L=0}^{\infty} \times\nonumber\\
& & \int_{0}^{\infty} dr\ u_{\sss l}(r) \left\{ 
 (\hbar\wfin + \hbar\winit ) (2\kappa_p+1) j_{\sss L}(\frac{qr}{2}) Y_{LM}^{\ast}(\hat{q})  + \right. \nonumber\\
& & \left. 2\kappa_n [\hbar\wfin - (-1)^L \hbar\winit] j_{\sss L}(\frac{Kr}{2}) 
Y_{LM}^{\ast}(\hat{K}) \right\}u_{\sss l'}(r) \left( \epspr \times \eps \right)_{-j} \times\nonumber\\
& &  \threej{L}{1}{\tilde{J}}{-j+M_f-M_i}{j}{M_i-M_f}
\threej{1}{\tilde{J}}{1}{-M_f}{M_f-M_i}{M_i} \times\nonumber\\
& & i^L (-1)^{j-M_i} Y_{LM}^{\ast}(\hat{q}) \sqrt{2\tilde{J}+1}\mxredba{[Y_L \otimes S]_{\tilde{J}} } \label{eq:rc5}.
\end{eqnarray}
We now examine the terms in equations~(\ref{eq:rc1})-(\ref{eq:rc4}) which contain commutators.  Adding 
these eight terms together gives
\begin{eqnarray}
\matrixm{rel, \mathrm{comm}} & = & \frac{\hbar^2}{2m_p} \mx{d_f}{ (\nabla^2 \hat{\Psi}_f) \hat{\Phi}_i + (\nabla^2\hat{\Phi}_i)  \hat{\Psi}_f +
\hat{\Phi}_i (\nabla^2 \hat{\Psi}_f) + \hat{\Psi}_f  (\nabla^2\hat{\Phi}_i)  }{d_i} + \nonumber\\
& & \{\hat{\Phi}_i, \hat{\Psi}_f\}  \leftrightarrow \{\hat{\Phi}_f, \hat{\Psi}_i\} .
\end{eqnarray}
Since we are working in the transverse gauge where $\vec{\nabla}\cdot\vec{A} = 0 $, the $\nabla^2 \hat{\Psi}$ terms can be 
neglected .  Also, looking back at the definitions of $\hat{\Psi}$, we notice that
\begin{equation}
\nabla^2 \hat{\Psi}_{i(f)} = \frac{\omega_{i(f)}^2}{4} \hat{\Psi}_{i(f)},
\end{equation}
Therefore, the commutator terms become
\begin{equation}
\matrixm{rel, \mathrm{comm}}  =  \frac{\hbar^2}{2m_p} \mx{d_f}{ \wfin^2 \left\{ 2\hat{\Phi}_i\hat{\Psi}_f + [\hat{\Psi}_f,\hat{\Phi}_i] \right\} +
\winit^2 \left\{ 2\hat{\Phi}_f\hat{\Psi}_i + [\hat{\Psi}_i,\hat{\Phi}_f] \right\} }{d_i}
\end{equation}
This is a correction to the contact term.  We can account for the ``non-commutator'' part of the above equation by making
the substitutions
\begin{eqnarray}
\hbar\winit & \rightarrow & \hbar\winit  - \frac{ (\hbar\winit)^2}{2m_p} \\
\hbar\wfin  & \rightarrow & \hbar\wfin + \frac{ (\hbar\wfin )^2}{2m_p}
\end{eqnarray}
in equation~(\ref{eq:rc5}).  This amounts to a correction of order $(\omega/m_d)^2$ smaller than the main term,
so it will be discarded.  We add the (previously neglected) first term of equation~(\ref{eq:rc6})
to the remaining part of the commutator to get
\begin{equation}
\matrixm{rel, \mathrm{comm}} = \left[\hbar\wfin - \hbar\winit + \frac{(\hbar\winit)^2}{m_p} \right]\mx{d_f}{\hat{\Phi}_i\hat{\Psi}_f + 
\hat{\Phi}_f\hat{\Psi}_i}{d_i}.
\end{equation}
This is the next-to-leading-order term, along with the $\kappa_n$ piece.  It is assumed to be small at energies
below 100 MeV and is not evaluated further.


\chapter{Pion Terms \label{app:pion} }
\begin{figure}
\centering
\parbox{75mm}{\centering\epsfig{file=pions1} }
\parbox{30mm}{(a)}
\parbox{75mm}{\centering\epsfig{file=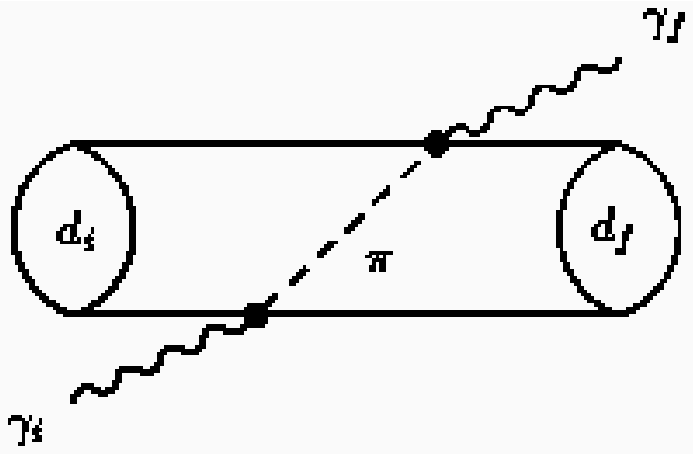} }
\parbox{30mm}{(a)}
\parbox{75mm}{\centering\epsfig{file=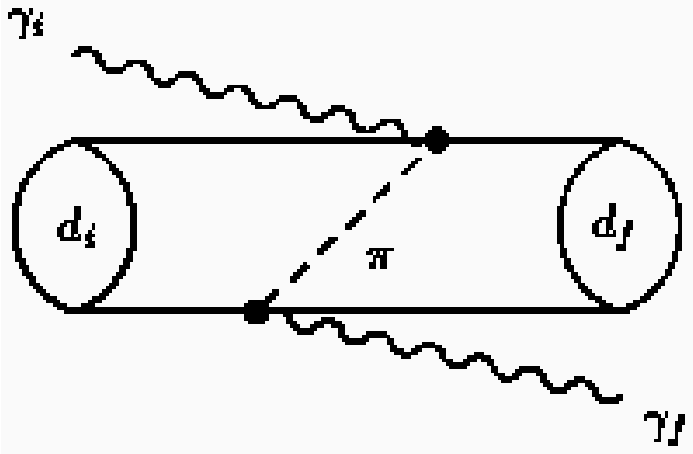} }
\parbox{30mm}{(a)}
\parbox{75mm}{\centering\epsfig{file=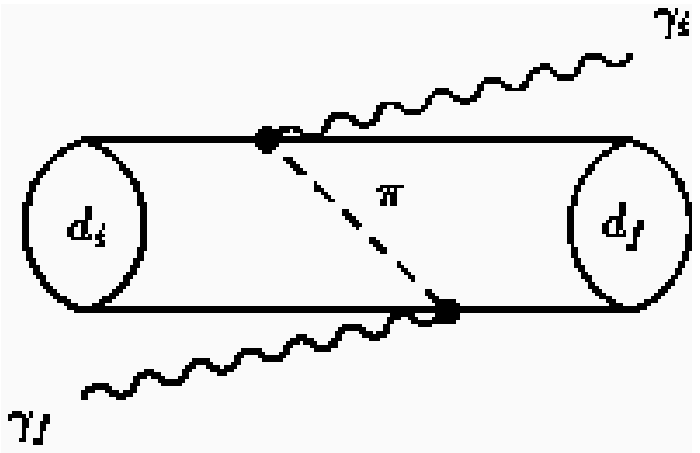} }
\parbox{30mm}{(a)}
\caption[Time-ordered Feynman diagrams with two $\gamma\pi N$ vertices]{Pion-exchange diagrams 
\label{fig:pion1} (a) $\matrixt{\pi 1,\mathrm{uncr}}$
(b) $\matrixt{\pi 2,\mathrm{uncr}}$ (c) $\matrixt{\pi 1,\mathrm{cr}}$ 
(d) $\matrixt{\pi 2,\mathrm{cr}}.$ All four diagrams are equal under 
the assumptions used here.}
\end{figure}

\noindent We now consider diagrams, such as those in Figure~\ref{fig:pion1},
 where a pion is exchanged between the two $\gamma N$ vertices.  
The first terms to be calculated are those which are needed to satisfy the low-energy
theorem described in Section 3.6.  We then calculate the potential
energy portion of the double commutator term in equation~(\ref{eq:mainterm9});
this exactly cancels the other pion-exchange diagrams at low energy.
Lastly, a correction to the $\gamma N$ vertex is presented.

The Hamiltonian for a point interaction between a pion, photon, and nucleon is
\begin{equation}
H^{\gamma\pi N}   = \sum_{j=n,p}-\frac{ i f e_{\pi} }{m_{\pi}} \left( \vec{\sigma}_j\cdot\vec{A}(\vec{x}_j) \right) 
\left( \iso{\tau}_j\cdot\iso{\phi}(\vec{x}_j) \right),
\end{equation}
\noindent where $\frac{f^2}{\hbar c} \approx 0.075$ and  $\iso{\tau}_j$ is the isospin operator for the $j$th
nucleon (the tilde indicates a vector in isospin space). $\vec{A}(\vec{x}_j)$ is given by equation (\ref{eq:seagull2}) 
and a similar expansion for the pion field is
\begin{equation}
\phi_{\pm}(\vec{x}_j) = \sum_{\vec{q}} \frac{\hbar}{\sqrt{V}} \sqrt{ \frac{2\pi}{E_{\pi}} }
\left( a_{\mp, \vec{q} } e^{i\vec{q}\cdot\vec{x}_j} + 
a_{\pm, \vec{q} }^{\dagger} e^{-i\vec{q}\cdot\vec{x}_j} \right), \label{eq:pion20}
\end{equation}
\noindent where $a_{\pm, \vec{q} }^{\dagger}$ creates a $\pi^{\pm}$ with
momentum $\vec{q}$.  Inserting the above equations into equation
(\ref{eq:mainterm1}) gives
\begin{equation}
\matrixt{\pi} = \matrixt{\pi 1,\mathrm{uncr}} + \matrixt{\pi 2,\mathrm{uncr}} + 
\matrixt{\pi 1,\mathrm{cr}} + \matrixt{\pi 2,\mathrm{cr}},
\end{equation}
\noindent where
\begin{eqnarray}
\matrixt{\pi 1,\mathrm{uncr}} & = & \sum_{C,\vec{P}_C}  \sum_{\pm} 
\int \frac{V d^3q}{(2\pi)^3}  \frac{f^2 e^2}{m_{\pi}^2 }
\frac{\hbar^3}{V^2} \frac{2\pi}{\sqrt{\wfin\winit}} \frac{2\pi}{E_{\pi}} \times\label{eq:pion1}\\
& &\bra{d_f, \vec{P}_f} (\vec{\sigma}_2\cdot\epspr) e^{-i\vec{k}_f\cdot\vec{x}_2} \frac{1}{\sqrt{2}}
\tau_{2\pm} e^{i\vec{q}\cdot\vec{x}_2} \ket{C, \vec{P}_C} \times\nonumber\\
& & \frac{1}{\hbar\winit - E_{\pi} + E_{d_i} - E_C -P_C^2/2m_D + i\varepsilon} \times\nonumber\\
& & \bra{C, \vec{P}_C} (\vec{\sigma}_1\cdot\eps) \nonumber
 e^{i\vec{k}_i\cdot\vec{x}_1} \frac{1}{\sqrt{2}}  \tau_{1\mp} e^{-i\vec{q}\cdot\vec{x}_1} \ket{d_i, \vec{P}_i} ,\\
\matrixt{\pi 2,\mathrm{uncr}} & = & \sum_{C, \vec{P}_C}  \sum_{\pm} 
\int \frac{V d^3q}{(2\pi)^3}  \frac{f^2 e^2}{m_{\pi}^2 }
\frac{\hbar^3}{V^2} \frac{2\pi}{\sqrt{\wfin\winit}} \frac{2\pi}{E_{\pi}} \times\label{eq:pion2}\\
& & \bra{d_f, \vec{P}_f} (\vec{\sigma}_1\cdot\epspr) e^{-i\vec{k}_f\cdot\vec{x}_1} \frac{1}{\sqrt{2}}
\tau_{1\pm} e^{i\vec{q}\cdot\vec{x}_1} \ket{C, \vec{P}_C} \times\nonumber\\
& & \frac{1}{\hbar\winit - E_{\pi} + E_{d_i} - E_C  -P_C^2/2m_D+ i\varepsilon} \times\nonumber\\
& & \bra{C, \vec{P}_C} (\vec{\sigma}_2\cdot\eps) \nonumber
e^{i\vec{k}_i\cdot\vec{x}_2} \frac{1}{\sqrt{2}}  \tau_{2\mp} e^{-i\vec{q}\cdot\vec{x}_2} \ket{d_i, \vec{P}_i} ,\\
\matrixt{\pi 1,\mathrm{cr}} & = & \sum_{C, \vec{P}_C} \sum_{\pm} 
\int \frac{V d^3q}{(2\pi)^3}  \frac{f^2 e^2}{m_{\pi}^2 }
\frac{\hbar^3}{V^2} \frac{2\pi}{\sqrt{\wfin\winit}} \frac{2\pi}{E_{\pi}} \times\label{eq:pion3}\\
& & \bra{d_f, \vec{P}_f} (\vec{\sigma}_2\cdot\eps) e^{i\vec{k}_i\cdot\vec{x}_2} \frac{1}{\sqrt{2}}
\tau_{2\pm} e^{i\vec{q}\cdot\vec{x}_2} \ket{C, \vec{P}_C} \times\nonumber\\
& & \frac{1}{-\hbar\wfin - E_{\pi} + E_{d_i} - E_C  -P_C^2/2m_D + i\varepsilon} \times\nonumber\\
& & \bra{C, \vec{P}_C} (\vec{\sigma}_1\cdot\epspr) \nonumber
 e^{-\vec{k}_f\cdot\vec{x}_1} \frac{1}{\sqrt{2}}  \tau_{1\mp} e^{-i\vec{q}\cdot\vec{x}_1} \ket{d_i, \vec{P}_i}, \\
\matrixt{\pi 2,\mathrm{cr}} & = & \sum_{C, \vec{P}_C} \sum_{\pm} 
\int \frac{V d^3q}{(2\pi)^3}  \frac{f^2 e^2}{m_{\pi}^2 }
\frac{\hbar^3}{V^2} \frac{2\pi}{\sqrt{\wfin\winit}} \frac{2\pi}{E_{\pi}} \times\label{eq:pion4}\\
& & \bra{d_f, \vec{P}_f} (\vec{\sigma}_1\cdot\eps) e^{i\vec{k}_i\cdot\vec{x}_1} \frac{1}{\sqrt{2}}
\tau_{1\pm} e^{i\vec{q}\cdot\vec{x}_1} \ket{C, \vec{P}_C} \times\nonumber\\
& & \frac{1}{-\hbar\wfin - E_{\pi} + E_{d_i} - E_C  -P_C^2/2m_D + i\varepsilon} \times\nonumber\\
& & \bra{C, \vec{P}_C} (\vec{\sigma}_2\cdot\epspr) \nonumber
e^{-\vec{k}_f\cdot\vec{x}_2} \frac{1}{\sqrt{2}}  \tau_{2\mp} e^{-i\vec{q}\cdot\vec{x}_2} \ket{d_i, \vec{P}_i}.
\end{eqnarray}
\noindent These four terms correspond to the four time-ordered Feynman diagrams of Figure~\ref{fig:pion1}.
We begin with $\matrixt{\pi 1,\mathrm{uncr}}$. 
After removing the center-of-mass states from $\matrixt{\pi 1,\mathrm{uncr}}$ and
inserting complete sets of $\vec{r}$ and $\vec{r}'$ states, we get
\begin{eqnarray}
\matrixm{\pi 1,\mathrm{uncr}} & = & \sum_C \sum_{\pm} \int d^3r\ d^3r' \frac{d^3q}{(2\pi)^3}  
\frac{\pi f^2 e^2}{m_{\pi}^2 } \frac{\hbar^2}{E_{\pi}} \times\nonumber\\
& & \bra{d_f} (\vec{\sigma}_2\cdot\epspr) 
e^{\frac{i}{2} \vec{k}_f\cdot\vec{r}} \tau_{2\pm} e^{-\frac{i}{2} \vec{q}\cdot\vec{r}} \ket{\vec{r}'} \times\nonumber\\
& & \mxemp{\vec{r}'}{C} \frac{1}{\hbar\winit - E_{\pi} + E_{d_i} - E_C 
 -P_C^2/2m_D + i\varepsilon} \mxemp{C}{\vec{r}}\nonumber\times\\
& & \bra{\vec{r}} (\vec{\sigma}_1\cdot\eps) e^{\frac{i}{2}\vec{k}_i\cdot\vec{r}} 
\tau_{1\mp} e^{-\frac{i}{2} \vec{q}\cdot\vec{r}} \ket{d_i} .
\end{eqnarray}
\noindent We now make the assumption that 
\begin{equation}
E_{\pi} = \sqrt{m_{\pi}^2 + \hbar^2 q^2 }\gg \hbar\winit + E_{d_i} - E_C -P_C^2/2m_D.  
\label{eq:pion9}
\end{equation}
\noindent This collapses the sum over $C$ and the integrals over $r'$ so that
\begin{eqnarray}
\matrixm{\pi 1,\mathrm{uncr}} & = & \sum_{\pm} \int d^3r \frac{\pi\hbar^2 f^2 e^2}{m_{\pi}^2 }
\mxemp{d_f}{\vec{r}}\int \frac{d^3q}{(2\pi)^3} \frac{ -e^{-i\vec{q}\cdot\vec{r}} }{m_{\pi}^2 +\hbar^2 q^2}\times\\
& & \bra{\vec{r}} (\vec{\sigma}_2\cdot\epspr) e^{\frac{i}{2} \vec{k}_f\cdot\vec{r}}\tau_{2\pm}
(\vec{\sigma}_1\cdot\eps) e^{\frac{i}{2}\vec{k}_i\cdot\vec{r}}
\tau_{1\mp}  \ket{d_i}. \nonumber
\end{eqnarray}
\noindent Performing the three-dimensional integral over $q$ yields:
\begin{eqnarray}
\matrixm{\pi 1,\mathrm{uncr}} & = & -\int d^3r \frac{\pi f^2 e^2}{m_{\pi}^2 }
\mxemp{d_f}{\vec{r}} \frac{e^{-m_{\pi}r/\hbar}}{4\pi r}\times\\
& &\bra{\vec{r}} (\vec{\sigma}_2\cdot\epspr) 
(\vec{\sigma}_1\cdot\eps)
\left( \tau_{2-}\tau_{1+} + \tau_{2+}\tau_{1-} \right) e^{\frac{i}{2}\vec{K}\cdot\vec{r}} \ket{d_i},\nonumber
\end{eqnarray}
\noindent where $\vec{K} \equiv \vec{k}_f + \vec{k}_i$.
\newline\newline
\noindent  The isospin operators are evaluated next.  We can write
\begin{eqnarray}
\tau_{2-}\tau_{1+} + \tau_{2+}\tau_{1-} & = & 2 \left( \iso{\tau}_1 \cdot \iso{\tau}_2 + 1 \right) \\
& = & 4 T^2 - \tau_1^2 - \tau_2^2 + 2 \\
& = & 4 T^2 - 4, \label{eq:pion7}
\end{eqnarray}
\noindent where we have defined the total isospin operator $\iso{T} \equiv ( \iso{\tau}_1 + \iso{\tau}_2 )/2$. 
Since the deuteron has zero isospin, $\matrixm{}$ becomes
\begin{equation}
\matrixm{\pi 1,\mathrm{uncr}}  = \int d^3r\frac{ f^2 e^2}{m_{\pi}^2 }
\mxemp{d_f}{\vec{r}} \frac{e^{-m_{\pi}r/\hbar}}{ r}
\bra{\vec{r}}(\vec{\sigma}_2\cdot\epspr)
(\vec{\sigma}_1\cdot\eps)  e^{\frac{i}{2}\vec{K}\cdot\vec{r}} \ket{d_i}.
\end{equation}
\noindent Inserting the partial wave expansion for the exponential (equation \ref{eq:seagull1}) and rewriting 
the dot products as sums over spherical components yields
\begin{eqnarray}
\matrixm{\pi 1,\mathrm{uncr}} & = & \sum_{L=0}^{\infty} \sum_{M=-L}^{L} \sum_{l=0,2} \sum_{l'=0,2}\sum_{j=-1,0,1} \sum_{j'=-1,0,1}
\times\\
& & \frac{4\pi f_{\pi}^2 e^2}{m_{\pi}^2 }
i^L (-1)^{j+j'} Y_{LM}^{\ast}(\hat{K}) 
\int dr u_{\sss l'}(r) \frac{e^{-m_{\pi}r/\hbar}}{ r} j_{\sss L}(\frac{Kr}{2}) u_{\sss l}(r) 
\nonumber\times\\
& & (\epspr)_{-j} (\eps)_{j'} \mxba{\sigma_{2,j} \sigma_{1,-j'} Y_{LM} } .\nonumber
\end{eqnarray}
\noindent We evaluate the general matrix element
\begin{equation}
\mx{L_1 S_1 J_1 M_1}{\sigma_{2,j} \sigma_{1,-j'} Y_{lm} Y_{l'm'} }{L_2 S_2 J_2 M_2},
\end{equation}
since this form appears in all of the pion terms to follow. The first
step is the formation of tensor products using equation~(\ref{eq:formula1}): 
\begin{equation}
\sigma_{2,j} \sigma_{1,-j'}  =  \sum_{S'M'} (-1)^{M'} \sqrt{2S'+1} \threej{1}{1}{S'}{j}{-j'}{-M'}
\left( \sigma_2 \otimes \sigma_1 \right)_{S'M'}
\end{equation} 
\noindent and 
\begin{eqnarray}
\lefteqn{ \sigma_{2,j} \sigma_{1,-j'} Y_{lm} Y_{l'm'}  =  
\sum_{\tilde{J}\tilde{M}} \sum_{S'M'} \sum_{LM} (-1)^{M'+S'-L+\tilde{M}+M} } \\
& & \threej{l}{l'}{L}{0}{0}{0} \threej{l}{l'}{L}{m}{m'}{-M} \threej{1}{1}{S'}{j}{-j'}{-M'} 
 \threej{S'}{L}{\tilde{J}}{M'}{M}{-\tilde{M}} \nonumber\times\\
& & \sqrt{ \frac{(2l+1)(2l'+1)(2L+1)(2S'+1)(2\tilde{J}+1)}{4\pi} }
\left[ \left( \sigma_2 \otimes \sigma_1 \right)_{S'}
\otimes Y_L \right]_{\tilde{J}\tilde{M}}. \nonumber
\end{eqnarray}
\noindent The sums run over all possible values allowed by the properties
of the $3j$ symbols.  Next, we use the Wigner-Eckart theorem.  This gives
\begin{eqnarray}
\lefteqn { \mx{L_1 S_1 J_1 M_1}{\sigma_{2,j} \sigma_{1,-j'} Y_{lm} Y_{l'm'} }{L_2 S_2 J_2 M_2} } \\ 
& = & \sum_{\tilde{J}\tilde{M}} \sum_{S'M'} \sum_{LM}
(-1)^{M'+S'-L+\tilde{M} + J_1 - M_1 + M } 
\threej{1}{1}{S'}{j}{-j'}{-M'} \threej{l}{l'}{L}{0}{0}{0} \nonumber\times\\
& &  \threej{l}{l'}{L}{m}{m'}{-M} 
\threej{S'}{L}{\tilde{J}}{M'}{M}{-\tilde{M}} \threej{J_1}{\tilde{J}}{J_2}{-M_1}{\tilde{M}}{M_2} \nonumber\times\\
& & \sqrt{ \frac{(2l+1)(2l'+1)(2L+1)(2S'+1)(2\tilde{J}+1)}{4\pi} } \nonumber\times\\
& & \mxred{L_1 S_1 J_1}{ \left[ \left( \sigma_2 \otimes \sigma_1 \right)_{S'} 
\otimes Y_L \right]_{\tilde{J}} }{L_2 S_2 J_2}. \nonumber
\end{eqnarray}
\noindent This tensor product can be separated into
spin and orbital angular momentum parts using equation (\ref{eq:formula3}). Thus, the matrix element becomes
\begin{eqnarray}
\lefteqn { \mx{L_1 S_1 J_1 M_1}{\sigma_{2,j} \sigma_{1,-j'} Y_{lm} Y_{l'm'} }{L_2 S_2 J_2 M_2} } \nonumber \\
& = & \sum_{\tilde{J}\tilde{M}} \sum_{S'M'} \sum_{LM} (-1)^{M'+S'-L+\tilde{M} + J_1 - M_1 + M}\label{eq:pion35}  \times\\
& & (2\tilde{J}+1)\sqrt{ \frac{(2l+1)(2l'+1)(2L+1)(2J_1+1)(2J_2+1)(2S'+1)}{4\pi} }\nonumber\times\\
& & \threej{1}{1}{S'}{j}{-j'}{-M'} \threej{l}{l'}{L}{0}{0}{0} \threej{l}{l'}{L}{m}{m'}{-M}
\threej{S'}{L}{\tilde{J}}{M'}{M}{-\tilde{M}} \nonumber\times\\
& & \threej{J_1}{\tilde{J}}{J_2}{-M_1}{\tilde{M}}{M_2}
\ninej{S_1}{L_1}{J_1}{S_2}{L_2}{J_2}{S'}{L}{\tilde{J}} \mxred{S_1}{ \left( \sigma_2 
\otimes \sigma_1 \right)_{S'}}{S_2} \mxred{L_1}{ Y_L }{L_2} \nonumber \\
& = & \sum_{\tilde{J}\tilde{M}} \sum_{S'M'} \sum_{LM} (-1)^{M'-L+\tilde{M} + J_1 - M_1 + M} (2\tilde{J}+1)(2S'+1)
\threej{l}{l'}{L}{0}{0}{0} \label{eq:pion5}\times\\
& & \threej{1}{1}{S'}{j}{-j'}{-M'} \threej{l}{l'}{L}{m}{m'}{-M}
\threej{S'}{L}{\tilde{J}}{M'}{M}{-\tilde{M}} \threej{J_1}{\tilde{J}}{J_2}{-M_1}{\tilde{M}}{M_2} \nonumber\times\\
& & \ninej{S_1}{L_1}{J_1}{S_2}{L_2}{J_2}{S'}{L}{\tilde{J}}
\sqrt{ \frac{(2l+1)(2l'+1)(2L+1)(2J_1+1)(2J_2+1)}{4\pi} } \nonumber\times\\
& &  \rbra{L_1} Y_L \rket{L_2} 
\sum_{s'} \sixj{1}{1}{S'}{S_2}{S_1}{s'} \mxred{S_1}{ S - t }{s'} \mxred{s'}{ S + t }{S_2}, \nonumber
\end{eqnarray}
\noindent where $\vec{S} \equiv (\vec{\sigma}_1 + \vec{\sigma}_2)/2$
and $\vec{t} \equiv (\vec{\sigma}_1 - \vec{\sigma}_2)/2$. 
The orbital angular momentum matrix elements can be evaluated using equation~(\ref{eq:formula5}),
while the spin matrix elements are listed below:
\begin{eqnarray}
\rbra{S_1}  S \rket{s'} & = & \sqrt{6} \delta_{s'1} \delta_{S_11}, \\
\rbra{S_1} t \rket{s'} & = & \sqrt{3} ( \delta_{s'0} \delta_{S_11} - \delta_{s'1} \delta_{S_10} ). 
\end{eqnarray}
\noindent Using this formula for the specific case at hand, 
 we see that $\matrixm{\pi 1,\mathrm{uncr}}$ becomes
\begin{eqnarray}
\matrixm{\pi 1,\mathrm{uncr} } & = & \sum_{LM} \sum_{ll'} \sum_{jj'} \sum_{\tilde{J}\tilde{M}} \sum_{S'M'} \sum_{s'}
\frac{12\pi f^2 e^2}{m_{\pi}^2 } (-1)^{j+j'+M'-L+\tilde{M} + 1 - M_f} Y_{LM}^{\ast}(\hat{K})  \times\nonumber\\
& & i^L (2\tilde{J}+1)(2S'+1)  (\epspr)_{-j} (\eps)_{j'}
\int dr\ u_{\sss l'}(r) \frac{e^{-m_{\pi}r/\hbar}}{ r} j_{\sss L}(\frac{Kr}{2}) u_{\sss l}(r)
\nonumber\times\\
& & \threej{1}{1}{S'}{j}{-j'}{-M'}\threej{S'}{L}{\tilde{J}}{M'}{M}{-\tilde{M}} 
\threej{1}{\tilde{J}}{1}{-M_f}{\tilde{M}}{M_i}\nonumber\times\\
& &\ninej{1}{l'}{1}{1}{l}{1}{S'}{L}{\tilde{J}} \sixj{1}{1}{S'}{1}{1}{s'}
\rbra{l'} Y_L \rket{l} (6\delta_{s'1}+3\delta_{s'0}). 
\end{eqnarray}
\noindent The $M$ sums are removed in the final result: 
\begin{eqnarray}
\matrixm{\pi 1,\mathrm{uncr} } & = & \sum_{l=0,2} \sum_{l'=0,2} \sum_{j=-1,0,1} \sum_{j=-1,0,1}
\sum_{S'=0,2} \sum_{\tilde{J} = 0,1,2} \sum_{L=0}^{\infty} \times\\  
& & \frac{12\pi f^2 e^2}{m_{\pi}^2 } i^L (-1)^{-L-M_i}
(2\tilde{J}+1)(2S'+1) (\epspr)_{-j} (\eps)_{j'}\nonumber\times\\
& & Y_{L,M_f-M_i-j+j'}^{\ast}(\hat{K})\int dr u_{\sss l'}(r) \frac{e^{-m_{\pi}r/\hbar}}{ r} 
j_{\sss L}(\frac{Kr}{2}) u_{\sss l}(r) \nonumber\times\\
& & \threej{1}{1}{S'}{j}{-j'}{j'-j}\threej{S'}{L}{\tilde{J}}{j-j'}{M_f-M_i-j+j'}{M_i-M_f}\nonumber\times\\
& & \threej{1}{\tilde{J}}{1}{-M_f}{M_f-M_i}{M_i} \ninej{1}{l'}{1}{1}{l}{1}{S'}{L}{\tilde{J}}
\rbra{l'} Y_L \rket{l} (-\delta_{S'0} + 2\delta_{S'2}). \nonumber
\end{eqnarray}
\noindent We now  turn  to $\matrixm{\pi 2,\mathrm{uncr} }$.  
Comparing equations (\ref{eq:pion1}) and (\ref{eq:pion2}),
we see the only difference is that the nucleon labels 1 and 2 have been switched.  
Since the numbering of the nucleons can't matter,
we must have $\matrixm{\pi 1,\mathrm{uncr} } = \matrixm{\pi 2,\mathrm{uncr} }.$ 
  In fact, under the approximation that the pion energy dominates
the energy denominators, all four time-orderings given by equations (\ref{eq:pion1})-(\ref{eq:pion4}) are equal. 
Therefore, we get
\begin{eqnarray}
\lefteqn{ \matrixm{\pi 1,\mathrm{uncr} } + \matrixm{\pi 2,\mathrm{uncr} } + \matrixm{\pi 1,\mathrm{cr} }
+ \matrixm{\pi 2,\mathrm{cr} } = }  \\ 
& &  \sum_{l=0,2} \sum_{l'=0,2} \sum_{j=-1,0,1} \sum_{j=-1,0,1}
\sum_{S'=0,2} \sum_{\tilde{J} = 0,1,2} \sum_{L=0}^{\infty} \frac{48\pi f^2 e^2}{m_{\pi}^2 } i^L (-1)^{-L-M_i} \times\nonumber\\
& & (2\tilde{J}+1)(2S'+1) Y_{L,M_f-M_i-j+j'}^{\ast}(\hat{K})\int_0^{\infty} dr\ u_{\sss l'}(r) \frac{e^{-m_{\pi}r/\hbar }}{ r} 
j_{\sss L}(\frac{Kr}{2}) u_{\sss l}(r) \nonumber\times\\
& &  (\epspr)_{-j} (\eps)_{j'}
\threej{1}{1}{S'}{j}{-j'}{j'-j}\threej{S'}{L}{\tilde{J}}{j-j'}{M_f-M_i-j+j'}{M_i-M_f}\nonumber\times\\
& & \threej{1}{\tilde{J}}{1}{-M_f}{M_f-M_i}{M_i} \ninej{1}{l'}{1}{1}{l}{1}{S'}{L}{\tilde{J}}
\rbra{l'} Y_L \rket{l} (-\delta_{S'0} + 2\delta_{S'2}). \nonumber
\end{eqnarray}
\begin{figure}
\centering
\epsfig{file=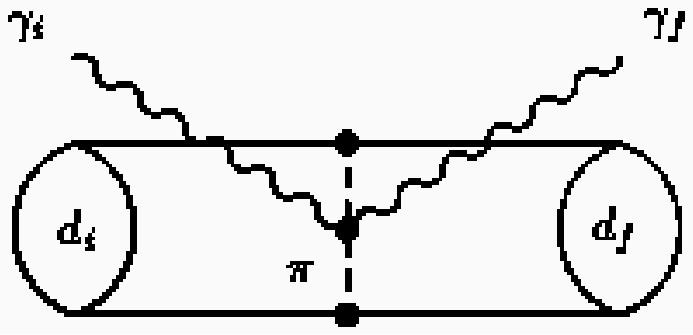}
\caption{Feynman Diagram with one $\gamma\gamma\pi\pi$ vertex \label{fig:pion13} }
\end{figure}
\noindent Next, we will consider the diagram in Figure~\ref{fig:pion13}.
The Hamiltonian for the photon/pion vertex is
\begin{equation}
H^{\gamma\gamma\pi\pi}  =  \frac{1}{4\pi\hbar^2}\int d^3xe^2 A^2(\vec{x}) \left[ \phi_+(\vec{x})\phi_-(\vec{x}) +
\phi_-(\vec{x})\phi_+(\vec{x}) \right],
\end{equation}
and the Hamiltonian for the $\pi N$ vertex is
\begin{eqnarray}
H^{\pi N} & = & \frac{ \hbar f }{ m_{\pi} }\sum_{j=1,2} \left( \vec{\sigma}_j \cdot
\vec{\nabla}_j \right) \left( \iso{\tau}_j \cdot \iso{\phi}(\vec{x}_j) \right) .
\end{eqnarray}
\begin{figure}
\centering
\parbox{75mm}{ \centering\epsfig{file=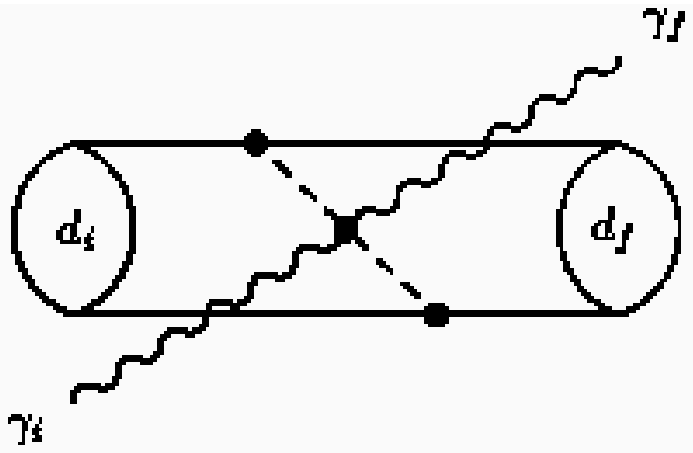} }
\parbox{30mm}{(a)}
\parbox{75mm}{ \centering\epsfig{file=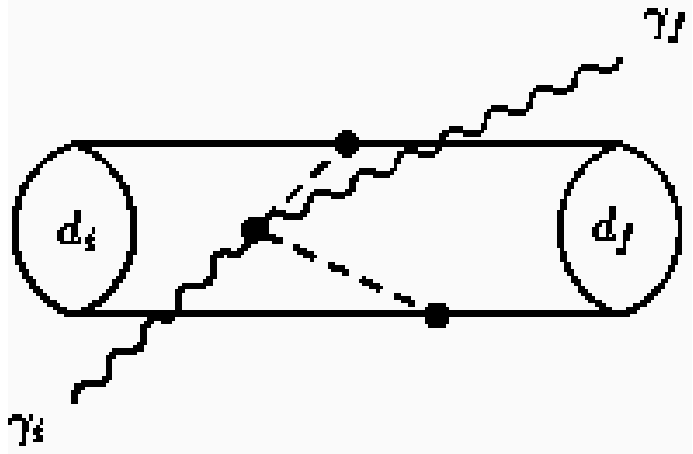} }
\parbox{30mm}{(b)}
\parbox{75mm}{ \centering\epsfig{file=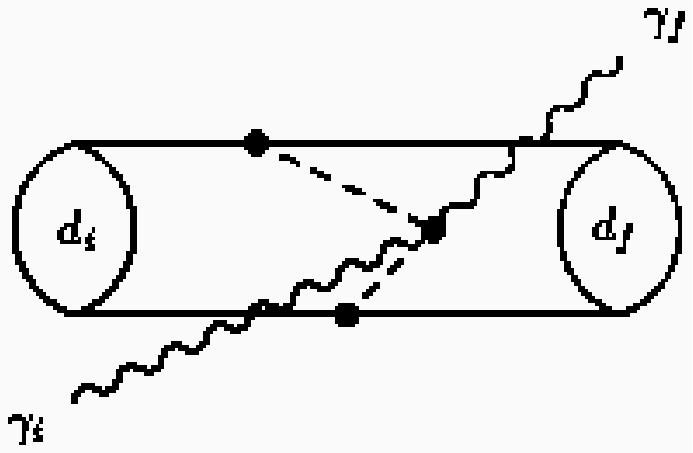} }
\parbox{30mm}{(c)}
\caption{ \label{fig:pion14} Time-ordered diagrams for Figure~\protect\ref{fig:pion13} } 
\end{figure}
\noindent The different time-orderings for this diagram are shown in Figure~\ref{fig:pion14}.  
Careful analysis shows that the sum of Figures~\ref{fig:pion14}(b) and (c) equals Figure~\ref{fig:pion14}(a).
There are three other time-orderings not pictured which are identical to Figure~\ref{fig:pion14}.
The net result of this counting is that Figure~\ref{fig:pion14}(a) needs to be multiplied by a factor of $4$.  
Calling the intermediate $NN$ state between
the two $\pi N$ vertices $C$,  the diagram in Figure~\ref{fig:pion14}(a) 
can be written in third-order perturbation theory as:
\begin{eqnarray}
\matrixt{\pi 5 } & = & 4 \sum_{C, \vec{P}_C}\int \frac{Vd^3q_1}{(2\pi)^3}  \frac{Vd^3q_2}{(2\pi)^3}
\frac{1}{E_{d_i} - E_C +\hbar\winit - \hbar\wfin - P_C^2/2m_d - E_{\pi_2} + i\varepsilon} \times\nonumber \\
& & \bra{d_f, \vec{P}_f, \gamma_f} \frac{\hbar  f }{ m_{\pi} }\sum_{j=n,p}\left( \vec{\sigma}_j \cdot
\vec{\nabla}_j \right) \left( \iso{\tau}_j \cdot \iso{\phi}(\vec{x}_j) \right)
\ket{C, \vec{P}_C, \pi_2, \gamma_f} \times\nonumber \\
& & \bra{C, \vec{P}_C, \pi_2, \gamma_f} 
\frac{1}{4\pi\hbar^2}\int d^3x\ e^2 A^2(\vec{x}) \left[ \phi_+(\vec{x})\phi_-(\vec{x}) + \nonumber\right.\\
& & \left. \phi_-(\vec{x})\phi_+(\vec{x}) \right]
 \ket{C, \vec{P}_C, \pi_1, \gamma_i}  \frac{1}{E_{d_i} - E_C - P_C^2/2m_d - E_{\pi_1} + i\varepsilon} \times\nonumber \\
& & \bra{C, \vec{P}_C, \pi_1, \gamma_i}
\frac{\hbar  f }{ m_{\pi} }\sum_{j'=n,p}\left( \vec{\sigma}_{j'} \cdot
\vec{\nabla}_{j'} \right) \left( \iso{\tau}_{j'} \cdot \iso{\phi}(\vec{x}_{j'}) \right)
\ket{d_i, \vec{P}_i, \gamma_i}.
\end{eqnarray}
\noindent Now, we insert equation~(\ref{eq:pion20}) for $\phi$ and equation~(\ref{eq:seagull2}) for $\vec{A}$.  
Under the usual assumption that the energy denominators are dominated by the pion energies (which
eliminates the sum over $C$), $\matrixt{}$ becomes
\begin{eqnarray} 
\matrixt{\pi 5} & = & 4 \sum_{\pm} \int d^3x\frac{Vd^3q_1}{(2\pi)^3}  \frac{Vd^3q_2}{(2\pi)^3}
\frac{1}{- E_{\pi_2} } \bra{d_f, \vec{P}_f}  \frac{\hbar}{\sqrt{V}} \sqrt{\frac{\pi}{E_{\pi_2}}}
\frac{ if\hbar }{ m_{\pi} } \times\nonumber\\
& & \left\{ \left( \vec{\sigma}_1\cdot\vec{q}_2 \right)
e^{i\vec{q}_2\cdot\vec{x}_1} \tau_{1\pm} +\left( \vec{\sigma}_2\cdot\vec{q}_2 \right)
e^{i\vec{q}_2\cdot\vec{x}_2} \tau_{2\pm} \right\} \times\nonumber\\
& & \frac{e^2 }{4\pi\hbar^2 V} \frac{2\pi\hbar}{\sqrt{\wfin\winit}}
e^{i(\vec{k}_i-\vec{k}_f)\cdot\vec{x} } 2 (\epsdp) \frac{\hbar^2}{V} \frac{2\pi}{\sqrt{E_{\pi_1}E_{\pi_2} }}
e^{i(\vec{q}_1-\vec{q}_2)\cdot\vec{x} }\times\nonumber\\
& & \frac{1}{ - E_{\pi_1} } 
\frac{\hbar}{\sqrt{V}} \sqrt{\frac{\pi}{E_{\pi_1}}}\frac{ -if\hbar }{ m_{\pi} } 
\left\{ \left( \vec{\sigma}_1\cdot\vec{q}_1 \right)
e^{-i\vec{q}_1\cdot\vec{x}_1} \tau_{1\mp} + \right.\nonumber \\
& & \left.\left( \vec{\sigma}_2\cdot\vec{q}_1 \right)
e^{-i\vec{q}_1\cdot\vec{x}_2} \tau_{2\mp} \right\} \ket{d_i, \vec{P}_i}.
\end{eqnarray}
\noindent Next we calculate the effect of the isospin operators.  Combinations involving only
one of the nucleons (such as $\tau_{1+}\tau_{1-}$) are already included in the polarizability terms and therefore are 
neglected here.  The remaining terms can be evaluated using equation~(\ref{eq:pion7}); the result 
after removing the center-of-mass states is
\begin{eqnarray}
\lefteqn{ \matrixm{\pi 5}  =  -\int d^3r\ d^3x\frac{d^3q_1}{(2\pi)^3}
\frac{d^3q_2}{(2\pi)^3} \frac{16\pi e^2f^2 \hbar^4}{m_{\pi}^2}
(\epsdp)\frac{1}{m_{\pi}^2+\hbar^2 q_1^2}
\frac{1}{m_{\pi}^2+\hbar^2q_2^2} }\\
& & \mxemp{d_f}{\vec{r}} \bra{\vec{r}} \left\{ (\vec{\sigma}_1\cdot\vec{q}_2)(\vec{\sigma}_2\cdot\vec{q}_1)
e^{i\vec{q}_2\cdot(\vec{r}/2 - \vec{x} )} e^{i\vec{q}_1\cdot(\vec{r}/2 + \vec{x})}  
e^{-i\vec{q}\cdot\vec{x}} + ( 1 \leftrightarrow 2 ) \right\} \ket{d_i} , \nonumber
\end{eqnarray}
\noindent where $\vec{q} \equiv \vec{k}_f - \vec{k}_i $.  To simplify the calculation, we 
shift the variable $\vec{x}$ by $-\vec{r}/2$.  The integrations over $\vec{q}_1$ and $\vec{q}_2$
can then easily be performed. 
Altogether, we get 
\begin{eqnarray}
\matrixm{\pi 5} & = & \int d^3r\ d^3x \frac{16 \pi e^2f^2}{m_{\pi}^2}
(\epsdp) \mxemp{d_f}{\vec{r}}  e^{-i\vec{q}\cdot(\vec{x}-\vec{r}/2)} \times\\
& &  \bra{\vec{r}}  (\vec{\sigma}_1\cdot\vec{\nabla}_r)
\left[ \vec{\sigma}_2\cdot(\vec{\nabla}_r+\vec{\nabla}_x) \right] 
\frac{e^{-m_{\pi} x/\hbar}}{4\pi x} \frac{e^{-m_{\pi} |\vec{r}-\vec{x}|/\hbar}}{4\pi |\vec{r}-\vec{x}|} \ket{d_i}.\nonumber
\end{eqnarray} 
\noindent The gradient with respect to $x$ can most easily be evaluated by integrating by parts.
Before evaluating the other gradient, we expand the exponentials into partial waves, and also use
\begin{equation}
\frac{e^{-m_{\pi}|\vec{r}-\vec{x}| /\hbar}}{|\vec{r}-\vec{x}|} =
-\frac{4\pi  m_{\pi} }{\hbar} \sum_{L=0}^{\infty} \sum_{M=-L}^L i_{\sss L}(\frac{m_{\pi} r_<}{\hbar})
k_{\sss L}(\frac{m_{\pi} r_>}{\hbar}) Y_{LM}(\hat{r}) Y_{LM}^{\ast}(\hat{x}),
\end{equation}
\noindent where $r_{<(>)}$ is the lesser(greater) of $r$ and $x$, and $i_{\sss L}$ and
$k_{\sss L}$ are the
spherical Bessel functions with imaginary arguments.  Also, the integration over the
angular part of $\vec{x}$ can be carried out because of the orthogonality of
the spherical harmonics.  Thus, $\matrixm{\pi 5}$ becomes
\begin{eqnarray}
\matrixm{\pi 5} & = & -\int d^3r\frac{64 \pi^2  e^2f^2}{\hbar m_{\pi}} (\epsdp) 
\sum_{LM} \sum_{L'M'} i^{L+L'} (-1)^{L'} \mxemp{d_f}{\vec{r}} \times\\
& & \bra{\vec{r}}Y_{LM}^{\ast}(\hat{q}) Y_{L'M'}^{\ast}(\hat{q}) Y_{LM}(\hat{r}) j_{\sss L}(\frac{qr}{2})
\int dx\ x^2 j_{\sss L'}(qx)  (\vec{\sigma}_1\cdot\vec{\nabla}_r) \times\nonumber\\
& & \left[ \vec{\sigma}_2\cdot(\vec{\nabla}_r+i\vec{q}) \right] i_{\sss L'}(\frac{m_{\pi} r_<}{\hbar})
k_{\sss L'}(\frac{m_{\pi} r_>}{\hbar}) \frac{e^{-m_{\pi} x/\hbar}}{x} Y_{L'M'}(\hat{r}) \ket{d_i} \nonumber .
\end{eqnarray}
\noindent The $\sigma$ dot products are expanded into spherical components:
\begin{eqnarray}
\matrixm{\pi 5} & = & -\int d^3r\frac{64 \pi^2 e^2f^2}{\hbar m_{\pi}} (\epsdp)
\sum_{LM} \sum_{L'M'} \sum_{jj'}  i^{L+L'} (-1)^{L'+j+j'} \times\\
& & \mxemp{d_f}{\vec{r}}\bra{\vec{r}}\sigma_{1,j'} \sigma_{2,-j} Y_{LM}^{\ast}(\hat{q}) Y_{L'M'}^{\ast}(\hat{q}) Y_{LM}(\hat{r}) 
j_{\sss L}(\frac{qr}{2}) \times\nonumber\\
& & (\nabla_{-j'}\nabla_j + iq_j\nabla_{-j'}) Y_{L'M'}(\hat{r}) f_{L'}(q,r) \ket{d_i} , \nonumber
\end{eqnarray}
\noindent where
\begin{equation}
 f_{L'}(q,r) \equiv \int_0^{\infty} dx\ x^2  j_{\sss L'}(qx) i_{\sss L'}(\frac{m_{\pi} r_<}{\hbar})
k_{\sss L'}(\frac{m_{\pi} r_>}{\hbar}) \frac{e^{-m_{\pi} x/\hbar}}{x}.
\end{equation}
\noindent Two applications of the gradient formula are needed to evaluate this expression.  
We write the final result as
\begin{eqnarray}
\matrixm{\pi 5} & = & -\frac{64 \pi^2 e^2f^2}{\hbar m_{\pi}} (\epsdp)
\sum_{L=0}^{\infty} \sum_{M=-L}^L \sum_{L'=0}^{\infty} \sum_{M'=-L'}^{L'} \sum_{j=-1,0,1} \sum_{j'=-1,0,1} 
\sum_{n = -2}^{2}  \times\\
& & i^{L+L'} (-1)^{L'+j+j'} \mxba{ \sigma_{1,j'} \sigma_{2,-j} Y_{L''M''} } Y_{LM}^{\ast}(\hat{q})
Y_{L'M'}^{\ast}(\hat{q}) \times\nonumber\\
& & \int_0^{\infty} dr\ u_{\sss l}(r)\ u_{\sss l'}(r) j_{\sss L}(\frac{qr}{2})
 C_{L'M'jj'}^{(n)} O_{L'}^{(n)}(r)  f_{L'}(q,r). \nonumber
\end{eqnarray}
\noindent where
\begin{eqnarray}
 C_{L'M'jj'}^{(-2)} & \equiv & (-1)^{1-j} \sqrt{L'(L'+1)}
\threej{L'-1}{L'}{1}{M'-j'}{-M'}{j'} \times\label{eq:pion11}\\
& & \threej{L'-2}{L'-1}{1}{M'-j'+j}{-M'+j'}{-j}, \nonumber \\
C_{L'M'jj'}^{(-1)} & \equiv & iq_j(-1)^{L'+M'-j'} \sqrt{L'} \threej{L'-1}{L'}{1}{M'-j'}{-M'}{j'}, \\
C_{L'M'jj'}^{(0)} O_{L'}^{(0)}(r) & \equiv &  (-1)^{-j} (L'+1)
\threej{L'+1}{L'}{1}{M'-j'}{-M'}{j'} \times\\
& & \threej{L'}{L'+1}{1}{M'-j'+j}{-M'+j'}{-j}  \times\nonumber\\
& & \ \ \ \ \left( \frac{d}{dr} + \frac{L'+2}{r} \right) 
\left(  \frac{d}{dr} - \frac{L'}{r} \right) + \nonumber\\
& & \threej{L'-1}{L'}{1}{M'-j'}{-M'}{j'}
\threej{L'}{L'-1}{1}{M'-j'+j}{-M'+j'}{-j} \times\nonumber\\
& & (-1)^{-j} L' \left( \frac{d}{dr} - \frac{L'-1}{r} \right)
\left(  \frac{d}{dr} + \frac{L'+1}{r} \right), \nonumber\\
C_{L'M'jj'}^{(1)}  & \equiv &  iq_j(-1)^{L'+M'-j'+1}  \sqrt{L'+1} \threej{L'+1}{L'}{1}{M'-j'}{-M'}{j'}, \\
C_{L'M'jj'}^{(2)} & \equiv &    (-1)^{1-j} \sqrt{(L'+1)(L'+2)}
\threej{L'+1}{L'}{1}{M'-j'}{M'}{j'} \times\\
& & \threej{L'+2}{L'+1}{1}{M'-j'+j}{-M'+j'}{-j},\nonumber \\
 O_{L'}^{(-2)}(r)  & \equiv &  \left( \frac{d}{dr} +\frac{L'}{r} \right)
\left(  \frac{d}{dr} + \frac{L'+1}{r} \right), \\
 O_{L'}^{(-1)}(r)  & \equiv &  \left(  \frac{d}{dr} + \frac{L'+1}{r} \right), \\
O_{L'}^{(1)}(r)  & \equiv & \left(  \frac{d}{dr} - \frac{L'}{r} \right), \\
O_{L'}^{(2)}(r)  & \equiv &\left( \frac{d}{dr} - \frac{L'+1}{r} \right)
\left(  \frac{d}{dr} - \frac{L'}{r} \right).  \label{eq:pion12}
\end{eqnarray}
\begin{figure}
\centering
\parbox{75mm}{\centering\epsfig{file=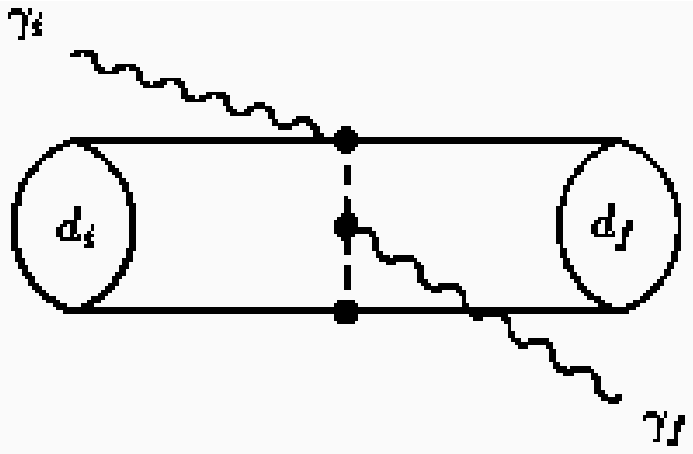} }
\caption{\label{fig:pion10} Feynman diagram corresponding to $\matrixt{\pi 6, \mathrm{uncr}}$. }
\end{figure}
We also consider diagrams such as the one shown in Figure~\ref{fig:pion10}.  Just as in the 
previous term, there are 6 time orderings; this contributes a factor of 4 to the calculation.
The Hamiltonian for the single photon/pion interaction is
\begin{equation}
H^{\pi\pi\gamma} = \frac{e}{4\pi\hbar} \int d^3x (\vec{q}_1 + \vec{q}_2)\cdot\vec{A}(\vec{x})
\left[ \phi_{-}(\vec{q}_1)\phi_{+}(\vec{q}_2) - \phi_{+}(\vec{q}_1)\phi_{-}(\vec{q}_2) \right] ,
\end{equation}
\noindent which means that this term is
\begin{eqnarray}
\lefteqn{ \matrixt{\pi 6, \mathrm{uncr}}  =   4 \sum_{C, \vec{P}_C}
\int \frac{Vd^3q_1}{(2\pi)^3}  \frac{Vd^3q_ 2}{(2\pi)^3} \times}  \\
& & \frac{1}{E_{d_i} - E_C +\hbar\winit - \hbar\wfin - P_C^2/2m_d - E_{\pi_2} + i\varepsilon} \times\nonumber \\
& & \bra{d_f, \vec{P}_f, \gamma_f} \frac{\hbar  f }{ m_{\pi} }\sum_{j=1,2}\left(
\vec{\sigma}_j \cdot \vec{\nabla}_j \right) \left( \iso{\tau}_j \cdot \iso{\phi}(\vec{x}_j) \right)
\ket{C, \vec{P}_C, \pi_2, \gamma_f} \times\nonumber \\
 & & \bra{C, \vec{P}_C, \pi_2, \gamma_f} \frac{e}{4\pi\hbar}\int d^3x  (\vec{q}_1 + \vec{q}_2)\cdot\vec{A}(\vec{x})
\left[ \phi_{-}(\vec{q}_1)\phi_{+}(\vec{q}_2) + \right. \nonumber\\
& & \left. \phi_{+}(\vec{q}_1)\phi_{-}(\vec{q}_2) \right]
\ket{C, \vec{P}_C, \pi_1, \gamma_f}  
\frac{1}{\hbar\winit + E_{d_i} - E_C - P_C^2/2m_d - E_{\pi_1} + i\varepsilon} \times\nonumber \\
& & \bra{C, \vec{P}_C, \pi_1, \gamma_i} \frac{i e f }{ m_{\pi} }
\sum_{n=1,2}\left( \vec{\sigma}_n \cdot
\vec{A}(\vec{x}_n) \right) \left( \iso{\tau}_n \cdot \iso{\phi}(\vec{x}_n) \right)
\ket{d_i, \vec{P}_i, \gamma_i}. \nonumber
\end{eqnarray}
\noindent This calculation follows the same steps as the previous one.  We first
simplify the energy denominators, and insert the expressions for $\vec{A}$ and $\tilde{\phi}$:
\begin{eqnarray}
\matrixt{\pi 6, \mathrm{uncr}} & = & 16\pi \sum_{\pm}\int d^3x\frac{Vd^3q_1}{(2\pi)^3}  \frac{Vd^3q_2}{(2\pi)^3}
\frac{1}{- E_{\pi_2} } \bra{d_f, \vec{P}_f}  \frac{\hbar}{\sqrt{V}} \sqrt{\frac{\pi}{E_
{\pi_2}}}
\frac{ if\hbar }{ m_{\pi} } \\
& & \left\{ \left( \vec{\sigma}_1\cdot\vec{q}_2 \right)
e^{i\vec{q}_2\cdot\vec{x}_1} \tau_{1\pm} +\left( \vec{\sigma}_2\cdot\vec{q}_2 \right)
e^{i\vec{q}_2\cdot\vec{x}_2} \tau_{2\pm} \right\} \times\nonumber\\
& & \frac{2\pi\hbar}{V\sqrt{\wfin\winit}}
\frac{e}{\hbar} (\vec{q}_1 + \vec{q}_2)\cdot\epspr e^{-i\vec{k}_f\cdot\vec{x}} 
 \frac{\hbar^2}{V} \frac{2\pi}{\sqrt{ E_{\pi_1}E_{\pi_2} }}
2e^{i(\vec{q}_1-\vec{q}_2)\cdot\vec{x} }\times\nonumber\\
& & \frac{1}{ - E_{\pi_1} }
\frac{\hbar}{\sqrt{V}} \sqrt{\frac{\pi}{E_{\pi_1}}}\frac{ i e f }{ m_{\pi} }
\left\{ \left( \vec{\sigma}_1\cdot\eps \right)
e^{-i\vec{q}_1\cdot\vec{x}_1}  e^{i\vec{k}_i\cdot\vec{x}_1} \tau_{1\mp} + \right.\nonumber \\
& & \left.\left( \vec{\sigma}_2\cdot\eps \right)
e^{-i\vec{q}_1\cdot\vec{x}_2} \tau_{2\mp}  e^{i\vec{k}_i\cdot\vec{x}_2}\right\} \ket{d_i, \vec{P}_i}.\nonumber
\end{eqnarray}
\noindent The only isospin factors which appear are $\sum_{\pm}\tau_{1\pm}\tau_{2\mp} = -4$.  
Removing the center-of-mass contributions and adding in the crossed term then gives
\begin{eqnarray}
\matrixm{\pi 6} & = &-\int d^3r\ d^3x\frac{d^3q_1}{(2\pi)^3}
\frac{d^3q_2}{(2\pi)^3} \frac{16\pi e^2f^2 \hbar^4}{m_{\pi}^2}
\frac{1}{m_{\pi}^2+\hbar^2 q_1^2}
\frac{1}{m_{\pi}^2+\hbar^2q_2^2} \mxemp{d_f}{\vec{r}}\\
& &\bra{\vec{r}} \left\{ (\vec{\sigma}_1\cdot\vec{q}_2)(\vec{\sigma}_2\cdot\eps)
(\vec{q}_1+\vec{q}_2)\cdot\epspr e^{i\vec{q}_1\cdot (\vec{x}+\vec{r}/2) }
e^{i\vec{q}_2\cdot (\vec{r}/2 - \vec{x})} e^{-i\vec{k_f}\cdot\vec{x}} e^{-i\vec{k}_i\cdot\vec{r}/2}
+ \right. \nonumber\\
& &  (\vec{\sigma}_2\cdot\vec{q}_2)(\vec{\sigma}_1\cdot\eps)
(\vec{q}_1+\vec{q}_2)\cdot\epspr e^{i\vec{q}_1\cdot (\vec{x}-\vec{r}/2) }
e^{i\vec{q}_2\cdot (-\vec{r}/2 - \vec{x})} e^{-i\vec{k_f}\cdot\vec{x}} e^{i\vec{k}_i\cdot\vec{r}/2}
+ \nonumber\\
& & (\vec{\sigma}_1\cdot\vec{q}_2)(\vec{\sigma}_2\cdot\epspr)
(\vec{q}_1+\vec{q}_2)\cdot\eps  e^{i\vec{q}_1\cdot (\vec{x}+\vec{r}/2) }
e^{i\vec{q}_2\cdot (\vec{r}/2 - \vec{x})} e^{i\vec{k_i}\cdot\vec{x}} e^{i\vec{k}_f\cdot\vec{r}/2}
+ \nonumber\\
& & \left. (\vec{\sigma}_2\cdot\vec{q}_2)(\vec{\sigma}_1\cdot\epspr)      
(\vec{q}_1+\vec{q}_2)\cdot\eps  e^{i\vec{q}_1\cdot (\vec{x}-\vec{r}/2) }
e^{i\vec{q}_2\cdot (-\vec{r}/2 - \vec{x})} e^{i\vec{k_i}\cdot\vec{x}} e^{-i\vec{k}_f\cdot\vec{r}/2}
\right\} \ket{d_i}.\nonumber
\end{eqnarray} 
\noindent We again shift $\vec{x}$ by $-\vec{r}/2$ for convenience, and
rewrite the momentum operators as gradients.  The $\vec{q}_1$ and $\vec{q}_2$
integrals can then be performed to yield 
\begin{eqnarray}
\lefteqn{ \matrixm{\pi 6}  = \int d^3r\ d^3x \frac{16\pi e^2f^2 }{m_{\pi}^2} 
 \mxemp{d_f}{\vec{r}}\times}\\
& &\bra{\vec{r}} \left\{ e^{-i\vec{k}_f\cdot (\vec{x}-\vec{r}/2) } e^{-i\vec{k}_i\cdot\vec{r}/2}
(\vec{\sigma}_1\cdot\vec{\nabla}_r)(\vec{\sigma}_2\cdot\eps)
(\vec{\nabla}_x + 2\vec{\nabla}_r)\cdot\epspr \right. +\nonumber\\
& &  \left.  e^{i\vec{k}_f\cdot\vec{r}/2} e^{i\vec{k}_i\cdot (\vec{x}-\vec{r}/2) }
(\vec{\sigma}_1\cdot\vec{\nabla}_r)(\vec{\sigma}_2\cdot\epspr)
(\vec{\nabla}_x + 2\vec{\nabla}_r)\cdot\eps \right\} \frac{e^{-m_{\pi} x/\hbar}}{4 \pi x} 
\frac{e^{-m_{\pi} |\vec{r}-\vec{x}|/\hbar}}{4\pi |\vec{r}-\vec{x}|} \ket{d_i}. \nonumber
\end{eqnarray}
\noindent We integrate by parts to evaluate the $\vec{x}$ gradient, and find it
to be zero.  Next, we
expand the exponentials into partial waves and break up the dot products into spherical
components.  After integrating over $\hat{x}$ to remove two of the spherical harmonics,
the matrix element becomes
\begin{eqnarray}
\matrixm{\pi 6} & = &-\int d^3r \frac{64\pi^2 e^2f^2 }{\hbar m_{\pi}}  \sum_{L=0}^{\infty} \sum_{M=-L}^L
\sum_{L'=0}^{\infty} \sum_{M'=-L'}^{L'} \sum_{\kappa=-1,0,1} \sum_{\mu = -1,0,1} \sum_{\nu=-1,0,1} \times\\
& & i^{L+L'} (-1)^{\kappa+\mu+\nu}  (\eps)_{-\nu} (\epspr)_{-\kappa} 
Y_{LM}^{\ast}(\hat{q})  j_{\sss L}(\frac{qr}{2}) \mxemp{d_f}{\vec{r}} \times\nonumber\\
&  & \bra{\vec{r}} Y_{LM}(\hat{r}) [(-1)^{L'} 2Y_{L'M'}^{\ast}(\hat{k}_f) \sigma_{1,\mu} \sigma_{2,\nu} (\nabla_r)_{-\mu} 
(\nabla_r)_{\kappa} Y_{L'M'}(\hat{r})  g_{L'}(\wfin, r) +  \nonumber\\
& & Y_{L'M'}^{\ast}(\hat{k}_i)\sigma_{1,\mu} \sigma_{2,\kappa} (\nabla_r)_{-\mu} [2(\nabla_r)_{\nu}
+ i (\vec{k}_f)_{\nu} ]Y_{L'M'}(\hat{r}) g_{L'}(\winit, r) ] \ket{d_i},  \nonumber
\end{eqnarray}
\noindent where
\begin{equation}
 g_{L'}(\omega, r) \equiv \int dx\ x^2 i_{\sss L'}(\frac{m_{\pi} r_<}{\hbar})
k_{\sss L'}(\frac{m_{\pi} r_>}{\hbar}) \frac{e^{-m_{\pi} x/\hbar}}{x} j_{\sss L'}(\omega x).
\end{equation}
\noindent The gradients can be evaluated using the gradient formula to give the final result
\begin{eqnarray}
\lefteqn{ \matrixm{\pi 6}  =  - \sum_{l=0,2} \sum_{l'=0,2}
\sum_{L=0}^{\infty} \sum_{M=-L}^L 
\sum_{L'=0}^{\infty} \sum_{M'=-L'}^{L'} \sum_{\kappa=-1,0,1} \sum_{\mu = -1,0,1} \sum_{\nu=-1,0,1}
\sum_{n = -2,0,2}  \times} \\
& & \frac{128 \pi^2 e^2f^2 }{\hbar m_{\pi}}  i^{L+L'} (-1)^{\kappa+\mu+\nu}  (\eps)_{-\nu} (\epspr)_{-\kappa} 
 Y_{LM}^{\ast}(\hat{q}) \int_0^{\infty} dr\ u_{\sss l}(r) j_{\sss L}(\frac{qr}{2}) u_{\sss l'}(r) \times\nonumber\\
& &  [ \mxba{(-1)^{L'} \sigma_{1,\mu} \sigma_{2,\nu} Y_{LM}(\hat{r}) Y_{L'+n,M'+\nu-\mu}(\hat{r}) }\times\nonumber\\
& &  \ \ \ \ \ \ \ \ Y_{L'M'}^{\ast}(\hat{k}_f) C_{L'M'\kappa\mu}^{(n)} O_{L'}^{(n)}(r)  g_{L'}(\wfin, r) +\nonumber\\
& & \mxba{\sigma_{1,\mu} \sigma_{2,\kappa}  Y_{LM}(\hat{r}) Y_{L'+n,M'+\nu-\mu}(\hat{r}) } \times\nonumber\\
& &  \ \ \ \ \ \ \ \ Y_{L'M'}^{\ast}(\hat{k}_i) C_{L'M'\nu\mu}^{(n)} O_{L'}^{(n)}(r)  g_{L'}(\winit, r) ] . \nonumber
\end{eqnarray}
\noindent The $C^{(n)}$ and $O^{(n)}$ factors are defined by equations~(\ref{eq:pion11})-(\ref{eq:pion12}),
and the angular element can be evaluated using equation~(\ref{eq:pion5}).
\begin{figure}
\centering
\parbox{75mm}{\centering\epsfig{file=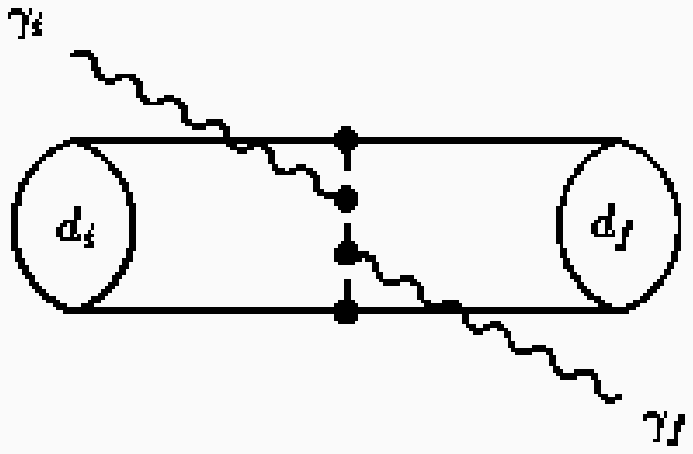} }
\caption{ Feynman diagram corresponding to $\matrixm{\pi 7}$.  \label{fig:pion11} }
\end{figure}

We now calculate the diagram pictured in Figure~\ref{fig:pion11}.  There are 12 separate time-orderings contributing 
a total factor of 4, as well as a factor of 2 for the term where nucleons 1 and 2 are switched,
 so the matrix element is
\begin{eqnarray}
\matrixt{\pi 7} & = & 8 \sum_{C, \vec{P}_C } \sum_{ij=n,p} \sum_{\pm} \int d^3x\ d^3y  \frac{ V d^3q_1}{(2\pi)^3}
\frac{ V d^3q_2}{(2\pi)^3} \frac{ V d^3q_3}{(2\pi)^3} \times \\
& & \bra{d_f, \vec{P}_f, \gamma_f }
\frac{\hbar f }{m_{\pi} } ( \vec{\sigma}_j\cdot\vec{\nabla}_j ) (\tilde{\tau}_j\cdot\tilde{\phi}(\vec{x}_j) )
\ket{C, \vec{P}_C, \gamma_f, \pi_3}  \times\nonumber \\
& & \frac{1}{ E_{d_i} + \hbar\winit - E_C - P_C^2/2m_d - E_{\pi_3} - \hbar\wfin + i\varepsilon} \times\nonumber\\
& & \mx{ C, \vec{P}_C, \gamma_f, \pi_3}{ \frac{ \pm e }{4\pi\hbar} (\vec{q}_2 + \vec{q}_3)\cdot\vec{A}(\vec{y})
\phi_{\mp}(\vec{q}_2) \phi_{\pm}(\vec{q}_3) }{C, \vec{P}_C, \pi_2} \times\nonumber\\
& & \frac{1}{ E_{d_i} + \hbar\winit - E_C - P_C^2/2m_d - E_{\pi_2}   + i\varepsilon} \times\nonumber\\
& & \mx{C, \vec{P}_C, \pi_2}{ \frac{ \pm e }{4\pi\hbar} (\vec{q}_1 + \vec{q}_2)\cdot\vec{A}(\vec{x})
\phi_{\mp}(\vec{q}_1) \phi_{\pm}(\vec{q}_2) }{C, \vec{P}_C, \pi_1, \gamma_i} \times\nonumber\\
& & \frac{1}{ E_{d_i}   - E_C - P_C^2/2m_d - E_{\pi_1}   + i\varepsilon} \times\nonumber\\
& & \mx{C, \vec{P}_C, \pi_1, \gamma_i}{ \frac{\hbar f }{m_{\pi} } ( \vec{\sigma}_j\cdot\vec{\nabla}_j ) 
(\tilde{\tau}_j\cdot\tilde{\phi}(\vec{x}_j) ) }{ d_i, \vec{P}_i, \gamma_i}.  \nonumber
\end{eqnarray}
The energy denominators can be simplified through the assumption that the pion energy is the
dominant factor.    After inserting the expressions for $\tilde{\phi}$ and $\vec{A}$ and removing the
overall center-of-mass factors, we get
\begin{eqnarray}
\matrixm{\pi 7} & = & -  \sum_{\pm} \int d^3x d^3y  \frac{d^3q_1}{(2\pi)^3}
\frac{d^3q_2}{(2\pi)^3} \frac{d^3q_3}{(2\pi)^3} d^3r\times \\
& & \mxemp{d_f}{\vec{r}} \bra{\vec{r}}  \frac{ 4 \pi e^2 f^2 \hbar^6}{m_{\pi}^2 } 
\tau_{2,\mp} \tau_{1,\pm}
e^{i\vec{k}_i\cdot\vec{x} } e^{-i\vec{k}_f\cdot\vec{y} }\times\nonumber \\
& & (\vec{\sigma}_1\cdot\vec{q}_1) (\vec{\sigma_2}\cdot\vec{q}_3 (\vec{q}_1 + \vec{q}_2)\cdot\eps
(\vec{q}_2 + \vec{q}_3)\cdot\epspr \times\nonumber\\
& & \frac{ e^{i\vec{q}_1\cdot(\vec{x}-\vec{r}/2)} }{m_{\pi}^2 + \hbar^2 q_1^2 }
\frac{ e^{i\vec{q}_2\cdot(\vec{y}-\vec{x})} }{m_{\pi}^2 + \hbar^2 q_2^2 }
\frac{ e^{-i\vec{q}_3\cdot(\vec{y}+\vec{r}/2)} }{m_{\pi}^2 + \hbar^2 q_3^2 } \ket{d_i} + \left\{(\vec{k}_i, \eps)
\leftrightarrow (-\vec{k}_f, \epspr) \right\} . \nonumber
\end{eqnarray}
In the calculations of the other pion terms, the next step was to perform the integrals over the
pion momenta.  However, in this case, this turns out to be more difficult because of (what will
become) the four gradients.   We will therefore integrate first over the dummy variables $x$ and $y$.
This generates two momentum delta functions, allowing two of the remaining integrals to be 
performed immediately.  The result of these integrations is
\begin{eqnarray}
\matrixm{\pi 7} & = & -  \sum_{\pm} \int\frac{  d^3q_2}{(2\pi)^3} d^3r \mxemp{d_f}{\vec{r}}  \bra{\vec{r}}
\frac{ 4\pi e^2 f^2 \hbar^6}{m_{\pi}^2 } \tau_{2,\mp} \tau_{1,\pm}
e^{i \vec{K}\cdot\vec{r}/2 } \times\\
& & \left[ \vec{\sigma}_2\cdot (\vec{q}_2 - \vec{k}_f) \right]
\left[ \vec{\sigma}_1\cdot (\vec{q}_2 - \vec{k}_i) \right]
2 (\vec{q}_2\cdot\eps) 2 (\vec{q}_1\cdot\epspr) e^{-i\vec{q}_2\cdot\vec{r} } \times\nonumber\\
& & \frac{1}{ m_{\pi}^2 + \hbar^2 q_2^2 } \frac{1}{ m_{\pi}^2 + \hbar^2 (\vec{q}_2 - \vec{k}_i)^2 }
\frac{1}{ m_{\pi}^2 + \hbar^2 (\vec{q}_2 - \vec{k}_f)^2 } \ket{d_i}
 + \nonumber\\
& & \left\{ (\vec{k}_i, \eps)\leftrightarrow (-\vec{k}_f, \epspr) \right\}. \nonumber
\end{eqnarray}
The isospin factor is the usual $ \sum_{\pm} \tau_{2,\mp} \tau_{1,\pm} = -4$.  Rewriting the
$\vec{q}_2$ vectors as gradients, the matrix element becomes
\begin{eqnarray}
\matrixm{\pi 7} & = &-\frac{ e^2 f^2}{m_{\pi}^2 } 64\pi  \hbar^6 \int d^3r \mxemp{d_f}{\vec{r}} 
e^{i \vec{K}\cdot\vec{r}/2 } \times\label{eq:pion14}\\
& & \bra{\vec{r}}  \left[ \vec{\sigma}_2\cdot (\vec{\nabla}_r  +i \vec{k}_f) \right]
\left[ \vec{\sigma}_1\cdot (\vec{\nabla}_r +i \vec{k}_i) \right]
 (\vec{\nabla}_r\cdot\eps)  (\vec{\nabla}_r\cdot\epspr)
\mathcal{I}(\kinit,\kfin; \vec{r}) \ket{d_i} \nonumber\\
& &  + \left\{ (\vec{k}_i, \eps)\leftrightarrow (-\vec{k}_f, \epspr) \right\} \nonumber,
\end{eqnarray}
where
\begin{equation}
\mathcal{I}(\kinit,\kfin; \vec{r}) \equiv \int \frac{  d^3q_2}{(2\pi)^3} 
\frac{e^{-i\vec{q}_2\cdot\vec{r} } }{[  m_{\pi}^2 + \hbar^2 q_2^2 ][m_{\pi}^2 + \hbar^2 (\vec{q}_2 - 
\vec{k}_i)^2][ m_{\pi}^2 + \hbar^2 (\vec{q}_2 - \vec{k}_i)^2] },
\end{equation}
and where the gradient only acts on $\mathcal{I}(\kinit,\kfin; \vec{r})$. Defining 
$Z(q,\omega) \equiv (m_{\pi}^2 + \hbar^2 q^2 + \hbar^2\omega^2)/2\hbar^2q\omega$ (as in \cite{ar80})
and expanding the exponential,
this integral becomes
\begin{eqnarray}
\mathcal{I}(\kinit,\kfin; \vec{r}) & = & \frac{1}{\hbar^4}
\int \frac{dq_2\ d\hat{q_2} }{(2\pi)^3} \sum_{L=0}^{\infty} \sum_{M=-L}^{L} \frac{(-i)^L \pi}{\winit\wfin} \times\\
& & \ \ \ \ \ \ \ \ \frac{ j_{\sss L}(q_2r) Y_{LM}(\hat{r}) Y_{LM}^{\ast}(\hat{q_2}) }{ [Z(q_2,\winit) - \hat{q_2}\cdot\hat{k}_i ]
[Z(q_2,\wfin) - \hat{q_2}\cdot\hat{k}_f ][m_{\pi}^2 + \hbar^2 q_2^2 ] }. \nonumber
\end{eqnarray} 
We now use the identity
\begin{equation}
\frac{1}{t-\hat{a}\cdot\hat{b} } = \sum_{\alpha=0}^{\infty} \sum_{\beta=-\alpha}^{\alpha} 4\pi Q_{\alpha}(t) Y_{\alpha\beta}(\hat{a})
Y_{\alpha\beta}^{\ast}(\hat{b}), 
\end{equation}
which is true for $t>1$, where $Q(t)$ is the irregular solution of Legendre's equation. 
This gives
\begin{eqnarray}
\mathcal{I}(\kinit,\kfin; \vec{r}) & = & \frac{1}{\hbar^4}\sum_{\alpha\beta} \sum_{\alpha'\beta'} \sum_{LM}
\int d\hat{q_2}  Y_{LM}^{\ast}(\hat{q_2}) Y_{\alpha\beta}^{\ast}(\hat{q_2})  Y_{\alpha'\beta'}^{\ast}(\hat{q_2}) \times\nonumber\\
& & \ \ \ \ \ \ \int dq_2  \frac{2 (-i)^L}{\winit\wfin(m_{\pi}^2 + \hbar^2 q_2^2)}
j_{\sss L}(q_2r) Y_{LM}(\hat{r})  Y_{\alpha\beta}(\hat{k}_i)  Y_{\alpha'\beta'}(\hat{k}_f) \times\nonumber\\
& & \ \ \ \ \ \  Q_{\alpha}[Z(q_2,\winit)] Q_{\alpha'}[Z(q_2,\wfin)].
\end{eqnarray}
The angular integral can be performed by combining two of the spherical harmonics and then using 
the orthogonality relation.  The result is
\begin{eqnarray}
\lefteqn{ \mathcal{I}(\kinit,\kfin; \vec{r})  = \sum_{\alpha\beta} \sum_{\alpha'\beta'} \sum_{LM}
\threej{\alpha}{\alpha'}{L}{0}{0}{0} \threej{\alpha}{\alpha'}{L}{\beta}{\beta'}{M} \times}\\
& & \sqrt{ \frac{(2\alpha+1) (2\alpha'+1)(2L+1) }{4\pi} }
\frac{2 (-i)^L}{\winit\wfin} Y_{LM}(\hat{r})  Y_{\alpha\beta}(\hat{k}_i) 
 Y_{\alpha'\beta'}(\hat{k}_f) \Lambda^{(0)}_{L\alpha\alpha'}(\winit,\wfin;r), \nonumber
\end{eqnarray}
where
\begin{equation}
\Lambda^{(n)}_{L\alpha\alpha'}(\winit,\wfin;r) \equiv
\frac{1}{\hbar^4} \int_0^{\infty} dq_2\ q_2^n \frac{  j_{\sss L}(q_2r) }{m_{\pi}^2 + \hbar^2 q_2^2 }
Q_{\alpha}[Z(q_2,\winit)] Q_{\alpha'}[Z(q_2,\wfin)].
\end{equation}
This notation allows a more compact formulation of the gradient formula:
\begin{eqnarray}
 \nabla_{\mu} Y_{LM}(\hat{r}) \Lambda^{(n)}_{L\alpha\alpha'}(\winit,\wfin;r) & = &
\sum_{a = \pm 1} (-1)^{\mu+L+M} \threej{L+a}{L}{1}{M+\mu}{-M}{-\mu} \times \label{eq:pion13}\\
& & \ \ \ \ \ \ \sqrt{L + \frac{1}{2}(1+a) } Y_{L+1,M+a}(\hat{r}) \Lambda^{(n+1)}_{L+a, \alpha\alpha'}(\winit,\wfin;r).  \nonumber
\end{eqnarray}
The final expression is cumbersome and therefore will not be explicitly written.  
Instead, we use equation~(\ref{eq:pion14}) and either 2, 3, or 4 applications of the gradient formula
(equation~\ref{eq:pion13}).
Switching the initial and final photons simply causes the function $\mathcal{I}$ to be
multiplied by a factor of $(-1)^{\alpha+\alpha'} = (-1)^L$.

We would now like to  calculate the contribution to the double commutator term of
equation~(\ref{eq:mainterm9}) from the pion-exchange potential.  This potential can be written as
\begin{equation}
V_{\pi}(\vec{r}) = \frac{f^2m_{\pi}}{\hbar} (\tilde{\tau}_1 \cdot \tilde{\tau}_2) \left[ S_{12} \left( \frac{1}{x^3} +
\frac{1}{x^2} + \frac{1}{3x} \right) e^{-x} + (\vec{\sigma}_1 \cdot \vec{\sigma}_2) \frac{1}{3} \left( \frac{e^{-x}}{x} -
\frac{4\pi}{m_{\pi}^3}\delta({\vec{r}}) \right) \right],
\end{equation}
where
\begin{eqnarray}
S_{12} & \equiv & 3 (\vec{\sigma}_1\cdot\hat{r})(\vec{\sigma}_2\cdot\hat{r}) -  (\vec{\sigma}_1 \cdot \vec{\sigma}_2),\\
x & \equiv & \frac{m_{\pi}r}{\hbar}.
\end{eqnarray}
The only operators in the double commutator
\begin{equation}
\left[ [V_{\pi}, \hat{\Phi}_i], \hat{\Phi}_f \right]
\end{equation}
which are non-commuting are the isospin operators.  We must therefore explicitly
indicate the isospin dependence of $\hat{\Phi}$, which is really a function of $\vec{x}_p$.  
Thus, the commutator is rewritten as
\begin{equation}
 \frac{e^2}{4\hbar^2} \sum_{l,m=n,p} \left[ [(\tilde{\tau}_1 \cdot \tilde{\tau}_2), \tau_{z,l}], \tau_{z,m} \right]
V_{\pi}^{\mathrm{no} \tau}(\vec{r})  \Phi_i(\vec{x}_l) \Phi_f(\vec{x}_m).
\end{equation}
Using the well-known commutator relation
\begin{equation}
[\tau_i, \tau_j] = 2i\varepsilon_{ijk}\tau_k,
\end{equation}
this becomes
\begin{equation}
-\frac{8e^2}{\hbar^2} V_{\pi}^{ \mathrm{no} \tau } (\vec{r})  \Phi_i(\vec{x}_p) \Phi_f(\vec{x}_p).
\end{equation}
The term we are interested in is just this commutator sandwiched between the initial and final deuteron
states.    By inserting the definitions of $\Phi$ (equations~\ref{eq:mainterm10} and \ref{eq:mainterm11}), 
we find the scattering amplitude:
\begin{eqnarray}
\lefteqn{ \matrixm{\pi \mathrm{comm}} = \sum_{L=0}^{\infty} \sum_{M=-L}^L \sum_{L'=-M'} \sum_{M'} \sum_{l=0,2} \sum_{l'=0,2} 
\frac{16\pi e^2 f^2 m_{\pi}}{(\hbar\winit)(\hbar\wfin)}
(-1)^{L'-\mufin} \times} \\
& & \wignerdi \wignerdf
i^{L+L'} \sqrt{\frac{(2L+1)(2L'+1)}{LL'(L+1)(L'+1)} }\times\nonumber\\
& & \bra{l' 1 1 M_f}  \left\{ (\vec{\sigma}_1\cdot\hat{r})(\vec{\sigma}_2\cdot\hat{r}) 
\mathcal{I}^{\pi 1}_{LL'll'}(\wfin,\winit) - \right. \nonumber\\
& & \left. (\vec{\sigma}_1 \cdot \vec{\sigma}_2) \mathcal{I}^{\pi 2}_{LL'll'}(\wfin,\winit) \right\} Y_{LM} Y_{L'M'} \ket{l 1 1 M_i}, \nonumber
\end{eqnarray}
where
\begin{eqnarray}
\mathcal{I}^{\pi 1}_{LL'll'}(\wfin,\winit) & \equiv & \int dr\ u_{\sss l}(r)\ u_{\sss l'}(r) 
(1 + r\frac{d}{dr}) j_{\sss L}(\frac{\winit r}{2}) \times\\
& & \ \ \ \ \ \ \ \ (1 + r\frac{d}{dr}) j_{\sss L'}(\frac{\wfin r}{2}) \frac{e^{-x}}{x^3}(3+3x+x^2), \nonumber \\
\mathcal{I}^{\pi 2}_{LL'll'}(\wfin,\winit) & \equiv & \int dr\ u_{\sss l}(r)\ u_{\sss l'}(r) 
(1 + r\frac{d}{dr}) j_{\sss L}(\frac{\winit r}{2}) \times\\
& & \ \ \ \ \ \ \ \ (1 + r\frac{d}{dr}) j_{\sss L'}(\frac{\wfin r}{2}) \frac{e^{-x}}{x^3}(1+x), \nonumber
\end{eqnarray}
with $x$ defined above.

\begin{figure}
\centering
\parbox{75mm}{\centering\epsfig{file=pions5} }
\parbox{30mm}{(a)}
\parbox{75mm}{\centering\epsfig{file=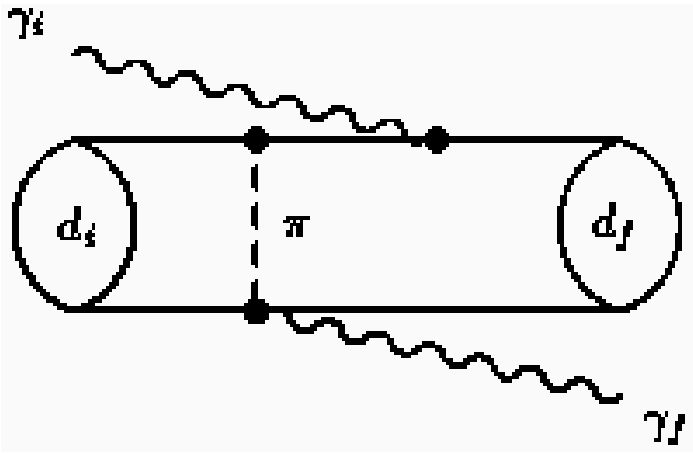} }
\parbox{30mm}{(b)}
\parbox{75mm}{\centering\epsfig{file=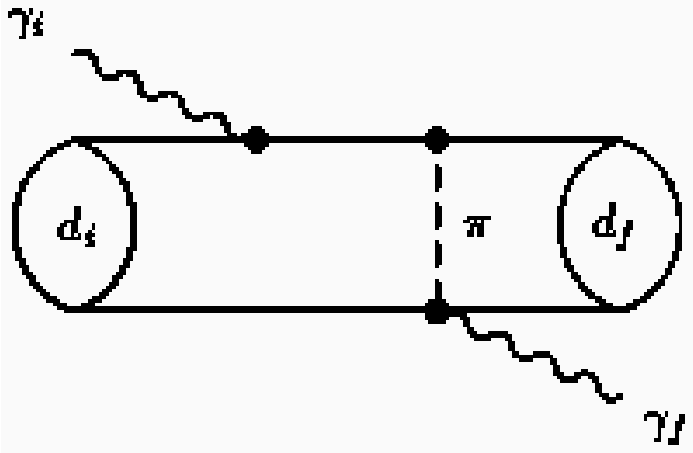} }
\parbox{30mm}{(c)}
\parbox{75mm}{\centering\epsfig{file=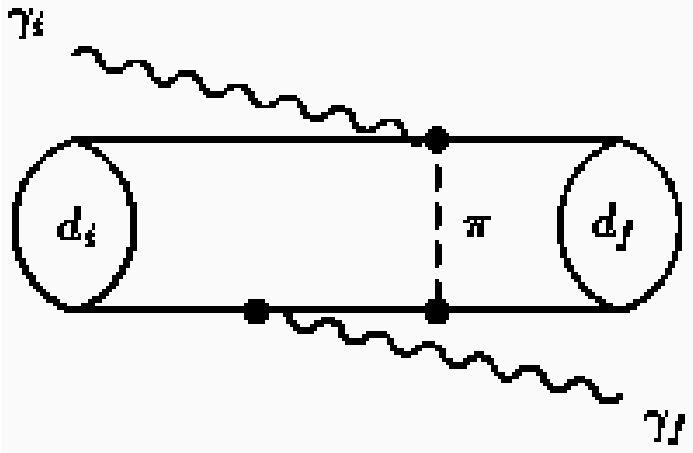} }
\parbox{30mm}{(d)}
\caption[Feynman diagrams for vertex corrections]{Vertex correction
diagrams \label{fig:pion5}(a)$\matrixt{\pi 3, \mathrm{uncr} }$
(b) $\matrixt{\pi 3, \mathrm{cr} }$ (c) $\matrixt{\pi 4, \mathrm{uncr} }$
(d) $\matrixt{\pi 4, \mathrm{cr} }$ }
\end{figure}

Finally, we calculate diagrams such as the one shown            
in Figure~\ref{fig:pion5}. These are vertex corrections to the ``ordinary'' second-order diagrams.
\begin{figure} 
\centering
\epsfig{file=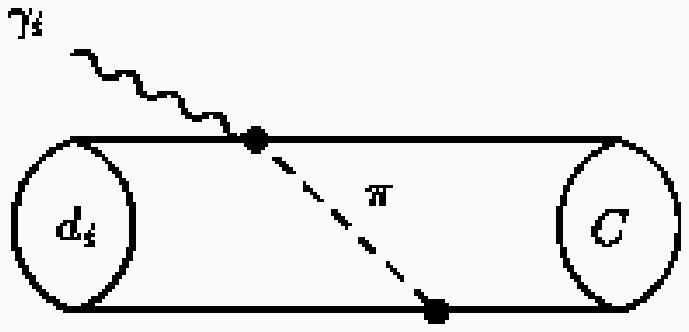}
\epsfig{file=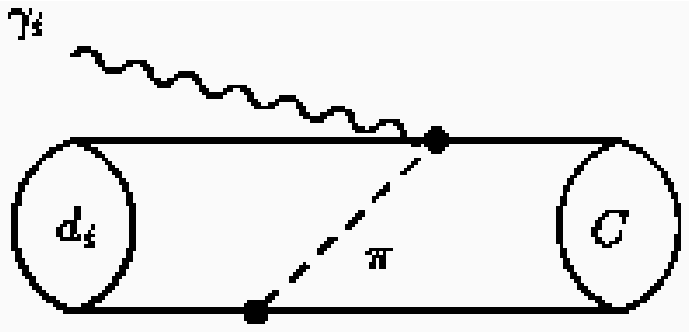}
\epsfig{file=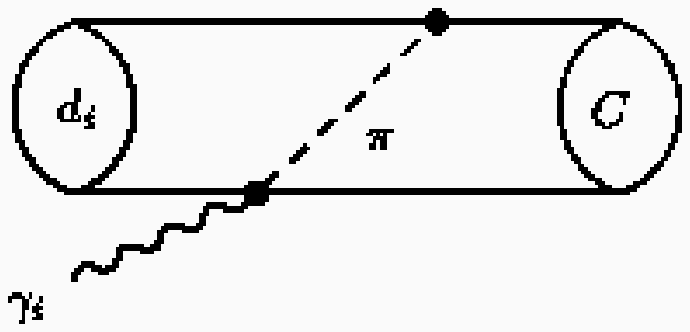}
\epsfig{file=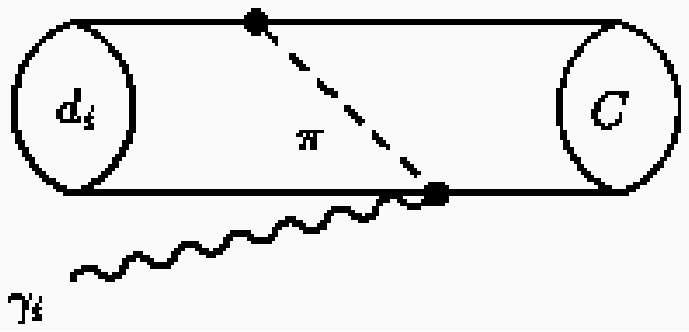}
\caption[Time-ordered vertex correction diagrams]{\label{fig:pion6} 
Possible time-orderings for the first half of the meson-exchange diagram
pictured in Figure~\ref{fig:pion5}(a). All are equal.}
\end{figure}
Figure~\ref{fig:pion6} shows the first part of Figure~\ref{fig:pion5}, up to the intermediate state $C$.
The plan is to find an effective Hamiltonian for this half of the diagram (the actual vertex correction), and then
use it to calculate the full diagram. Calling the new intermediate $NN$ state $B$, 
the first of the diagrams pictured in Figure~\ref{fig:pion6} is
\begin{eqnarray}
\matrixt{\pi \frac{1}{2}} & = & \sum_{B,\vec{P}_B} \sum_{\pm} \int \frac{V d^3q}{(2\pi)^3} \frac{\hbar^2}{V}
 \frac{2\pi}{ E_{\pi} } \bra{C, \vec{P}_C} \frac{ f\hbar }{ m_{\pi} }     
\left( \vec{\sigma}_2 \cdot \vec{\nabla}_2  \right) \frac{1}{\sqrt{2}} \tau_{2\pm} e^{i\vec{q}\cdot\vec{x}_2}
\ket{B, \vec{P}_B}  \nonumber\times \\  
& & \frac{1}{\hbar\winit + E_{d_i} - E_B - P_B^2/2m_d - E_{\pi} + i\varepsilon} \times\nonumber\\
& & \bra{B, \vec{P}_B} 
\frac{-i f} { m_{\pi} }      
\left( \vec{\sigma}_1 \cdot \vec{A}(\vec{x}_1)  \right)
\frac{\pm e}{\sqrt{2}} \tau_{1 \mp} e^{-i\vec{q}\cdot\vec{x}_1} \ket{d_i, \vec{P}_i, \gamma_i}.
\end{eqnarray}
\noindent Next, we make the assumption of equation~(\ref{eq:pion9}) to simplify
the denominator, and then remove the complete set of $B$ states to get
\begin{eqnarray}
\matrixt{\pi \frac{1}{2} } & = &  \frac{\pi e \hbar^3  f^2 }{  m_{\pi}^2 }      
\bra{C, \vec{P}_C } ( - \tau_{2+}\tau_{1-} + \tau_{2-}\tau_{1+} )   \times \\
& & \ \ \ \ \ \int \frac{ d^3q}{(2\pi)^3} \left( \vec{\sigma}_2 \cdot \vec{q}
\right) \frac{ e^{i\vec{q}\cdot(\vec{x}_2-\vec{x}_1)}  }{m_{\pi}^2 + \hbar^2 q^2}
\left( \vec{\sigma}_1 \cdot \vec{A}(\vec{x}_1)  \right) \ket{d_i, \vec{P}_i, \gamma_i}. \nonumber
\end{eqnarray}
\noindent Performing the $q$-integration yields
\begin{eqnarray}
\matrixt{\pi \frac{1}{2} } & = & \frac{e  f^2 }{  m_{\pi}^2 }      
\bra{C, \vec{P}_C }  \left[ \iso{\tau}_1 \times \iso{\tau}_2
\right]_z \vec{\sigma}_2\cdot \vec{\nabla} \left( \frac{e^{-m_{\pi} | \vec{x}_2-\vec{x}_1 | / \hbar}}{
2 | \vec{x}_2-\vec{x}_1 | } \right) \label{eq:pion6}\times\\
& & \ \ \ \ \ \ \ \ 
\left( \vec{\sigma}_1 \cdot \vec{A}(\vec{x}_1)  \right) \ket{d_i, \vec{P}_i, \gamma_i}, \nonumber
\end{eqnarray}
\noindent where the gradient is with respect to $\vec{x}_2-\vec{x}_1$ (which will become $\vec{r}$).  The term
where the pion is destroyed, rather than created, at the photon vertex, can be calculated similarly, and turns
out to be identical to equation~(\ref{eq:pion6}).  Adding these two terms to the two terms where nucleons 1 and
2 are exchanged (see Figure~\ref{fig:pion6} for the Feynman diagrams) gives  
\begin{eqnarray}
\matrixt{\pi \frac{1}{2}, \mathrm{tot} } & = & \frac{e\hbar   f^2 }{  m_{\pi}^2 } \bra{C, \vec{P}_C }      
\left[ \iso{\tau}_1 \times \iso{\tau}_2 \right]_z  \left\{ \left( \vec{\sigma}_1 \cdot \vec{A}(\vec{x}_1)\right)
\left( \vec{\sigma}_2\cdot \vec{\nabla}  \frac{e^{-m_{\pi} | \vec{x}_2-\vec{x}_1 | / \hbar}}{
| \vec{x}_2-\vec{x}_1 | } \right)\right. +\nonumber\\
& & \left.\left( \vec{\sigma}_2 \cdot \vec{A}(\vec{x}_2)\right) \left( \vec{\sigma}_1\cdot \vec{\nabla}
 \frac{e^{-m_{\pi} | \vec{x}_2-\vec{x}_1 | / \hbar}}{| \vec{x}_2-\vec{x}_1 | } \right)\right\}
\ket{d_i, \vec{P}_i, \gamma_i}.
\end{eqnarray}
\noindent Therefore, we can write the effective Hamiltonian which includes all terms where one of the photons interacts
with a pion as
\begin{eqnarray}
H^{\pi\mathrm{vc} } & = & \frac{e\hbar  f^2 }{  m_{\pi}^2 }
\left[ \iso{\tau}_1 \times \iso{\tau}_2 \right]_z  \left\{ \left( \vec{\sigma}_1 \cdot \vec{A}(\vec{x}_1)\right)
\left( \vec{\sigma}_2\cdot \vec{\nabla}  \frac{e^{-m_{\pi} | \vec{x}_2-\vec{x}_1 | / \hbar}}{
| \vec{x}_2-\vec{x}_1 | } \right)\right. +\nonumber\\
& & \left.\left( \vec{\sigma}_2 \cdot \vec{A}(\vec{x}_2)\right) \left( \vec{\sigma}_1\cdot \vec{\nabla}
\frac{e^{-m_{\pi} | \vec{x}_2-\vec{x}_1 | / \hbar}}{| \vec{x}_2-\vec{x}_1 | } \right)\right\}. \label{eq:pion25}
\end{eqnarray}
\noindent This formula applies to either incoming or outgoing photons, since no assumptions
about the photons have been made.  Since this Hamiltonian changes the deuteron isospin,
we must use second-order perturbation theory with another isospin-changing
Hamiltonian.  The only operator available for this is $\vec{t} \equiv (\vec{\sigma}_p - \vec{\sigma}_n)/2$.
The largest contribution comes from $H^{\mathrm{int}} = -\int J^{(\sigma)}(\vec{\dum})\cdot A^{(1)}(\vec{\dum}) d^3\dum$ 
(equation \ref{eq:nextterm10}).
Therefore, the vertex corrections (vc) to the scattering amplitude are
\begin{eqnarray}
\matrixt{\pi\mathrm{vc} 1 } & = & - \sum_{C,\vec{P}_C}  \bra{d_f, \vec{P}_f }
\int J^{(\sigma)}(\vec{\dum})\cdot A^{(1)}(\vec{\dum}) d^3\dum
\ket{C,\vec{P}_C} \times \\
& & \frac{1}{\hbar\winit + E_{d_i} - E_C - P^2_C/2m_d + i\varepsilon}
\bra{C,\vec{P}_C} H^{\pi\mathrm{vc} } \ket{d_i, \vec{P}_i,
\gamma_i} ,\nonumber\\ 
\matrixt{\pi\mathrm{vc} 2 } & = & - \sum_{C,\vec{P}_C} \bra{d_f, \vec{P}_f }
\int J^{(\sigma)}(\vec{\dum})\cdot A^{(1)}(\vec{\dum}) d^3\dum
\ket{C,\vec{P}_C,\gamma_i, \gamma_f}  \times \\
& & \frac{1}{-\hbar\wfin + E_{d_i} - E_C  - P^2_C/2m_d+ i\varepsilon}\bra{C,\vec{P}_C,\gamma_i, \gamma_f}
H^{\pi\mathrm{vc} }\ket{d_i, \vec{P}_i, \gamma_i} \nonumber\\
\matrixt{\pi\mathrm{vc} 3} & = & - \sum_{C,\vec{P}_C}  \bra{d_f, \vec{P}_f }
 H^{\pi\mathrm{vc} } \ket{C,\vec{P}_C} 
\frac{1}{\hbar\winit + E_{d_i} - E_C  - P^2_C/2m_d+ i\varepsilon} \times\nonumber\\
& & \bra{C,\vec{P}_C}
\int J^{(\sigma)}(\vec{\dum})\cdot A^{(1)}(\vec{\dum}) d^3\dum
\ket{d_i, \vec{P}_i, \gamma_i} ,  \\
\matrixt{\pi\mathrm{vc}  4 } & = & - \sum_{C,\vec{P}_C} \bra{d_f, \vec{P}_f }
H^{\pi\mathrm{vc} } \ket{C,\vec{P}_C,\gamma_i, \gamma_f}  \times\nonumber\\
& & \frac{1}{-\hbar\wfin + E_{d_i} - E_C  - P^2_C/2m_d+ i\varepsilon}\nonumber\times \\
& & \bra{C,\vec{P}_C,\gamma_i, \gamma_f} \int J^{(\sigma)}(\vec{\dum})\cdot A^{(1)}(\vec{\dum}) d^3\dum
\ket{d_i, \vec{P}_i, \gamma_i} . 
\end{eqnarray}
The calculations are very similar to the ones in Appendix G, so we give just one intermediate
step in the calculation of $\matrixt{\pi \mathrm{vc} 1 }$.  After inserting the expressions for
the Hamiltonians (from equations~\ref{eq:pion25} and  \ref{eq:nextterm10}) and performing some simplifications, we get
\begin{eqnarray}
\lefteqn{ \matrixt{\pi\mathrm{vc} 1 }  =  -\sum_{L_C S_C J_C M_C}  \sum_{ll'} \sum_{LM} \sum_{L'M'}
\frac{4\pi\mufin e^2\hbar^2f^2\wfin}{m_p m_{\pi}^2 } i^{L+L'} (-1)^{1-M_f+L'-\mufin} Y_{LM}^{\ast}(\hat{k}_i) \times}\nonumber\\
& & \left[ \mu_p + (-1)^{L'} \mu_n \right]  \wignerdf \sqrt{2\pi(L'+1)}
\threej{1}{L'}{J_C}{-M_f}{M'}{M_C}  \times\nonumber\\
& & \int r dr r' dr' j_{\sss L'-1}(\frac{\wfin r'}{2}) u_{\sss l'}(r') \green j_{\sss l}(\frac{\winit r}{2})
\frac{d}{dr} \frac{e^{-m_{\pi}r/\hbar} }{r} u_{\sss l}(r) \times\nonumber\\
& & \mxredbc{ [ Y_{L'-1} \otimes t ]_{L'} } \bra{l' 1 1 M_f} \left\{(\vec{\sigma}_1\cdot\eps)(\vec{\sigma}_2\cdot\hat{r}) + \right.\nonumber\\
& & \left. (-1)^L (\vec{\sigma}_2\cdot\eps)(\vec{\sigma}_1\cdot\hat{r}) \right\} Y_{LM} \ket{l 1 1 M_i}.
\end{eqnarray}
We have used the fact that
\begin{equation}
\mx{T=0}{ [\tilde{\tau}_1\times\tilde{\tau}_2]_z }{T=1} = -2i.
\end{equation}
Equation~(\ref{eq:pion35}) can be used to help the evaluate the second matrix element above.  The final result is
\begin{eqnarray}
\lefteqn{ \matrixm{\pi \mathrm{vc} 1 }  =  -\sum_{l=0,2} \sum_{l'=0,2} \sum_{\tilde{J}=|1-J_C|}^{1+J_C} 
\sum_{J_C=|1-L''|}^{1+L''} \sum_{L''=|L-L'|}^{L+L'}
\sum_{L=0}^{\infty} \sum_{L'=1}^{\infty} \sum_{i=-1,0,1} \sum_{j=-1,0,1} \times}\\
& & \frac{72\pi e^2 f^2 \hbar^2 \mufin\wfin}{m_p m_{\pi}^2}
i^{L+L'} (-1)^{L'-\mufin-M_f-M_i+j+J_C-L''} \left[ \mu_p + (-1)^{L'} \mu_n \right]  \times\nonumber\\
& & \wignerdf (\eps)_{-i} (2\tilde{J}+1)(2L+1) \sqrt{ \frac{(L'+1)(2L''+1)(2J_C+1)}{4\pi} }
\times\nonumber\\
& & \threej{1}{L'}{J_C}{-M_f}{M_f-M_i-i}{M_i+i} \threej{1}{1}{1}{j}{i}{-i-j}  \threej{1}{L}{L''}{0}{0}{0} \times\nonumber\\
& & \threej{1}{L}{L''}{-j}{0}{j} \threej{1}{L''}{\tilde{J}}{i+j}{-j}{-i} \threej{J_C}{\tilde{J}}{1}{-M_i-i}{i}{M_i} \times\nonumber\\
& & \ninej{0}{J_C}{J_C}{1}{l}{1}{1}{L''}{\tilde{J}} \mxred{l' 1 1}{  [ Y_{L'-1} \otimes t ]_{L'} }{J_C 0 J_C}
\mxred{J_C}{ Y_{L''} }{l} \times\nonumber\\
& & \int_0^{\infty} \int_0^{\infty} r\ dr\ r'\ dr'\ j_{\sss L'-1}(\frac{\wfin r'}{2}) u_{\sss l'}(r') \green j_{\sss L}(\frac{\winit r}{2})
\frac{d}{dr} \frac{e^{-m_{\pi}r/\hbar} }{r} u_{\sss l}(r).\nonumber
\end{eqnarray}
The other correction terms are:
\begin{eqnarray}
\lefteqn{ \matrixm{\pi \mathrm{vc} 2 }  =  -\sum_{l=0,2} \sum_{l'=0,2} \sum_{\tilde{J}=|1-J_C|}^{1+J_C}
\sum_{J_C=|1-L''|}^{1+L''} \sum_{L''=|L-L'|}^{L+L'}
\sum_{L=0}^{\infty} \sum_{L'=1}^{\infty} \sum_{i=-1,0,1} \sum_{j=-1,0,1} \times}\\
& &  \frac{72\pi e^2 f^2 \hbar^2 \muinit\winit}{m_p m_{\pi}^2}
i^{L+L'} (-1)^{L'+i+j-L''-M_f+J_C-M_i-\muinit} \left[ \mu_p + (-1)^{L} \mu_n \right]\times\nonumber\\
& &   Y_{L'M'}^{\ast}(\hat{k}_f) (\epspr)_{-i}
(2\tilde{J}+1) \sqrt{ (L+1)(2L'+1)(2L''+1)(2J_C+1) } \times\nonumber\\
& & \threej{1}{L'}{L''}{0}{0}{0}\threej{1}{L'}{L''}{-j}{M_f-M_i-\muinit-i}{M_i+\muinit+i+j-M_f} \times\nonumber\\
& & \threej{1}{1}{1}{j}{i}{-i-j}\threej{1}{L''}{\tilde{J}}{i+j}{M_f-M_i-\muinit-i-j}{\muinit+M_i-M_f} \times\nonumber\\
& & \threej{J_C}{L}{1}{-M_i-\muinit}{\muinit}{M_i}\threej{1}{\tilde{J}}{J_C}{-M_f}{M_f-M_i-\muinit}{M_i+\muinit} \times\nonumber\\
& & \ninej{1}{l'}{1}{0}{J_C}{J_C}{1}{L''}{\tilde{J}} \mxred{J_C 0 J_C}{  [ Y_{L-1} \otimes t ]_{L} }{l 1 1}
\mxred{l'}{ Y_{L''} }{J_C} \times\nonumber\\
& & \int_0^{\infty} \int_0^{\infty} r\ dr\ r'\ dr'\ j_{\sss L-1}(\frac{\winit r}{2}) u_{\sss l}(r) \green j_{\sss L'}(\frac{\wfin r'}{2})
\frac{d}{dr'} \frac{e^{-m_{\pi}r'/\hbar} }{r'} u_{\sss l'}(r')\nonumber ,
\end{eqnarray}
\begin{eqnarray} 
\lefteqn{\matrixm{\pi\mathrm{vc} 3 }  =  -\sum_{l=0,2} \sum_{l'=0,2} \sum_{\tilde{J}=|1-J_C|}^{1+J_C}
\sum_{J_C=|1-L''|}^{1+L''} \sum_{L''=|L-L'|}^{L+L'}
\sum_{L=0}^{\infty} \sum_{L'=1}^{\infty} \sum_{i=-1,0,1} \sum_{j=-1,0,1} \times}\\
& & \frac{72\pi e^2 f^2 \hbar^2 \mufin\wfin}{m_p m_{\pi}^2}
i^{L+L'} (-1)^{L'-\mufin+J_C+j+L''}  \left[ \mu_p + (-1)^{L'} \mu_n \right]\times\nonumber\\
& & \wignerdf (\eps)_{-i}
(2\tilde{J}+1)(2L+1) \sqrt{ \frac{(L'+1)(2L''+1)(2J_C+1)}{4\pi} } \times\nonumber\\
& & \threej{J_C}{L'}{1}{i-M_f}{M_f-M_i-i}{M_i} \threej{1}{1}{1}{j}{i}{-i-j}  \threej{1}{L}{L''}{0}{0}{0} \times\nonumber\\
& & \threej{1}{L}{L''}{-j}{0}{j} \threej{1}{L''}{\tilde{J}}{i+j}{-j}{-i}
\threej{1}{\tilde{J}}{J_C}{-M_f}{i}{M_f-i} \times\nonumber\\
& & \ninej{1}{l'}{1}{0}{J_C}{J_C}{1}{L''}{\tilde{J}} \mxred{J_C 0 J_C}{  [ Y_{L'-1} \otimes t ]_{L'} }{l 1 1}
\mxred{l'}{ Y_{L''} }{J_C} \times\nonumber\\
& & \int_0^{\infty} \int_0^{\infty} r\ dr\ r'\ dr'\ j_{\sss L'-1}(\frac{\wfin r'}{2}) u_{\sss l}(r') \greenpr j_{\sss L}(\frac{\winit r}{2})
\frac{d}{dr} \frac{e^{-m_{\pi}r/\hbar} }{r} u_{\sss l'}(r) \nonumber, 
\end{eqnarray}
\begin{eqnarray}
\lefteqn{ \matrixm{\pi \mathrm{vc} 4 }  =  -\sum_{l=0,2} \sum_{l'=0,2} \sum_{\tilde{J}=|1-J_C|}^{1+J_C}
\sum_{J_C=|1-L''|}^{1+L''} \sum_{L''=|L-L'|}^{L+L'}
\sum_{L=0}^{\infty} \sum_{L'=1}^{\infty} \sum_{i=-1,0,1} \sum_{j=-1,0,1} \times}\\
& & \frac{72\pi e^2 f^2 \hbar^2 \muinit\winit}{m_p m_{\pi}^2}
i^{L+L'} (-1)^{L'+i+j-L''+J_C-\muinit} \left[ \mu_p + (-1)^{L} \mu_n \right]\times\nonumber\\
& &   Y_{L'M'}^{\ast}(\hat{k}_f) (\epspr)_{-i}
(2\tilde{J}+1) \sqrt{ (L+1)(2L'+1)(2L''+1)(2J_C+1) } \times\nonumber\\
& & \threej{1}{L'}{L''}{0}{0}{0}\threej{1}{L'}{L''}{-j}{M_f-M_i-\muinit-i}{M_i+\muinit+i+j-M_f} \times\nonumber\\
& & \threej{1}{1}{1}{j}{i}{-i-j}\threej{1}{L''}{\tilde{J}}{i+j}{M_f-M_i-\muinit-i-j}{\muinit+M_i-M_f}\times\nonumber\\
& & \threej{1}{L}{J_C}{-M_f}{\muinit}{M_f-\muinit}\threej{J_C}{\tilde{J}}{1}{\muinit-M_f}{M_f-M_i-\muinit}{M_i} \times\nonumber\\
& & \ninej{0}{J_C}{J_C}{1}{l'}{1}{1}{L''}{\tilde{J}} \mxred{l'1 1}{  [ Y_{L-1} \otimes t ]_{L} }{J_C 0 J_C}
\mxred{J_C}{ Y_{L''} }{l} \times\nonumber\\
& & \int_0^{\infty} \int_0^{\infty} r\ dr\ r'\ dr'\ j_{\sss L-1}(\frac{\winit r}{2}) u_{\sss l}(r') \greenpr j_{\sss L'}(\frac{\wfin r'}{2})
\frac{d}{dr'} \frac{e^{-m_{\pi}r'/\hbar} }{r'} u_{\sss l}(r').\nonumber
\end{eqnarray}


\chapter{Useful Formulas \label{app:form} }

\noindent Tensor product definition:
\begin{equation}
[T_{k_1} \otimes T_{k_2}]_{km} = \sum_{m_1 m_2} \sqrt{2k+1} (-1)^{-k_1+k_2-m} \threej{k_1}{k_2}{k}{m_1}{m_2}{-m}
T_{k_1m_1} T_{k_2m_2} \label{eq:formula1} .
\end{equation}
\noindent Wigner-Eckart theorem:
\begin{equation}
\mx{J_f M_f}{T_{JM}}{J_i}{M_i} = (-1)^{J_f-M_f} \threej{J_f}{J}{J_i}{-M_f}{M}{M_i}
\mxred{J_f}{T_J}{J_i} \label{eq:formula2} .
\end{equation}
Formulas for uncoupling tensor products: 
\begin{eqnarray}
\lefteqn{\mxred{ j_1 j_2 j  }{ [T_{k_1} \otimes T_{k_2}]_{k} }{ {j_1}' {j_2}' j' }  =} \label{eq:formula3} \\
& & \sqrt{(2j+1)(2j'+1)(2k+1)}  \ninej{j_1}{j_2}{j}{{j_1}'}{{j_2}'}{j'}{k_1}{k_2}{k}
\mxred{ j_1 }{T_{k_1}}{ {j_1}' }\mxred{ j_2 }{ T_{k_2} }{ {j_2}' }, \nonumber
\end{eqnarray}
\begin{eqnarray}
\lefteqn{\mxred{ j  }{ [T_{k_1} \otimes T_{k_2}]_{k} }{  j' }  =} \label{eq:formula4} \\
& & (-1)^{k+j+j'} \sqrt{2k+1} \sum_{\tilde{\j}} \sixj{k_1}{k_2}{k}{j'}{j}{\tilde{\j}}
\mxred{j}{T_{k_1}}{\tilde{\j}} \mxred{\tilde{\j}}{T_{k_2}}{j'} \nonumber.
\end{eqnarray}
Useful reduced matrix elements:
\begin{equation}
\mxred{l'}{Y_L}{l} = (-1)^{l'} \sqrt{\frac{(2l'+1)(2L+1)(2l+1)}{4\pi }} \threej{l}{L}{l'}{0}{0}{0} \label{eq:formula5}.
\end{equation}
\begin{eqnarray}
\lefteqn{\mxred{ L_f S_f J_f }{ Y_{L}  }{L_i S_i J_i} = }\label{eq:formula6}\\
& & \sqrt{ \frac{ (2L_f+1)(2L_i+1)(2J_i+1)(2J_f+1)(2L+1)(2L'+1)(2S_f+1)}{4\pi} }\times\nonumber\\
& & (-1)^{L_f+S_f} \delta_{S_f,S_i} \ninej{L_f}{S_f}{J_f}{L_i}{S_i}{J_i}{L}{0}{L} \threej{L_f}{L}{L_i}{0}{0}{0}, \nonumber\\
\lefteqn{\mxred{ L_f S_f J_f }{ \left[ Y_{L'} \otimes S \right]_L }{L_i S_i J_i} = } \label{eq:formula7}\\
& & (-1)^{L_f} \sqrt{ \frac{3(2L_f+1)(2L_i+1)(2J_i+1)(2J_f+1)(2L+1)(2L'+1)}{2\pi} } \nonumber\times\\
& & \ninej{L_f}{S_f}{J_f}{L_i}{S_i}{J_i}{L'}{1}{L} \threej{L_f}{L'}{L_i}{0}{0}{0} \delta_{S_f,S_i} \delta_{S_i,1} \nonumber ,\\
\lefteqn{\mxred{ L_f S_f J_f }{ \left[ Y_{L'} \otimes t \right]_L }{L_i S_i J_i} = } \label{eq:formula8}\\
& & (-1)^{L_f} \sqrt{ \frac{3(2L_f+1)(2L_i+1)(2J_i+1)(2J_f+1)(2L+1)(2L'+1)}{16\pi} } \times\nonumber\\
& & \left[ (-1)^{S_i} - (-1)^{S_f} \right]
\ninej{L_f}{S_f}{J_f}{L_i}{S_i}{J_i}{L'}{1}{L} \threej{L_f}{L'}{L_i}{0}{0}{0} \nonumber,\\
\lefteqn{\mxred{ L_f S_f J_f }{ \left[ Y_{L'} \otimes L \right]_L }{L_i S_i J_i} = } \label{eq:formula9}\\
& & \sqrt{ \frac{L_i(L_i+1)(2L_i+1)^2 (2L_f+1)(2J_i+1)(2J_f+1)(2L+1)(2L'+1)}{4\pi} } \times\nonumber\\
& &  (-1)^{L_i+L_f+J_i+S_f}
\delta_{S_i,S_f} \sixj{J_f}{L_f}{S_f}{L_i}{J_i}{L} \threej{L_f}{L'}{L_i}{0}{0}{0} \sixj{L'}{1}{L}{L_i}{L_f}{L_i} \nonumber.
\end{eqnarray}
\noindent Addition of two spherical harmonics:
\begin{eqnarray}
\lefteqn{ Y_{jm} Y_{j'm'} = }  \label{eq:formula10}\\
& & \sum_{JM} (-1)^M \sqrt{\frac{(2j+1)(2j'+1)(2J+1)}{4\pi}} \threej{j}{j'}{J}{0}{0}{0}
\threej{j}{j'}{J}{m}{m'}{-M} Y_{JM} \nonumber.
\end{eqnarray}
Spherical Bessel recursion relations:
\begin{eqnarray}
\frac{2l+1}{x} j_{\sss l}(x) & = & j_{\sss l-1}(x) + j_{\sss l+1}(x) \label{eq:formula11}, \\
(2l+1)j_{\sss l}'(x) & = & lj_{\sss l-1}(x) - (l+1)j_{\sss l+1}(x) \label{eq:formula12}.
\end{eqnarray}
\noindent The gradient formula:
\begin{eqnarray}
\vec{\nabla}\Phi (r) Y_{LM}(\hat{r}) & = & -\sqrt{\frac{L+1}{2L+1}} \left(
\frac{d}{dr} - \frac{L}{r} \right) \Phi(r)\vsh{L}{L+1}{M} +  \label{eq:formula13} \\
& & \sqrt{\frac{L}{2L+1}} \left( \frac{d}{dr} + \frac{L+1}{r}\right) \Phi(r)\vsh{L}{L-1}{M} \nonumber,\\
\vec{\nabla}\times \left[ f(r) \vsh{L}{L}{M} \right]  & = &
i\left( \frac{d}{dr} - \frac{L}{r} \right) f(r) \sqrt{ \frac{L}{2L+1} } \vsh{L}{L+1}{M} +  \label{eq:formula14} \\
& & i\left( \frac{d}{dr} + \frac{L+1}{r} \right) f(r) \sqrt{ \frac{L+1}{2L+1} } \vsh{L}{L-1}{M}  \nonumber,\\
 \vec{\nabla}\times \left[ f(r) \vsh{L}{L+1}{M} \right] & = &
i\left( \frac{d}{dr} + \frac{L+2}{r} \right) f(r) \sqrt{ \frac{L}{2L+1} } \vsh{L}{L}{M} \label{eq:formula15},\\
\vec{\nabla}\times \left[ f(r) \vsh{L}{L-1}{M} \right] & = &
 i\left( \frac{d}{dr} -\frac{L}{r} \right) f(r) \sqrt{\frac{L+1}{2L+1} } \vsh{L}{L}{M} \label{eq:formula16},\\
\vec{\nabla}\cdot \left[ f(r) \vsh{L}{L}{M} \right] & = & 0 \label{eq:formula17},\\
\vec{\nabla}\cdot \left[ f(r) \vsh{L}{L+1}{M} \right] & = &
-\sqrt{\frac{L+1}{2L+1} } \left( \frac{d}{dr} + \frac{L+2}{r} \right) f(r) Y_{LM}(\hat{r}) \label{eq:formula18},\\
\vec{\nabla}\cdot \left[ f(r) \vsh{L}{L-1}{M} \right] & = &
\sqrt{\frac{L}{2L+1} } \left( \frac{d}{dr} + \frac{L-1}{r} \right) f(r) Y_{LM}(\hat{r}) \label{eq:formula19}.
\end{eqnarray}
Vector spherical harmonic definition: 
\begin{equation}
\vsh{J}{L}{M}  = \sum_{\nu=-1,0,1}  (-1)^{L-M-1} \sqrt{2J+1} \threej{1}{L}{J}{-\nu}{M+\nu}{-M}
{\hat{r}}_{-\nu} Y_{L,M+\nu}(\hat{r}). \label{eq:formula20}
\end{equation}



\end{document}